\documentclass[10pt,notitlepage,pra,superscriptaddress,preprintnumbers,aps,nofootinbib,tightenlines,floatfix]{revtex4-1}

\usepackage{hyperref}
\usepackage{mathtools}
\usepackage{amsmath,amssymb}
\usepackage{cleveref}
\usepackage{graphicx}
\usepackage{xcolor}
\usepackage[utf8]{inputenc}
\usepackage{bm}  
\usepackage{csquotes}
\usepackage[section]{placeins}
\usepackage{csquotes}
\usepackage{scalerel}
\usepackage{comment}
\usepackage[export]{adjustbox}
\usepackage{booktabs}
\usepackage{overpic}
\usepackage{youngtab}
\usepackage{setspace}
\usepackage{multirow}
\usepackage{longtable}
\usepackage{array}

\allowdisplaybreaks

\newcommand{\gsim}{\raisebox{-0.7ex}{$\stackrel{\textstyle >}{\sim}$ }}
\newcommand{\lsim}{\raisebox{-0.7ex}{$\stackrel{\textstyle <}{\sim}$ }}

\newcommand\scalemath[2]{\scalebox{#1}{\mbox{\ensuremath{\displaystyle #2}}}}

\newcount\hour \newcount\hourminute \newcount\minute
\hour=\time \divide \hour by 60
\hourminute=\hour \multiply \hourminute by 60
\minute=\time \advance \minute by -\hourminute
\newcommand{\mydate}{\ \today \ - \number\hour :\number\minute}

\makeatletter
\renewcommand*\env@matrix[1][*\c@MaxMatrixCols c]{%
  \hskip -\arraycolsep
  \let\@ifnextchar\new@ifnextchar
  \array{#1}}
\makeatother

\usepackage[braket, qm]{qcircuit}
\usepackage{tikz}
\newcommand*\qcontrolcolor[2]{\push{\scriptstyle \tikz[baseline=(char.base)]{
                              \node[shape=circle,draw,inner sep=0.6pt,color=#1] (char) {#2};}}
                              \qw}  
\newcommand*\qcontrolcolordouble[2]{\push{ \tikz[baseline=(char.base)]{
                              \node[shape=circle,draw,inner sep=0.6pt,color=#1,double] (char) {#2};}}
                              \qw}  
\newcommand*\qcontrol[1]{\qcontrolcolor{black}{#1}}
\newcommand*\qcontroldouble[1]{\qcontrolcolordouble{black}{#1}}

\newcommand\controlsector[4]{\ensuremath{
\begin{pmatrix}
    \mathbf{#1} \qquad \ \  \mathbf{#3} \\
    \ \hat{\Box} + \hat{\Box}^\dagger \\
    \mathbf{#2} \qquad \ \  \mathbf{#4}
\end{pmatrix}}}
\newcommand\controlsectorc[4]{\ensuremath{
\begin{pmatrix}
    \mathbf{#1} \qquad \ \  \mathbf{#3} \\
    \ (\hat{\Box},\hat{\Box}^\dagger) \\
    \mathbf{#2} \qquad \ \  \mathbf{#4}
\end{pmatrix}}}
\newcommand\plaquettestate[4]{\ensuremath{\begin{pmatrix}
      \bullet, \mathbf{#3}, \bullet \\
      \mathbf{#4}, \mathbf{#2}\\
      \bullet, \mathbf{#1}, \bullet
    \end{pmatrix}}}
\newcommand\plaquettestateket[4]{\ensuremath{\Bigg|\plaquettestate{#1}{#2}{#3}{#4} \Bigg\rangle}}
\newcommand\plaquettestatebra[4]{\ensuremath{\Bigg\langle \plaquettestate{#1}{#2}{#3}{#4} \Bigg|}}

\newcommand\CG[6]{\ensuremath{\langle #1, #2, #3,#4 | #5, #6\rangle }}
\newcommand\CGstar[6]{\ensuremath{\langle #5, #6 | #1, #2, #3, #4\rangle }}

\newcommand\nineR[9]{\ensuremath{ \begin{Bmatrix}
    \mathbf{#1} & \mathbf{#2} & \mathbf{#3} \\
    \mathbf{#4} & \mathbf{#5} & \mathbf{#6} \\
    \mathbf{#7} & \mathbf{#8} & \mathbf{#9}
  \end{Bmatrix}}}

\begin{document}

\title{A Trailhead for Quantum Simulation of SU(3) Yang-Mills Lattice Gauge Theory in the Local Multiplet Basis}

\author{Anthony Ciavarella}
\email{aciavare@uw.edu}
\affiliation{InQubator for Quantum Simulation (IQuS), Department of Physics, University of Washington, Seattle, WA 98195, USA}
\author{Natalie Klco}
\email{natklco@caltech.edu}
\affiliation{Institute for Quantum Information and Matter (IQIM) and Walter Burke Institute for Theoretical Physics, California Institute of Technology, Pasadena CA 91125, USA}
\author{Martin J.~Savage}
\email{mjs5@uw.edu}
\affiliation{InQubator for Quantum Simulation (IQuS), Department of Physics, University of Washington, Seattle, WA 98195, USA}

\date{\mydate}
\preprint{IQuS@UW-21-001}

\begin{abstract}
Maintaining local interactions in the quantum simulation of gauge field theories relegates most states in the Hilbert space to be unphysical---theoretically benign, but experimentally difficult to avoid.
Reformulations of the gauge fields can modify the ratio of physical to gauge-variant states often through classically preprocessing the Hilbert space and modifying the representation of the field on qubit degrees of freedom.  This paper considers the implications of representing SU(3) Yang-Mills gauge theory on a lattice of irreducible representations in both
  a global basis of projected global quantum numbers and
  a local basis in which controlled-plaquette operators support efficient time evolution.
  Classically integrating over the internal gauge space at each vertex (e.g., color isospin and color hypercharge) significantly
  reduces both the qubit requirements and the dimensionality of the unphysical Hilbert space.
  Initiating tuning procedures that may inform future calculations at scale,   the time evolution of one- and two-plaquettes are implemented on one of IBM's superconducting quantum devices, and early benchmark quantities are identified.
  The potential advantages of qudit environments, with either constrained 2D hexagonal or 1D nearest-neighbor internal state connectivity, are discussed for future large-scale calculations.
\end{abstract}

\maketitle

{
\small
\twocolumngrid
\tableofcontents
\onecolumngrid
}

\section{Introduction}
\label{sec:intro}
\noindent
Precision calculations of Standard Model (SM) processes
where quantum chromodynamics (QCD)~\cite{Politzer:1973fx,Politzer:1974fr,Gross:1973id}
 plays a role
are important for a number of critical applications in high-energy physics (HEP) and nuclear physics (NP).
These range from highly-inelastic hadron collisions and fragmentation at the energy-frontier,
to neutrino-nucleus processes essential in understanding the nature of neutrinos,
to searches for  physics beyond the SM,
to predicting the behavior of matter under extreme conditions as found in supernova and collisions of binary neutron stars.
Since the discovery of QCD in the early 1970's, substantial progress has been made in developing analytic and numerical techniques to establish predictive capabilities for observables,
along with a complete quantification of uncertainties, for SM processes that are beyond reach of experiment or observation.
Effective field theories (EFTs) have played a central role in analytic progress,
while lattice QCD~\cite{Wilson:1974sk,Creutz:1980zw} has been central to numerical progress.
Generally, these techniques are frameworks for making predictions that converge to those of QCD in particular limits, and that provide systematically reducible uncertainties (both systematic and statistical), in a limited range of kinematics or SM parameters.
Lattice QCD has enjoyed great success in performing precision calculations of hadronic spectra, electroweak matrix elements and form factors,
properties of high-temperature low-density systems
and simple multi-hadron systems (for a review of quantities, see Ref.~\cite{Aoki:2019cca}).
Further, it is making inroads towards light nuclei  and reactions, and higher-density systems
(for recent reviews, see, for example,  Refs.~\cite{Bazavov:2019lgz,Joo:2019byq}).
Stochastic sampling at the heart of lattice QCD calculations of
finite density systems, including nuclei, encounters
\enquote{sign} problems
and signal-to-noise problems (also a sign problem)
in computing observables.
Further, real-time dynamics  are not practical in such calculations (which are explicitly constructed in Euclidean space) at scale.
Establishing an algorithm with efficient scaling to compute SM matrix elements with dynamical electroweak fields (chiral gauge theories) has also proven elusive.
The limitations of classical computing, as demonstrated by these and  other examples,
were anticipated by Feynman and by Benioff in the early 1980s~\cite{Feynman1982,Benioff:1980},
and are now providing impediments to advancing important scientific applications.

Advances in the control of coherence and entanglement in the laboratory have led to
the availability of first quantum devices and inspired theoretical developments leveraging the computational potential of quantum systems.
Due to their natural capability of coherently manipulating quantum wavefunctions, quantum devices are expected to more naturally perform real-time dynamics of non-equilibrium and dense quantum systems at scale.
One challenge in utilizing this envisioned advancement is in devising  encodings of physical systems of theoretical interest onto the controllable subspace of  quantum devices.
As is typical, the choice of basis used to represent the structure of information impacts
practical computational difficulty at  subsequent steps.
For the quantum simulation of lattice gauge theories, the bases that are chosen to spatially latticize the field and to digitize, or capture through finite quantum resources, the continuous local group manifold affects the processes of initialization,  the complexity of the time evolution operator, creation of localized wavepackets, and the resulting distribution of information required to be captured in the final measurement process.
Due to its ubiquity and potential relevance to the initialization process, the efficiency of the time evolution operator is often prioritized, and a strategy of spatially local distributions of quantum registers is employed~\cite{Jordan:2011ne,Jordan:2011ci}.
With this tactic, algorithms for the implementation of large lattices can be defined through the design of a small number of local operators, commonly acting in two conjugate bases, that,
when parallelized, allow for temporal propagation of the field with computation times that are independent of the volume.
Thus, constructing an algorithm for the quantum simulation of  fields begins with studies of the chosen basis or mapping of the field to quantum hardware.

Considerable effort has been devoted to developing quantum algorithms for the design and time evolution of lattice gauge theories on quantum  devices~\cite{Byrnes:2005qx,Zohar:2012ay,Zohar:2012xf,Banerjee:2012pg,Banerjee:2012xg,Zohar:2013zla,Zohar:2014qma,Marcos:2014lda,Brennen:2015pgn,Zohar:2015hwa,Martinez2016,Muschik:2016tws,Zohar:2016iic,Banuls:2017ena,Bender:2018rdp,Alaeian:2018egk,Kaplan:2018vnj,Zache:2018jbt,Stryker:2018efp,Raychowdhury:2018osk,Davoudi:2019bhy,Alexandru:2019nsa,Raychowdhury:2019iki,Lamm:2019bik,Klco:2019evd,Banuls:2019bmf,Lamm:2019bik,Halimeh:2019svu,Luo:2019vmi,Altman:2019vbv,Bender:2020ztu,Haase:2020kaj,Buser:2020cvn,Paulson:2020zjd,Lamm:2020jwv,Gustafson:2020yfe,Davoudi:2020yln,Shaw:2020udc,Halimeh:2020djb,Briceno:2020rar,Ikeda:2020agk,Kharzeev:2020kgc,Buser:2020cvn,Mueller:2020vha,Barata:2020jtq,Kasper:2020owz,Atas:2021ext},
often through the Kogut-Susskind Hamiltonian
formulation~\cite{KogutSusskind1975,Kogut:1979wt,Robson1980,Ligterink:2000wf,Bronzan:1984xb}.
Consequently, there has been a wide range of explorations of quantum simulation basis design for fields from the scalar field to gauge theories e.g.,
on a position-space lattice in the eigenbasis of the field operator~\cite{Jordan:2011ne,Jordan:2011ci}, in
a basis of the local free-field eigenstates~\cite{Klco:2018zqz},
on a lattice of momentum modes~\cite{Yeter-Aydeniz:2018mix},
in the magnetic basis~\cite{Bender:2020ztu,Haase:2020kaj},
through gauge field integration in low-dimensional spaces~\cite{Martinez2016,Muschik:2016tws},
on an orbifold lattice~\cite{Kaplan:2002wv,Buser:2020cvn},
in a prepotential framework or basis of gauge invariant loop, string, and hadron excitations~\cite{Schwinger:1965,Anishetty:2009nh,Anishetty:2009ai,Mathur:2010wc,Mathur:2011sr,Raychowdhury:2013rwa,Raychowdhury:2018osk,Raychowdhury:2019iki,Davoudi:2020yln},
through the use of a spin system producing the desired continuous fields approaching a critical point~\cite{HORN1981149,ORLAND1990647,Chandrasekharan:1996ih,Brower:2003vy,Wiese:2006kp,Singh:2019uwd,Bhattacharya:2020gpm},
through discrete subgroups and group space decimation~\cite{Zohar:2014qma,Zohar:2016iic,Lamm:2019bik,Ji:2020kjk}, through mesh digitization~\cite{Hackett:2018cel},
using light-front formulations of lattice field theory~\cite{Kreshchuk:2020dla,Kreshchuk:2020kcz},
and
in hybrid and analog approaches leveraging natural properties of trapped ions or ultracold atoms in optical lattices~\cite{Banerjee:2012xg,Zohar:2012xf,Tagliacozzo:2012df,Zohar:2013zla,Zohar:2015hwa,Davoudi:2019bhy}.
These strategies are important as optimal design is likely to depend on the physical properties of specific quantum architectures, which continue to be developed.
Furthermore, this range of formulations can provide  robustness in
evaluating systematic uncertainties
(from the performance of quantum hardware and algorithms)
for observables that are inaccessible to classical computation.

In this paper, the multiplet basis utilized in the work of
Byrnes and Yamamoto~\cite{Byrnes:2005qx} is integrated over the local gauge space at each vertex of the lattice, reducing the
Hilbert space describing the system down to the local SU(N) irreducible representations below a chosen truncation.
This approach has been previously used to
explore $(1+1)$-dim SU(2) lattice gauge theory~\cite{Banuls:2017ena}, further implemented for a $1$-dim chain
of plaquettes in SU(2) lattice gauge theory~\cite{Klco:2019evd}, and is here developed for application to SU(3) lattice gauge theory.

Quantum simulations of Yang-Mills theories and QCD are in their infancy.
Precision calculations  of quantities that can  be directly compared with experiment are far in the future, and
are expected to require major advances in quantum devices, algorithms and formalism.
However, in starting along the path to this ultimate objective,
explorations of simple systems, establishing informative benchmarks, analyzing features of profitable mappings, observing natural structures, quantifying truncation sensitivity, and identifying amenable architectures are all important  steps.
We focus on understanding the behavior of simple systems, one- and two-plaquette systems,
with regard to coupling, truncations in color space, the scaling of global and local basis states and operators, and the mapping of color irreps onto qubits, qutrits and qudits.
We perform quantum simulations of low-truncation one- and two-plaquette systems using IBM's
QExperience superconducting quantum devices.
Further, we examine a framework (that appears to scale amiably) for the use of controlled-plaquette operators on qudit systems
as a  way to perform simulations of SU(3) Yang-Mills gauge field theory.

\section{The SU(3) Yang-Mills Hamiltonian}
\label{sec:hamiltonian}
\noindent
Quantum simulations of Yang-Mills gauge theory can be performed
by discretizing the gauge fields in the spatial directions using a cubic lattice of sites
and defining link variables connecting adjacent sites of this underlying grid.
These link variables are parallel transporters that connect,
for SU(3), color vectors at one site to those at an adjacent site.
The Hamiltonian is a sum over the chromo-electric and chromo-magnetic contributions,
as first discussed by Kogut and Susskind~\cite{KogutSusskind1975},
\begin{equation}
\hat H  =
\frac{g^2}{2 a^{d-2}} \
\sum_{ b, {\rm links}}
\ |  \hat {\bf E}^{(b)} |^2
\ +\
\frac{1}{2 a^{4-d} g^2} \sum_{{\rm plaquettes} }
\left[\
6\ -\ \hat { \Box}(\mathbf{x}) \ -\ \hat {\Box}^\dagger(\mathbf{x})
\ \right]
\ \ \ ,
\label{eq:QCDham}
\end{equation}
where $g$ is the strong coupling constant,
$a$ is the lattice spacing between adjacent sites,
and
$d$ is the number of spatial dimensions.
In the irrep basis of tensor indices that are
labeled by $(p,q)$, the number of (fundamental, anti-fundamental) indices with total dimension
\begin{equation}
  \dim(p,q) = \frac{(p+1)(q+1)(p+q+2)}{2} \ \ \ ,
\end{equation}
the electric Hamiltonian is diagonal with eigenvalues determined by the Casimir operator,
\begin{equation}
	 \sum_{b} |\hat{\mathbf{E}}^{(b)}|^2\ket{p,q} = \frac{p^2 + q^2 + pq +3p + 3q}{3} \ket{p,q} \ \ \ .
\label{eq:casimir}
\end{equation}
The plaquette operator, $\hat{\Box}(\mathbf{x})$,  is defined as
\begin{equation}
\hat {\Box} (\mathbf{x})  =
{\rm Tr}\left[\
\hat U^{\bf 3}({\bf x}, {\bf x} + a {\bm \mu})\
\hat U^{\bf 3}({\bf x} + a {\bm \mu}, {\bf x} + a {\bm \mu}+ a {\bm \nu})\
\hat U^{\bf 3}({\bf x} + a {\bm \mu}+ a {\bm \nu}, {\bf x} + a {\bm \nu})\
\hat U^{\bf 3}( {\bf x} + a {\bm \nu}, {\bf x})\
\right]
\ \ \ ,
\label{eq:PlaqOp}
\end{equation}
where  $\hat U^{\bf 3}({\bf x},{\bf y})$ are $3\times 3$ unitary matrices,
and ${\bm\mu}$ and ${\bm\nu}$ are unit vectors that define the orientation of the plaquette.
In the electric basis,
links are defined by states of the color irrep to which they belong, ${\bf R}$, and
the (uncorrelated) orientations in the two color spaces they connect,
$\alpha$ and  $\beta$,
$| {\bf R}, \alpha , \beta \rangle$.
The electric contribution from each link  is proportional to the Casimir operator
acting on the link without changing the color irrep, while the plaquette operators,
$\hat{\Box} + \hat{\Box}^\dagger$,
add color fluxes to the links in the plaquette, ${\bf 3}$ and $\overline{\bf 3}$, which
change the irrep of each link, subject to Gauss's law.
Constraints imposed to define physically allowed states of the system are included through
additional conditions.
In the absence of external color charges and quarks, Gauss's law
is satisfied by the product of link irreps
at each vertex combining to a color singlet.

\subsection{The Plaquette Operator}
\label{subsec:plaquetteop}
\noindent
In the standard formulation of Hamiltonian lattice gauge
theory~\cite{KogutSusskind1975}, wavefunctions carry Clebsch-Gordon (CG) factors
at each vertex with the effect of enforcing local gauge invariance.
\begin{figure}[!ht]
  \centering
  \includegraphics[width = 0.5\textwidth]{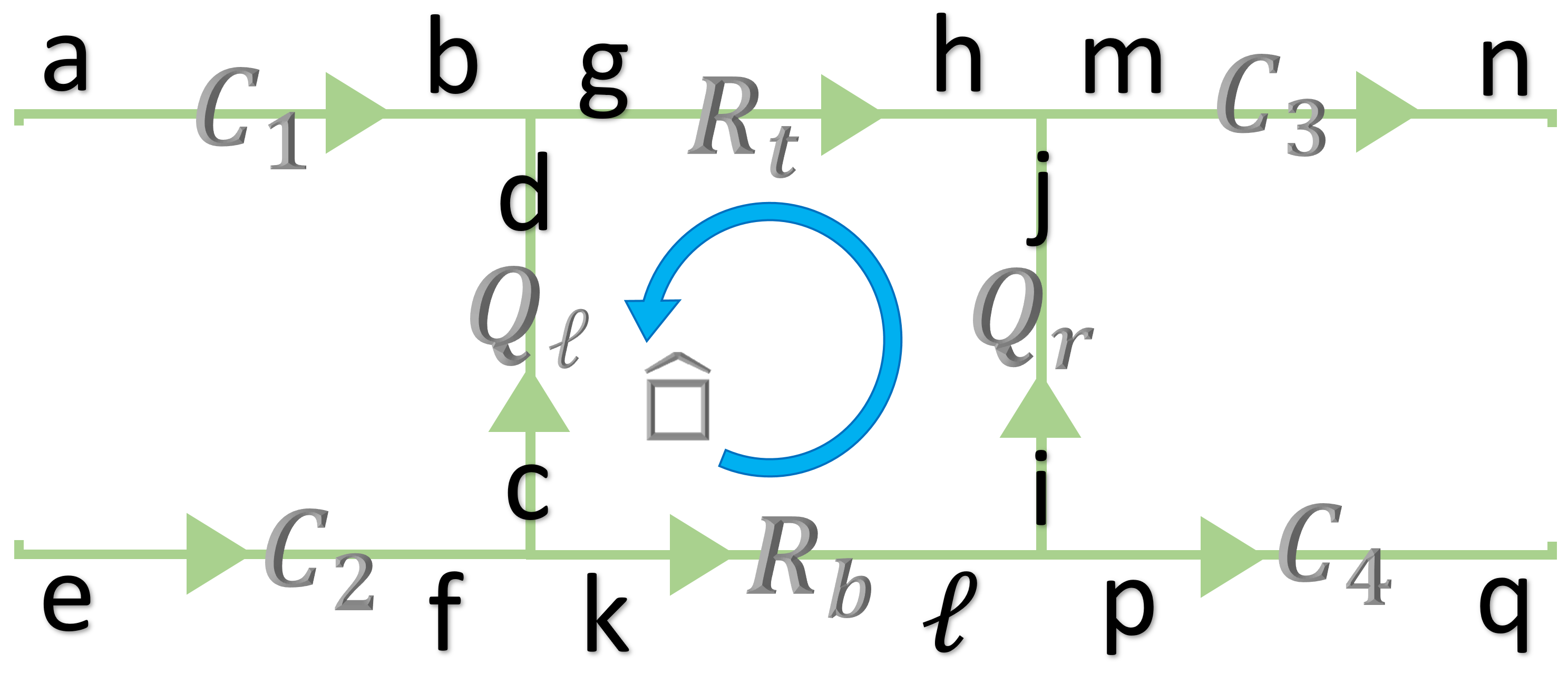}
  \caption{
  Following an arrow convention generalizable to higher dimension, the above link labels will be employed.
  Indices local to one end of each link represent a set of indices characterizing the local gauge space
  e.g., the color spin and color hypercharge in SU(3).
  }
  \label{fig:labels}
\end{figure}
Using the notation of Fig.~\ref{fig:labels}, an example of the local vertex structure is (upper-left vertex)
\begin{equation}
  |\psi_{3pt}\rangle \sim \sum_{ b,g,d,\Gamma}
  \CG{\mathbf{C}_1}{b}{\overline{\mathbf{R}}_t}{g}{\overline{\mathbf{Q}}_\ell}{d}_{\Gamma}
   \ |\mathbf{C}_1, a, b\rangle |\mathbf{Q}_\ell, c, d\rangle |\mathbf{R}_t, g, h\rangle
   \ \ \ ,
  \label{eq:vertexCGs}
\end{equation}
where the sum is over the quantum numbers internal to the links at the vertex.
The subscript, $\Gamma$, on the SU(3) CG coefficient indexes the multiplicity of combined irreps achieved through  tensor contractions.
An example of this multiplicity is in the product $\mathbf{8} \otimes \mathbf{8}$ that can be combined to produce the 8-dimensional irrep in two distinct ways,
symmetric and antisymmetric contractions,
with  distinct CGs.
These multiplicities mildly complicate the calculation of plaquette matrix elements, but are otherwise benign with respect to the structure of the quantum simulation.

With a truncation including only up to the single-index irreps, the vertices that contain a singlet (and are thus gauge invariant) are $\mathbf{1}\otimes\mathbf{1}\otimes\mathbf{1}, \mathbf{1}\otimes\overline{\mathbf{3}}\otimes\mathbf{3},
\mathbf{3}\otimes \mathbf{3} \otimes \mathbf{3}$, and those related under global conjugation and permutation symmetries. With a truncation including the $\mathbf{8}$ irrep, described by the two index tensor with one upper and one lower index, the number of gauge invariant vertices rises to include the $\mathbf{1}\otimes\mathbf{8} \otimes \mathbf{8}, \mathbf{3} \otimes \overline{\mathbf{3}}\otimes \mathbf{8}$ and  $\mathbf{8}\otimes \mathbf{8} \otimes \mathbf{8}$.

A key role of the vertex CGs is to allow a  localization of the plaquette operator, determining the magnetic Hamiltonian, as the minimal contracted loop of local link operators (directionality as in Fig.~\ref{fig:labels}),
\begin{align}
  \hat{\Box} &= \hat U_{\alpha, \beta}^{\mathbf{3}}
  \hat U_{\beta, \gamma}^{\mathbf{3}}
  \left(\hat U_{\gamma, \delta}^{\mathbf{3}} \right)^\dagger
  \left(\hat U_{\delta, \alpha}^{\mathbf{3}}\right)^\dagger
  \ \ \ ,
  \label{eq:boxop} \\
  \hat U_{\alpha, \beta}^{\mathbf{r}} |\mathbf{R},a, b\rangle
  &= \sum_{\oplus {\bf R}', \vec{\Gamma}} \sum_{a' b'} \sqrt{\frac{\dim( \mathbf{R})}{\dim(\mathbf{R}')}}  |\mathbf{R}', a', b'\rangle \
  \CG{\mathbf{R}}{a}{\mathbf{r}}{\alpha}{\mathbf{R}'}{a'}_{\Gamma_1}
  \CGstar{\mathbf{R}}{b}{\mathbf{r}}{\beta}{\mathbf{R}'}{b'}_{\Gamma_2}
  \label{eq:linkop} \ \ \ ,
\end{align}
where $\mathbf{r}$ indicates the representation of the link operator, the $a(b)$ label
states within an irrep in the left(right) spaces, and the primes denote final state properties generated by the application of the link operator.
For SU(2),
the internal state labels are naturally identified with half integers capturing the third component projection of the total angular momentum.
For SU(3), these labels may be the three-component vector of color isospin(T)-hypercharge(Y) rational numbers
$(T,T^z,Y)$ with $T^z,Y$ additive as utilized in Ref.~\cite{Byrnes:2005qx} or, more abstractly, the Gelfand–Tsetlin patterns.

While this four-link operator is naively capable of producing transitions outside the gauge-invariant subspace,
the vertex CGs prevent such transitions.
To be concrete, consider the application of a plaquette operator impacting two links of a three-point vertex in an initial state of $\mathbf{C}_1 = \mathbf{R}_t = \mathbf{3}$ and $\mathbf{Q}_\ell = \mathbf{8}$. Schematically,
\begin{equation}
  \hat{\Box}^\dagger|\mathbf{3}\rangle_{C_1} |\mathbf{8}\rangle_{Q_\ell} |\mathbf{3}\rangle_{R_t} \rightarrow  \begin{gathered}\includegraphics[width=0.2\textwidth]{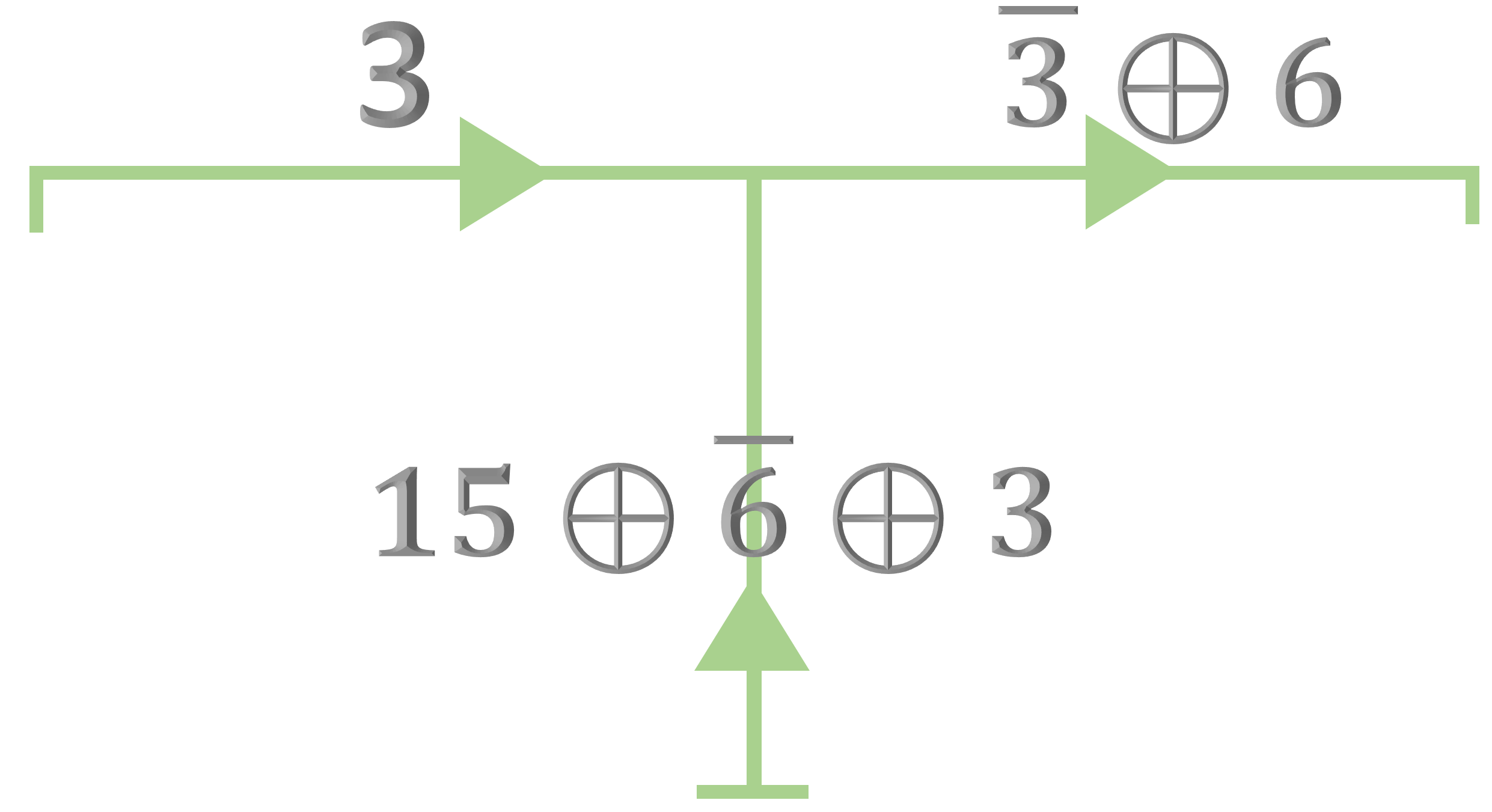}\end{gathered} \rightarrow \begin{gathered}\begin{cases}
    |\mathbf{3}\rangle |\overline{\mathbf{6}} \rangle|\overline{\mathbf{3}}\rangle & \langle \mathbf{3}, \ , \mathbf{3}, \ | \mathbf{6} , \ \rangle \\
    |\mathbf{3}\rangle |\mathbf{3}\rangle |\overline{\mathbf{3}}\rangle & \langle \mathbf{3}, \ , \mathbf{3}, \ | \overline{\mathbf{3}}, \ \rangle \\
    |\mathbf{3} \rangle |\mathbf{3}\rangle |\mathbf{6}\rangle & \langle \mathbf{3}, \ , \overline{\mathbf{6}}, \ | \overline{\mathbf{3}}, \ \rangle  \\
    |\mathbf{3} \rangle |\mathbf{15} \rangle |\mathbf{6} \rangle & \langle \mathbf{3}, \ , \overline{\mathbf{6}}, \ | \overline{\mathbf{15}}, \ \rangle
  \end{cases} \end{gathered} \ \ \ , \label{eq:vertexneedslegs}
\end{equation}
where the right shows the physical irrep configurations populated by the plaquette operator application and the associated CGs that would appear in the vertex factor.
When applying the $\hat{\Box}^\dagger$ operator, $\mathbf{3}$'s will be applied, according to Eq.~\eqref{eq:linkop}, to $\mathbf{R}_t$ and $\mathbf{Q}_\ell$.
Some combinations of the irreps generated by the plaquette operator are disallowed by Gauss's law, requiring information of the state of the neighboring link $\mathbf{C}_1$ stored in the vertex CG to maintain gauge invariance.
An example of such a configuration disallowed by the neighboring link is $|\mathbf{3}\rangle_{C_1} |\mathbf{15}\rangle_{Q_\ell} |\bar{\mathbf{3}}\rangle_{R_t}$ as $\mathbf{3}\otimes \mathbf{3}$ does not produce a $\overline{\mathbf{15}}$ or, equivalently, there is no singlet present in $\mathbf{3} \otimes \mathbf{3} \otimes \mathbf{15}$ tensor product.

As detailed in Appendix~\ref{app:plaquetteMEs}, the vector components at each vertex can be captured analytically through the calculation of composite CG factors.
As a result, the basis for
quantum simulation can be simplified to expressing
an SU(3) irrep on each link, leaving internal quantum numbers to impact the matrix elements comprising the local plaquette operator calculated classically.
This formulation extends the observations previously made in SU(2) lattice gauge theory for a one-dimensional string of links~\cite{Banuls:2017ena} and plaquettes~\cite{Klco:2019evd}.
Defining notation through the 9-R symbol,
\begin{equation}
  \begin{Bmatrix}
    \mathbf{A} & \mathbf{B} & \mathbf{C} \\
    \mathbf{3} & \mathbf{1} & \mathbf{3} \\
    \mathbf{D} & \mathbf{B} & \mathbf{E}
  \end{Bmatrix} = \sum \langle \mathbf{D}, y', \mathbf{B}, x| \mathbf{E}, q' \rangle_{\Gamma_1} \langle \mathbf{A}, y, \mathbf{B}, x | \mathbf{C}, q\rangle_{\Gamma_2} \langle \mathbf{A}, y, \mathbf{3}, c | \mathbf{D}, y'\rangle_{\Gamma_3} \langle \mathbf{C}, q, \mathbf{3}, c| \mathbf{E}, q'\rangle_{\Gamma_4} \label{eq:9R} \ \ \ ,
\end{equation}
where the sum is over all local vector and multiplicity indices,
the plaquette matrix elements may be expressed as
\begin{multline}
  \Bigg\langle \begin{pmatrix}\mathbf{C}_{1}, \mathbf{R}_{t}',\mathbf{C}_{3} \\  \mathbf{Q}_{\ell}', \mathbf{Q}_r' \\
  \mathbf{C}_{2}, \mathbf{R}_{b}', \mathbf{C}_{4}\end{pmatrix} \Bigg| \hat{\Box} \Bigg| \begin{pmatrix}\mathbf{C}_{1}, \mathbf{R}_{t},\mathbf{C}_{3} \\  \mathbf{Q}_{\ell}, \mathbf{Q}_r \\
  \mathbf{C}_{2}, \mathbf{R}_{b}, \mathbf{C}_{4}\end{pmatrix} \Bigg\rangle = \\ \sqrt{\frac{\dim(\mathbf{R}_{t}) \dim(\mathbf{R}_{b}) }{\dim(\mathbf{R}'_{t}) \dim(\mathbf{R}'_{b}) \dim(\mathbf{Q}_\ell) \dim(\mathbf{Q}_r) \dim(\mathbf{Q}'_\ell)^3 \dim(\mathbf{Q}'_r)^3}} \\
  \begin{Bmatrix}
    \overline{\mathbf{R}}_{t} & \mathbf{C}_{1} & \overline{\mathbf{Q}}_\ell \\
    \mathbf{3} & \mathbf{1} & \mathbf{3} \\
    \overline{\mathbf{R}}_{t}' & \mathbf{C}_{1} & \overline{\mathbf{Q}}_\ell'
  \end{Bmatrix}_{\includegraphics[width=0.03\textwidth]{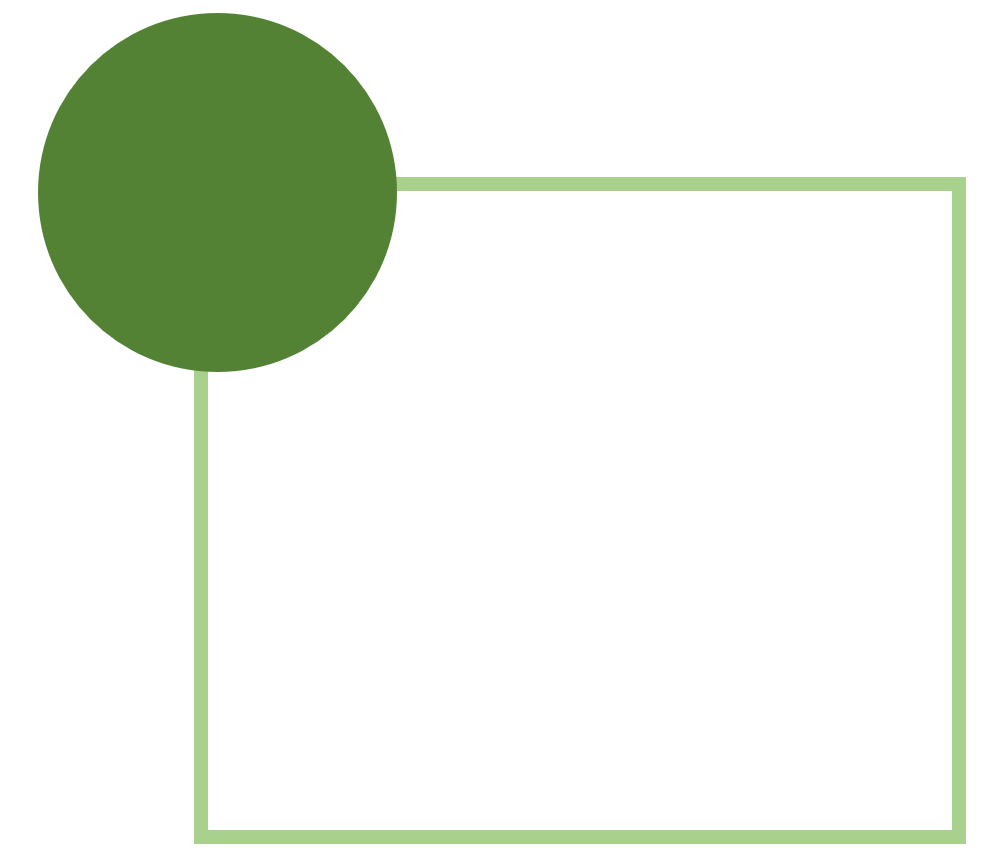}}
  \begin{Bmatrix}
    \mathbf{R}_{t} & \overline{\mathbf{C}}_{3} & \overline{\mathbf{Q}}_r \\
    \overline{\mathbf{3}} &  \mathbf{1} & \overline{\mathbf{3}} \\
    \mathbf{R}_{t}' & \overline{\mathbf{C}}_{3} & \overline{\mathbf{Q}}_r'
  \end{Bmatrix}_{\includegraphics[width=0.03\textwidth]{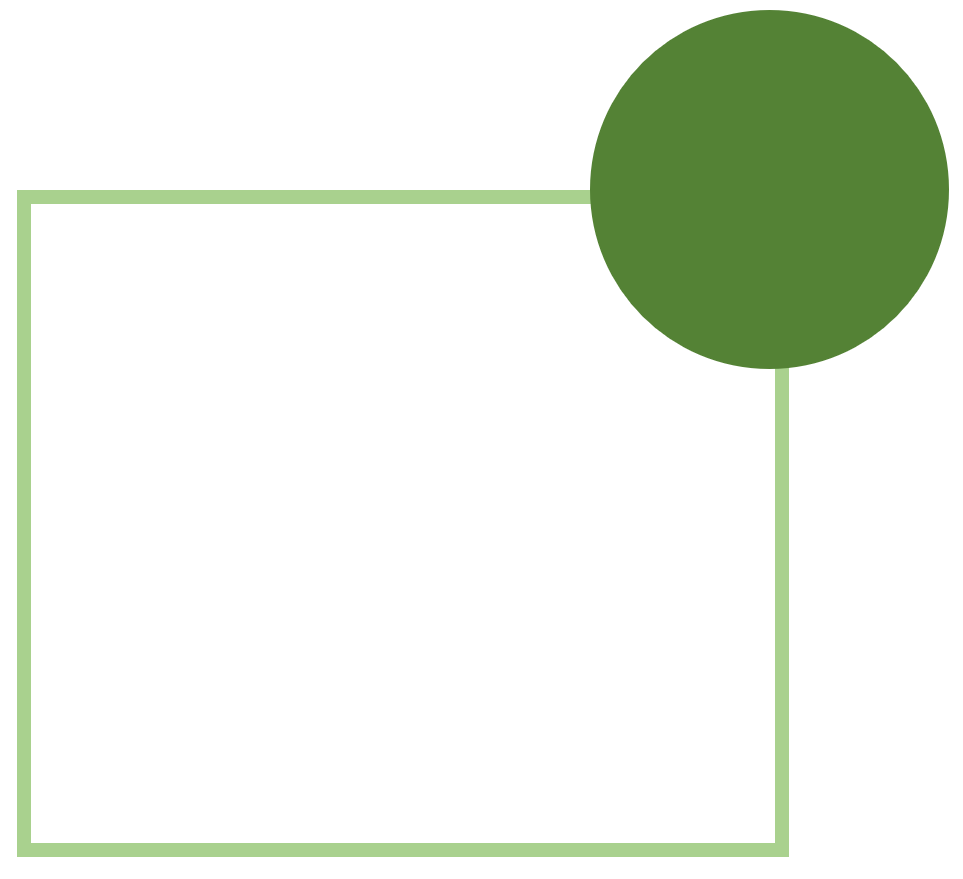}}
  \begin{Bmatrix}
    \overline{\mathbf{R}}_{b} & \mathbf{C}_{2} & \mathbf{Q}_\ell \\
    \overline{\mathbf{3}} & \mathbf{1} & \overline{\mathbf{3}} \\
    \overline{\mathbf{R}}_{b}' & \mathbf{C}_{2} & \mathbf{Q}_\ell'
  \end{Bmatrix}_{\includegraphics[width=0.03\textwidth]{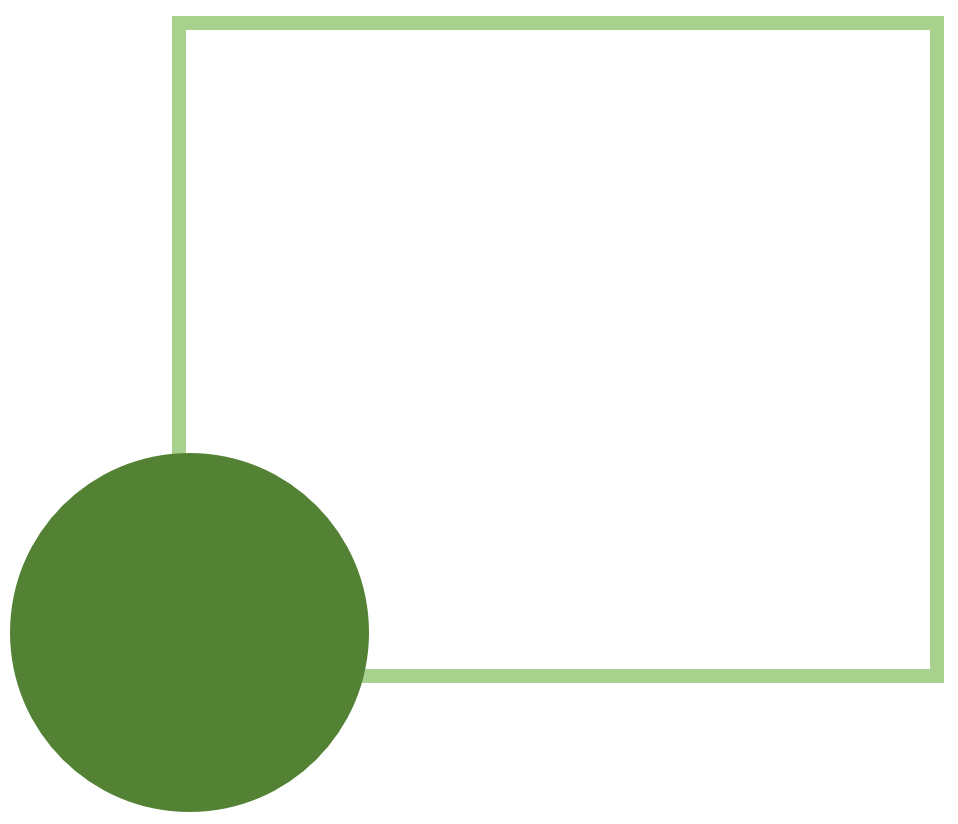}}
  \begin{Bmatrix}
    \mathbf{R}_{b} & \overline{\mathbf{C}}_{4} & \mathbf{Q}_r \\
    \mathbf{3} & \mathbf{1} & \mathbf{3} \\
    \mathbf{R}_{b}' & \overline{\mathbf{C}}_{4} & \mathbf{Q}_r'
  \end{Bmatrix}_{\includegraphics[width=0.03\textwidth]{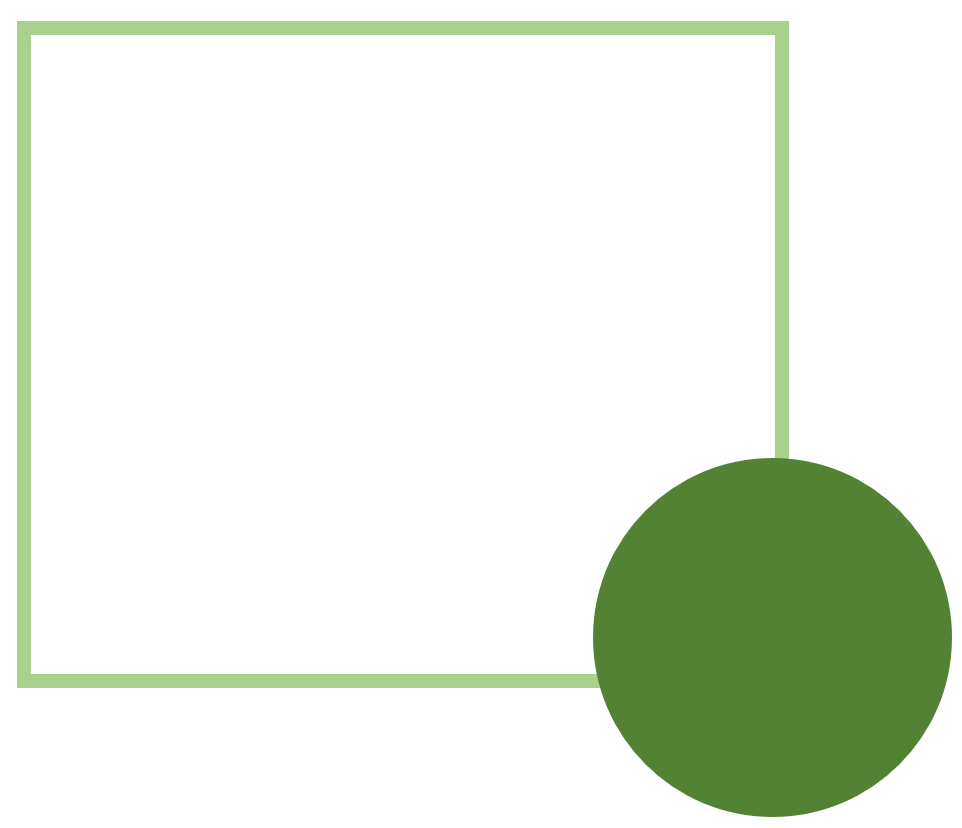}}
  \ \ \ ,
  \label{eq:plaquetteArbMEs}
\end{multline}
where the subscipts on the 9-R symbols graphically denote the corresponding vertex in Fig.~\ref{fig:labels}.
The delocalization of information necessary to consider in the application of a plaquette operator,
depicted in Eq.~\eqref{eq:vertexneedslegs},
is  set at the distance of neighboring links
and does not grow beyond this locality for larger lattices, nor in higher dimension.
Because the simple Hilbert space structure of qubit degrees of freedom
will not provide the vertex CGs necessary to retain a four-link local plaquette operator,
the CGs, usually separately relegated to the vertex and the operator, have been included here in their entirety.
Two methods of implementation will be explored below.
In Section~\ref{sec:globalbasis2p}, the vertex CGs of Eq.~\eqref{eq:vertexCGs} will
be manually captured through symmetry-projected, global wavefunctions of the lattice mapped to quantum states of a quantum device.
In Section~\ref{sec:localbasis}, the vertex CGs will be  captured in the structure of
an eight-link operator controlled on the quantum states of the four neighboring links.

\subsection{Connectivity in Multiplet Space}
\label{subsec:connectivity}
\noindent
When designing  operations for the  implementation of dynamical processes within a Hilbert space,
it is helpful to understand the natural connectivity between states.
This basis-dependent feature will affect the efficiency of digital formulations of time evolution
as well as their ease of implementation on quantum architectures with limited connectivity.
Naturally, designing quantum hardware with connectivity matching that of the field Hilbert space (or vice versa) is expected to be advantageous.

\begin{figure}[!ht]
\centering
  {\Large \textbf{SU(2):}} \qquad
  \begin{minipage}[c]{0.7\textwidth} \includegraphics[width=1\textwidth]{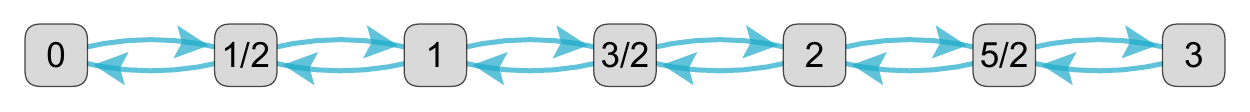}
  \end{minipage}
  \begin{minipage}[c]{0.05\textwidth}
  $\mathbf{\cdots}$
  \end{minipage}
  \\
  \vspace{0.3cm}
  {\Large \textbf{SU(3):}} \qquad
  \begin{minipage}{0.8\textwidth}
  \includegraphics[width=1.0\textwidth]{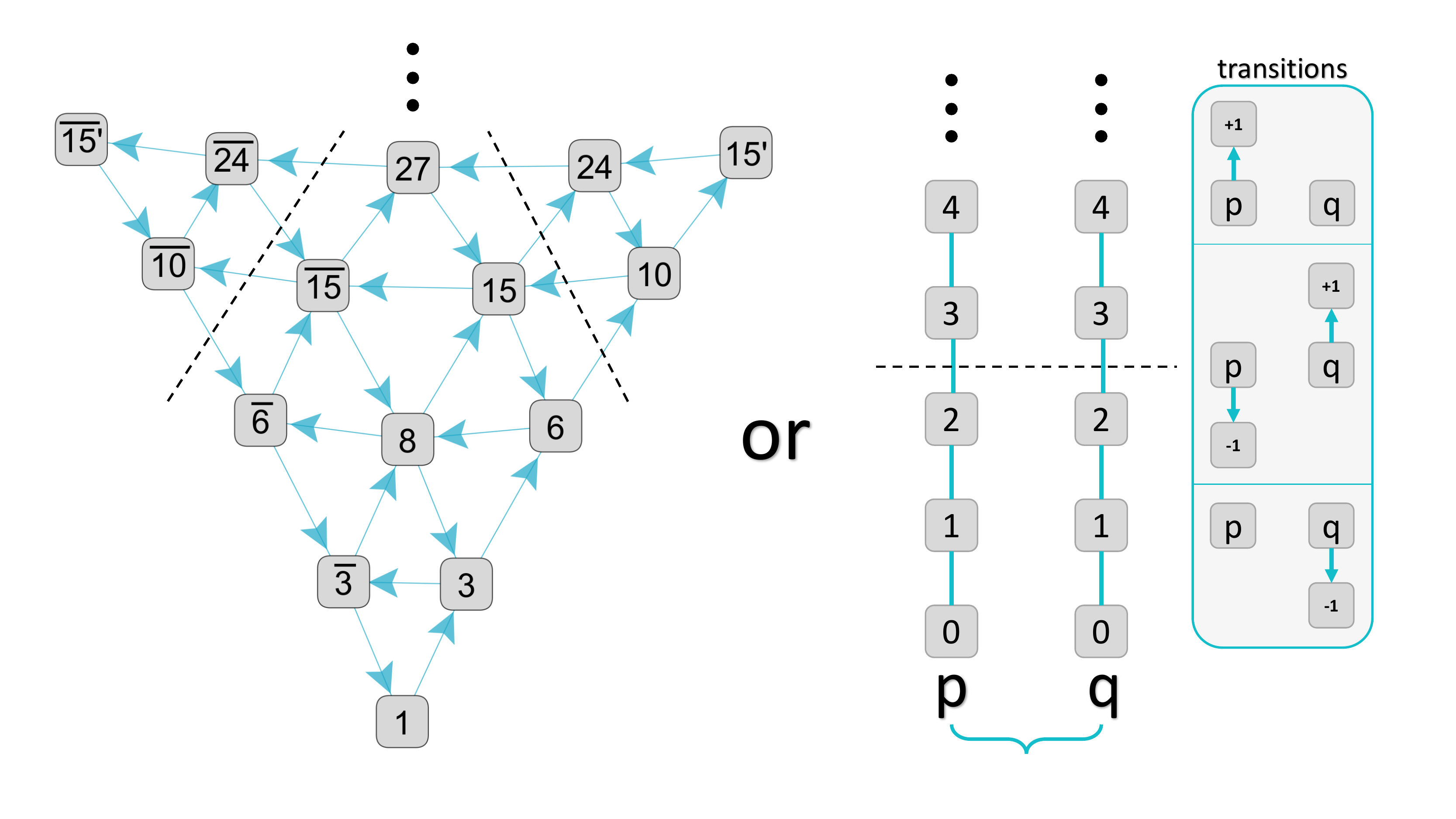}
  \end{minipage}
  \caption{
 Connectivity diagrams for the low-Casimir irreps in SU(2) and SU(3) gauge theory upon application of
 the plaquette operator.
 In SU(2), connections are bidirectional.
 In SU(3),
 connections between multiplets are directional, shown here for the application of the fundamental representation.
 The link Hilbert space can be captured through the connectivity of a single constrained
 hexagonal lattice of quantum states (lower-left panel) or through a pair of correlated one dimensional lattices (lower-right panel).
  }
  \label{fig:connectivitydiagram}
\end{figure}

When an SU(2) link operator in the fundamental representation acts, it is capable of raising
or lowering the total angular momentum $j$ value of the link state by $\pm\frac{1}{2}$.
When the vector components of the link Hilbert space are classically incorporated into the matrix elements of the plaquette operator, as discussed above, these $j$ values are sufficient to describe the state of the local link degree of freedom.
Thus, in a basis of multiplets, the relevant connectivity of quantum states within an SU(2) gauge link is in the form of a simple ladder, as shown in Fig.~\ref{fig:connectivitydiagram}.
While the coefficients associated with connections between these states
depend on the surrounding links and the associated local CG factors,  states  interact
with maximally two neighboring states.

For SU(3) lattice gauge theory in the multiplet basis,
the connectivity among states within the local gauge link Hilbert space is only slightly more elaborate, and is well known from group theory.
For the link operator in the $\mathbf{3}$ or  $\overline{\mathbf{3}}$,
the tensor indices become,
\begin{align}
  (p,q) \otimes (1,0) &= (p+1,q) \oplus (p-1,q+1) \oplus (p,q-1)
  \ \ \ ,
  \nonumber\\
    (p,q) \otimes (0,1) &= (p,q+1) \oplus (p+1,q-1) \oplus (p-1,q)
  \ \ \ .
  \label{eq:linkop3generatedirreps}
\end{align}
A connectivity diagram of the transitions described by Eq.~\eqref{eq:linkop3generatedirreps} is shown at the
bottom of Fig.~\ref{fig:connectivitydiagram}, with black dashed lines indicating a possible truncation of irreps with up to two fundamental and two antifundamental indices.
For each irrep not affected by the lower boundary of zero indices or the upper truncation,
three connections exit the irrep associated with the three transitions described in Eq.~\eqref{eq:linkop3generatedirreps}.
Connections into any irrep through the fundamental link operator also appear with maximal number three
and along distinct paths from those exiting the irrep.
In this sense, the one-dimensional nearest-neighbor locality of the
SU(2) link-operator-generated gauge space is promoted in SU(3)
to a two-dimensional hexagonal lattice in the bulk of high irrep truncation on each link.
Importantly, these connections remain local upon the two-dimensional manifold.

\subsection{Embeddings of the Gauge Space}
\label{subsec:embeddings}

Due to their role in defining the Hilbert space,
the basis used to digitize gauge fields impacts  many aspects of quantum simulations of gauge-field theories.
It is for this reason that understanding the practical consequences of  basis choice,
or distributions of the field content onto the degrees of freedom,
is expected to play a central role in optimizing simulations on different  quantum architectures.

Before discussing the bases explored in this manuscript for the digitization of the SU(3) gauge field, it is worth
pausing to reflect upon the basic assumption that continuous gauge fields be digitized at all.
There is an alternative to digitizing the gauge field directly that retains a spatially local distribution of qubit degrees of freedom.  Rather than implementing the gauge field continuum limit and then the thermodynamic and spatial continuum limits, it has been proposed, under the names of link models or qubit regularization, that this can be replaced with a one-step process by devising a spin system of appropriate local symmetries with a critical point in the same universality class as the field of interest~\cite{HORN1981149,ORLAND1990647,Chandrasekharan:1996ih,Brower:2003vy,Wiese:2006kp,Singh:2019uwd,Bhattacharya:2020gpm}.
Tuning to the lattice continuum limit at a phase transition of the latticized spin system produces an emergent relativistic field theory.
One way to interpret success with this approach is through spatial blocking of the lattice producing effectively continuous field values in the continuum limit at the critical point.
Just as the correlation length of the field determines the lattice volume needed to approach the thermodynamic limit for the spatial continuum with continuous fields, a second correlation length will be present in the spin lattice expressing the blocked volume necessary to capture effectively continuous fluctuations in the local field degrees of freedom.
For a scalar field, it has been shown that the local digitization of the field, and the efficiency of the local quantum Fourier transform (QFT), allows the effectively continuous fluctuations in the local field degrees of freedom to be captured with double exponential convergence in local qubit number~\cite{Jordan:2011ne,Jordan:2011ci,Macridin:2018oli,Klco:2018zqz}.
Benefitting from this application of the Nyquist-Shannon (NS) sampling theorem, the number of qubits per site relevant to foreseeable applications is expected to be $\lesssim 5$~\cite{Macridin:2018oli,Klco:2018zqz}.
With current methodologies, the spatial continuum of the scalar field retains polynomial lattice artifacts as the volume-sized QFT is not expected to be efficient and thus modifications to the lattice dispersion relation will appear polynomially with the lattice spacing~\cite{Symanzik:1983dc,Klco:2018zqz}.
When working in a spin system, both the field continuum and the spatial continuum limits are na\"ively expected to be of the latter type, allowing neither to enjoy the rapid NS convergence.
It is for this reason, with an expectation that local NS
convergence will retain some relevance for digitized quantum fields beyond the scalar field, that the current manuscript
will focus on the local digitization of the gauge field rather than the identification of a UV completion structured as
a local spin system sharing a universiality class with QCD. However, the rapidly evolving quantum ecosystem and
exploratory nature of current development encourages thorough investigation of all possible avenues for embedding
gauge fields into controllable quantum architectures.

Multiple embeddings of the lattice Hilbert space through the basis of SU(3) irreps will be considered in the following.
The first distinction that can be made is whether the embedding is \emph{global} or \emph{local}.
A local embedding of the Hilbert space assigns a qubit register to each link of the lattice, thus storing local information of the field in locally-distributed quantum systems.
Distributing qubits/qudits across the lattice in this way is not always the most efficient use of Hilbert space.
In particular there are a number of symmetries present in the gauge theory creating correlations between the link states e.g., vertex gauge symmetries, spatial parity and color-parity.
Being  good symmetries of the Hamiltonian,
a state that begins in one symmetry sector will remain within the sector throughout its dynamical evolution.
Transferring to a \emph{global} embedding allows manual projections into these symmetry sectors, as utilized in Refs.~\cite{Klco:2018kyo,Lu:2018pjk}.
The resulting efficient use of Hilbert space, beyond reducing the total qubit resource requirements
of the calculation, protects the calculation from gauge-variant or symmetry-breaking errors.
Each state of the quantum hardware is associated with a full lattice configuration with the desired projected symmetry.
As long as an error maintains the computational basis, all errors maintain global and vertex symmetries of the lattice in the global embedding.

While the global embedding is advantageous for small lattices on noisy hardware,
the classical computational demands for pre-conditioning the Hilbert space and computing symmetry-projected matrix elements
scale poorly with the volume of the lattice.
For this reason, we further explore two link-local embeddings of the lattice Hilbert space in the irrep basis.
In the first approach, a single quantum register or qudit is used to capture the gauge space of each link, the operations of Section~\ref{sec:localbasis} describe how the internal modes must interact in order to express time evolution of a quantum state with respect to the magnetic Hamiltonian.
While full connectivity has been permitted within a qudit, with rotations mixing populations between any two modes, it can be seen from Fig.~\ref{fig:connectivitydiagram} that the required mode connectivity does contain a sense of locality.
In particular, the dynamical mode connectivity, reflecting that produced by the plaquette operator between irreps, will have a structure of nearest neighbor locality in a truncated and constrained two-dimensional hexagonal lattice.
While this locality is an improvement over arbitrarily non-local interactions, this connectivity is potentially sub-optimal, unless a qudit architecture is designed that naturally reflects this two-dimensional structure.
In particular, for a one-dimensional embedding of the irrep modes into a ladder-structured qudit, necessary mode rotations delocalize as the irrep truncation increases, reflected by the growing number of irreps per row in Fig.~\ref{fig:connectivitydiagram}.

One natural way to address the growing two-dimensional link structure in gauge space, inspiring the second local embedding explored in this work, is to introduce a number of qudits on every link equal to the rank of the gauge group, two for SU(3), as introduced by Byrnes and Yamamoto~\cite{Byrnes:2005qx}.
These two qudits will reflect the tensor index structure of the irrep, in SU(3) denoted as $(p,q)$ in Section~\ref{sec:hamiltonian} above, with one qudit specifying the value of $p$ and the second indicating the value of $q$.
In this local $(p,q)$ representation of the local irreps, the $p$ and $q$ registers are simply integers from zero to the maximum number of tensor indices considered.
The plaquette operator
produces correlated transitions by $\pm 1,0$ within the $p$ and $q$ qudits at each active link of a plaquette.
In this way,
further separating the link space into a pair of qudits naturally simplifies the two-dimensional connectivity
shown in the left panel of Fig.~\ref{fig:connectivitydiagram} into a correlated pair of qudits shown in the right panel
of Fig.~\ref{fig:connectivitydiagram},
each requiring only a raising/lowering operator within a one-dimensional embedded space.
With this splitting of the link space into two qudits, however, operators of up to 8 qudits controlled on another 8 qudits will be required for constructing the local time evolution operators for the one-dimensional plaquette string.
As hardware-specific strategies evolve for implementing mode-isolated multi-qudit unitary rotations, tradeoffs in the fidelity of intra- and inter-qudit operations will inspire a decision: implementing a two-dimensional gauge space within a single qudit at each link versus implementing one-dimensional correlated gauge spaces within each of two qudits per link, or some spatially dependent combination of the two.

As will become clear, the choice of basis inspires different ways to perform the gauge field truncations, resulting in different convergence properties and resource requirements for simulation.
Local truncations at the level of the representations in the link Hilbert spaces readily scale to larger systems, requiring resources that scale with the spatial volume of the simulation.
While a global truncation connects well to intuition based on globally conserved quantities, the implementation of such a basis does not scale well with increasing system size.
It would appear that an adaptive local truncation scheme constrained by a global truncation
may be required for optimal use of available resources in future simulations, though constructing such a scheme is beyond the scope of this work.

\section{The Single Plaquette}
\label{sec:singleplaquette}
\noindent
A single plaquette is one of the simplest gauge-invariant objects that can be constructed
within a lattice gauge field theory.
For Yang-Mills gauge theory, 
the Hamiltonian responsible for its dynamics is a special case of that given in Eq.~(\ref{eq:QCDham}),
given by
\begin{equation}
	\hat{H} =
\frac{g^2}{2} \ \sum_{ b, {\rm links}}\ |  \hat {\bf E}^{(b)} |^2
 + \frac{1}{2g^2}\left(6 - \hat { \Box}-  \hat { \Box}^\dagger\right)
 \ \ \ ,
\end{equation}
where the lattice spacing is set to $a=1$, and $b$ is an adjoint color index.
The Hilbert space of a single link has been defined previously,
spanned by the eigenstates of the electric-field  strength operator,
$ | {\bf R}, \alpha , \beta \rangle =  | {\bf R}, \alpha  \rangle_L  | {\bf R},  \beta \rangle_R$.
The Gauss's law constraint allows the state of the one-plaquette system to be expressed in terms of
basis states of the form
\begin{equation}
	| {\bf R} \rangle = \frac{1}{\text{dim}({\bf R})^2}
	\sum_{\alpha, \beta, \gamma, \delta}
	| {\bf R}, \alpha , \beta \rangle_1
	| {\bf R}, \beta ,  \gamma \rangle_2
	| {\bf R}, \gamma , \delta  \rangle_3
	| {\bf R}, \delta ,  \alpha \rangle_4
 \ \ \ ,
\label{eq:oneplaqwf}
\end{equation}
where ${\bf R}$ is the representation of each link.
The irreducible representations of $SU(3)$ with tensor representation $T^{a_1,a_2,\cdots,a_p}_{b_1,b_2,\cdots,b_q}$
can be specified by $p$ and $q$,
and the global basis states for the one plaquette system 
is conventionally denoted by $\ket{p,q}$.
The electric energy of a basis state is determined  by the action of the Casimir operator (Eq.~\eqref{eq:casimir}), and is equal to four times the
value of the Casimir operator on a single link multiplied by a factor of $g^2/2$.
The plaquette operator is defined in Eq.~(\ref{eq:PlaqOp}), and the subsequent section.
For the one-plaquette system, matrix elements between basis states
$\langle {\bf R}_f | \hat \Box | {\bf R}_i \rangle = 1$
if ${\bf R}_f$ is present in the decomposition of ${\bf R}_i \otimes {\bf 3}$,
and $0$ otherwise, by the completeness of CG coefficients.

\FloatBarrier

\subsection{Color Space Truncation Errors}
\label{subsec:oneplaqexpsup}
\noindent
As any computational framework is comprised of a finite number of controllable degrees of freedom, numerical explorations of gauge theories require spatial latticization as well as a form of digitization and truncation of the continuous field.
As discussed above for SU(3) gauge theory on a single plaquette, digitization will here be accomplished in a gauge invariant way by truncating the number of tensor indices $(p,q)$ at values $(\Lambda_p,\Lambda_q)$ with $\Lambda_p = \Lambda_q $  a choice following the natural color parity symmetry of the system.
Any truncation of the field will introduce controlled, systematic errors that must be quantified.
Previous explorations of digitized scalar fields on a spatial lattice found that the field-space digitization converged double-exponentially in the number of qubits per lattice site describing the local field~\cite{Jordan:2011ci,Jordan:2011ne,Macridin:2018gdw,Macridin:2018oli,Klco:2018zqz}.
This convergence is attributed to the Nyquist-Shannon sampling theorem when the field and conjugate momentum bases are distributed appropriately.
In Appendix~\ref{app:onesu2plaquette},
it is shown analytically that the asymptotic form of the
\enquote{color} space wavefunction for a single plaquette in SU(2) gauge theory is Gaussian with respect to the irrep dimensionality or number of tensor indices.
In the case of $1+1$ dim. SU(2) lattice gauge theory, the exponential convergence of low-lying
quantities with increasing truncation has been identified previously as discussed, for example,  in Ref.~\cite{Davoudi:2020yln}.
In this section, it is shown numerically that this Gaussian convergence is also present in the color space of an SU(3) plaquette ground state, as well as its static and dynamical observables.

\begin{figure}[!ht]
  \centering
  \includegraphics[width=0.265\textwidth]{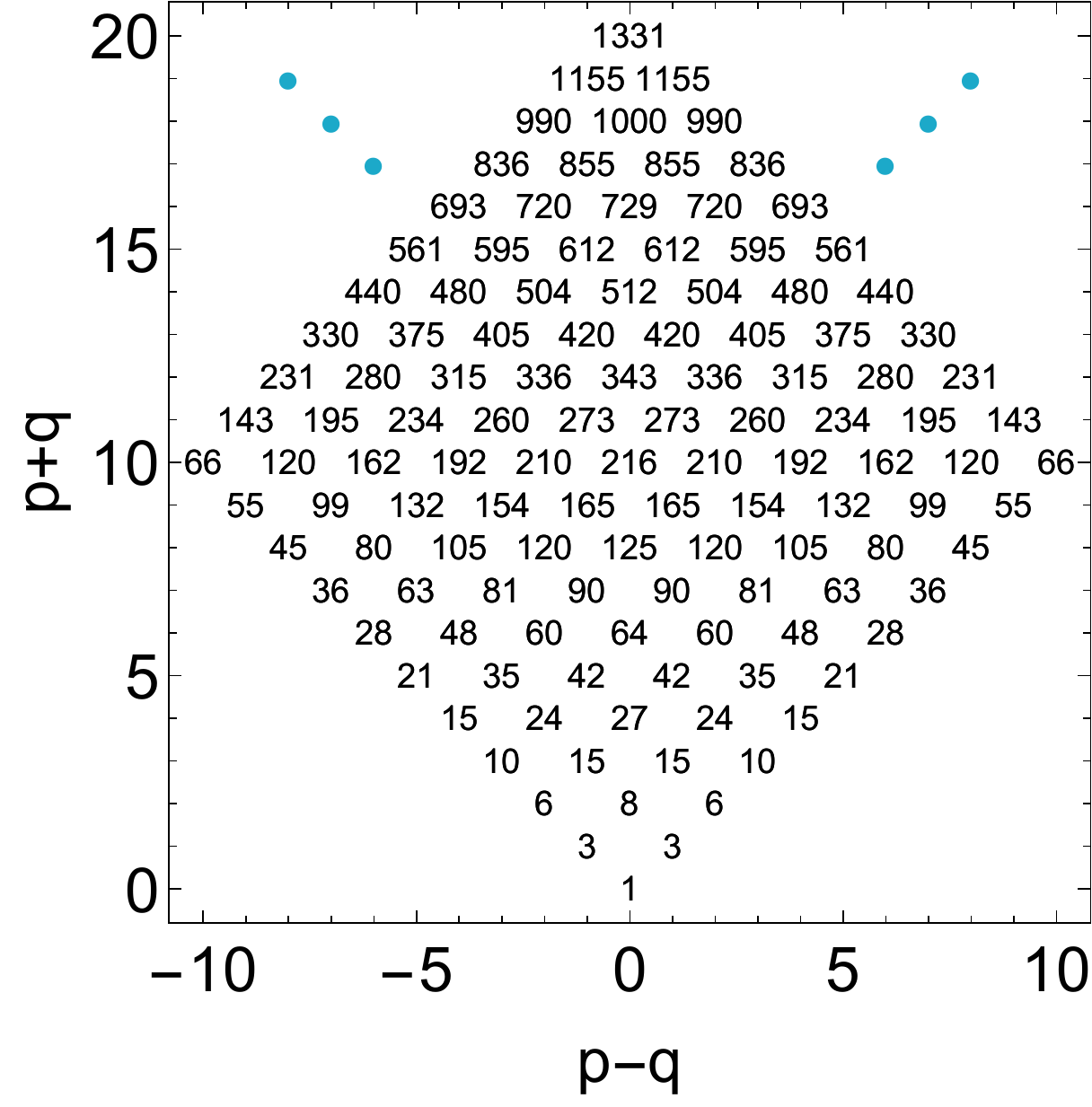}
  \begin{overpic}[width=0.3\textwidth,percent]{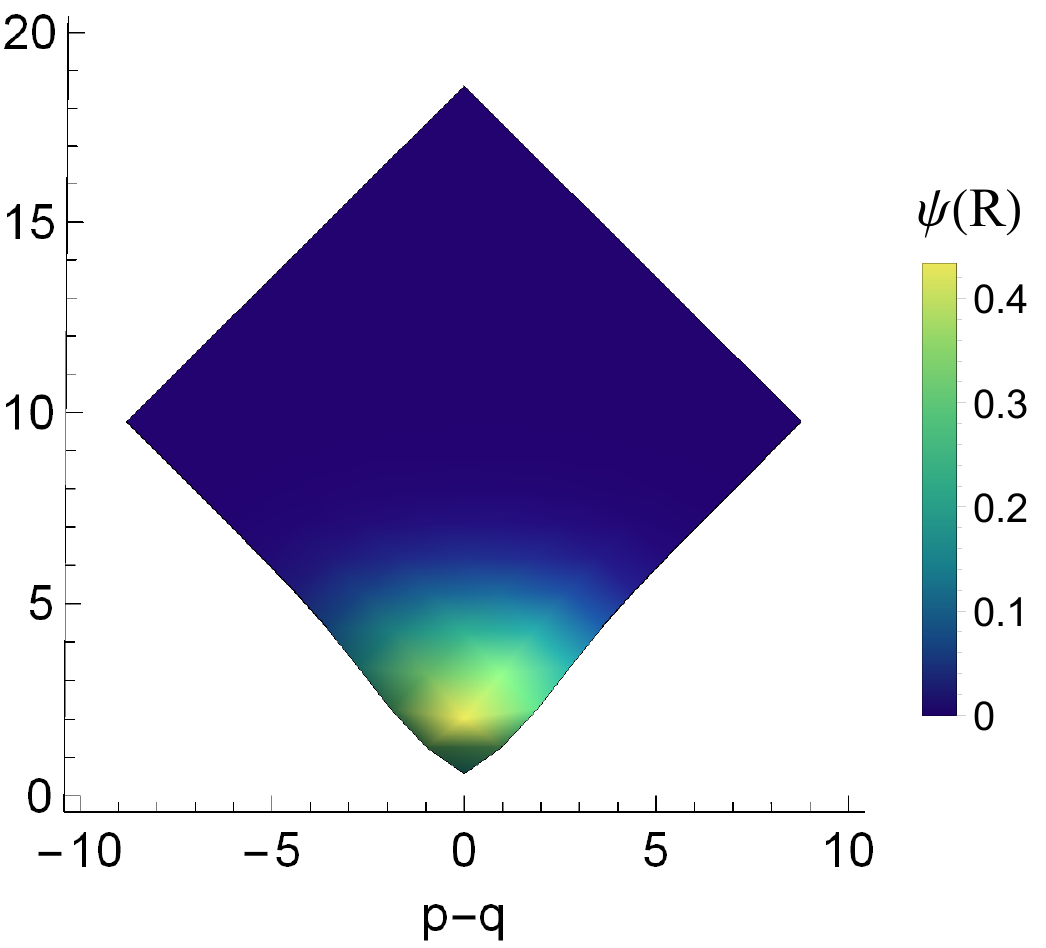}
    \put(84,72){\includegraphics[width=0.04\textwidth]{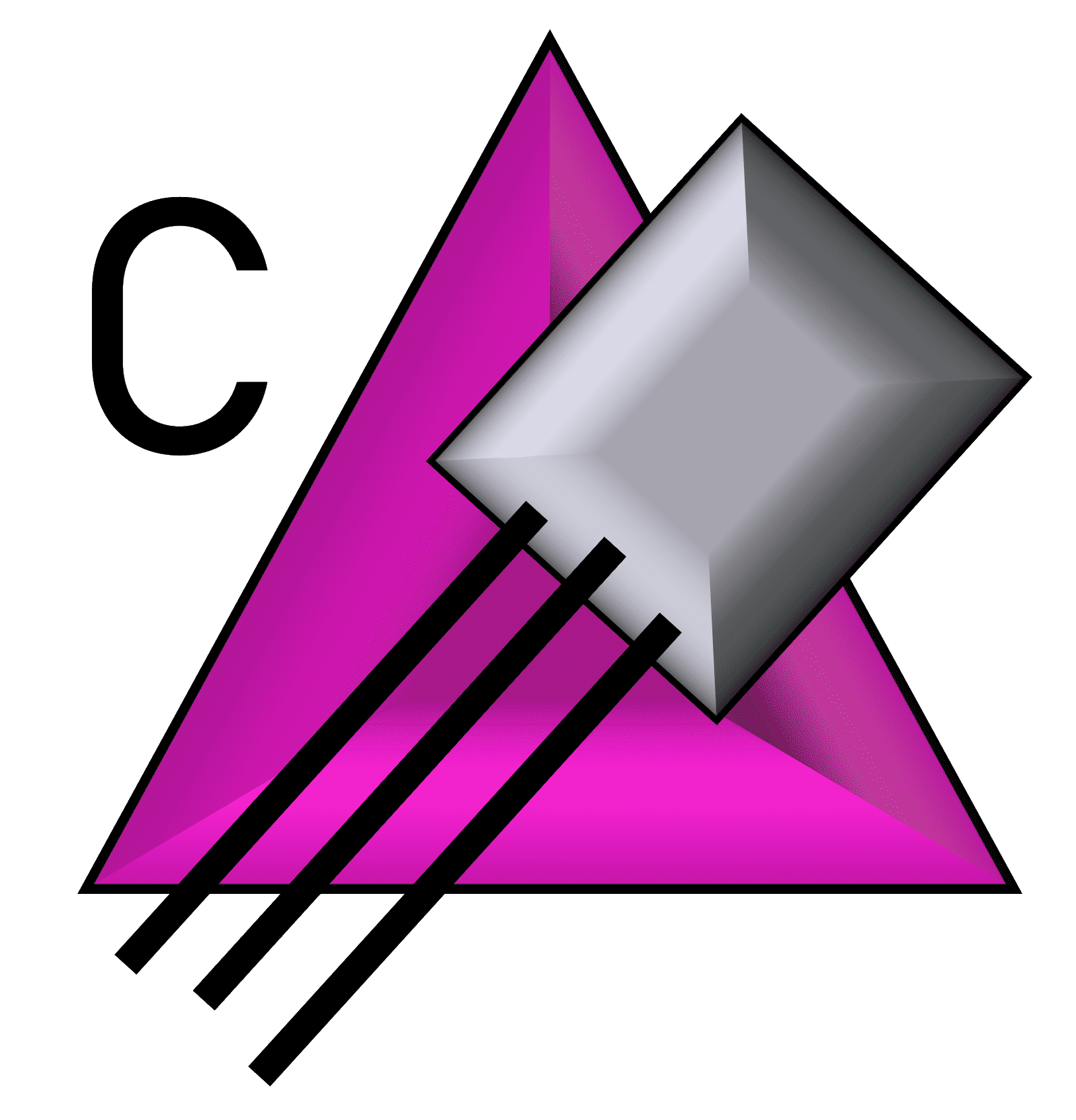}}
\end{overpic}
\begin{overpic}[width=0.325\textwidth,percent]{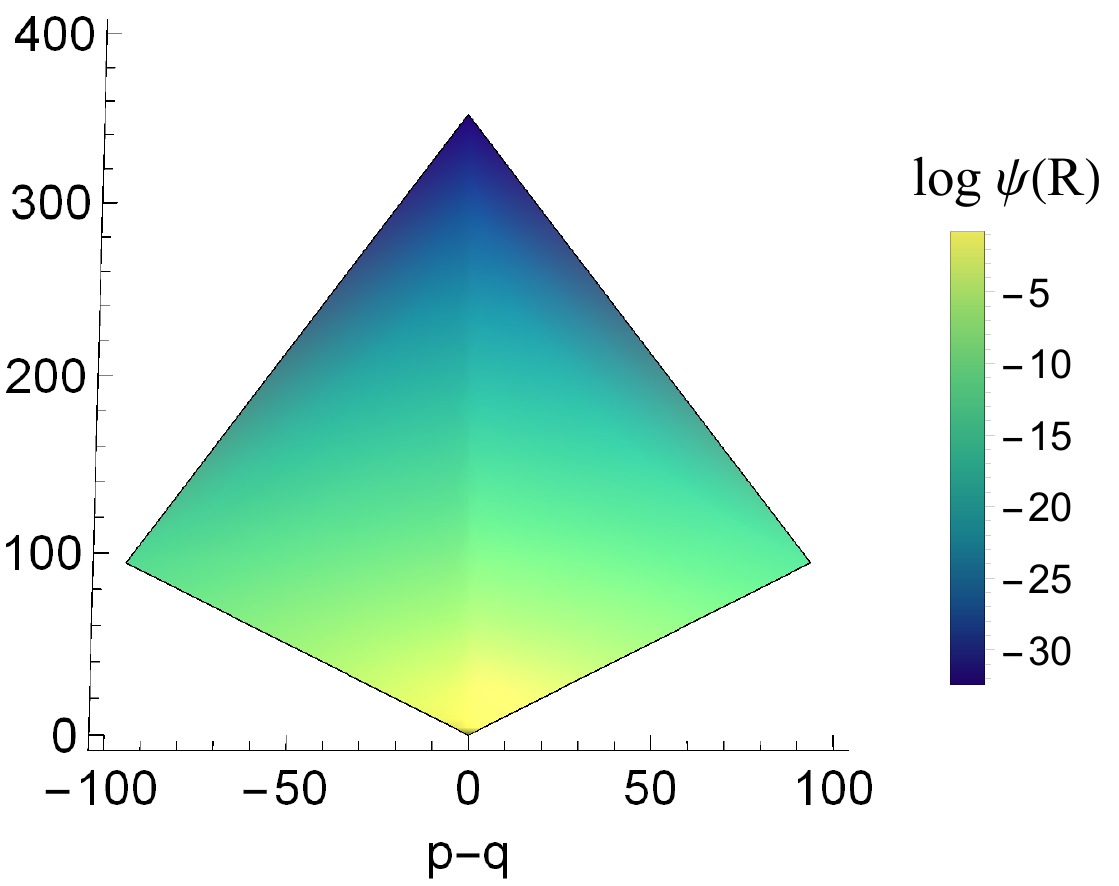}
    \put(82,65.5){\includegraphics[width=0.04\textwidth]{iconC1_bf.png}}
\end{overpic}
  \caption{
  On a grid (left panel) of irreducible representations organized by their dimensionality and plaquette connectivity (as shown in Fig.~\ref{fig:connectivitydiagram}), support of the the ground state wavefunction $\psi(\mathbf{R})$, shown for $g = 0.5$, is localized to low irrep dimensionalities (center panel).  Conjugate irreps appear on the left half of the grid with real irreps appearing along the center vertical.  The right panel shows $\log \psi(\mathbf{R})$ on a scaled quadratic grid for visual clarity of the convergence structure.   }
  \label{fig:oneplaq_wavefunction}
\end{figure}

Focusing upon the ground state wavefunction in a basis of irreducible representations,
Fig.~\ref{fig:oneplaq_wavefunction} shows an exponentially localized distribution of amplitudes.
In the left panel of Fig.~\ref{fig:oneplaq_wavefunction}, a grid of irrep dimensionalities is established akin to that of Fig.~\ref{fig:connectivitydiagram}.
Neighboring points on this grid are connected through the magnetic plaquette operator and thus experience a sense of locality.
The $x$-axis of this space is the difference between the number of fundamental and anti-fundamental indices in the tensor representation of each irrep, resulting in conjugate irreps residing at negative values and real irreps residing along the $p-q=0$ axis.
The dimensionalities of irreps below a truncation of $\Lambda_p = \Lambda_q = 10$ are shown explicitly, and higher index irreps would appear in the upper triangles.
In this space, the ground state amplitudes of the SU(3) single plaquette wavefunction at coupling $g = 0.5$ are shown in the center and right panels.
In the center panel, it is seen that support of this wavefunction is highly localized to the low-index regime.
Of course, the extent of localization is $g$-dependent and becomes dispersed as $g$ is lowered toward the weak coupling limit.
The right panel presents the same wavefunction as in the center panel, but with logarithmic and quadratic functional distortions on the wavefunction amplitudes and the tensor indices, respectively.
From this perspective, the asymptotic Gaussian structure of the irrep-space wavefunction is visually clear.

\begin{figure}
\centering
\begin{overpic}[width = 0.46\textwidth ,percent]{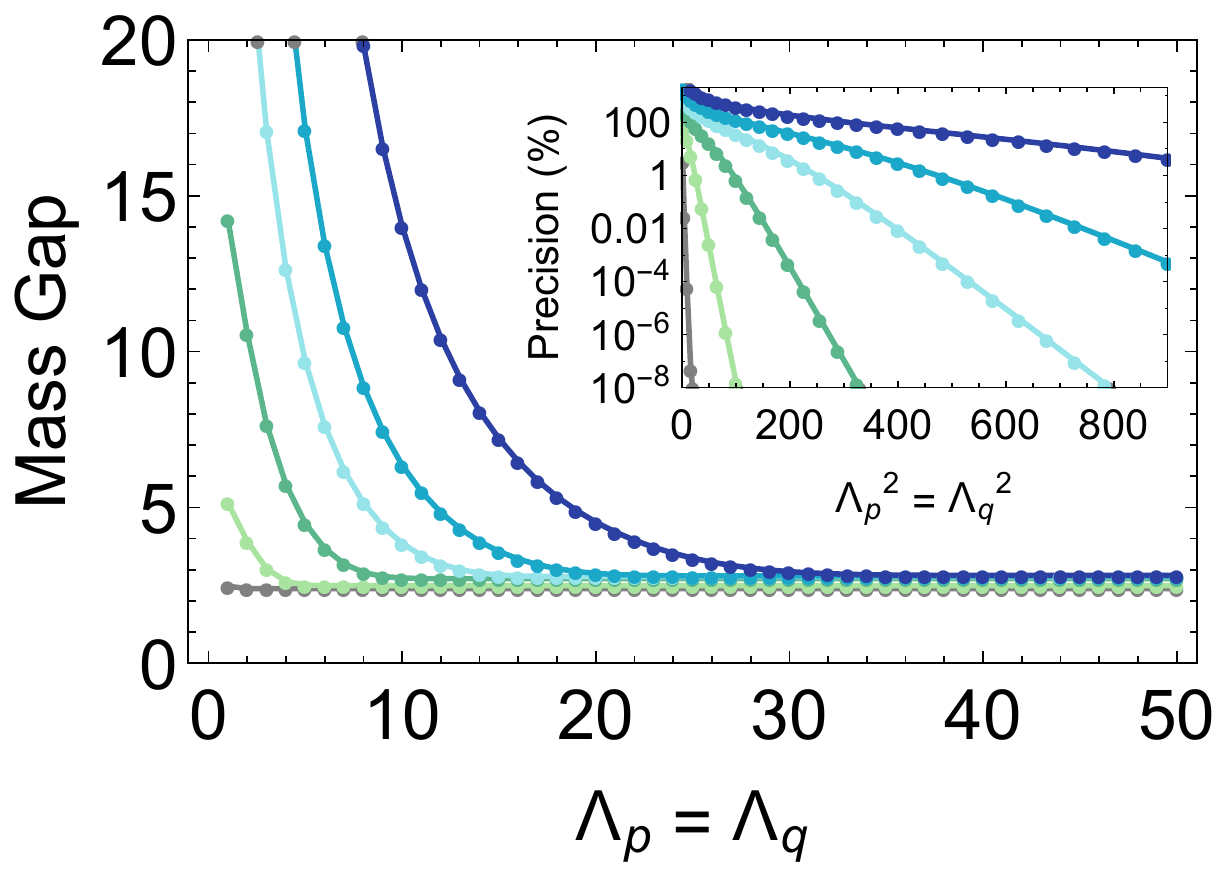}
    \put(93,62){\includegraphics[width=0.06\textwidth]{iconC1_bf.png}}
\end{overpic}
\begin{overpic}[width=0.46\textwidth,percent]{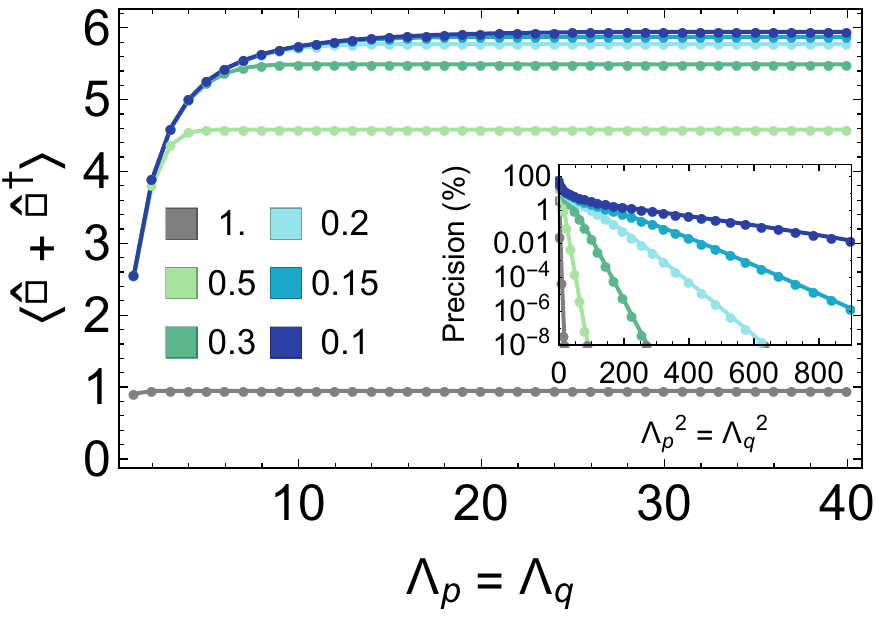}
    \put(93,63){\includegraphics[width=0.06\textwidth]{iconC1_bf.png}}
\end{overpic}
\caption{Mass gap (left panel) and vacuum expectation value of the Hermitian magnetic plaquette operator
$\hat{\Box}+\hat{\Box}^\dagger$ (right panel) for one plaquette  in SU(3) gauge theory as a function of $\Lambda_p$, the irrep tensor index truncation.
Convergence is demonstrated for six different values of the coupling ($g = 0.1$ to $1$). Inset panels show the percent deviation in observables from their values without truncation. The inset $x$-axes are squared for visual clarity of the convergence structure.}
	\label{fig:oneplaq_staticconvergence}
\end{figure}

The exponential localization of the single plaquette wavefunction extends this profitable convergence also to static and dynamic observables.
Figure~\ref{fig:oneplaq_staticconvergence} shows the convergence of the mass gap and the magnetic plaquette operator expectation value at a range of couplings.
\begin{figure}
\begin{overpic}[width = 0.46\textwidth,percent]{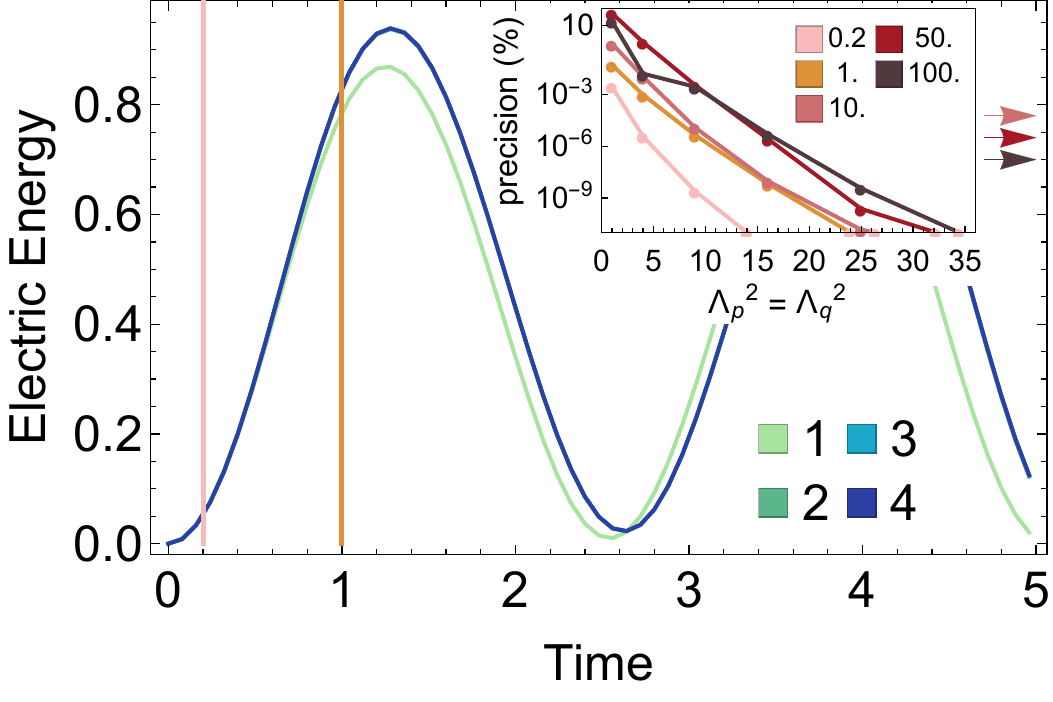} \put(92,61){\includegraphics[width=0.06\textwidth]{iconC1_bf.png}}
	\end{overpic}
\begin{overpic}[width = 0.46\textwidth,percent]{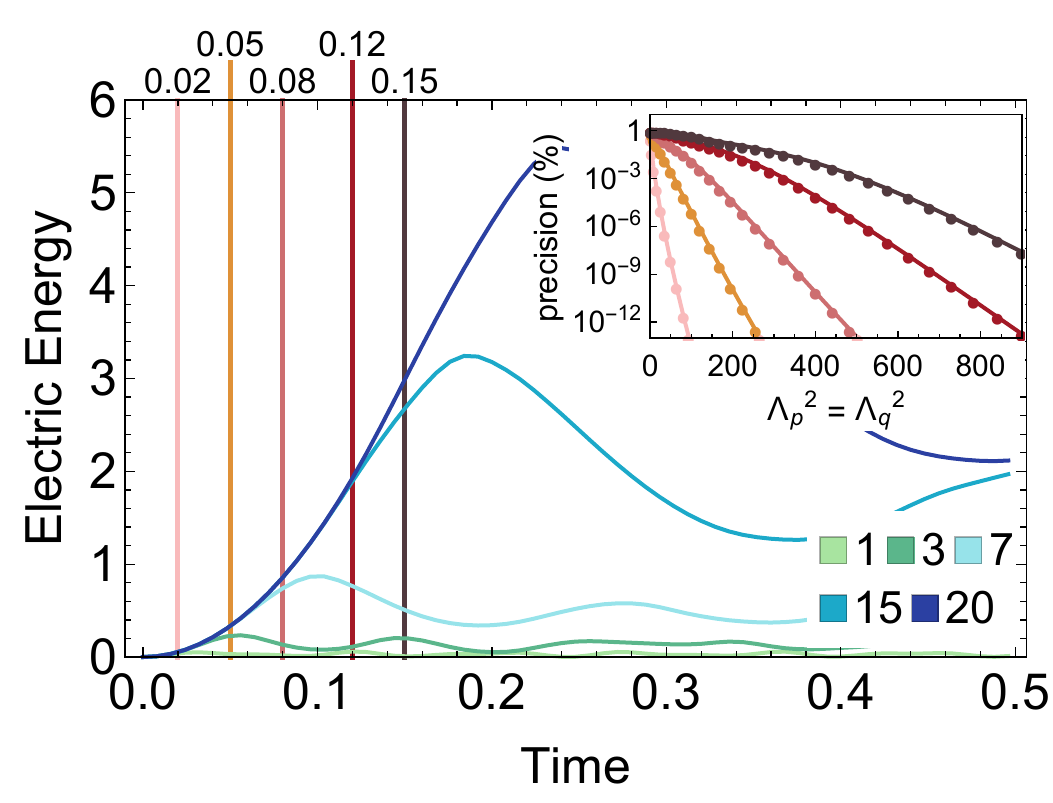} \put(90,61){\includegraphics[width=0.06\textwidth]{iconC1_bf.png}}
	\end{overpic}
	\caption{Expectation of the electric energy as a function of time for the single plaquette beginning in the strong coupling vacuum. Evolution is shown for tensor index truncations $\Lambda_p\leq4~(20)$ in the left (right) panel where $g =1~(0.1)$.
Insets show the percent deviation of the electric energy from its value without truncation at fixed times indicated by vertical lines in the main panels ($t = 10, 50, 100$ are beyond the domain of left panel).
The inset $x$-axes are squared for visual clarity of the convergence structure.  }
	\label{fig:one_plaq_he_convergence}
\end{figure}
Static observables for the unit coupling are found to converge to $10^{-8}$ percent of their asymptotic values at a low irrep truncation of $\Lambda_p = 4$  up to and including tensor irreps with four fundamental and four anti-fundamental indices.
As $g$ is lowered and the wavefunction disperses in irrep space, truncation errors naturally become more dramatic.
Interestingly, the mass gap demonstrates low $g$-dependence at high truncation, $\Lambda_p$, throughout the shown coupling range.
The insets of Fig.~\ref{fig:oneplaq_staticconvergence} provide convergence information with tensor index truncations scaled quadratically, as in the right panel of Fig.~\ref{fig:oneplaq_wavefunction}, such that the linear trajectories experienced at large tensor index truncations express Gaussian-type convergence structure.
From these insets, one can connect necessary quantum resources to the attainable precision of local observables as the weak-coupling limit is approached.
For example, percent-level precision for these quantities at couplings $g\geq0.3$ is expected to be achievable with $\Lambda_p \leq 10$ or equivalently 3-4 qubits per index register.
These features are expected to apply to the link-space localization and convergence on larger lattices of SU(3) gauge theory.
This suggests that SU(3)  Yang-Mills simulations in a cubic spatial lattice of extent $10\times 10\times 10$
could be performed with $\lsim 10^4$ qubits at this coupling.

It is important to keep in mind that our analysis has been performed in the electric basis, and requires increasing resources with decreasing lattice spacing to achieve the same level of precision for any given quantity.
Therefore, there is a minimum lattice spacing (coupling)  below which computations are inaccessible to the electric basis for a given quantum device and available classical computing resources.
Recent work by Haase {\it et al}~\cite{Haase:2020kaj} has shown in the context of QED that working instead with the magnetic basis is potentially more effective in calculations at small lattice spacings, making it an interesting area for further investigations.

As a final demonstration, Fig.~\ref{fig:one_plaq_he_convergence} shows similar exponential convergence properties also for dynamic observables.
Calculating the time evolution of the electric energy as a function of time at increasing field truncations, $\Lambda_j$,  similar Gaussian precision improvements are observed.
Related to the fact that the $g = 1$ mass gap is well captured at low truncation, the $g = 1$ time evolution in the left panel of Fig.~\ref{fig:one_plaq_he_convergence} is well captured at low truncations even at long times.
For example, the expectation value of the electric energy at time $t = 100$ is achievable at single precision with just 3 qubits per index register.
As the coupling is reduced and the wavefunction experiences reduced locality,
larger truncations are demanded to achieve precise calculations of long-time observables.
The right panel of Fig.~\ref{fig:one_plaq_he_convergence} quantifies this scenario for a coupling of $g = 0.1$.
Convergence calculations such as these inform estimations of  resource requirements for future lattice gauge theory simulations that will be implemented at a selection of coupling strengths and extrapolated to inform the continuum limit with a complete quantification of uncertainties.
In subsequent subsections, we show the results of simulations of these systems performed on IBM's
superconducting architecture.

\FloatBarrier

\subsection{Global Basis for One Plaquette}
\subsubsection{\texorpdfstring{ $\mathbf{8}$ Truncation }{ 8 Truncation}}
\label{sec:OnePlaq8}
\noindent
It is informative to study a simple basis truncation of
$p,q \le 1$, containing the irreps $\{{\bf 1}, {\bf 3} , \overline{\bf 3} , {\bf 8}\}$.
These can be straightforwardly mapped to two qubits as $\{|00\rangle, |10\rangle, |01\rangle, |11\rangle \} = \{ |\mathbf{1}\rangle, |\mathbf{3}\rangle, |\overline{\mathbf{3}}\rangle, |\mathbf{8}\rangle\}$.
In this basis, the Hamiltonian is
\begin{eqnarray}
\hat{H} =
{ \frac{g^2}{ 2} }
\left(
\begin{array}{cccc}
0 & 0 & 0 & 0   \\
0 &  {\frac{16}{ 3}} & 0 & 0   \\
0 & 0 & {\frac{16}{3}} & 0   \\
0 & 0 & 0 & 12    \\
\end{array}
\right)
\ +\
{\frac{1}{ g^2}}
\left(
3  \ \hat{\mathbb{I}}  \ -\
{\frac{1}{ 2}}
\left(
\begin{array}{cccc}
0 & 1 & 1 & 0  \\
1 &   0  & 1 & 1  \\
1 &   1  & 0 & 1  \\
0 &   1  & 1 & 0  \\
\end{array}
\right)
\right)
\ \ \ ,
\label{eq:Ham1p8}
\end{eqnarray}
consistent with the results provided in Ref.~\cite{Ligterink:2000wf} (when truncated at $f_1\le 2$ and removing the contribution from the ${\bf 6}$ and $\overline{\bf 6}$).
In terms of operators acting on a two-qubit system, the electric Hamiltonian operator can be decomposed as
\begin{equation}
	\hat{H}_E = \frac{g^2}{6}\left(17 \ \hat{\mathbb{I}}\otimes \hat{\mathbb{I}}
	- 9\  \hat{Z}\otimes \hat{\mathbb{I}} - 9\ \hat{\mathbb{I}}\otimes  \hat{Z}
	+  \hat{Z}\otimes  \hat{Z}\right)
	\ \ \ ,
\end{equation}
where $\hat{\mathbb{I}}$ is the identity operator.
The magnetic Hamiltonian can similarly be decomposed as
\begin{equation}
	\hat{H}_B =
	\frac{3}{g^2}\hat{\mathbb{I}}\otimes \hat{\mathbb{I}}
	-\frac{1}{2g^2}\left( \hat{X}\otimes\hat{\mathbb{I}} + \hat{\mathbb{I}}\otimes  \hat{X} + \frac{1}{2}\left( \hat{X}\otimes  \hat{X} +  \hat{Y} \otimes \hat{Y}\right)\right)
	\ \ \ .
\end{equation}
While there is a wide range of tactics being explored for the time evolution of quantum systems~\cite{berry2015simulating,low2017optimal,low2018hamiltonian,low2019hamiltonian,childs2019nearly}, the method of Trotterization~\cite{Trotter1959OnTP,doi:10.1063/1.529425} is a qubit-efficient approach introducing zero auxiliary qubits.
Focusing on this latter method, time evolution through Trotterization~\cite{Lloyd1073,zalka1998simulating,berry2007efficient,su2020nearly,wiebe2011simulating}
for a time $\Delta t$ of a general quantum wavefunction is approximated at first and second orders as
\begin{equation}
	e^{-i \Delta t \sum\limits_k \hat{H}_k}
	\ \sim \
	\prod_k e^{-i \Delta t \hat{H}_k}+ O(\Delta t^2)
	\ \sim \
	\prod_{k=N}^1 e^{-i \frac{\Delta t}{2} \hat{H}_k} \prod_{k=1}^N e^{-i \frac{\Delta t}{2} \hat{H}_k}+ O(\Delta t^3)
	\ \ \ .
\end{equation}
Neglecting terms proportional to the identity, the one-plaquette Hamiltonian can be separated into Trotterized operators,
$\hat{H} = \hat{H}_1 + \hat{H}_2$,
with
\begin{align}
	\hat{H}_1 & =  \left(\frac{17 g^2}{6} + \frac{3}{g^2} \right)\hat{\mathbb{I}} \otimes \hat{\mathbb{I}} -\frac{g^2}{6}\left(9 \  \hat{Z}\otimes \hat{\mathbb{I}} +9\  \hat{\mathbb{I}}\otimes  \hat{Z}\right)  -\frac{1}{2g^2}\left( \hat{X}\otimes\hat{\mathbb{I}} + \hat{\mathbb{I}}\otimes \hat{X} \right) \nonumber \\
	\hat{H}_2 & = \frac{g^2}{6}  \hat{Z} \otimes  \hat{Z} - \frac{1}{4g^2}\left( \hat{X}\otimes  \hat{X} +  \hat{Y} \otimes  \hat{Y}\right)
	\ \ \ .
\end{align}
The matrix exponential of the first Hamiltonian contribution can be implemented with single-qubit gates,
while that of the second Hamiltonian contribution can be implemented as
\begin{equation}
  e^{i(a \hat{X}\otimes\hat{X} + b \hat{Y}\otimes\hat{Y} + c \hat{Z}\otimes\hat{Z})} = \begin{gathered}
    \Qcircuit @R=1.em @C=1.2em {
    &\ctrl{1} & \gate{e^{i a X }} & \gate{H} & \ctrl{1} & \gate{S} & \gate{H} & \ctrl{1} & \gate{e^{i \frac{\pi}{4} X}} & \qw \\
    & \targ & \gate{e^{i c Z}} & \qw & \targ & \gate{e^{-i b Z}} & \qw & \targ & \gate{e^{-i \frac{\pi}{4} X}} & \qw
    }
  \end{gathered} \ \ \ ,
  \label{eq:XX_circuit}
\end{equation}
using the decomposition of the SU(4) Cartan subalgebra~\cite{PhysRevA.69.010301,PhysRevA.77.066301}.

\begin{figure}
\begin{overpic}[width = 0.47\textwidth,percent]{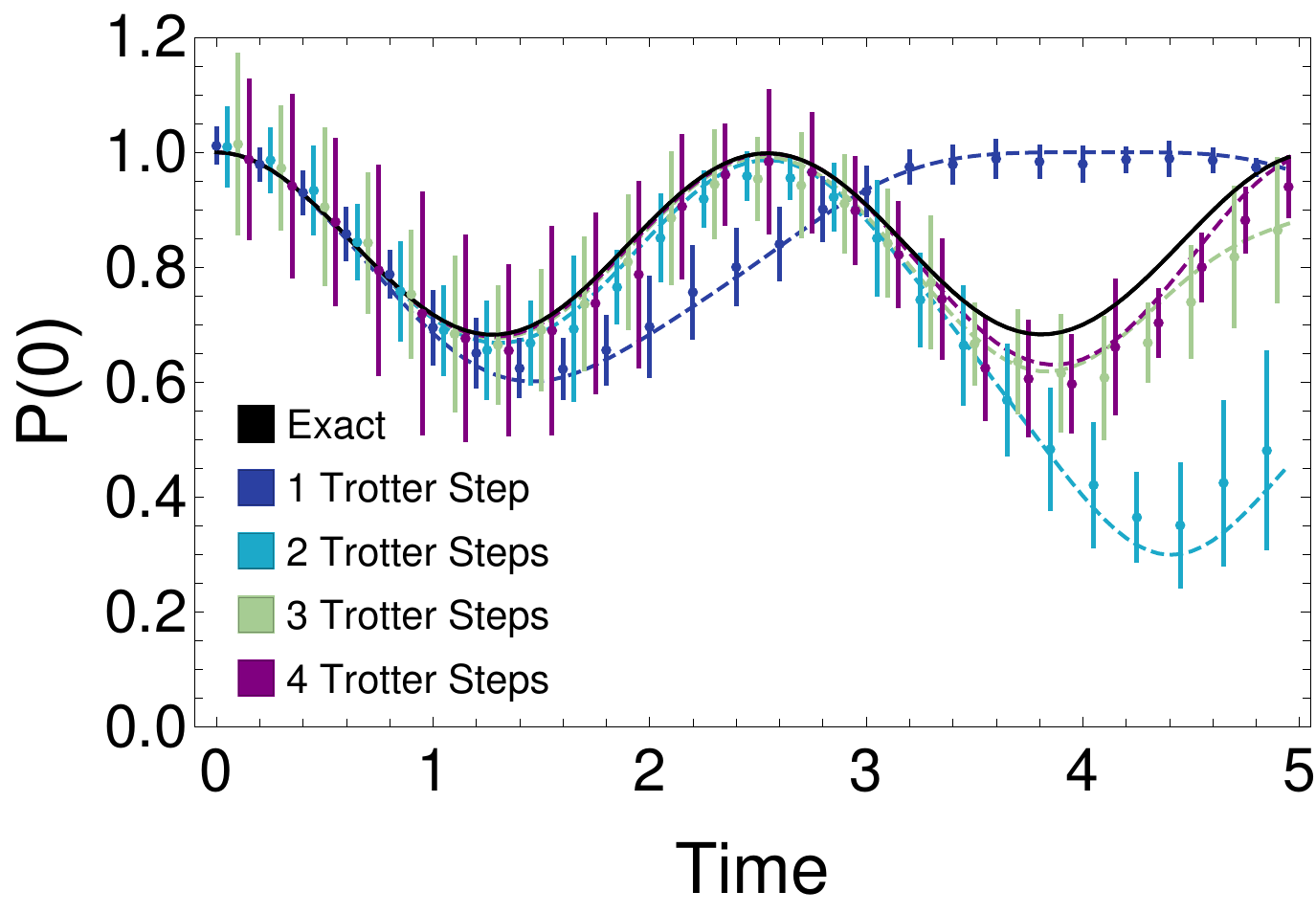}
    \put(93,60){\includegraphics[width=0.06\textwidth]{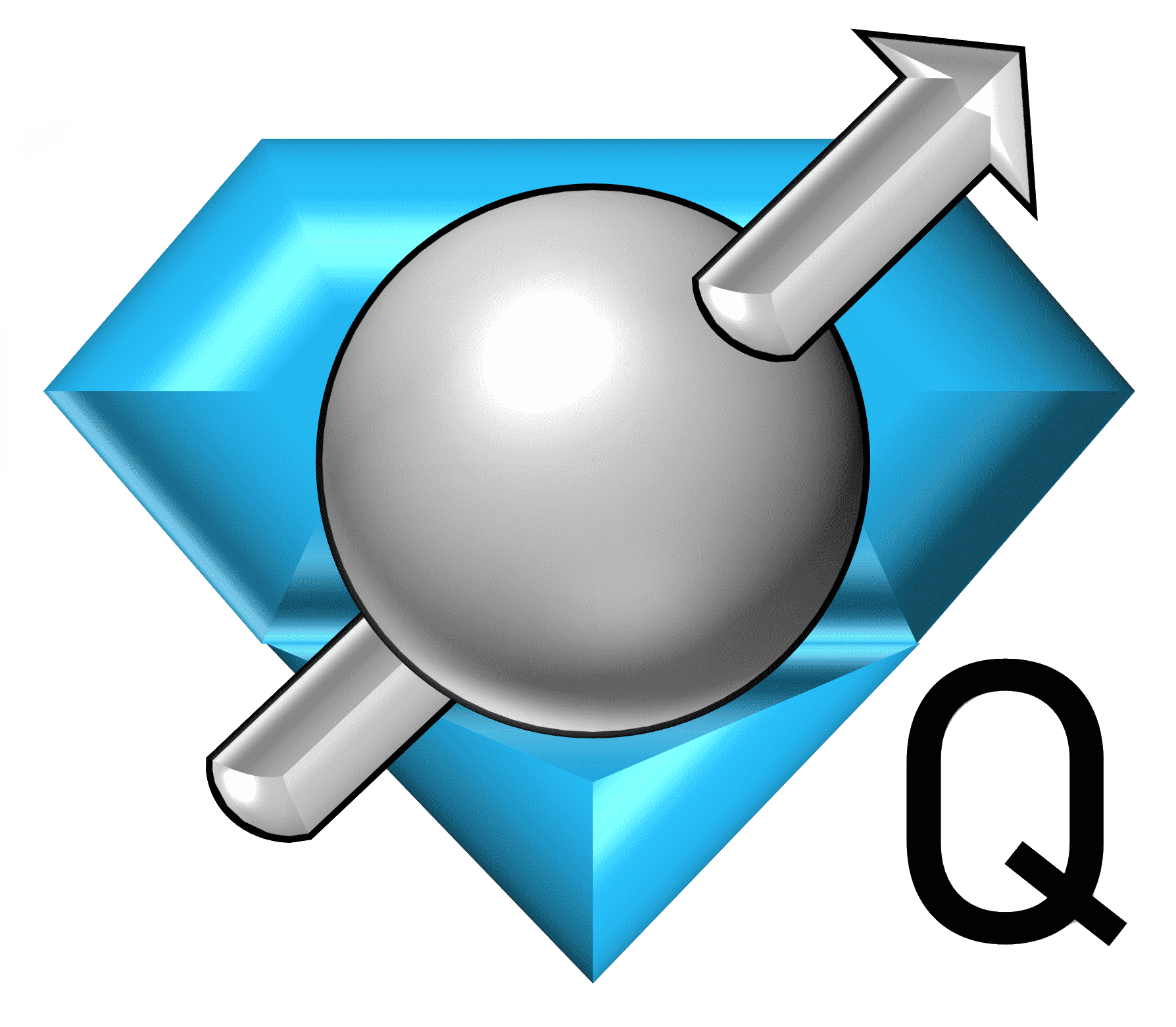}}
	\end{overpic}
\quad
\begin{overpic}[width = 0.47\textwidth,percent]{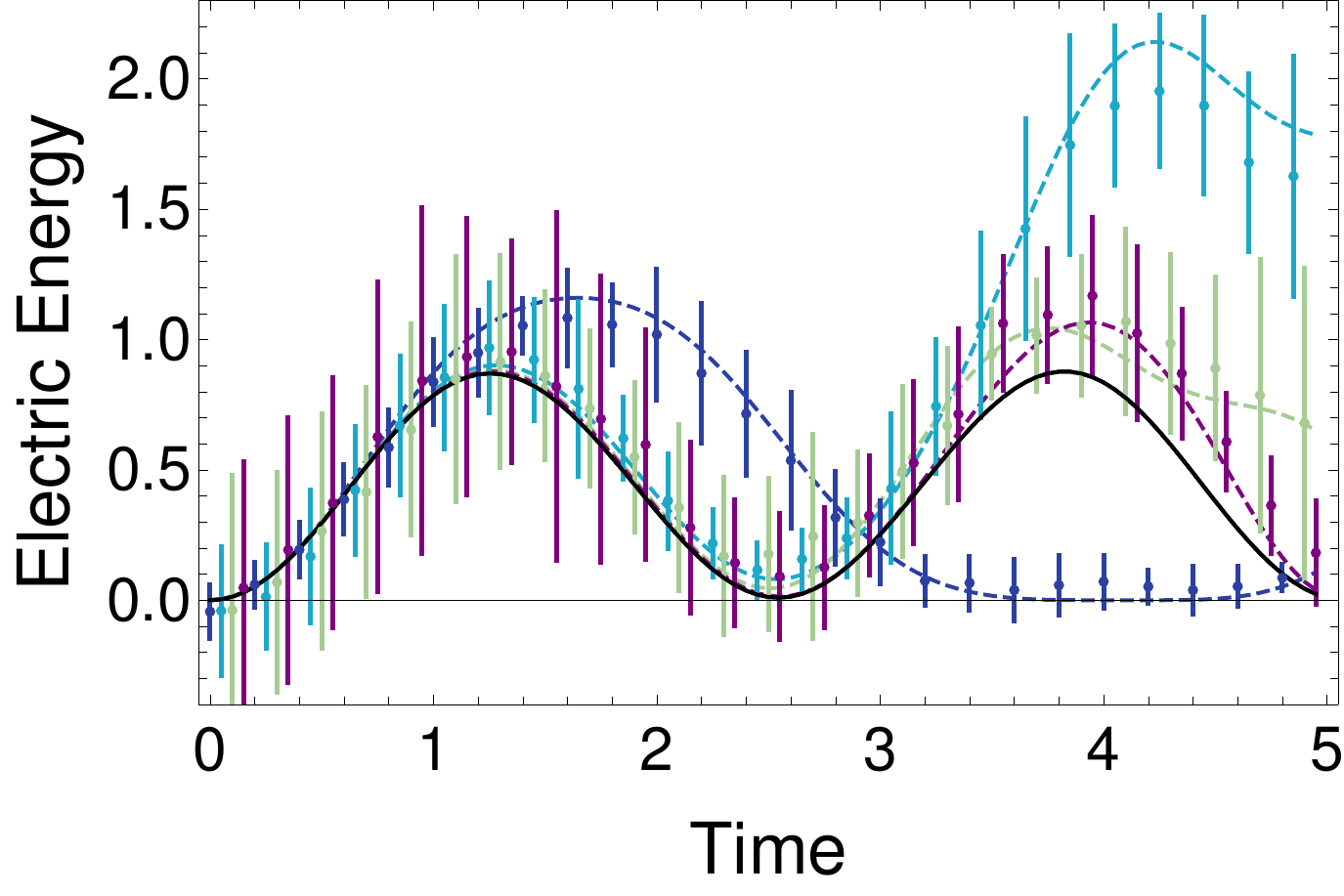}
    \put(93,60.5){\includegraphics[width=0.06\textwidth]{iconQ1_bf.png}}
	\end{overpic}
	\caption{
	The (trivial-) vacuum-to-vacuum persistence probability
	$|\langle 00 |\ \hat U(t)\ | 00\rangle|^2$ (left panel)
	and the energy in the electric field (right panel)
	of the one-plaquette system
	derived from the Hamiltonian given in Eq.~(\ref{eq:Ham1p8}) for color irreps ${\bf 1}, {\bf 3} , \overline{\bf 3} , {\bf 8}$.
	Dashed lines correspond to the exact results for 2nd-order Trotterization given in Eq.~(\ref{eq:Global8Trotter2})	with $\Delta t = t, t/2, t/3, t/4, 0$. 	Points correspond to quadratic extrapolations of results obtained from IBM's {\tt Athens} quantum processor, with systematic and statistical uncertainties combined in quadrature. }
\label{fig:one_plaq_p0_athens}
\end{figure}

The panels of Fig. \ref{fig:one_plaq_p0_athens} show the probability of a single plaquette
remaining in the trivial vacuum, $|00\rangle$, and its electric energy fluctuations for a color irrep basis truncated to
$\{{\bf 1}, {\bf 3} , \overline{\bf 3} , {\bf 8}\}$.
Up to four $2^{\text{nd}}$-order Trotter steps of the form,
\begin{equation}
	\hat{U}(\Delta  t) = e^{-i \frac{\Delta t}{2} \hat{H}_1} e^{-i \Delta t \hat{H}_2} e^{-i \frac{\Delta t}{2} \hat{H}_1} \ \ \ ,
	\label{eq:Global8Trotter2}
\end{equation}
are implemented for a coupling of $g=1$.
Beyond this number, it is found that the increased gate fidelity and coherence demands of the extended quantum circuit do not allow controlled mitigation of noise.
The dashed curves correspond to exact classical calculations of each Trotterization, with the limit of continuous time evolution shown by the solid black curve.
The data points correspond to results of the circuits discussed above implemented on IBM's {\tt Athens} quantum processor~\cite{ibmAthens},
a superconducting qubit system
in the lineage of IBM's devices using Qiskit~\cite{Qiskit}.
The connectivity of this device is linear across five superconducting qubits~\cite{ibmAthens} and
two of the middle qubits were used to store the wavefunction of the truncated SU(3) plaquette.

The largest sources of systematic uncertainty in simulating with current quantum devices are measurements and CNOT gates, with the former dominating by a factor of $\sim 3$ in basic benchmarking of the ${\tt Athens}$ device.
To account for systematic errors associated with CNOT gates, previously employed extrapolation
procedures~\cite{PhysRevX.7.021050,PhysRevLett.119.180509} have been utilized.
Mitigation of this CNOT-gate error combines the results obtained by replacing each CNOT in a circuit with an odd number, r (for r = 3, 5, 7), of CNOTs
and extrapolating to $r=0$ (as CNOT.CNOT$=\hat{\mathbb{I}}$).
Linear and quadratic extrapolations to $r=0$
in the number of CNOTs-per-circuit-CNOT were performed,
and
$\frac{1}{2} | { O}(\text{linear})- { O}(\text{quadratic}) |$
was used as an estimate of the systematic uncertainty in the extrapolation of a quantity ${O}$.
In many cases, the linear fit was of relatively poor quality and gate fidelity limited the reliable extraction of sufficient samples in $r$ to estimate the systematic uncertainty at a comparison of higher polynomials.
Hence, this comparison provides only an estimate and should not be considered a complete quantification of CNOT errors.

Measurement errors were mitigated in two ways.
The first was implementing Qiskit's {\tt measurement filter} subroutine~\cite{ibmmeaserror}
during production, which removes the leading order measurement errors by optimizing an approximate inverse of the calculated all-to-all measurement matrix.
When the error introduced by application of a single CNOT gate is small compared to those of the measurement procedure, it is viable to mitigate measurement errors through the use of auxiliary qubits by implementing a majority- or unanimous-vote for the measurement result.
In this democratic approach, each auxiliary qubit is connected as the CNOT target controlled on a qubit in the plaquette Hilbert space and provides one correlated measurement to inform post-selected voting.
After calibration, the typical single qubit measurement error rate on the {\tt Athens} processor is approximately $3\%$ and the typical CNOT error rate is approximately $0.9\%$ \cite{ibmAthens}.
As a result, the unanimous voting criterion provides an improvement that is found in some cases to be comparable to that of the measurement filter, with degradation for circuits implemented at times distant from a calibration procedure.
The initially positive results observed in this work, along with the scalability of the voting procedure, inspire future exploration of the device-dependent tuning necessary to optimize this measurement error mitigation strategy.

In addition to a choice of measurement error mitigation, the calculation shown in Fig.~\ref{fig:one_plaq_p0_athens} was implemented with both a 3- and 4-CNOT gate version of $e^{i \left(a \hat{X} \otimes \hat{X} + b \hat{Y} \otimes \hat{Y} + c \hat{Z} \otimes \hat{Z} \right)}$, the time evolution of the Cartan subalgebra.
In the absence of noise, these two implementations should give the same results.
While additional noise would be reasonably expected for the 4-CNOT calculations, temporal fluctuations in error rates of the device instead produced lower noise fluctuations for the 4-CNOT calculations.
Thus, in order to express most accurately the uncertainties associated with this calculation on quantum hardware, the four
implementations
(3-CNOT Cartan subalgebra with unanimous voting, 4-CNOT Cartan subalgebra with the measurement filter, and others) after $r$-extrapolation have been combined.
The uncertainty is a quadrature combination of the extrapolation errors and standard deviations of the four implementations.
As a result, the uncertainties presented in Fig.~\ref{fig:one_plaq_p0_athens} and throughout this manuscript are not statistical confidence intervals, but also capture the systematic errors associated with gate fidelities and temporal fluctuations of the device between calibrations that produce dominant error contributions.

\subsubsection{\texorpdfstring{ $\mathbf{6}$ Truncation }{ 6 Truncation}}
\label{sec:three_qubit_6}
\noindent
The next lowest-lying irreps beyond the $\{\mathbf{1},\mathbf{3},\mathbf{\bar{3}},\mathbf{8}\}$
to be included in the one-plaquette basis are the $\mathbf{6}$ and $\mathbf{\bar{6}}$.
With six basis states, three qubits are required with two remaining unphysical states in the Hilbert space.
As was leveraged in Ref.~\cite{Klco:2019evd}, the freedom of gauge-variant completion allows couplings and interactions within the unphysical subspace to be chosen to simplify the implementation of the Hamiltonian on the quantum device.
The one-plaquette basis can be embedded into the Hilbert space of three qubits with the encoding, $\{|0\rangle, |1\rangle, |2\rangle, |3\rangle, |4\rangle, |7\rangle\} = \{\ket{\mathbf{1}}, \ket{\mathbf{3}}, \ket{\mathbf{\bar{3}}}, \ket{\mathbf{8}}, \ket{\mathbf{6}}, \ket{\mathbf{\bar{6}}}\}$, leaving the states $\ket{101}$ and $\ket{110}$ to be unphysical.
With this mapping, the Hamiltonian can be gathered into seven terms, $\hat{H} = \sum\limits_{i=0}^7 \hat{H}_i $, with
\begin{align}
	 \hat{H}_0 &= g^2 \left( \frac{14}{3} \hat{\mathbb{I}} \otimes \hat{\mathbb{I}} \otimes \hat{\mathbb{I}} -\frac{11}{6}  \hat{Z} \otimes \hat{\mathbb{I}} \otimes \hat{\mathbb{I}}-\frac{3}{2} \hat{\mathbb{I}} \otimes \hat{\mathbb{I}} \otimes  \hat{Z}-\frac{3}{2}  \hat{Z} \otimes  \hat{Z} \otimes \hat{\mathbb{I}} + \frac{1}{6} \hat{\mathbb{I}} \otimes  \hat{Z} \otimes  \hat{Z} \right)  \ \ \ ,\nonumber \\
	 \hat{H}_1 &= -\frac{1}{4 g^2} \hat{\mathbb{I}} \otimes \hat{\mathbb{I}} \otimes  \hat{X}  \ \ \ ,\nonumber \\
	 \hat{H}_2 &= -\frac{1}{4 g^2} \left(\hat{\mathbb{I}} +  \hat{Z} \right)\otimes  \hat{X} \otimes \hat{\mathbb{I}}  \ \ \ , \nonumber \\
	 \hat{H}_3 &= -\frac{1}{4g^2} \left(\hat{\mathbb{I}} \otimes \hat{X} \otimes  \hat{X} + \hat{\mathbb{I}} \otimes  \hat{Y} \otimes  \hat{Y} \right)  \ \ \ ,\nonumber \\
	 \hat{H}_4 &= -\frac{1}{4g^2} \left( \hat{X} \otimes \hat{\mathbb{I}} \otimes  \hat{X} +  \hat{Z} \otimes \hat{\mathbb{I}}\otimes  \hat{X} \right)  \ \ \ ,\nonumber \\
	 \hat{H}_5 &= -\frac{1}{4 g^2}  \hat{Y} \otimes  \hat{Z} \otimes  \hat{Y}  \ \ \ ,\nonumber \\
	 \hat{H}_6 &= -\frac{1}{2g^2} \left(\hat{b}^\dagger \otimes \hat{b} \otimes \hat{b} + \hat{b} \otimes \hat{b}^\dagger \otimes \hat{b}^\dagger \right)  \ \ \ ,\nonumber \\
	 \hat{H}_7 &= -\frac{1}{8g^2}  \hat{X} \otimes \left(\hat{\mathbb{I}}- \hat{Z} \right) \otimes \left(\hat{\mathbb{I}}- \hat{Z} \right) \ \ \ ,
	 \label{eq:H17}
\end{align}
where $\hat{b} = ( \hat{X} + i  \hat{Y})/2$.

Middle qubits on the {\tt Athens} quantum processor were chosen to represent the state of the system,
while the two remaining qubits were used to mitigate the measurement errors of the
second and fourth qubits when employing voting protocols for measurement error mitigation.
A single application of the Trotterized time evolution operator
acted on the trivial vacuum $|000\rangle$,
and the CNOT error extrapolation procedure described in
Sec.~\ref{sec:OnePlaq8} was applied.
Due to the nearest neighbor couplings of the device, the Trotterized time evolution operator
is decomposed into $38$ CNOT gates and $37$ single qubit gates.
However, many of the CNOT gates in the circuit were required to compensate for the linear nearest neighbor coupling
of qubits;
on a device with all-to-all couplings between the qubits,
this Trotterized time evolution operator could be implemented with $20$ CNOT gates.

Unfortunately, implementation of this three-qubit calculation is found to exceed the capabilities of the {\tt Athens} architecture with controllable systematic errors.
The $r\geq3$ measurements show clear signs of coherence time saturation and the $r=1$ experiences already large deviations.  This combination results in an inability to perform an $r$ extrapolation and thus an inability to mitigate the CNOT gate errors.
It is possible that an extrapolation at fractional $r$, introducing error-exacerbating CNOT pairs at stochastically-chosen fractions of the CNOTs in the circuit as implemented in Ref.~\cite{Klco:2019evd}, could be reliably implemented, though the errors experienced already at $r = 1$ remain daunting anchors for extrapolation.

An additional interesting quantity to inform development is the survival probability in the physical subspace.
As gauge theories are commonly designed for quantum simulation by embedding locally-interacting gauge-invariant spaces within larger Hilbert spaces, maintaining symmetry subspaces to high fidelity will be an important property of future quantum devices.
These subspace fidelity demands also reside at the heart of many quantum error correction protocols with the space of logical quantum information embedded non-locally in a low-energy Hilbert space satisfying local symmetries.
In Ref.~\cite{Klco:2019evd}, a gauge invariant survival probability of approximately 40\% was calculated for a physical/unphysical Hilbert space ratio of $4/12$ at the peak of the first oscillation in the electric energy of two plaquettes in SU(2) lattice gauge theory utilizing a circuit of 6 CNOTs on the IBM ${\tt Tokyo}$ 20-qubit quantum device.
In the current application, a physical/unphysical Hilbert space ratio of $6/2$ was explored with a time evolution operator of 38 CNOTs, demonstrating a survival probability of approximately 90\% at the minimum of the first oscillation in the electric energy of the SU(3) global basis plaquette.
While the latter represents a possible improvement in survival probability per CNOT, the former, being within both the coherence time and gate-fidelity coherence time of the device, was found to allow reliable extrapolation to a survival probability of approximately 60\% and accurately capture the time evolution of electric-basis observables at the accuracy of a single Trotter step.
These observations further support the necessity of multi-dimensional optimization in the design of quantum architectures.
In the next subsection, the flexibility of the global basis is leveraged to perform a projection into the color parity symmetric space respected by the SU(3) Hamiltonian that is shown to allow reliable exploration of the $\{\mathbf{1}, \mathbf{3}, \overline{\mathbf{3}},\mathbf{8}, \mathbf{6}, \overline{\mathbf{6}}\}$ dynamics through reduction onto two qubits with no unphysical subspace.

\subsection{Color Parity Bases}
\noindent
To simulate the time evolution of the trivial electric vacuum,
only states that are connected by repeated applications of the Hamiltonian are required to
be included in the simulated basis.
Observing the structure of the plaquette operator, color parity is a symmetry of SU(3) lattice Hamiltonian.
Thus, time evolution will only couple states of positive color parity,
\begin{equation}
	\ket{R^+} = {\frac{1}{\sqrt{2}}} \left[\ \ket{R} + \ket{\bar{R}} \ \right]
	\ \ \ ,
\end{equation}
to the electric strong-coupling vacuum.
By including only states of the form of $\ket{R^+}$ in the basis, the evolution of the trivial electric
vacuum can be simulated with a reduced Hilbert space, or a higher precision calculation can be performed
using the same quantum register.

\subsubsection{\texorpdfstring{ $(\mathbf{1},\mathbf{3^+})$ }{ (1,3 plus)}}

The lowest non-trivial truncation in the color parity basis consists of the states
$\ket{\mathbf{1}}$ and $\ket{\mathbf{3^+}} = \frac{1}{\sqrt{2}} (\ket{\mathbf{3}} + \ket{\mathbf{\bar{3}}})$.
This can be mapped onto a single qubit with the basis choice
$\ket{0} = \ket{\mathbf{1}}$ and $\ket{1} = \ket{\mathbf{3^+}}$,
and the Hamiltonian becomes,
\begin{equation}
	\hat{H} = \left(\frac{4}{3}g^2 + \frac{11}{4g^2} \right) \hat{\mathbb{I}} + \left(-\frac{4}{3}g^2 + \frac{1}{4g^2} \right)\hat{Z} - \frac{1}{\sqrt{2} g^2}  \hat{X} \ \ \ .
	\label{eq:Color3Ham}
\end{equation}
With the availability of arbitrary single qubit gates, the associated time evolution can be implemented with a single unitary rotation without Trotterization.
\begin{figure}
\begin{overpic}[width = 0.4\textwidth,percent]{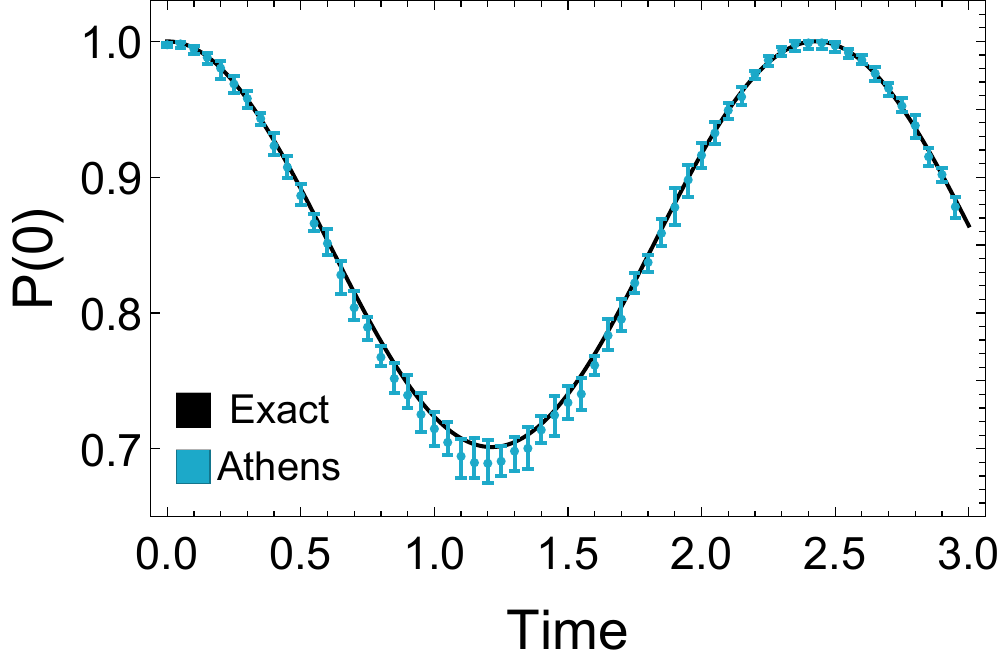} \put(91,58){\includegraphics[width=0.06\textwidth]{iconQ1_bf.png}}
\end{overpic}
\qquad
\begin{overpic}[width = 0.4\textwidth,percent]{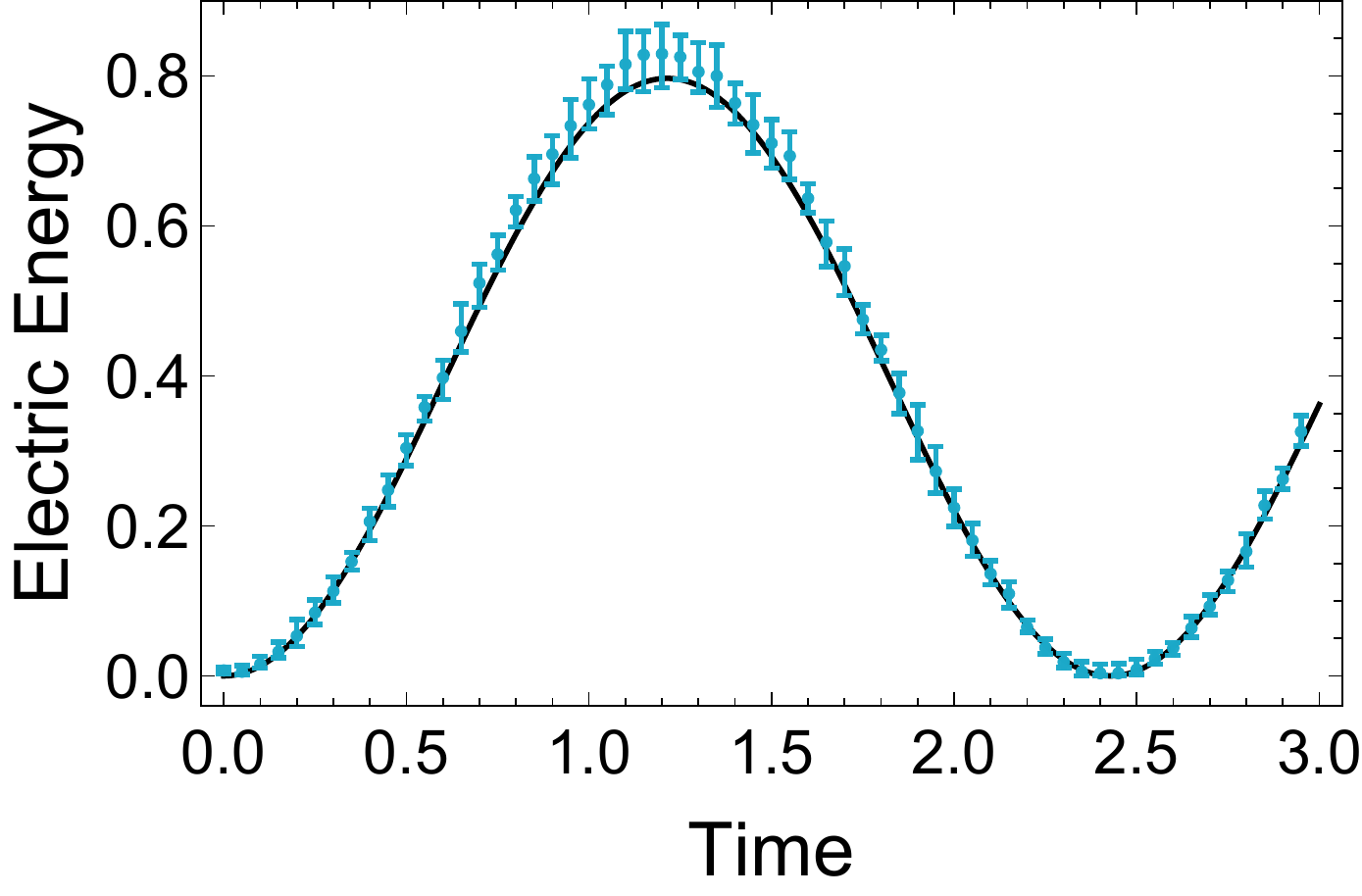} \put(91,58){\includegraphics[width=0.06\textwidth]{iconQ1_bf.png}}
	\end{overpic}
	\caption{
The (trivial-) vacuum-to-vacuum persistence probability
	$|\langle 00 |\ \hat U(t)\ | 00\rangle|^2$ (left panel)
	and the energy in the electric field (right panel)
	of the one-plaquette system in the color parity basis truncated at
	$\mathbf{3^+}$ evolved according to the Hamiltonian in Eq.~\eqref{eq:Color3Ham}.
	The points correspond to the average value and the maximal extent of 68\% binomial confidence intervals across four implementations on IBM's {\tt Athens} quantum processor, expressing both statistical and systematic uncertainties.
}
	\label{fig:one_plaq_color_3}
\end{figure}
Figure~\ref{fig:one_plaq_color_3} shows the results of performing the $\mathbf{3}^+$ time evolution on the {\tt Athens} quantum processor
with $g=1$ beginning in the electric vacuum.
The combinations of measurement error and CNOT extrapolations have been employed as
described in Sec.~\ref{sec:OnePlaq8}.
As this calculation does not require CNOT gates, there is significantly less noise relative to the associated two-qubit calculation performed in the global basis without color parity projection.

\subsubsection{\texorpdfstring{ $(\mathbf{1},\mathbf{3^+},\mathbf{8},\mathbf{6^+})$ }{ (1,3 plus,8,6 plus) }}
With two qubits,
the color parity basis can be extended to include the $\ket{\mathbf{8}}$ and $\ket{\mathbf{6^+}}$ states in a basis encoding of the form $\{\ket{00}, \ket{01}, \ket{10}, \ket{11} \} = \{\ket{\mathbf{1}}, \ket{\mathbf{3^+}}, \ket{\mathbf{6^+}}, \ket{\mathbf{8}} \}$,
leading to a Hamiltonian of the form,
\begin{multline}
	\hat{H} =  g^2 \left(\frac{23}{6} \hat{\mathbb{I}} \otimes \hat{\mathbb{I}} - \frac{5}{2} \hat{Z} \otimes \hat{\mathbb{I}} - \frac{1}{2}
	\hat{\mathbb{I}} \otimes \hat{Z} - \frac{5}{6} \hat{Z} \otimes \hat{Z} \right)  \\
	- \frac{1}{2g^2} \left( \sqrt{2}\ \hat{\mathbb{I}}\otimes \hat{X}+\sqrt{2} \ \hat{X}\otimes \left(\frac{\hat{\mathbb{I}}-\hat{Z}}{2}\right) + \frac{1}{2} \hat{X}\otimes \hat{X} + \frac{1}{2} \hat{Y}\otimes \hat{Y} + \frac{1}{4}\left(\hat{\mathbb{I}}+\hat{Z}\right)\otimes\left(\hat{\mathbb{I}}-\hat{Z}\right) -6 \ \hat{\mathbb{I}} \otimes \hat{\mathbb{I}}  \right) \ \ \ .
	\label{eq:ColorParity6Ham}
\end{multline}
To Trotterize,
the single qubit terms can be grouped together, and the Cartan subalgebra $ (\hat{X}\otimes \hat{X}, \hat{Y}\otimes \hat{Y}, \hat{Z}\otimes \hat{Z})$ can be implemented as in the case of the global basis $\mathbf{8}$-truncation above.
When including the $\mathbf{6}$ irrep in the color parity projected basis, there is an additional $ \hat{X}\otimes \hat{Z}$ term in the Hamiltonian whose Trotterized time evolution can be decomposed with the following circuit,
\begin{equation}
  e^{i \alpha \hat{X} \otimes \hat{Z}} = \begin{gathered}
    \Qcircuit @R=1em @C=1em {
    & \gate{H} & \targ & \gate{e^{i \alpha Z}} & \targ & \gate{H} \\
    & \qw & \ctrl{-1} & \qw & \ctrl{-1} & \qw \\
    } \ \ \ .
  \end{gathered}
\end{equation}
Explicitly, the first order Trotterized time evolution operator is chosen to be implemented as
$\hat U(t) = e^{-i \hat{H}_3 t} e^{-i \hat{H}_2 t} e^{-i \hat{H}_1 t}$ with
\begin{align}
\hat{H}_1 & = \left(\frac{23 g^2}{6}+\frac{23}{8 g^2}\right) \hat{\mathbb{I}} \otimes \hat{\mathbb{I}} - \left(\frac{5g^2}{2} + \frac{1}{8g^2} \right)  \hat{Z} \otimes \hat{\mathbb{I}} -\left(\frac{g^2}{2} - \frac{1}{8g^2} \right) \hat{\mathbb{I}} \otimes \hat{Z} - \frac{1}{2\sqrt{2} g^2} \hat{X}\otimes\hat{\mathbb{I}} - \frac{1}{\sqrt{2} g^2} \hat{\mathbb{I}}\otimes \hat{X} \nonumber \ \ \ , \\
	\hat{H}_2 & = \frac{1}{2 \sqrt{2} g^2} \hat{X}\otimes \hat{Z} \nonumber \ \ \ , \\
	\hat{H}_3 & = -\frac{1}{4g^2}  \hat{X}\otimes \hat{X} -\frac{1}{4g^2} \hat{Y}\otimes \hat{Y} -  \left( \frac{5}{6}g^2 - \frac{1}{8g^2} \right) \hat{Z}\otimes \hat{Z} \ \ \ .
\end{align}
Implementing this Trotterized time evolution employs $10$ single qubit gates and $6$ CNOT gates. The accuracy of the simulation can be improved by using a second order Trotterized time evolution operator of the form,
\begin{equation}
	\hat{U}(t) = e^{-i \hat{H}_1 \frac{t}{2}} e^{-i \hat{H}_2 \frac{t}{2}} e^{-i \hat{H}_3 t} e^{-i \hat{H}_2 \frac{t}{2}} e^{-i \hat{H}_1 \frac{t}{2}} \ \ \ ,
	\label{eq:Color6Trotter2}
\end{equation}
which employs $15$ single qubit gates and $8$ CNOT gates.
Because the second order Trotter step requires fewer gates than performing two first order Trotter steps,  higher order Trotterizations may be capable of improving the calculation.
\begin{figure}
\begin{overpic}[width= 0.48\textwidth,percent]{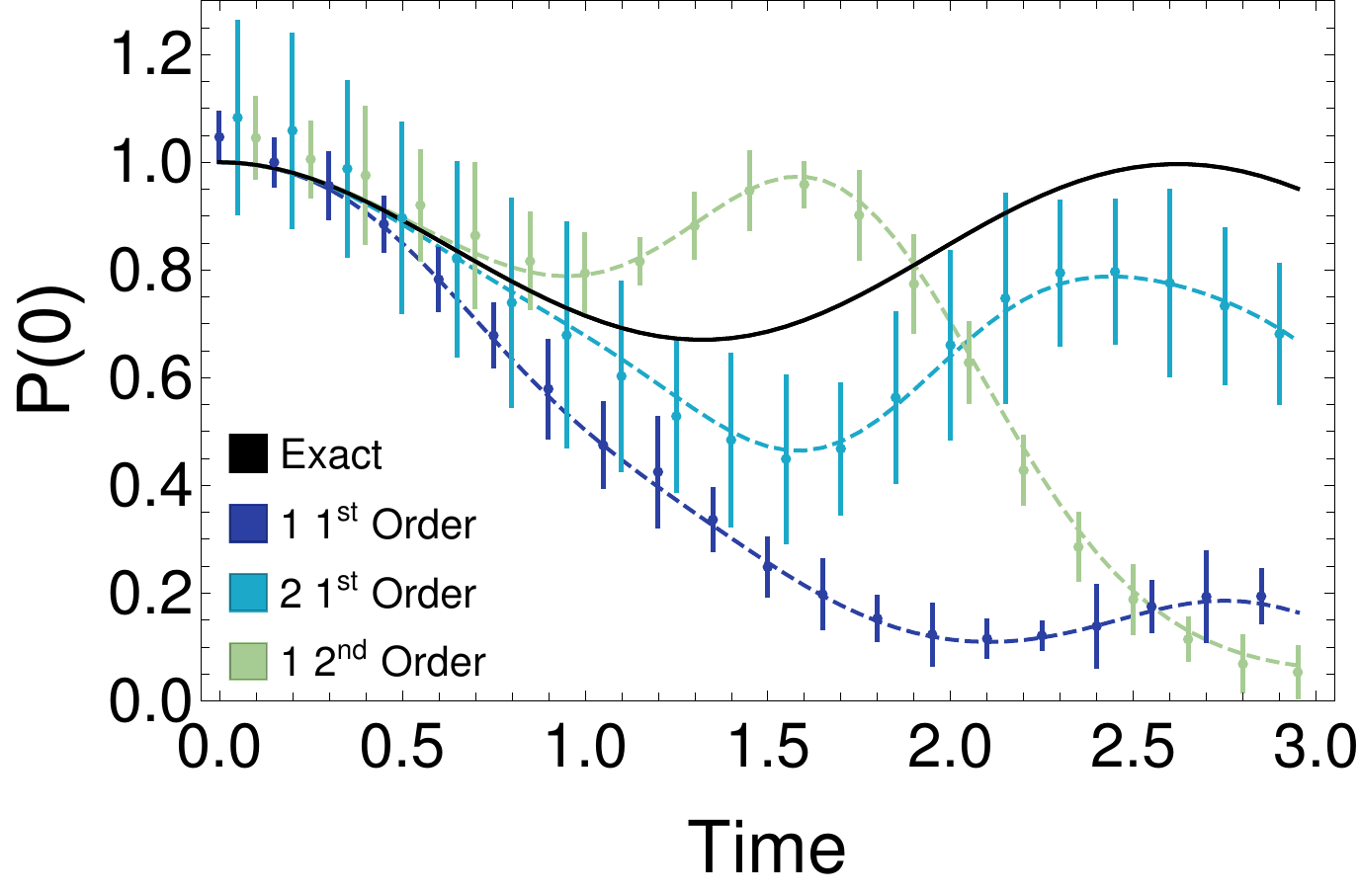} \put(91.5,59.5){\includegraphics[width=0.06\textwidth]{iconQ1_bf.png}}
	\end{overpic}
\quad
\begin{overpic}[width= 0.47\textwidth,percent]{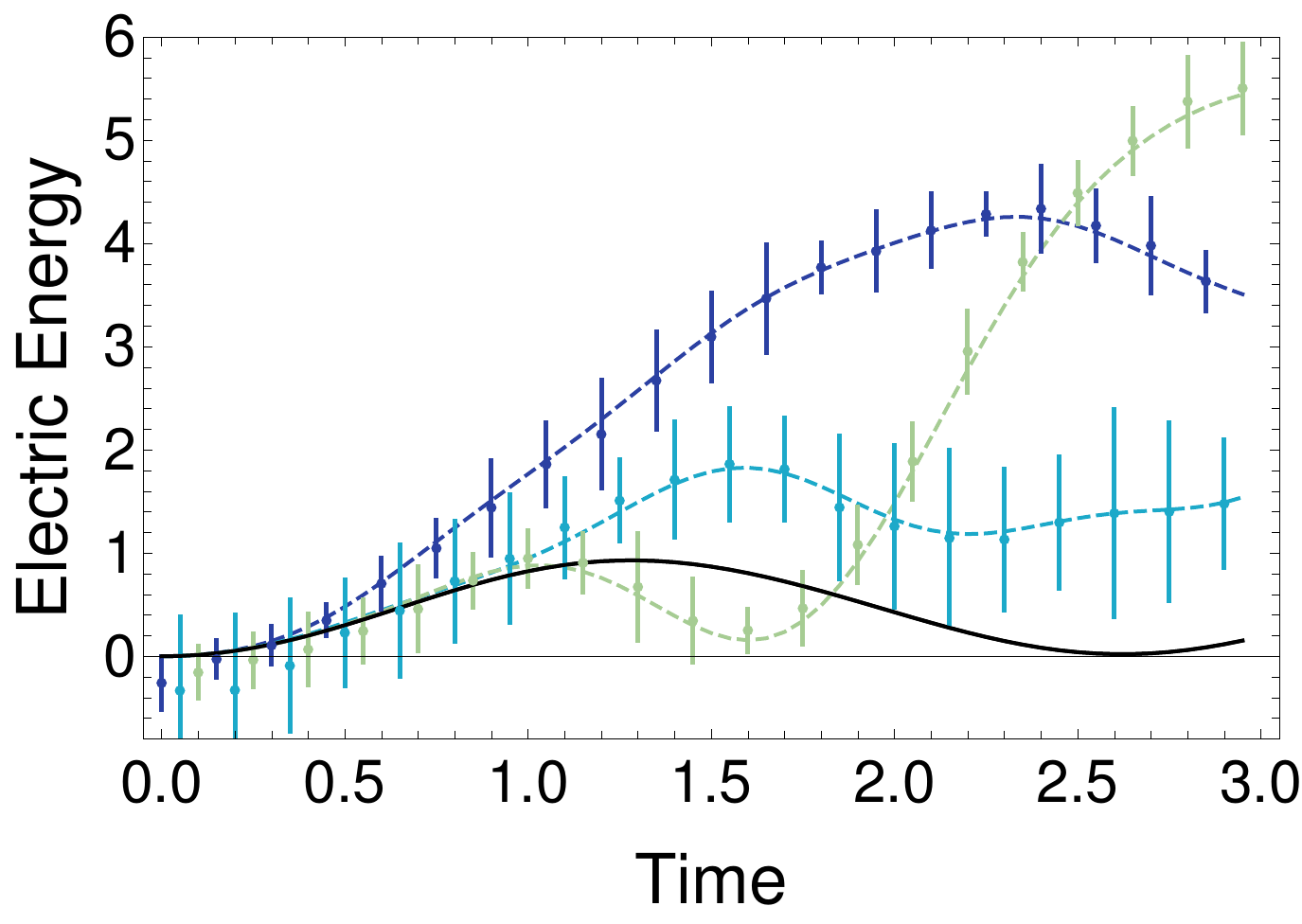} \put(65,61){\includegraphics[width=0.06\textwidth]{iconQ1_bf.png}}
	\end{overpic}
	\caption{
	The (trivial-) vacuum-to-vacuum persistence probability $|\langle 00 |\ \hat U(t)\ | 00\rangle|^2$ (left panel) and the energy in the electric field (right panel) of the one-plaquette system derived from the Hamiltonian given in Eq.~(\ref{eq:ColorParity6Ham}) in the color parity basis truncated at $\mathbf{6^+}$. The different curves correspond to the exact results for $1^{\text{st}}$-order Trotterizations with $\Delta t = t, t/2$ and for a single step of $2^{\text{nd}}$-order Trotterization with $\Delta t = t$. 	
The points correspond to quadratic extrapolations of results obtained from IBM's {\tt Athens} quantum processor, with systematic and statistical uncertainties combined in quadrature.
}
	\label{fig:one_plaq_color_6_trot1}
\end{figure}

Implementation of these two forms of Trotterization are presented in Fig.~\ref{fig:one_plaq_color_6_trot1}.
Due to the extra $ \hat{X} \otimes \hat{Z}$ term in the Hamiltonian, the time evolution circuit requires more CNOT gates and the additional gates in the circuit causes noise to dominate the calculation earlier than when using the global basis truncated at ${\bf 8}$.
Adding a majority choice mitigation of the measurement error in addition to the measurement filter does not significantly improve the results, indicating that the breakdown in the calculation is due to noise in the circuit rather than the measurement process.
As a result, only two steps of the $1^{\text{st}}$-order and one step of the $2^{\text{nd}}$ order Trotterizations were found to be reliable compared to the four $2^{\text{nd}}$ order steps achievable for the $\mathbf{8}$-truncated global basis.
This two qubit calculation of the $\mathbf{6}$-truncated single plaquette contains all of the states coupled to the vacuum present in the three qubit global basis calculation above~\ref{sec:three_qubit_6}. However, due to the more efficient mapping, reliable time evolution is achievable with the added color parity projection.

\FloatBarrier

\subsection{Rudimentary Single Plaquette Benchmarks}
\label{subsec:benchmarks}
\noindent
While the performance of many-body dynamics cannot be captured in a single metric, benchmarks for scientific application can provide useful information toward the simulation of dynamical lattice gauge theories as quantum devices develop.
Near term benchmarks are likely too rudimentary to survive into the production era, but may provide helpful guidance in the near term NISQ era.
Given the exacerbated noise experienced by many quantum devices at local extrema of time evolved observables, a succinct, yet meaningful, quantity expressing device performance in this area is the extrema of the electric energy fluctuations for a single plaquette of an SU(3) lattice.
Analogously to the array of hardware calibrations used to capture the high-dimensional optimization affecting the quality of operations and measurements across devices, the peaks and troughs in the fluctuation of the electric energy focuses on one informative aspect of the time evolution.

\begin{figure} \begin{overpic}[width=0.45\textwidth,percent]{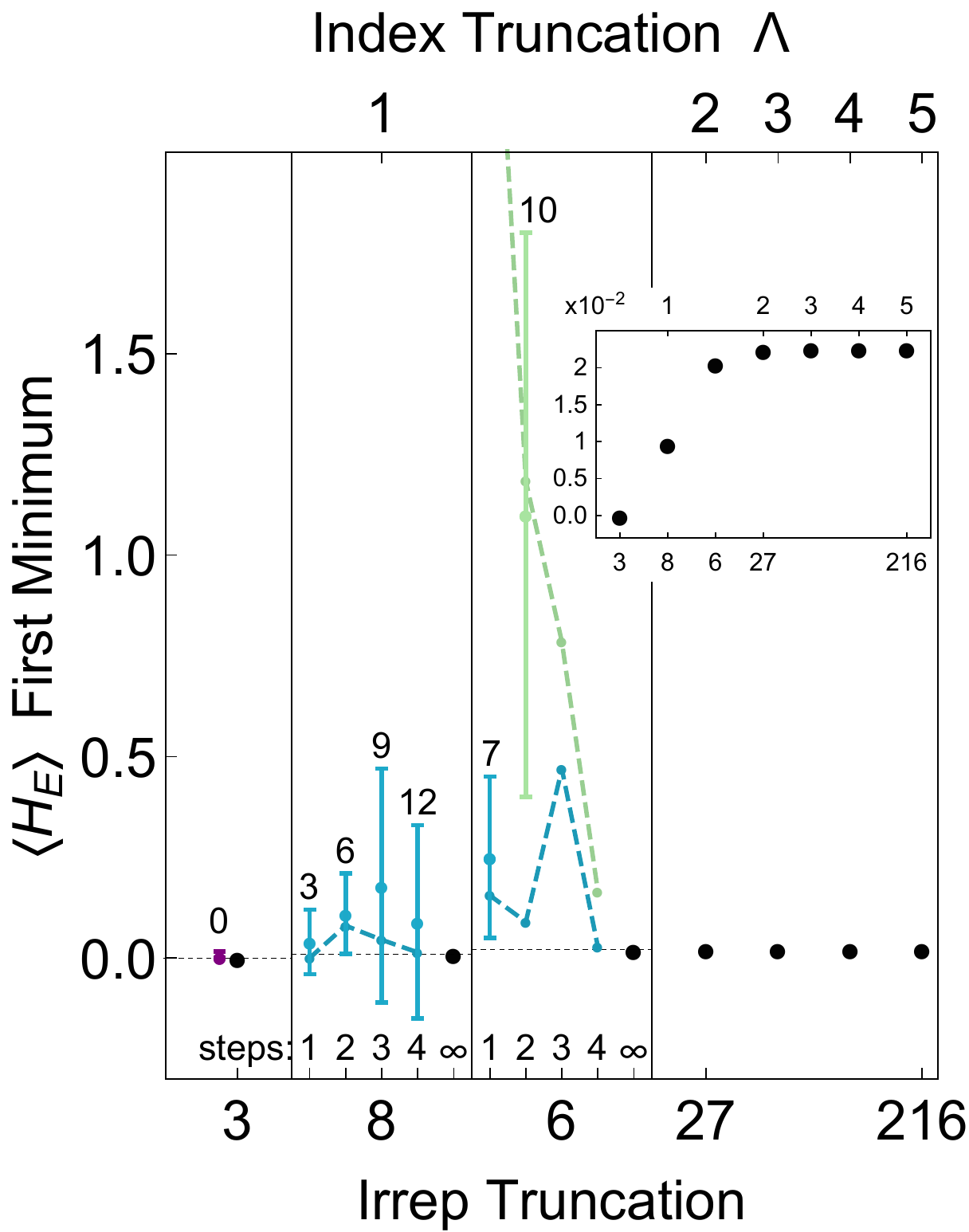}
    \put(8,82.5){\includegraphics[width=0.06\textwidth]{iconQ1_bf.png}}
	\end{overpic}
\begin{overpic}[width=0.43\textwidth,percent]{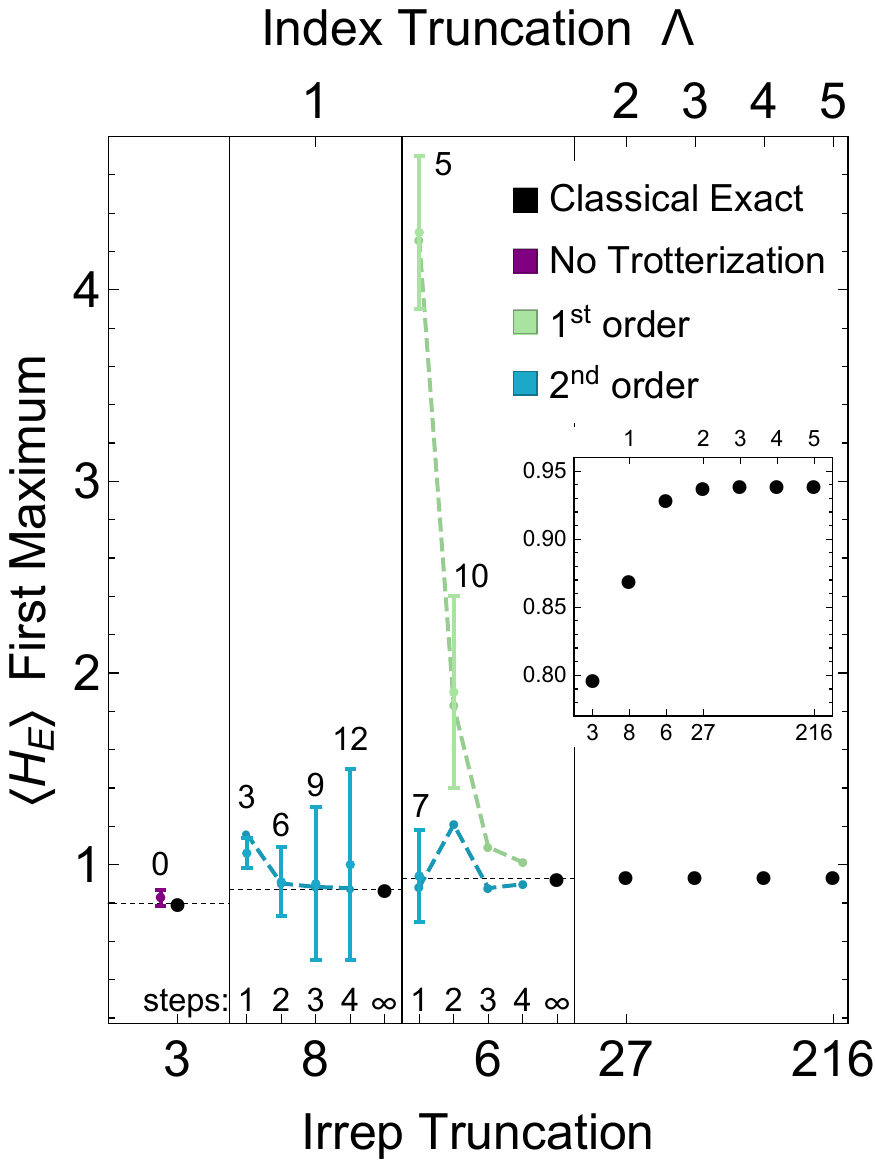} \put(4,83){\includegraphics[width=0.06\textwidth]{iconQ1_bf.png}}
	\end{overpic}
	\caption{Calculation of the first local minimum and maximum in the temporal fluctuations of the electric energy for one irrep-truncated SU(3) plaquette at unit coupling on the {\tt Athens} quantum processor beginning from the strong coupling vacuum state.  Expected theoretical approximations from Trotterization are shown as dashed lines.  Insets provide the exact values at each truncation with smaller vertical axis scale for perspective.  The irrep truncation of $\mathbf{8}$ was calculated in the global basis while the $\mathbf{3}$ and $\mathbf{6}$ were evaluated in color parity projected global bases.  Numbers associated with each point indicate the relevant number of CNOT gates.}
	\label{fig:minmaxbenchmark}
\end{figure}

The left (right) panel of Fig.~\ref{fig:minmaxbenchmark} shows the values of the first minimum (maximum) in the electric energy
time evolution performed on the {\tt Athens} quantum processor.  Numerical values for the data appearing in Fig.~\ref{fig:minmaxbenchmark} can be found in Tables~ \ref{table:ElectricMin} and~ \ref{table:ElectricMax} of Appendix~\ref{app:benchmarks}.
Being a single-qubit calculation and thus requiring no Trotterization, the data of the $\mathbf{3}^+$ truncation is well controlled as seen in Fig.~\ref{fig:one_plaq_color_3}.
Increasing the irreps included in the basis moving to the right also increases the number of qubits necessary to capture the Hilbert space.
Within a sub-panel at fixed irrep truncation, the number of Trotter steps used to time evolve to the local extrema is increased moving to the right, increasing the gate fidelity and coherence demanded of the quantum device.
Thus, from left to right each panel of Fig.~\ref{fig:minmaxbenchmark} trades the impact of theoretical approximations for the impact of hardware noise.
Ideally, this type of figure will show \emph{windows}, in which Trotter errors are reduced and hardware noise has not yet overwhelmed the calculation, for an array of irrep truncations in order to inform a systematic extrapolation to the limit of infinite truncation.

As discussed in Subsection~\ref{subsec:oneplaqexpsup}, the exploration of decreasing coupling increases the required Hilbert space of the ground state wavefunction in the basis of electric multiplets.
While the convergence of observables at fixed $g$ is subsequently exponential in the irrep truncation, finite computational resources, both in quantity and quality,
will limit the parameter regime that can be controllably explored with extrapolation to infinite $\Lambda$ irrep cut off.
This relationship has been  visually translated in
Fig.~\ref{fig:minmaxbenchmark}
to the presence of \emph{windows}, with the smallest $g$
reliably accessible being that for which a set of windows relevant for extrapolation are achievable.
In this preliminary exploration, reliable extraction of
results with increasing
gauge field truncation was  limited to $\Lambda_p=1$ with the $\{\mathbf{1}, \mathbf{3}, \overline{\mathbf{3}}, \mathbf{8}\}$ basis, where Fig.~\ref{fig:minmaxbenchmark} shows a crossing between the regime dominated by Trotter errors to the regime dominated by hardware noise with little extended intermediate window.
For this reason, a coupling of $g = 1$ was chosen for presentation, expressing the limited Hilbert space delocalization that can be accurately captured.
However, the implemented quantum circuits experience only modified rotation angles for different couplings.
For this reason, it is expected that uncertainties associated with the implementation of alternate $g$ values will be commensurate with those presented, though the angle dependence of current hardware performance supports a more thorough exploration.

\FloatBarrier

\subsection{Single Plaquette Operator Scalability}
\label{subsec:oneplaqscaling}
While exploring Pauli decompositions of operators are important first steps for lattice gauge theory time evolution on quantum architectures, construction of relevant operators can quickly become treacherous as the color space truncation is raised.
This was experienced in the quantum simulation of SU(2) lattice gauge theory and is seen to arise also in SU(3).
For this reason, exploring alternate compilation protocols for the representation of plaquette operators amenably for hardware implementation is of vital significance.
In this subsection, an approach based on a decomposition into two-level unitaries will be presented for the single plaquette global basis and similar methods will be used to describe the local plaquette operator for extended lattices in Section~\ref{sec:localbasis}.

As discussed in Subsection~\ref{subsec:embeddings}, the further splitting of the local irrep basis into two registers per link representing the fundamental and antifundamental indices is likely to be practically advantageous in requiring only nearest neighbor connectivity within the two Hilbert spaces representing the link as shown at the right of Fig.~\ref{fig:connectivitydiagram}.
Because the Hilbert space of the single plaquette lattice satisfying the local Gauss's law is structurally similar to that of one (unconstrained) link of a larger lattice, the one plaquette system can be represented in a global $|p,q\rangle$ basis with $p$ and $q$ digitized in a binary encoding on two separate qubit registers.
With this encoding, the operators capturing the $p$ and $q$ index values are diagonal,
\begin{equation}
	\hat{p} = \sum_{k=0}^{n-1} 2^{k} \frac{\mathbb{I} - \hat{Z}_{p,k} }{2}
\ \ \ \ , \ \ \ \
	\hat{q} = \sum_{k=0}^{n-1} 2^{k} \frac{\mathbb{I} - \hat{Z}_{q,k} }{2}
	\ \ \ \ ,
\end{equation}
where the subscripts on $ Z$ specify which register and qubit the operator acts upon.
With this representation, the electric term in the Hamiltonian becomes a sum over one- and two-qubit Pauli-Z operators as the Casimir of Eq.~\eqref{eq:casimir} is quadratic.

Because the connectivity of both the $p$- and $q$-registers is nearest neighbor, it is convenient to consider
the (non-unitary) operator, $\hat{B}_n$ that maps $\ket{p}$ to $\ket{p-1}$ and annihilates $|p\rangle = |0\rangle$
when $p$ is stored in the binary encoding with $n$ qubits.
The operator $\hat{B}_n$ can be constructed recursively according to
\begin{equation}
	\hat{B}_n = \mathbb{I}\otimes \hat{B}_{n-1} + \hat{b} \bigotimes_{k=1}^{n-1} \hat{b}^\dagger
	\ =\  \sum_{k=0}^{n-1} \left(\bigotimes_{j=1}^{n-k-1} \mathbb{I}\right)\otimes \hat{b} \otimes \left(\bigotimes_{i=1}^k \hat{b}^\dagger \right)
	\ \ \ ,
\end{equation}
expressed in $n$ contributing terms with an increasing number of $\hat{b}^\dagger$ operators in the Hilbert spaces.
The operator  $\hat b$ has been defined previously, just below Eq.~(\ref{eq:H17}).
Using $n$ qubits to represent each of the $p$ and $q$ registers,
the plaquette operator can be written as
\begin{equation}
	\hat{\square} = \hat{B}_n^\dagger \otimes \mathbb{I} + \hat{B}_n \otimes \hat{B}_n^\dagger + \mathbb{I}\otimes \hat{B}_n
	\ \ \ ,
\label{eq:oneplaqbox}
\end{equation}
where the first and second Hilbert spaces represent the $p$ and $q$ registers, respectively.
Note again that the single plaquette lattice shares a Hilbert space structure with that of the single (unconstrained) link of an extended lattice, leading to later connections between Eq.~\eqref{eq:oneplaqbox} and the local link operator.
As an explicit example, for a $|p\rangle$ or $|q\rangle$ register of four qubits, this non-unitary lowering operator would be decomposed as
\begin{equation}
  \hat{B}_4 = \mathbb{I}_8 \otimes \hat{b}  + \mathbb{I}_4 \otimes \hat{b} \otimes \hat{b}^\dagger  + \mathbb{I}_2\otimes \hat{b} \otimes \hat{b}^\dagger \otimes \hat{b}^\dagger +  \hat{b} \otimes \hat{b}^\dagger \otimes \hat{b}^\dagger \otimes \hat{b}^\dagger \ \ \ .
  \label{eq:B4example}
\end{equation}
If decomposed in the Pauli basis, the $\hat{B}_n$ operator would span $\sum\limits_{k = 1}^{n} 2^k = 2^{n+1}-2$ unique terms.
Decomposing the plaquette operator in the Pauli basis subsequently demands $2^{n+2}(2^n-1)$ unique operators, or $\mathcal{O}(\Lambda_p \Lambda_q)$ as the index truncation is exponential in the number of qubits per register.
The Pauli decomposition of the Hermitian combination $\hat{\square} + \hat{\square}^\dagger$ relevant for Trotterized time evolution presents a factor of two simplification to $2^{n+1}(2^n-1)$ unique operators, due to the Hermiticity of the Pauli matrices.
Ignoring simplifications for the implementation of terms sharing a basis, each of these exponentially numerous terms can be implemented with $\mathcal{O}(2n)$ or $\mathcal{O}(\log(\Lambda_p \Lambda_q))$ CNOT gates~\cite{Nielsen:2011:QCQ:1972505} resulting in the total number of gates to implement the plaquette operator time evolution scaling exponentially with $n$ or polynomially in $\Lambda_{p,q}$.

While the exponential suppression of wavefunction amplitudes discussed in Subsection~\ref{subsec:oneplaqexpsup} may allow this na\"ive approach to be practically fruitful, it is possible to restructure the plaquette time evolution decomposition for improved scaling.
Again allowing $p$ and $q$ to be represented by two quantum registers with $n$ qubits each,
the Hermitian combination of plaquette operators present in the magnetic Hamiltonian can be written as a sum of the form
\begin{equation}
	\hat{\square} + \hat{\square}^\dagger  = \left(\hat{B}_n + \hat{B}_n^\dagger\right) \otimes \mathbb{I} + \mathbb{I} \otimes \left( \hat{B}_n + \hat{B}_n^\dagger \right) + \hat{B}_n \otimes \hat{B}_n^\dagger + \hat{B}_n^\dagger \otimes \hat{B}_n = \sum_{j = 1}^{n^2 + 2n} \hat{O}_j + \hat{O}_j^\dagger \ \ \ ,
\label{eq:boxpboxdagB}
\end{equation}
where $\hat{O}_j$ is a tensor product operator of elements of the form $\{\mathbb{I}, \hat{b}, \hat{b}^\dagger\}^{\otimes 2n}$.
For example, the $\hat{O}_j$ operators associated with the first term may consist of $k$ identity operators, one $\hat{b}$ operator, $(n-k-1)$ conjugate $\hat{b}$ operators, and $n$ identity operators for the $q$-register.
Each of these Hermitian operators, $\hat{O} + \hat{O}^\dagger$, can be identified as a two-level unitary between computational basis states dictated by the Hilbert space locations of the $\hat{b}$ and $\hat{b}^\dagger$ operators.
As discussed in Ref.~\cite{Nielsen:2011:QCQ:1972505} (4.5.2), the time evolution associated with such operators can be implemented by first transforming the basis through a Gray code~\cite{gray1953}, implementing a controlled single-qubit rotation, and then inverting the Gray code transformation.
For example, implementing the time evolution associated with the last term of Eq.~\eqref{eq:B4example} in the $|p\rangle$ register, contributing to the first term of Eq.~\eqref{eq:boxpboxdagB}, connects the states $|1000\rangle$ and $|0111\rangle$ in the $p$-register for every state in the $q$-register.
The time evolution according to this term can then be implemented with the following circuit acting on the $p$-register
\begin{equation}
  e^{-i \alpha (\hat{b}\hat{b}^\dagger \hat{b}^\dagger \hat{b}^\dagger+\hat{b}^\dagger \hat{b}\hat{b}\hat{b})} =
  \begin{gathered}
  \Qcircuit @R = 0.8em @C=0.8em {
  & \targ & \ctrl{1} & \ctrl{1} & \ctrl{1} & \ctrl{1} & \ctrl{1} & \targ & \qw \\
  & \ctrl{-1} & \ctrl{1} & \ctrl{1} & \gate{e^{-i \alpha \hat{X}}} & \ctrl{1} & \ctrl{1} & \ctrl{-1} & \qw \\
  & \ctrl{-1} & \ctrl{1} & \targ & \ctrlo{-1} & \targ & \ctrl{1} & \ctrl{-1} & \qw \\
  & \ctrl{-1} & \targ & \ctrlo{-1} & \ctrlo{-1} & \ctrlo{-1} & \targ & \ctrl{-1} & \qw
  }
  \end{gathered} =
  \begin{gathered}
    \Qcircuit @R=0.8em @C = 0.8em {
    & \gate{\hat{X}} & \qw & \qw & \ctrl{1} & \gate{e^{-i \alpha \hat{X}}} & \ctrl{1} & \qw & \qw & \gate{\hat{X}} & \qw \\
    & \qw & \qw & \ctrl{1} & \targ & \ctrlo{-1} & \targ & \ctrl{1} & \qw & \qw & \qw \\
    & \qw & \ctrl{1} & \targ & \qw & \ctrlo{-1} & \qw & \targ & \ctrl{1} & \qw & \qw \\
    & \qw & \targ & \qw & \qw  & \ctrlo{-1} & \qw & \qw & \targ & \qw & \qw
    }
  \end{gathered}
   \ \ \ ,
\end{equation}
where $\alpha$ is both time and coupling dependent.
The second equality emphasizes the  simplifications often available in the practical application of Gray code techniques, though the generic implementation of the first equality will be momentarily convenient for the scaling discussion.
At the left, a Gray code is implemented in the order $0111 \rightarrow 1111 \rightarrow 1110 \rightarrow 1100 \rightarrow 1000$ through the first three multi-controlled-$\hat{X}$ operators and the location of the central controlled rotation operator.
With maximal Hamming distance of $2n$ between two bit strings spanning the $\{ |p\rangle , |q\rangle \}$ basis, the maximal depth of any Trotter contribution of the form $e^{-i \alpha (\hat{O}_j + \hat{O}_j^\dagger)}$ will be $2(2n)+1$ in terms of these maximally-$(2n-1)$-controlled $\hat{X}$ and rotation operators.
Decomposing each of these $C^kNOT$ operators into Toffoli, CNOT, and single qubit gates can be done with $\mathcal{O}(k)$ gates without the introduction of any auxiliary qubits~\cite{gidneycNnotBlog,Nielsen:2011:QCQ:1972505}.
In practice, however, the desire to avoid the demand of exponentially precise rotation gates may inspire the use of a single auxiliary qubit.
This efficiency of multi-controlled CNOT operators translates directly to an equivalent efficiency in the decomposition of the general multi-controlled SU(2) rotation at the center of this circuit~\cite{Barenco:1995na}.
With this identification of two-level unitaries treated through Gray code manipulation, the total number of gates to implement the plaquette operator time evolution is found to scale polynomially with $n = \log_2 (\Lambda_p+1)$, the number of qubits used to represent the tensor indices of irreducible representations composing the basis.

While the qubit decompositions of these two-level contributions to the plaquette time evolution are straightforward and technically efficient, later discussions in Section~\ref{sec:localbasis} of the local plaquette operator
will maintain this level of abstraction due to an expectation that currently-developing qudit frameworks may provide advantageous hardware-specific approaches for the implementation of two-level rotations.

\section{Global Basis: Two Plaquettes}
\label{sec:globalbasis2p}
\noindent
The results obtained for a single plaquette, detailed in Section~\ref{sec:singleplaquette}, have provided insights into the convergence of the color-representation truncation in a simple system.
Some other features that are required for QCD calculations at scale only first appear in more complex systems, such as
a two-plaquette system subject to spatial periodic boundary conditions (PBCs).
The SU(3) two-plaquette systems are similar to those of the SU(2) system explored in Ref.~\cite{Klco:2019evd},
but with additional structure associated with the SU(3) gauge group defining the link variables.
\begin{figure}
  \includegraphics[width = 0.4\textwidth]{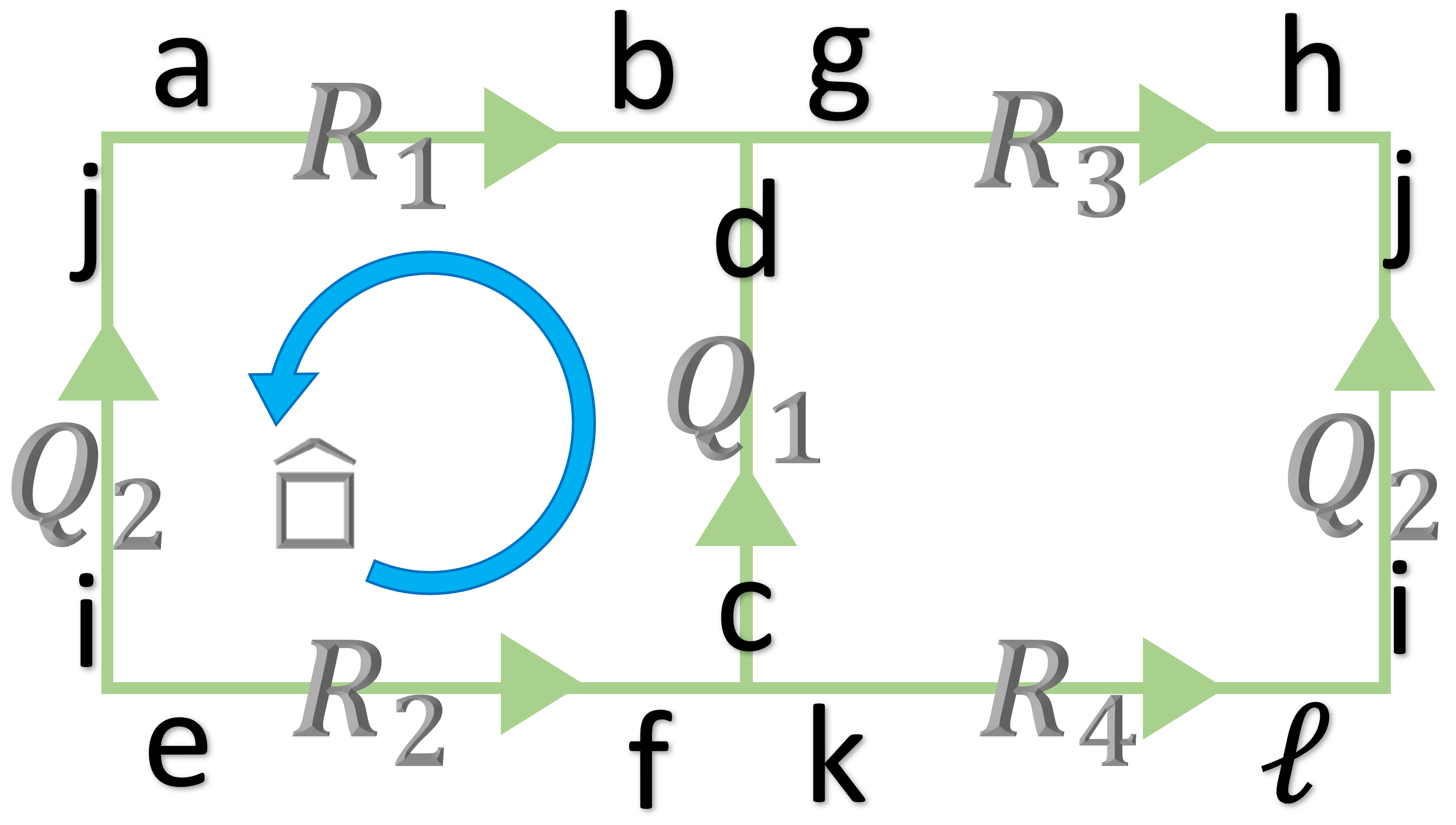}
  \caption{
Two plaquettes with periodic boundary conditions
 and an arrow convention amenable to infinite extension in the two-dimensional plane.
 Indices local to each end of each link characterize states in SU(3) e.g., the color isospin and hypercharge indices. }
\label{fig:twoplaqlabels}
\end{figure}
Figure~\ref{fig:twoplaqlabels}
shows the layout of the two plaquettes, along with our conventions that
define the action of  plaquette operators.
With an eye toward an efficient mapping of the problem onto quantum hardware,
we employ the techniques used in
Refs.~\cite{Banuls:2017ena,Klco:2019evd} to \enquote{integrate over} the gauge
group at each lattice site.
Local gauge invariance of the theory is used to eliminate redundancies associated with the local orientations
in color space, allowing the vertex amplitudes to be defined completely by
the dimensionality of irreducible representations of the intersecting links.
This process reduces the dimensionality of the Hilbert space and the associated
resources required for quantum simulation
compared with previous algorithms,
for example, Ref.~\cite{Byrnes:2005qx}.

Similar to the methods employed for the one-plaquette system, Gauss's law can be explicitly satisfied in the global wavefunctions by construction of the basis states.
Using the dimensionality of the color irrep of each link, as shown in Fig.~\ref{fig:twoplaqlabels},
the  basis states for the two-plaquette system are written as
$| \chi ({\bf R_1} , {\bf Q_1} , {\bf R_2} , {\bf R_3} , {\bf Q_2} , {\bf R_4}) \rangle$.
The gauge invariant lattice wavefunction for this two-plaquette system, as discussed in greater generality in Appendix~\ref{app:plaquetteMEs}, is
\begin{multline}
 | \chi ({\bf R_1} , {\bf Q_1} , {\bf R_2} , {\bf R_3} , {\bf Q_2} , {\bf R_4}) \rangle
 =
  \frac{1}{\dim(\mathbf{Q}_1) \dim (\mathbf{Q}_2)} \sum_{all} |\mathbf{R}_1, a, b\rangle |\mathbf{Q}_1, c, d\rangle | \mathbf{R}_2, e, f\rangle |\mathbf{R}_3, g, h\rangle |\mathbf{Q}_2, i, j\rangle  |\mathbf{R}_4, k, \ell\rangle \\
  \langle \mathbf{R}_3, h, \bar{\mathbf{R}}_1, a | \bar{\mathbf{Q}}_2, j\rangle_{\Gamma_{312}} \
  \langle \mathbf{R}_1, b, \bar{\mathbf{R}}_3, g| \bar{\mathbf{Q}}_1, d\rangle_{\Gamma_{131}} \\   \langle \mathbf{R}_4, \ell, \bar{\mathbf{R}}_2, e | \mathbf{Q}_2, i\rangle_{\Gamma_{422}} \
\langle \mathbf{R}_2, f, \bar{\mathbf{R}}_4, k | \mathbf{Q}_1, c\rangle_{\Gamma_{241}}
  \ \ \ ,
  \label{eq:twoplaqbasis}
\end{multline}
where $|\mathbf{R}, a, b\rangle$ is  a link-state in the electric basis
and $\langle \mathbf{R}_i, f, {\mathbf{R}_j}, k | \mathbf{Q}_k, c\rangle_{\Gamma_{ijk}}$ are SU(3) CG coefficients.

The global wavefunctions of the two-plaquette system are formed from combinations of these basis states, consistent with the global symmetries of the system such as: color-parity symmetry resulting from the sum of $\Box+\Box^\dagger$ in the Hamiltonian,
e.g., $ \{ {\bf R}_i , {\bf Q}_i  \} \leftrightarrow \{ \overline{\bf R}_i , \overline{\bf Q}_i  \}$,
translation invariance, and reflection symmetry.
These symmetries lead to a natural block-diagonalization of the Hamiltonian in these projected bases.
Quantum numbers may be assigned to the states in each block,
$\pm 1$ for each of the symmetries in the case of two-plaquettes.
In this section, we consider a global basis in which dynamical quantum states are mapped to symmetry-projected configurations of the full two-plaquette lattice.  Two related local truncations in color space are used to explore the convergence of both local and global truncations.

\subsection{\texorpdfstring{Two-Plaquette: $\{ {\bf 1}, {\bf 3} , \overline{\bf 3} \}$ Local Truncation}{Two-Plaquette: \{1, 3, 3\} Local Truncation}}
\label{subsec:133}
\noindent
In limiting the local link basis to color irreps $\{ {\bf 1}, {\bf 3} , \overline{\bf 3} \}$ for the two-plaquette system
without constraints and symmetries, there are $3^6$ independent basis states.
Imposing Gauss's law at each vertex reduces this number down to 27.  Further restricting to global singlet states, as is the strong coupling vacuum and preserved by the Hamiltonian, the dynamical Hilbert space becomes 9 dimensional, which decomposes into sectors of dimensions (4, 2, 2, 1) under the discrete symmetries of color parity and spatial translation.
Focusing on the sector that contains the trivial vacuum, the basis states  in the $++$ sector are,
\begin{eqnarray}
| \psi^{({\bf 1} {\bf 3}\overline{\bf 3};++)}_1 \rangle & = &
| \chi ({\bf 1} , {\bf 1} , {\bf 1} , {\bf 1} , {\bf 1} , {\bf 1}) \rangle
\nonumber\\
| \psi^{({\bf 1} {\bf 3}\overline{\bf 3};++)}_2 \rangle & = &
\frac{1}{2}\left[\
| \chi ({\bf 3} , \overline{\bf 3} , \overline{\bf 3}  , {\bf 1} , {\bf 3} , {\bf 1}) \rangle
+
| \chi (\overline{\bf 3} , {\bf 3} , {\bf 3}  , {\bf 1} , \overline{\bf 3} , {\bf 1}) \rangle
+
| \chi ({\bf 1} , {\bf 3} , {\bf 1}  , {\bf 3} , \overline{\bf 3} , \overline{\bf 3} ) \rangle
+
| \chi ({\bf 1} , \overline{\bf 3} , {\bf 1}  , \overline{\bf 3} , {\bf 3} , {\bf 3} ) \rangle
\ \right]
\nonumber\\
| \psi^{({\bf 1} {\bf 3}\overline{\bf 3};++)}_3 \rangle & = &
\frac{1}{\sqrt{2}}\left[\
| \chi ({\bf 3} , {\bf 1} , \overline{\bf 3}  , {\bf 3} , {\bf 1} , \overline{\bf 3} ) \rangle
+
| \chi (\overline{\bf 3} , {\bf 1} , {\bf 3}  , \overline{\bf 3} , {\bf 1} , {\bf 3} ) \rangle
\ \right]
\nonumber\\
| \psi^{({\bf 1} {\bf 3}\overline{\bf 3};++)}_4 \rangle & = &
\frac{1}{\sqrt{2}}\left[\
| \chi ({\bf 3} , {\bf 3} , \overline{\bf 3}  , \overline{\bf 3} , \overline{\bf 3} , {\bf 3} ) \rangle
+
| \chi (\overline{\bf 3} , \overline{\bf 3} , {\bf 3}  , {\bf 3} , {\bf 3} , \overline{\bf 3} ) \rangle
\ \right]
\ \ \ ,
\label{eq:133basispp}
\end{eqnarray}
where the superscript "$++$" denotes the transformation properties under color parity inversion and spatial translation, respectively.
The wavefunctions in the other sectors are
\begin{eqnarray}
| \psi^{({\bf 1} {\bf 3}\overline{\bf 3};-+)}_2 \rangle & = &
\frac{1}{2}\left[\
| \chi ({\bf 3} , \overline{\bf 3} , \overline{\bf 3}  , {\bf 1} , {\bf 3} , {\bf 1}) \rangle
-
|\chi (\overline{\bf 3} , {\bf 3} , {\bf 3}  , {\bf 1} , \overline{\bf 3} , {\bf 1}) \rangle
+
| \chi ({\bf 1} , {\bf 3} , {\bf 1}  , {\bf 3} , \overline{\bf 3} , \overline{\bf 3} ) \rangle
-
| \chi ({\bf 1} , \overline{\bf 3} , {\bf 1}  , \overline{\bf 3} , {\bf 3} , {\bf 3} ) \rangle
\ \right]
\nonumber\\
| \psi^{({\bf 1} {\bf 3}\overline{\bf 3};-+)}_3 \rangle & = &
\frac{1}{\sqrt{2}}\left[\
| \chi ({\bf 3} , {\bf 1} , \overline{\bf 3}  , {\bf 3} , {\bf 1} , \overline{\bf 3} ) \rangle
-
| \chi (\overline{\bf 3} , {\bf 1} , {\bf 3}  , \overline{\bf 3} , {\bf 1} , {\bf 3} ) \rangle
\ \right]
\ \ \ ,
\label{eq:133basismp}
\end{eqnarray}
in the $-+$ sector,
\begin{eqnarray}
| \psi^{({\bf 1} {\bf 3}\overline{\bf 3};+-)}_2 \rangle & = &
\frac{1}{2}\left[\
| \chi ({\bf 3} , \overline{\bf 3} , \overline{\bf 3}  , {\bf 1} , {\bf 3} , {\bf 1}) \rangle
+
|\chi (\overline{\bf 3} , {\bf 3} , {\bf 3}  , {\bf 1} , \overline{\bf 3} , {\bf 1}) \rangle
-
| \chi ({\bf 1} , {\bf 3} , {\bf 1}  , {\bf 3} , \overline{\bf 3} , \overline{\bf 3} ) \rangle
-
| \chi ({\bf 1} , \overline{\bf 3} , {\bf 1}  , \overline{\bf 3} , {\bf 3} , {\bf 3} ) \rangle
\ \right]
\ \ \ ,
\label{eq:133basispm}
\end{eqnarray}
in the $+-$ sector, and
\begin{eqnarray}
| \psi^{({\bf 1} {\bf 3}\overline{\bf 3};--)}_2 \rangle & = &
\frac{1}{2}\left[\
| \chi ({\bf 3} , \overline{\bf 3} , \overline{\bf 3}  , {\bf 1} , {\bf 3} , {\bf 1}) \rangle
-
| \chi (\overline{\bf 3} , {\bf 3} , {\bf 3}  , {\bf 1} , \overline{\bf 3} , {\bf 1}) \rangle
-
| \chi ({\bf 1} , {\bf 3} , {\bf 1}  , {\bf 3} , \overline{\bf 3} , \overline{\bf 3} ) \rangle
+
| \chi ({\bf 1} , \overline{\bf 3} , {\bf 1}  , \overline{\bf 3} , {\bf 3} , {\bf 3} ) \rangle
\ \right]
\nonumber\\
| \psi^{({\bf 1} {\bf 3}\overline{\bf 3};--)}_4 \rangle & = &
\frac{1}{\sqrt{2}}\left[\
| \chi ({\bf 3} , {\bf 3} , \overline{\bf 3}  , \overline{\bf 3} , \overline{\bf 3} , {\bf 3} ) \rangle
-
| \chi (\overline{\bf 3} , \overline{\bf 3} , {\bf 3}  , {\bf 3} , {\bf 3} , \overline{\bf 3} ) \rangle
\ \right]
\ \ \ ,
\label{eq:133basismm}
\end{eqnarray}
in the $--$ sector.

\begin{figure}
\begin{overpic}[width = 0.4\textwidth,percent]{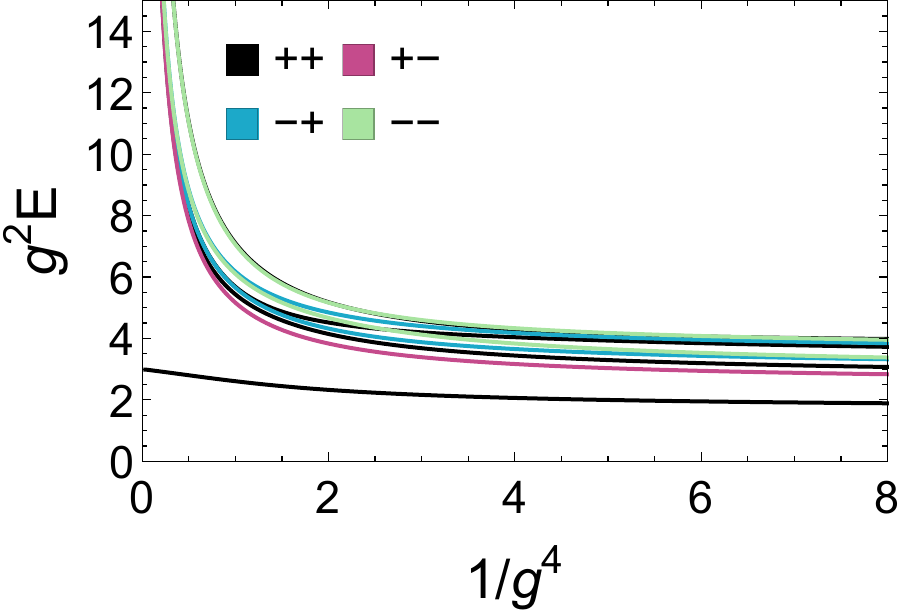}
    \put(90,59.5){\includegraphics[width=0.06\textwidth]{iconC1_bf.png}}
\end{overpic}
\begin{overpic}[width = 0.42\textwidth,percent]{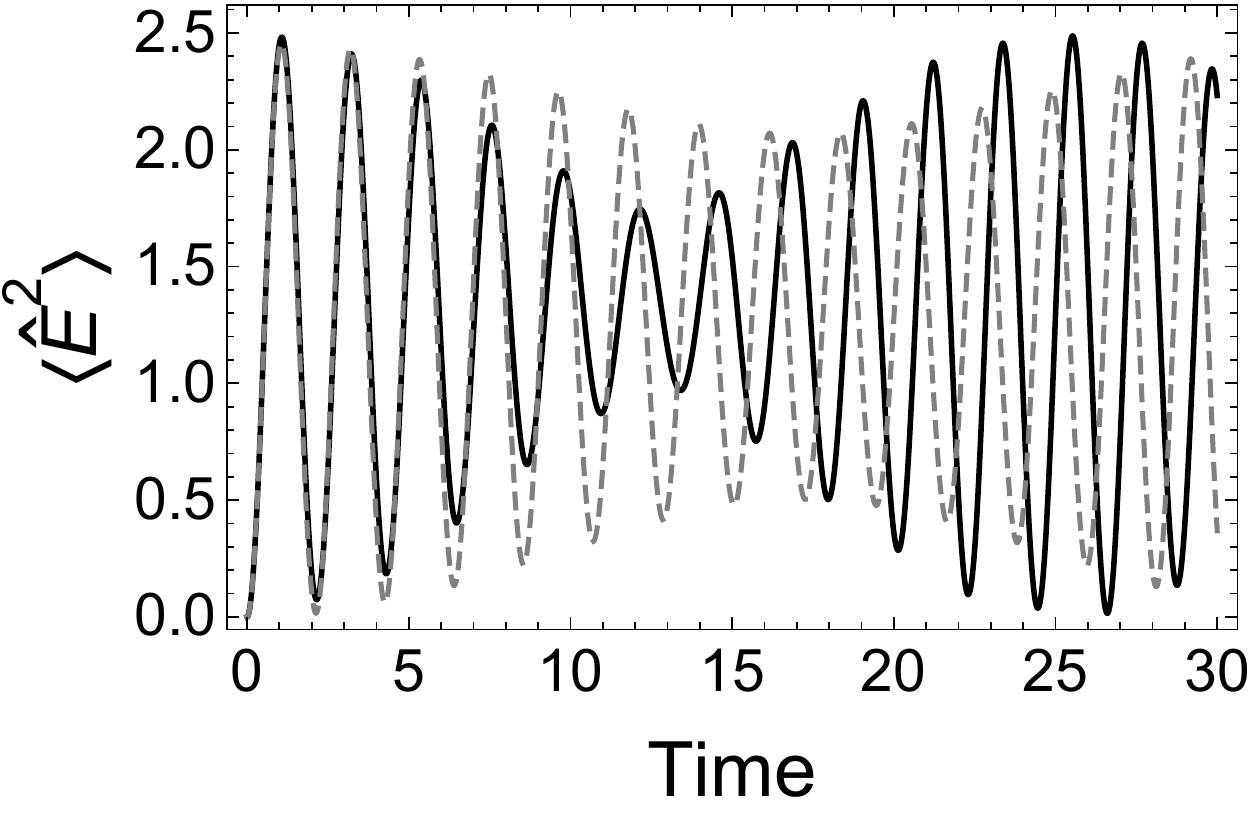}
    \put(90,57){\includegraphics[width=0.06\textwidth]{iconC1_bf.png}}
\end{overpic}
  \caption{ (left panel)
The  energy eigenvalues of the two-plaquette system as a function of coupling.
The vertical axis shows the scaled energy eigenvalues, $g^2 E_i$ versus $1/g^4$,
from each of the sectors,
$\hat H^{({\bf 1} {\bf 3}\overline{\bf 3};++)} $ (black),
$\hat H^{({\bf 1} {\bf 3}\overline{\bf 3};-+)} $ (blue),
$\hat H^{({\bf 1} {\bf 3}\overline{\bf 3};+-)} $ (pink),
and
$\hat H^{({\bf 1} {\bf 3}\overline{\bf 3};--)} $ (green),
given in Eqs.~(\ref{eq:HppLam3}) and (\ref{eq:HpmHmpHmmLam3}).
(right panel) Time evolution of $\sum\limits_a | {\bf E}^a |^2$ in the
$++$ two-plaquette system (with PBCs) locally truncated to $\{ {\bf 1}, {\bf 3} , \overline{\bf 3} \}$ and
globally truncated to basis Casimir's of $\frac{16}{3}$ (dashed gray curve) and of $8$ (solid black curve) for $g=1$.
The system is initially in the trivial vacuum.
}
\label{fig:133energies}
\end{figure}

By a direct calculation of the Hamiltonian matrix elements, both the Casimir  and  plaquette operators, we find Hamiltonian matrices of the following form in the $++$ sector,
\begin{eqnarray}
\hat H^{({\bf 1} {\bf 3}\overline{\bf 3};++)} & = &
\frac{g^2}{2}\
\left(
\begin{array}{cccc}
0 & 0 & 0 & 0   \\
0 & \frac{16}{3} & 0 & 0  \\
0 &0 & \frac{16}{3} & 0  \\
0 &0 &0 & 8
\end{array}
\right)
\ +\
\frac{1}{ 2 g^2}\
\left(
\begin{array}{cccc}
6 & -2 & 0 & 0   \\
-2 & 5 & -\frac{\sqrt{2}}{ 9} &  -\frac{\sqrt{2}}{ 3}   \\
0 & -\frac{\sqrt{2}}{ 9} & 6   & -\frac{2}{ 3}    \\
0 & -\frac{\sqrt{2}}{ 3} &  -\frac{2}{ 3} &   6
 \end{array}
\right)
\ \ \ ,
\label{eq:HppLam3}
\end{eqnarray}
and in the other sectors
\begin{eqnarray}
\hat H^{({\bf 1} {\bf 3}\overline{\bf 3};-+)} & = &
\frac{g^2}{ 2}\
\left(
\begin{array}{cc}
\frac{16}{ 3} & 0   \\
0 & \frac{16}{ 3}   \\
\end{array}
\right)
\ +\
\frac{1}{2 g^2}\
\left(
\begin{array}{cccc}
7 & -\frac{\sqrt{2}}{ 9} \\
-\frac{\sqrt{2}}{ 9} & 6
 \end{array}
\right)
\ \ \ ,
\nonumber\\
\hat H^{({\bf 1} {\bf 3}\overline{\bf 3};+-)} & = &
\frac{g^2}{ 2}\ \frac{16}{3} \ +\  \frac{1}{  g^2}\ \frac{5}{ 2}
\ \ \ ,
\nonumber\\
\hat H^{({\bf 1} {\bf 3}\overline{\bf 3};--)} & = &
\frac{g^2}{ 2}\
\left(
\begin{array}{cc}
\frac{16}{ 3} & 0   \\
0 & 8   \\
\end{array}
\right)
\ +\
 \frac{1}{  2 g^2}\
\left(
\begin{array}{cccc}
7 & -\frac{\sqrt{2}}{ 3} \\
-\frac{\sqrt{2}}{ 3} & 6
 \end{array}
\right)
\ \ \ .
\label{eq:HpmHmpHmmLam3}
\end{eqnarray}
The eigenvalues of these sectors are shown in the left panel of Fig.~\ref{fig:133energies} as a function of the coupling, $g$.
The axes have been re-scaled, according to their behavior in the strong and weak coupling limits,
to be $g^2 E$ vs $1/g^4$.
At the left of this panel resides the strong coupling limit where the electric contributions to the Hamiltonian dominate and the ground state is well separated.
At the right of this panel resides the weak coupling limit where the magnetic contributions to the Hamiltonian dominate and the ground state remains gapped below excitations.
For demonstration purposes, $g=1$ is chosen in what follows,
however the behavior as a function of coupling should be noted when considering the lattice continuum limit, where $g a \rightarrow 0$.

While the present basis is highly truncated, and we will explore a larger basis in subsequent sections,
it is orienting to see the effect of a global truncation.
In the right panel of Fig.~\ref{fig:133energies}, the time evolution  of the system initially in the trivial vacuum,
$| \psi^{({\bf 1} {\bf 3}\overline{\bf 3};++)}_1 \rangle$ for a coupling $g=1$ is displayed.
From this evaluation it is seen that the lowered global cutoff at a quadratic Casimir of $\frac{16}{3}$ has an impact that increases with evolution temporal extent, a natural observation considering the low-Casimir initialization.
Discrepancies first appear in magnitude at local extrema and build a significant phase shift over a few oscillations as the restricted Hilbert space of the added global truncation has effectively reduced the period.
Informed by understanding of the truncation dependence of the single-plaquette wavefunction discussed in Section~\ref{sec:singleplaquette}, it is not surprising that the presence of states at the global truncation boundary in this system are significant.

Though severely truncated to the local $\{ {\bf 1}, {\bf 3} , \overline{\bf 3} \}$ basis, this example parallels previous quantum simulations of a two plaquette  SU(2) system~\cite{Klco:2019evd} and naturally maps onto a qutrit device with each qutrit describing the color state of a link.
Further, the global wavefunctions discussed above can be simulated
with two qubits (embedding the four global states in the $++$ sector).
While current understanding indicates local bases to be advantageous at scale, global bases will continue to be valuable techniques (e.g., Refs.~\cite{Dumitrescu:2018njn,Klco:2018kyo,Lu:2018pjk}) for exploring the quantum simulation capabilities of available quantum architectures.

\subsubsection{Hardware Implementation}
\begin{figure} [!ht]
\begin{overpic}[width=0.46\textwidth,percent]{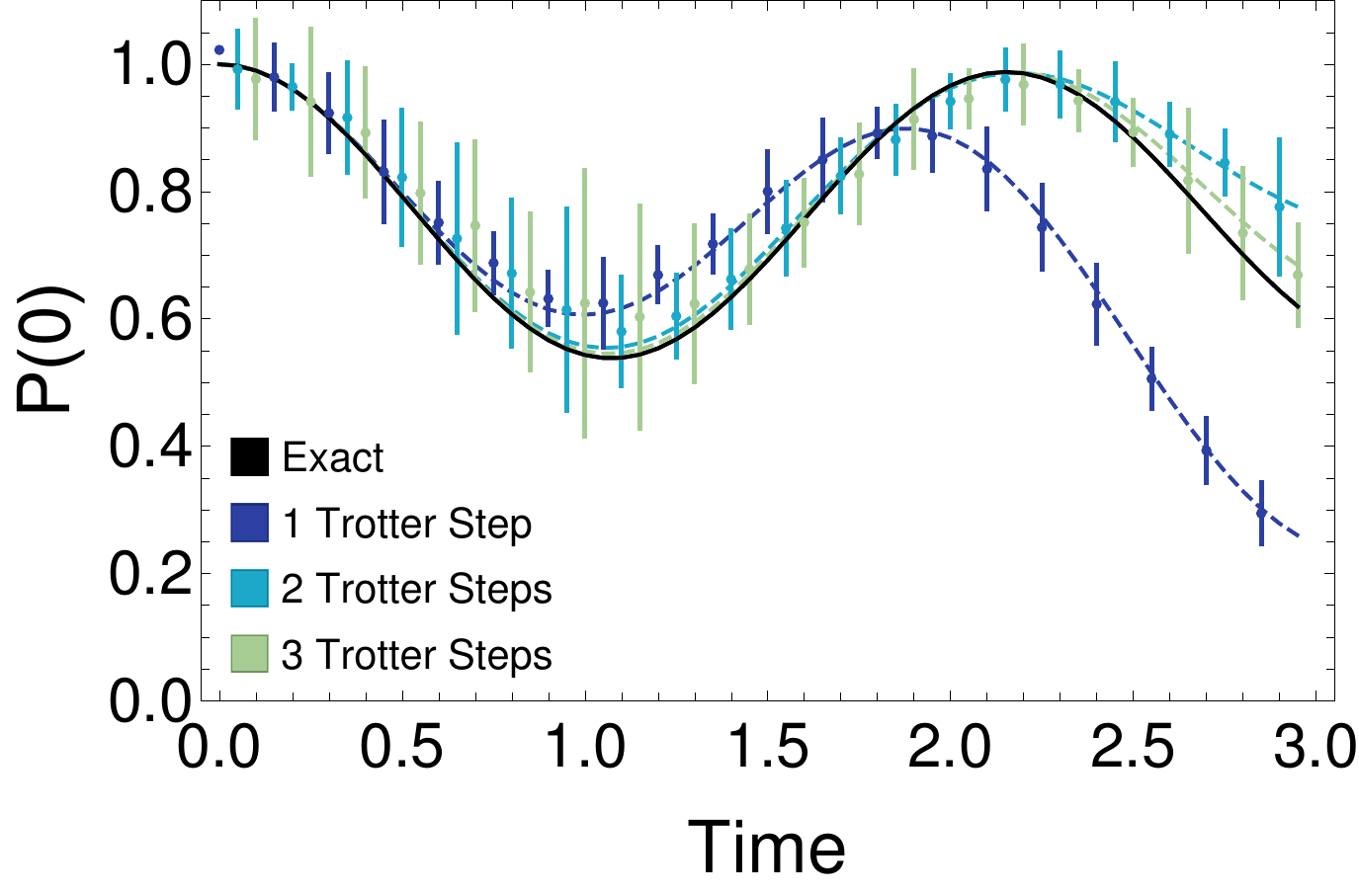} \put(91.5,58.5){\includegraphics[width=0.06\textwidth]{iconQ1_bf.png}}
	\end{overpic}
\quad
\begin{overpic}[width=0.46\textwidth,percent]{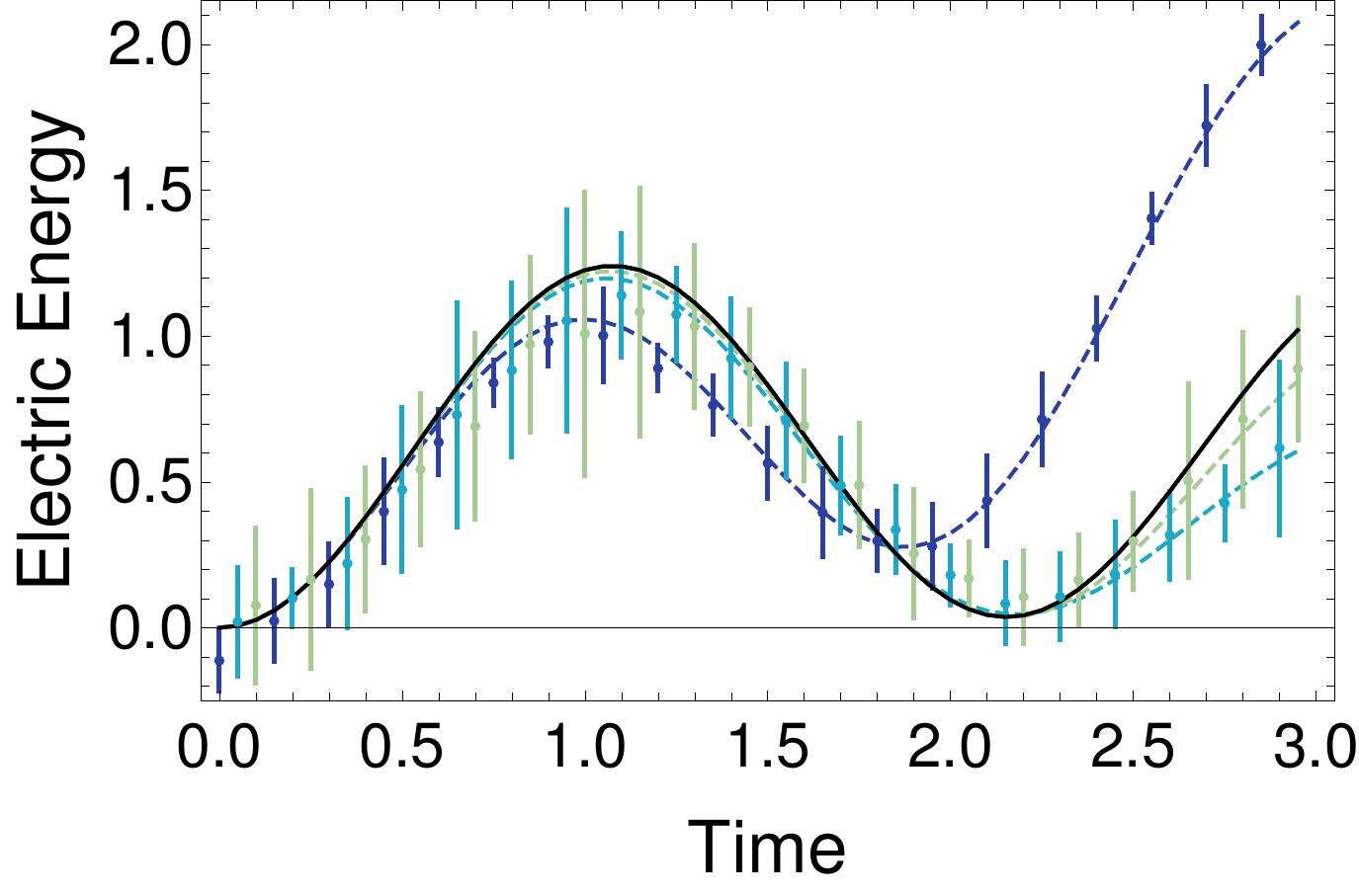} \put(92,59){\includegraphics[width=0.06\textwidth]{iconQ1_bf.png}}
	\end{overpic}
	\caption{
	The (trivial-) vacuum-to-vacuum persistence probability $|\langle 00 |\ \hat U(t)\ | 00\rangle|^2$ (left panel)
	and the energy in the electric field (right panel)
of the two plaquette system in the color parity basis truncated locally at $\mathbf{3}$ and $\overline{\mathbf{3}}$.
Evolution is a $1^{\text{st}}$-order Trotterization of the Hamiltonian in Eq.~(\ref{eq:HppLam3}).
Points correspond to quadratic extrapolations of results obtained from IBM's {\tt Athens} quantum processor, with systematic and statistical uncertainties combined in quadrature.}
	\label{fig:two_plaq_color_3_trot1}
\end{figure}

\noindent
Mapping the ++ sector of Eq.~\eqref{eq:133basispp} onto the computational basis of two qubits, $\left\{| \psi^{({\bf 1} {\bf 3}\overline{\bf 3};++)}_1 \rangle, \dots, | \psi^{({\bf 1} {\bf 3}\overline{\bf 3};++)}_4 \rangle \right\}  \rightarrow \left\{|0\rangle, \dots, |3\rangle\right\}$,
the Hamiltonian of Eq.~\eqref{eq:HppLam3} is decomposed in the Pauli basis as
\begin{multline}
 \hat H^{({\bf 1} {\bf 3}\overline{\bf 3};++)} =  g^2 \left( \frac{7}{3} \hat{\mathbb{I}} \otimes \hat{\mathbb{I}} - \hat{Z} \otimes \hat{\mathbb{I}} - \hat{\mathbb{I}} \otimes \hat{Z} - \frac{1}{3} \hat{Z} \otimes \hat{Z} \right)
	- \frac{1}{2g^2} \left( \frac{1}{4} \left(-23 \ \hat{\mathbb{I}} \otimes \hat{\mathbb{I}} + \hat{Z} \otimes \hat{\mathbb{I}} - \hat{\mathbb{I}} \otimes \hat{Z} - \hat{Z} \otimes \hat{Z} \right) \right. \\ \left.  + \frac{1}{3 \sqrt{2}} \hat{X} \otimes \left( \hat{\mathbb{I}} - \hat{Z}\right) + \frac{2}{3} \left(2 \ \hat{\mathbb{I}} + \hat{Z}\right) \otimes \hat{X} + \frac{1}{9 \sqrt{2}} \left( \hat{X} \otimes \hat{X} + \hat{Y} \otimes \hat{Y} \right) \right) \ \ \ .
\end{multline}
A Trotterized time evolution can be constructed by further decomposing into a sum of three terms,
\begin{align}
	\hat{H}_1 & = \left(\frac{7}{3}g^2 + \frac{23}{8g^2} \right) \hat{\mathbb{I}} \otimes \hat{\mathbb{I}} - \left(\frac{1}{8g^2} + g^2 \right) \hat{Z} \otimes \hat{\mathbb{I}} +  \left(\frac{1}{8g^2} - g^2 \right) \hat{\mathbb{I}} \otimes \hat{Z} - \frac{1}{6g^2 \sqrt{2}} \hat{X} \otimes \hat{\mathbb{I}} - \frac{2}{3g^2} \hat{\mathbb{I}} \otimes \hat{X}
	\ \  \ ,
	\nonumber \\
	\hat{H}_2 & = \left(\frac{1}{8 g^2} - \frac{g^2}{3}\right) \hat{Z} \otimes \hat{Z} - \frac{1}{18g^2 \sqrt{2}} \left( \hat{X} \otimes \hat{X} + \hat{Y} \otimes \hat{Y} \right)
		\ \  \ ,
\nonumber \\
	\hat{H}_3 & = \frac{1}{6g^2 \sqrt{2}} \hat{X} \otimes \hat{Z} - \frac{1}{3g^2} \hat{Z} \otimes \hat{X} \ \ \ .
\end{align}
The first order Trotterized time evolution operator used in the following implementation is
$\hat{U}(t) = e^{-i\hat{H}_3 t} e^{-i\hat{H}_2 t} e^{-i\hat{H}_1 t}$.
Application of
the first evolution contains only single qubit operators in $\hat{H}_1$,
which can be implemented by single qubit quantum gates without further Trotterization,
while the second evolution can be implemented using the quantum circuit in Eq.~\eqref{eq:XX_circuit},
and the third can be implemented with the following circuit relation
\begin{equation}
  e^{i \left(\alpha \hat{Z} \otimes \hat{X} + \beta \hat{X} \otimes \hat{Z} \right)} =
  \begin{gathered}
    \Qcircuit @R=0.6em @C=0.7em {
    & \gate{H} & \ctrl{1} & \gate{H} & \gate{e^{i \alpha \hat{Z}}} & \gate{H} & \ctrl{1} & \gate{H} & \qw \\
    & \qw & \targ & \qw & \gate{e^{i \beta \hat{Z}}} & \qw & \targ & \qw & \qw
    }
  \end{gathered} \ \ \ .
\end{equation}
The results of performing first order Trotter time steps with $g=1$ beginning in the electric vacuum are shown in Fig.~\ref{fig:two_plaq_color_3_trot1}.
Two middle  qubits were used to store the state of the system and, when the measurement error mitigation is implemented
through voting, the remaining three qubits were used to inform the post-selection described in Section~\ref{sec:OnePlaq8}.
As the results show, three Trotter steps are capable of reproducing the first maximum and minimum in the evolution of the electric energy and calculations on the {\tt Athens} quantum processor are in agreement with the exact calculation.
%

\subsection{\texorpdfstring{Two-Plaquette: $\{ {\bf 1}, {\bf 3} , \overline{\bf 3} , {\bf 8} \}$ Local Truncation}{Two-Plaquette: \{1, 3, 3, 8\} Local Truncation}}
\label{subsec:1338}
\noindent
To further explore global wavefunctions and also to demonstrate a further complexity in such calculations,
the discussion in Subsection~\ref{subsec:133} is here extended to include the ${\bf 8}$ in the local link basis.
The construction involves
an expanded basis that requires considering non-trivial multiplicities in the products of irreps, in particular in
${\bf 8}\otimes {\bf 8}={\bf 27}\oplus {\bf 10}\oplus \overline{\bf 10}\oplus{\bf 8}\oplus{\bf 8}\oplus{\bf 1}$.
Of the $4^6$ states in this local basis, 109 of them satisfy Gauss's law.
Projecting further to the global color singlet states---the global color charge being a quantum number conserved by the Hamiltonian---there are 41 distinct physical configurations potentially connected to the strong coupling vacuum.

These physical and global color singlet states combine into states with definite transformation properties under the discrete symmetries of color parity, translation, and reflection, which is no longer redundant in this larger basis as ${\bf 3}\otimes \overline{\bf 3} = {\bf 8}\oplus {\bf 1}$ leads to configurations that can be odd under reflection.
Focusing only on the $+++$ sector, the 15 independent states are,
\begin{eqnarray}
| \psi^{({\bf 1} {\bf 3}\overline{\bf 3}{\bf 8};+++)}_1 \rangle & = &
| \chi ({\bf 1} , {\bf 1} , {\bf 1} , {\bf 1} , {\bf 1} , {\bf 1}) \rangle
\ \ \ ,
\nonumber\\
| \psi^{({\bf 1} {\bf 3}\overline{\bf 3}{\bf 8};+++)}_{2a} \rangle & = &
\frac{1}{2}\left[\
| \chi ({\bf 3} , \overline{\bf 3} , \overline{\bf 3}  , {\bf 1} , {\bf 3} , {\bf 1}) \rangle
+
| \chi (\overline{\bf 3} , {\bf 3} , {\bf 3}  , {\bf 1} , \overline{\bf 3} , {\bf 1}) \rangle
+
| \chi ({\bf 1} , {\bf 3} , {\bf 1}  , {\bf 3} , \overline{\bf 3} , \overline{\bf 3} ) \rangle
+
| \chi ({\bf 1} , \overline{\bf 3} , {\bf 1}  , \overline{\bf 3} , {\bf 3} , {\bf 3} ) \rangle
\ \right]
\ \ \ ,
\nonumber\\
| \psi^{({\bf 1} {\bf 3}\overline{\bf 3}{\bf 8};+++)}_{2b} \rangle & = &
\frac{1}{\sqrt{2}}\left[\
| \chi ({\bf 3} , {\bf 1} , \overline{\bf 3}  , {\bf 3} , {\bf 1} , \overline{\bf 3} ) \rangle
+
| \chi (\overline{\bf 3} , {\bf 1} , {\bf 3}  , \overline{\bf 3} , {\bf 1} , {\bf 3} ) \rangle
\ \right]
\ \ \ ,
\nonumber\\
| \psi^{({\bf 1} {\bf 3}\overline{\bf 3}{\bf 8};+++)}_{3} \rangle & = &
\frac{1}{\sqrt{2}}\left[\
| \chi ({\bf 8} , {\bf 1} , {\bf 1}  , {\bf 8} , {\bf 1} , {\bf 1} ) \rangle
+
| \chi ({\bf 1} , {\bf 1} , {\bf 8}  , {\bf 1} , {\bf 1} , {\bf 8} ) \rangle
\ \right]
\ \ \ ,
\nonumber\\
| \psi^{({\bf 1} {\bf 3}\overline{\bf 3}{\bf 8};+++)}_{4} \rangle & = &
\frac{1}{\sqrt{2}}\left[\
| \chi ({\bf 3} , {\bf 3} , \overline{\bf 3}  , \overline{\bf 3} , \overline{\bf 3} , {\bf 3} ) \rangle
+
| \chi (\overline{\bf 3} , \overline{\bf 3} , {\bf 3}  , {\bf 3} , {\bf 3} , \overline{\bf 3} ) \rangle
\ \right]
\ \ \ ,
\nonumber\\
| \psi^{({\bf 1} {\bf 3}\overline{\bf 3}{\bf 8};+++)}_{5a} \rangle & = &
\frac{1}{2}\left[\
| \chi ({\bf 3} , {\bf 1} , \overline{\bf 3}  , {\bf 3} , {\bf 8} , \overline{\bf 3}) \rangle
+
| \chi (\overline{\bf 3} , {\bf 1} , {\bf 3}  , \overline{\bf 3} , {\bf 8} , {\bf 3}) \rangle
+
| \chi ({\bf 3} , {\bf 8} , \overline{\bf 3}  , {\bf 3} , {\bf 1} , \overline{\bf 3}) \rangle
+
| \chi (\overline{\bf 3} , {\bf 8} , {\bf 3}  , \overline{\bf 3} , {\bf 1} , {\bf 3}) \rangle
\ \right]
\ \ \ ,
\nonumber\\
| \psi^{({\bf 1} {\bf 3}\overline{\bf 3}{\bf 8};+++)}_{5b} \rangle & = &
\frac{1}{2\sqrt{2}}\left[\
| \chi ({\bf 3} , \overline{\bf 3} , \overline{\bf 3}  , {\bf 1} , {\bf 3} , {\bf 8}) \rangle
+
| \chi ({\bf 3} , \overline{\bf 3} , \overline{\bf 3}  , {\bf 8} , {\bf 3} , {\bf 1}) \rangle
+
| \chi (\overline{\bf 3} , {\bf 3} , {\bf 3}  , {\bf 1} , \overline{\bf 3} , {\bf 8}) \rangle
+
| \chi (\overline{\bf 3} , {\bf 3} , {\bf 3}  , {\bf 8} , \overline{\bf 3} , {\bf 1}) \rangle
\right.
\nonumber\\
&&
\qquad \left.
+
| \chi ( {\bf 1} , {\bf 3} , {\bf 8} , {\bf 3} , \overline{\bf 3} , \overline{\bf 3}) \rangle
+
| \chi ( {\bf 8} , {\bf 3} , {\bf 1} , {\bf 3} , \overline{\bf 3} , \overline{\bf 3}) \rangle
+
| \chi ( {\bf 1} , \overline{\bf 3} , {\bf 8} , \overline{\bf 3} , {\bf 3} , {\bf 3}) \rangle
+
| \chi ( {\bf 8} , \overline{\bf 3} , {\bf 1} , \overline{\bf 3} , {\bf 3} , {\bf 3}) \rangle
 \right]
\ \ \ ,
 \nonumber\\
 | \psi^{({\bf 1} {\bf 3}\overline{\bf 3}{\bf 8};+++)}_{6a} \rangle & = &
\frac{1}{\sqrt{2}}\left[\
| \chi ({\bf 3} , {\bf 8} , \overline{\bf 3}  , {\bf 3} , {\bf 8} , \overline{\bf 3} ) \rangle
+
| \chi (\overline{\bf 3} , {\bf 8} , {\bf 3}  , \overline{\bf 3} , {\bf 8} , {\bf 3} ) \rangle
\ \right]
\ \ \ ,
 \nonumber\\
 | \psi^{({\bf 1} {\bf 3}\overline{\bf 3}{\bf 8};+++)}_{6b} \rangle & = &
\frac{1}{2}\left[\
| \chi ({\bf 3} ,\overline{\bf 3} , \overline{\bf 3}  , {\bf 8} , {\bf 3} , {\bf 8} ) \rangle
+
| \chi (\overline{\bf 3} ,{\bf 3} , {\bf 3}  , {\bf 8} , \overline{\bf 3} , {\bf 8} ) \rangle
+
|  \chi ({\bf 8} , {\bf 3} , {\bf 8} ,{\bf 3} ,\overline{\bf 3} , \overline{\bf 3} ) \rangle
+
| \chi ( {\bf 8} , \overline{\bf 3} , {\bf 8} , \overline{\bf 3} ,{\bf 3} , {\bf 3} ) \rangle
\ \right]
\ \ \ ,
 \nonumber\\
| \psi^{({\bf 1} {\bf 3}\overline{\bf 3}{\bf 8};+++)}_{7a} \rangle & = &
| \chi ({\bf 8} , {\bf 1} , {\bf 8} , {\bf 8} , {\bf 1} , {\bf 8}) \rangle
\ \ \ ,
 \nonumber\\
 | \psi^{({\bf 1} {\bf 3}\overline{\bf 3}{\bf 8};+++)}_{7b} \rangle & = &
\frac{1}{\sqrt{2}}\left[\
| \chi ({\bf 8} , {\bf 8} , {\bf 8}  , {\bf 1} , {\bf 8} , {\bf 1} ) \rangle
+
| \chi ({\bf 1} , {\bf 8} , {\bf 1} , {\bf 8} , {\bf 8} , {\bf 8} ) \rangle
\ \right]
\ \ \ ,
 \nonumber\\
 | \psi^{({\bf 1} {\bf 3}\overline{\bf 3}{\bf 8};+++)}_{7c} \rangle & = &
\frac{1}{\sqrt{2}}\left[\
| \chi ({\bf 1} , {\bf 8} , {\bf 8}  , {\bf 8} , {\bf 8} , {\bf 1} ) \rangle
+
| \chi ({\bf 8} , {\bf 8} , {\bf 1} , {\bf 1} , {\bf 8} , {\bf 8} ) \rangle
\ \right]
\ \ \ ,
 \nonumber\\
 | \psi^{({\bf 1} {\bf 3}\overline{\bf 3}{\bf 8};+++)}_{8a} \rangle & = &
\frac{1}{\sqrt{2}}\left[\
| \chi ({\bf 8} , {\bf 1} , {\bf 8}  , {\bf 8} , {\bf 8} , {\bf 8} ) \rangle
+
| \chi ({\bf 8} , {\bf 8} , {\bf 8} , {\bf 8} , {\bf 1} , {\bf 8} ) \rangle
\ \right]
\ \ \ ,
 \nonumber\\
 | \psi^{({\bf 1} {\bf 3}\overline{\bf 3}{\bf 8};+++)}_{8b} \rangle & = &
\frac{1}{2}\left[\
| \chi ({\bf 1} , {\bf 8} , {\bf 8}  , {\bf 8} , {\bf 8} , {\bf 8} ) \rangle
+
| \chi ({\bf 8} , {\bf 8} , {\bf 8} , {\bf 1} , {\bf 8} , {\bf 8} ) \rangle
+
| \chi ({\bf 8} , {\bf 8} , {\bf 1}  , {\bf 8} , {\bf 8} , {\bf 8} ) \rangle
+
| \chi ({\bf 8} , {\bf 8} , {\bf 8} , {\bf 8} , {\bf 8} , {\bf 1} ) \rangle
\ \right]
\ \ \ ,
 \nonumber\\
| \psi^{({\bf 1} {\bf 3}\overline{\bf 3}{\bf 8};+++)}_9 \rangle & = &
| \chi ({\bf 8} , {\bf 8} , {\bf 8} , {\bf 8} , {\bf 8} , {\bf 8}) \rangle
\ \ \ .
\label{eq:1338basisppp}
\end{eqnarray}
States have been grouped together by the value of the Casimir operator, e.g.,
$| \psi^{({\bf 1} {\bf 3}\overline{\bf 3}{\bf 8};+++)}_{2a} \rangle$ and
$| \psi^{({\bf 1} {\bf 3}\overline{\bf 3}{\bf 8};+++)}_{2b} \rangle$ both have a
Casimir of $\sum\limits_a|{\bf E}^a|^2 = \frac{16}{3}$.
The basis states associated with $\{ {\bf 1}, {\bf 3} , \overline{\bf 3}  \}$ local truncation are,
of course, found in this basis,
$| \psi^{({\bf 1} {\bf 3}\overline{\bf 3};+++)}_{1} \rangle = | \psi^{({\bf 1} {\bf 3}\overline{\bf 3}{\bf 8};+++)}_{1} \rangle$,
$| \psi^{({\bf 1} {\bf 3}\overline{\bf 3};+++)}_{2} \rangle = | \psi^{({\bf 1} {\bf 3}\overline{\bf 3}{\bf 8};+++)}_{2a} \rangle$,
$| \psi^{({\bf 1} {\bf 3}\overline{\bf 3};+++)}_{3} \rangle = | \psi^{({\bf 1} {\bf 3}\overline{\bf 3}{\bf 8};+++)}_{2b} \rangle$,
$| \psi^{({\bf 1} {\bf 3}\overline{\bf 3};+++)}_{4} \rangle = | \psi^{({\bf 1} {\bf 3}\overline{\bf 3}{\bf 8};+++)}_{4} \rangle$.
Interestingly,
there is a configuration with one or more of the links in the ${\bf 8}$ that has a smaller Casimir, e.g.,
 $| \psi^{({\bf 1} {\bf 3}\overline{\bf 3}{\bf 8};+++)}_{3} \rangle$ has a smaller Casimir than
 $| \psi^{({\bf 1} {\bf 3}\overline{\bf 3}{\bf 8};+++)}_{4} \rangle$, further emphasizing a practical difference between global and local truncations.
The basis states in the other sectors can be constructed straightforwardly (by inspection from the $+++$ states).
\begin{figure}
\centering
\begin{overpic}[width = 0.7\textwidth,percent]{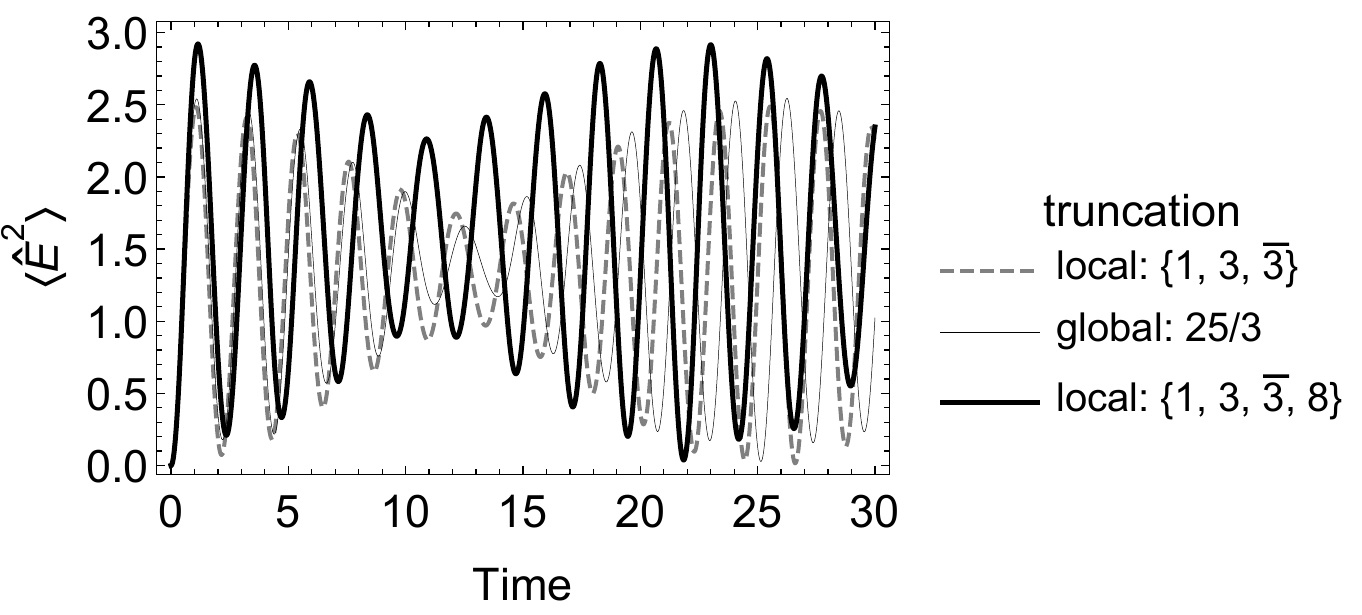}
    \put(62,39.5){\includegraphics[width=0.06\textwidth]{iconC1_bf.png}}
\end{overpic}
  \caption{
Time evolution of the electric Casimir operators $\sum\limits_a | {\bf E}^a |^2$ in the
symmetry-sector-$(+++)$ two-plaquette system
(with PBCs) for coupling $g = 1$ initialized in the trivial vacuum at three different truncations: a local truncation of irreps $\{ {\bf 1}, {\bf 3} , \overline{\bf 3}\}$ (gray, dashed), a global truncation to basis Casimirs of $\frac{25}{3}$ (thin solid gray line), and the local truncation of $\{ {\bf 1}, {\bf 3} , \overline{\bf 3} , {\bf 8}\}$ expressed in Eq.~\eqref{eq:1338basisppp} (thick black line).
}
\label{fig:1338truncgl}
\end{figure}

Using the same methods as described previously, the electric and magnetic matrix elements of the Kogut-Susskind Hamiltonian can be determined with this symmetry-projected basis, leading to a $15\times 15$ dimensional matrix for the $+++$ sector.
The Hamiltonian matrix containing the electric contributions is diagonal,
\begin{eqnarray}
\hat H_E^{({\bf 1} {\bf 3}\overline{\bf 3} {\bf 8};+++)} & = &
\frac{g^2}{2} {\rm diag}\left( 0, \frac{16}{3}, \frac{16}{3}, 6, 8, \frac{25}{3}, \frac{25}{3}, \frac{34}{3}, \frac{34}{3}, 12, 12, 12, 15, 15, 18 \right)
\ \ \ ,
\label{eq:1338E}
\end{eqnarray}
while the magnetic contributions are
\begin{align}
\hat H_{B,\alpha}^{({\bf 1} {\bf 3}\overline{\bf 3} {\bf 8};+++)}  &=
\frac{3}{g^2} \hat{\mathbb{I}}_{15} \ \ \ ,  \nonumber
\\
\hat H_{B,\beta}^{({\bf 1} {\bf 3}\overline{\bf 3} {\bf 8};+++)} &=
-\frac{1}{2 g^2}
\scalemath{0.8}{
\left(
\begin{array}{ccccccccccccccc}
 0 & 2 & 0 & 0 & 0 & 0 & 0 & 0 & 0 & 0 & 0 & 0 & 0 & 0 & 0 \\
 2 & 1 & \frac{\sqrt{2}}{9} & 0 & \frac{\sqrt{2}}{3} & \frac{4 \sqrt{2}}{9} & 0 & \frac{8 \sqrt{2}}{9} & 0 & 0 & \sqrt{2} & 0 & 0 & 0 & 0 \\
 0 & \frac{\sqrt{2}}{9} & 0 & 0 & \frac{2}{3} & 0 & \frac{4 \sqrt{2}}{9} & 0 & \frac{8 \sqrt{2}}{9} & 0 & 0 & 0 & 0 & 0 & 0 \\
 0 & 0 & 0 & 0 & 0 & 0 & \frac{1}{\sqrt{2}} & 0 & 0 & 0 & 0 & 0 & 0 & 0 & 0 \\
 0 & \frac{\sqrt{2}}{3} & \frac{2}{3} & 0 & 0 & \frac{2}{3} & \frac{\sqrt{2}}{3} & \frac{1}{3} & \frac{1}{3 \sqrt{2}} & 0 & 0 & 0 & 0 & 0 & 0 \\
 0 & \frac{4 \sqrt{2}}{9} & 0 & 0 & \frac{2}{3} & 0 & -\frac{2 \sqrt{2}}{9}  & 0 & \frac{1}{9 \sqrt{2}} & 0 & 0 & 0 & 0 & 0 & 0 \\
 0 & 0 & \frac{4 \sqrt{2}}{9} & \frac{1}{\sqrt{2}} & \frac{\sqrt{2}}{3} & -\frac{2 \sqrt{2}}{9}  & \frac{1}{4} & \frac{1}{9 \sqrt{2}} & 0 & 0 & 0 & \frac{1}{\sqrt{2}} & 0 & -\frac{1}{2 \sqrt{2}} & 0 \\
 0 & \frac{8 \sqrt{2}}{9} & 0 & 0 & \frac{1}{3} & 0 & \frac{1}{9 \sqrt{2}} & 0 & \frac{1}{288 \sqrt{2}} & 0 & 0 & 0 & 0 & 0 & 0 \\
 0 & 0 & \frac{8 \sqrt{2}}{9} & 0 & \frac{1}{3 \sqrt{2}} & \frac{1}{9 \sqrt{2}} & 0 & \frac{1}{288 \sqrt{2}} & \frac{1}{16} & \frac{1}{4} & \frac{1}{4 \sqrt{2}} & 0 & -\frac{1}{4} & -\frac{1}{4 \sqrt{2}} & \frac{1}{8} \\
 0 & 0 & 0 & 0 & 0 & 0 & 0 & 0 & \frac{1}{4} & 0 & 0 & 0 & 0 & 0 & 0 \\
 0 & \sqrt{2} & 0 & 0 & 0 & 0 & 0 & 0 & \frac{1}{4 \sqrt{2}} & 0 & 0 & 0 & 0 & 0 & 0 \\
 0 & 0 & 0 & 0 & 0 & 0 & \frac{1}{\sqrt{2}} & 0 & 0 & 0 & 0 & 0 & 0 & 0 & 0 \\
 0 & 0 & 0 & 0 & 0 & 0 & 0 & 0 & -\frac{1}{4} & 0 & 0 & 0 & 0 & 0 & 0 \\
 0 & 0 & 0 & 0 & 0 & 0 & -\frac{1}{2 \sqrt{2}} & 0 & -\frac{1}{4 \sqrt{2}} & 0 & 0 & 0 & 0 & 0 & 0 \\
 0 & 0 & 0 & 0 & 0 & 0 & 0 & 0 & \frac{1}{8} & 0 & 0 & 0 & 0 & 0 & 0 \\
\end{array}
\right)
}
\ \ \ ,
\label{eq:1338B}
\end{align}
where $\mathbb{I}_n$ is the $n$-dimensional identity matrix, and the full Hamiltonian
in the $+++$ sector is the sum
$\hat H^{({\bf 1} {\bf 3}\overline{\bf 3} {\bf 8};+++)} = \hat H_E^{({\bf 1} {\bf 3}\overline{\bf 3} {\bf 8};+++)} +
\hat H_{B,\alpha}^{({\bf 1} {\bf 3}\overline{\bf 3} {\bf 8};+++)} + \hat H_{B,\beta}^{({\bf 1} {\bf 3}\overline{\bf 3} {\bf 8};+++)}$.

The wavefunctions in Eq.~(\ref{eq:1338basisppp}) form a complete set of gauge invariant states that could
be accessed through applications of the plaquette operators to the trivial vacuum
with a local link truncation of $\{ {\bf 1}, {\bf 3} , \overline{\bf 3} , {\bf 8} \}$.
As previously mentioned,
the smallest Casimir for any state containing an ${\bf 8}$ is less than the maximum Casimir
associated with the $\{ {\bf 1}, {\bf 3} , \overline{\bf 3} \}$ local truncation.
Similarly, the smallest Casimir for a state containing a ${\bf 6}$ or $\overline {\bf 6}$ is 10, which lies below that
of the Casimir of
$| \psi^{({\bf 1} {\bf 3}\overline{\bf 3}{\bf 8};+++)}_{6a} \rangle$ and higher states.
Therefore,
there is a fixed number of states in the global basis beyond which changes
to observables from including higher-Casimir states
provide an estimate of the systematic uncertainty from the local link truncation, but do not improve the fidelity of
predictions.
For the $\{ {\bf 1}, {\bf 3} , \overline{\bf 3} , {\bf 8} \}$ local truncation,
the basis states $1, 2a, 2b, 3, 4, 5a, 5b$ have Casimirs below that of the state with the first appearance of the
${\bf 6}$ or $\overline {\bf 6}$.
Therefore computing observables from these wavefunctions provides a consistent prediction for the
contribution from $\{ {\bf 1}, {\bf 3} , \overline{\bf 3} , {\bf 8} \}$ link states.
Differences between  predictions from the $1, 2a, 2b, 3, 4, 5a, 5b$ states and those of part or all of the larger Hamiltonian matrix provide an estimate of the irrep truncation uncertainties,
i.e. the impact of omitting the ${\bf 6}, \overline {\bf 6}, {\bf 10}, \cdots $.
Figure~\ref{fig:1338truncgl} shows the time dependence of the energy in the electric field for three truncations.
The dashed gray curve corresponds to the truncation imposed by the $\{ {\bf 1}, {\bf 3} , \overline{\bf 3} \}$ local link truncation,
the thin solid gray curve corresponds to the global truncation at a Casimir of $25/3$ including the contribution from the ${\bf 8}$ that is below the threshold for the contribution from the
$({\bf 6}, \overline{\bf 6})$, and the solid black curve corresponds to the evolution from the complete matrices in
Eqs.~(\ref{eq:1338E}) and (\ref{eq:1338B}), locally truncated at with $\Lambda_p = \Lambda_q=1$.
The difference between the solid gray and black curves provides an estimate of the systematic uncertainty due to the truncation in color space.
This parallels naive dimensional analysis that is used to estimate the systematic uncertainty introduced by the omission of counterterms in low-energy EFTs.
We conclude from this analysis that
the color-space truncation defined at the link level has not fully converged at $g=1$,
and inclusion of the ${\bf 6}, \overline{\bf 6}$, followed by the three-index tensor representations,
${\bf 10}$, $\overline{\bf 10}$, ${\bf 15}$ and $\overline{\bf 15}$, will be required to
obtain a result that is converged at the percent level.

A technical detail related to  multiplicites in products of irreps appears
in calculations with the $\{ {\bf 1}, {\bf 3} , \overline{\bf 3} , {\bf 8} \}$ local truncated basis,
but absent in the $\{ {\bf 1}, {\bf 3} , \overline{\bf 3} \}$ truncation.
Specifically, as is well known, there are two distinct transitions to the ${\bf 8}$ irrep in the tensor product ${\bf 8}\otimes {\bf 8}$: the symmetric and anti-symmetric contractions,
with two
distinct sets of CG coefficients that contribute to amplitudes, for example, in
the fusion of states, $A_{\bf 8} + B_{\bf 8} \rightarrow  C_{\bf 8}$.
For the calculations in this section,
the high-lying states in the spectrum involving three ${\bf 8}$ links at one vertex,
require a coherent sum over amplitudes in their fusion.
Modifications to the local or global basis that would denote symmetrization or antisymmetrization at relevant vertices
are not required.
A consistently phased set of CG coefficients used to sum over color states at each vertex
is sufficient to arrive at amplitudes and matrix elements.

In the same way that time evolution in the two-plaquette global basis truncated to
$\{ {\bf 1}, {\bf 3} , \overline{\bf 3}  \}$ was performed above by mapping states of projected global symmetries to states in the quantum hardware,
one can contemplate an analogous computation for the $\{ {\bf 1}, {\bf 3} , \overline{\bf 3} , {\bf 8} \}$
local truncation from the $15\times 15$ matrices in Eqs.~(\ref{eq:1338E}) and
(\ref{eq:1338B}).   The first step in developing the qubit-based quantum circuit, using the ordering of states that
follows naturally from the basis of increasing global quadratic Casimir, is to project the Hamiltonian onto tensor products of Pauli and Identity operators, to give rise to coefficients of the form,
\begin{eqnarray}
\hat H
& \rightarrow &
c_{ijkl}\ \hat{\sigma}^i\otimes\hat{\sigma}^j \otimes\hat{\sigma}^k\otimes\hat{\sigma}^l
\ \ ,
\end{eqnarray}
where $\hat{\sigma}^\alpha = \{ \hat{\mathbb{I}}_2 , \hat{\sigma}^x , \hat{\sigma}^y , \hat{\sigma}^z\ \}$.
The coefficients $c_{ijkl}$ are
\begin{eqnarray}
c_{ijkl} & = &
\frac{1}{16}\ {\rm Tr} \left[\ \hat H  \ \hat{\sigma}^i \otimes \hat{\sigma}^j\otimes \hat{\sigma}^k\otimes \hat{\sigma}^l \ \right]
\ \ ,
\end{eqnarray}
where an extra row of zeros has been added to the matrices in Eqs.~(\ref{eq:1338E}) and
(\ref{eq:1338B}).
For the plaquette operator, there are 104 non-zero $c_{ijkl}$.
If the system is further truncated to eight states to be implemented on three qubits, then the number of non-zero coefficients is reduced to 30.
While Pauli decompositions do not always utilize quantum resources optimally, as demonstrated in Section~\ref{subsec:oneplaqscaling}, the lack of uniform Hilbert space organization (as is present in the local $(p,q)$ basis) leads to challenges in identifying scalable alternatives for global basis circuit decomposition.
Even for this small system, involving only two plaquettes, the anticipated limitations of working with the global basis are becoming evident.

\section{Local Basis: The Plaquette Operator}
\label{sec:localbasis}
\noindent
Unlike the space-efficient global basis, where classical pre-preprocessing identifies and isolates the physical sector of the gauge field and each symmetry projected configuration of the field is mapped onto quantum hardware, the local basis distributes local qubit registers uniformly across the lattice to express local quantum numbers of the field.
In this way, an operator acting on a limited number of quantum registers (dictated by its inherent spatial locality) can be developed on small lattices while subsequently retaining relevance even in the infinite volume limit.
In this section, a formulation of the local plaquette operator for the SU(3) magnetic Hamiltonian is presented in the language of digital circuit elements on an architecture comprised of a qudit ($d$-level quantum system) representing the gauge field on each link of a one-dimensional string of plaquettes.

As discussed in Section~\ref{subsec:plaquetteop}, the plaquette operator in a Hilbert space without naturally embedded CG factors will necessarily be controlled on the link registers neighboring the plaquette.
Consider the case of a local qutrit representing the irreps $\{0, 1, 2\} \leftrightarrow \{\mathbf{1}, \mathbf{3}, \overline{\mathbf{3}} \}$ on each link.
To define the structure of the associated magnetic time evolution operator is to characterize an approach for the evolution of an infinite volume lattice with a truncation on any local excitation of the field.
For each plaquette operator interacting with 8 qutrits in this system, there are 81 physical states out of the total $3^8$ that satisfy Gauss's law and contain a singlet at each vertex.
In order to design an operator implementing the correct quantum dynamics, it is necessary to accurately mix these physical states among themselves as well as to assure vanishing matrix elements between these states and the unphysical Hilbert space.
The remaining portion of the operator, mixing the unphysical states among themselves, is a source of flexibility for the intended scientific application of gauge theory simulation and can be optimized or chosen as desired to simplify the circuit implementation.
This freedom has been referred to as \emph{gauge variant completion} (GVC) and will be used in the following design of the plaquette time evolution operator.

Of the 81 physical states currently being considered for plaquette operator design, there are 27 unique external link configurations, as shown in Table~\ref{tab:controlsectors133bar}.
\begin{table}
  \begin{tabular}{c}
  \hline
  \hline
    $\{\mathbf{C}_{1}, \mathbf{C}_{2}, \mathbf{C}_{3}, \mathbf{C}_{4} \}$ \\
    \hline
  $\mathbf{1111}$\\
  $\mathbf{3333 \qquad \bar{3}\bar{3}\bar{3}\bar{3} } $\\
   $\mathbf{3\bar{3}3\bar{3}  \qquad  \bar{3}3\bar{3}3 }$ \\
   $\mathbf{3\bar{3}\bar{3}3  \qquad  \bar{3}33\bar{3}  }$\\
  $\mathbf{113\bar{3}  \qquad  11\bar{3}3  \qquad  3\bar{3}11  \qquad \bar{3}311} $  \\
   $\mathbf{1313  \qquad  1\bar{3}1\bar{3}  \qquad  3131  \qquad \bar{3}1\bar{3}1 } $ \\
   $\mathbf{1331  \qquad  1 \bar{3} \bar{3} 1  \qquad  3131  \qquad \bar{3}1\bar{3}1  }$\\
   $\mathbf{13\bar{3}\bar{3}  \qquad  1\bar{3}33  \qquad 31\bar{3}\bar{3}  \qquad  \bar{3}133  \qquad  33\bar{3}1  \qquad  331\bar{3}  \qquad  \bar{3}\bar{3}13  \qquad  \bar{3}\bar{3}31} $ \\
   \hline
   \hline
  \end{tabular}
  \caption{Physical control sectors of the $\{\mathbf{1}, \mathbf{3}, \overline{\mathbf{3}}\}$-truncated SU(3) plaquette operator.}
  \label{tab:controlsectors133bar}
\end{table}
These 27 control sectors are grouped in rows by vertical and horizontal spatial parity as well as by global conjugation.
The time evolution under the Hermitian plaquette operator can be implemented as a series of commuting operators in the control sectors of Table~\ref{tab:controlsectors133bar},
\begin{equation}
\begin{gathered}
  \Qcircuit @R=1.0em @C=0.3em {
  & \ctrl{1} & \qw \\
  & \ctrl{1} & \qw \\
  & \ctrl{1} & \qw \\
  & \ctrl{1} & \qw \\
  & \multigate{3}{\hat{\Box}+ \hat{\Box}^\dagger} & \qw \\
  & \ghost{\hat{\Box}+ \hat{\Box}^\dagger} & \qw \\
  & \ghost{\hat{\Box}+ \hat{\Box}^\dagger} & \qw \\
  & \ghost{\hat{\Box}+ \hat{\Box}^\dagger} & \qw
  }
\end{gathered}
=
\prod_{\vec{\mathbf{C}}}
\begin{gathered}
  \Qcircuit @R=0.2em @C=0.3em {
  & \qcontrol{$\mathbf{C}_1$} & \qw \\
  & \qcontrol{$\mathbf{C}_2$} \qwx & \qw \\
  & \qcontrol{$\mathbf{C}_3$} \qwx & \qw \\
  & \qcontrol{$\mathbf{C}_4$} \qwx & \qw \\
  & \multigate{3}{\controlsector{\mathbf{C}_1}{\mathbf{C}_2}{\mathbf{C}_3}{\mathbf{C}_4}} \qwx & \qw \\
  & \ghost{\controlsector{\mathbf{C}_1}{\mathbf{C}_2}{\mathbf{C}_3}{\mathbf{C}_4}} & \qw \\
  & \ghost{\controlsector{\mathbf{C}_1}{\mathbf{C}_2}{\mathbf{C}_3}{\mathbf{C}_4}} & \qw \\
  & \ghost{\controlsector{\mathbf{C}_1}{\mathbf{C}_2}{\mathbf{C}_3}{\mathbf{C}_4}} & \qw
  }
\end{gathered}
=
\begin{gathered}
  \Qcircuit @R=0.45em @C=0.3em {
  & \qcontrol{$\mathbf{1}$} & \qcontrol{$\mathbf{3}$} & \qcontrol{$\overline{\mathbf{3}}$} & \qw & \qcontrol{$\overline{\mathbf{3}}$} & \qw\\
  & \qcontrol{$\mathbf{1}$} \qwx & \qcontrol{$\mathbf{3}$} \qwx & \qcontrol{$\overline{\mathbf{3}}$} \qwx & \qw & \qcontrol{$\overline{\mathbf{3}}$} \qwx & \qw \\
  & \qcontrol{$\mathbf{1}$} \qwx & \qcontrol{$\mathbf{3}$} \qwx  & \qcontrol{$\overline{\mathbf{3}}$} \qwx & \qw & \qcontrol{$\mathbf{3}$} \qwx & \qw \\
  & \qcontrol{$\mathbf{1}$} \qwx & \qcontrol{$\mathbf{3}$} \qwx  & \qcontrol{$\overline{\mathbf{3}}$} \qwx & \qw & \qcontrol{$\mathbf{1}$}  \qwx & \qw \\
  & \multigate{3}{\controlsector{1}{1}{1}{1}}  \qwx & \multigate{3}{\controlsector{3}{3}{3}{3}} \qwx & \multigate{3}{\controlsector{\overline{3}}{\overline{3}}{\overline{3}}{\overline{3}}} \qwx  & \cds{3}{\cdots} &  \multigate{3}{\controlsector{\overline{3}}{\overline{3}}{3}{1}} \qwx & \qw
  \\
  & \ghost{\controlsector{1}{1}{1}{1}}  & \ghost{\controlsector{1}{1}{1}{1}} & \ghost{\controlsector{1}{1}{1}{1}} & \qw &  \ghost{\controlsector{1}{1}{1}{1}} & \qw
  \\
  & \ghost{\controlsector{1}{1}{1}{1}}  & \ghost{\controlsector{1}{1}{1}{1}} & \ghost{\controlsector{1}{1}{1}{1}} & \qw &  \ghost{\controlsector{1}{1}{1}{1}} & \qw
  \\
  & \ghost{\controlsector{1}{1}{1}{1}}  & \ghost{\controlsector{1}{1}{1}{1}} & \ghost{\controlsector{1}{1}{1}{1}} & \qw &  \ghost{\controlsector{1}{1}{1}{1}} & \qw
  \\
  }
\end{gathered}
\label{eq:magneticCircuit}
\end{equation}
with a total of 27 controlled operators, mixing three physical states each, that are clearly mutually commuting.

To implement the above magnetic interaction, an architecture of qutrits is natural and amenable to generalization when higher truncations of the gauge space are designed.
Consider the placement of a qutrit on each link degree of freedom.
The  Pauli operations in the qutrit space flip pairs of states,
\begin{align}
  X_{01} &= \begin{pmatrix}
    0 & 1 & 0 \\
    1 & 0 & 0 \\
    0 & 0 & 1 \\
  \end{pmatrix}
  \qquad
  &X_{02} &= \begin{pmatrix}
    0 & 0 & 1 \\
    0 & 1 & 0 \\
    1 & 0 & 0 \\
  \end{pmatrix}
  \qquad
  &X_{12} &= \begin{pmatrix}
    1 & 0 & 0 \\
    0 & 0 & 1 \\
    0 & 1 & 0 \\
  \end{pmatrix}
  \label{eq:qutritXs}\\
  Y_{01} &= \begin{pmatrix}
    0 & -i & 0 \\
    i & 0 & 0 \\
    0 & 0 & 1 \\
  \end{pmatrix} 
  \nonumber
  \qquad
  &Y_{02} &= \begin{pmatrix}
    0 & 0 & -i \\
    0 & 1 & 0 \\
    i & 0 & 0 \\
  \end{pmatrix}
  \qquad
  &Y_{12} &= \begin{pmatrix}
    1 & 0 & 0 \\
    0 & 0 & -i \\
    0 & i & 0 \\
  \end{pmatrix} \ \ \ .
\end{align}
The natural rotation operator generalizing those available on current quantum architectures is the Givens rotation that  transfers population between two levels within the qudit,
\begin{equation}
  G_{jk}^\phi(t) = \exp \left[ -i t \left( e^{i \phi} |j\rangle \langle k | + e^{-i \phi} |k\rangle \langle j| \right) \right]
  \ \ \ .
\end{equation}
Specific rotation angles of 0 and $\pi/2$ correspond to natural extensions of the Pauli-basis rotations on qubits,
\begin{align}
   G_{jk}^0(t) &= \begin{gathered}\Qcircuit @C 0.2em @R 0.6em { & \gate{G_{jk}^{\mathcal{X}}(t)} & \qw }\end{gathered} = \exp\left[ -i t \left( |j\rangle \langle k | + |k \rangle \langle j|\right)\right] =  \exp \left[ -i t \mathcal{X}_{jk}\right] \ \ \ ,
    \\
   G_{jk}^{\frac{\pi}{2}}(t) &= \begin{gathered}\Qcircuit @C 0.2em @R 0.6em { & \gate{G_{jk}^{\mathcal{Y}}(t)} & \qw }\end{gathered} = \exp\left[ -i t \left(i |j\rangle \langle k | - i |k \rangle \langle j|\right)\right] =  \exp \left[ -i t \mathcal{Y}_{jk}\right] \ \ \ ,
\end{align}
where the calligraphic $\mathcal{X}, \mathcal{Y}$ structures are Hermitian but not unitary.
In terms of these operators, one GVC of the magnetic Hamiltonian in each control sector can be constructed through the following Hermitian combinations, with coefficients and transitions determined by the plaquette matrix elements discussed is Appendix~\ref{app:plaquetteMEs}.  With the plaquette Hilbert space designated in the linearized basis of $|\mathbf{R}_b\rangle|\mathbf{Q}_r\rangle |\mathbf{R}_t\rangle |\mathbf{Q}_{\ell}\rangle$, the active space rotations are,
\begin{align}
     \begin{gathered}\Qcircuit @R=0.2em @C=0.3em { & \multigate{3}{\controlsector{\mathbf{1}}{\mathbf{1}}{\mathbf{1}}{\mathbf{1}}} & \qw \\
     & \ghost{\controlsector{\mathbf{1}}{\mathbf{1}}{\mathbf{1}}{\mathbf{1}}} & \qw \\
     & \ghost{\controlsector{\mathbf{1}}{\mathbf{1}}{\mathbf{1}}{\mathbf{1}}} & \qw \\
     & \ghost{\controlsector{\mathbf{1}}{\mathbf{1}}{\mathbf{1}}{\mathbf{1}}} & \qw}\end{gathered} \quad &= \quad \exp\left[-i \alpha \left( \mathcal{X}_{01}\mathcal{X}_{01}\mathcal{X}_{02}\mathcal{X}_{02} + \mathcal{X}_{02}\mathcal{X}_{02}\mathcal{X}_{01}\mathcal{X}_{01} + \mathcal{X}_{12}\mathcal{X}_{12}\mathcal{X}_{12}\mathcal{X}_{12} \right) \right]
     \ \ \ ,
    \label{eq:boxboxD1}\\
  \begin{gathered}\Qcircuit @R=0.2em @C=0.3em { & \multigate{3}{\controlsector{\mathbf{1}}{\mathbf{1}}{\mathbf{3}}{\overline{\mathbf{3}}}} & \qw \\
     & \ghost{\controlsector{\mathbf{1}}{\mathbf{1}}{\mathbf{3}}{\mathbf{3}}} & \qw \\
     & \ghost{\controlsector{\mathbf{1}}{\mathbf{1}}{\mathbf{3}}{\mathbf{3}}} & \qw \\
     & \ghost{\controlsector{\mathbf{1}}{\mathbf{1}}{\mathbf{3}}{\mathbf{3}}} & \qw}\end{gathered} \quad &= \quad   \exp \left[ -i \alpha \left(\frac{1}{3} \mathcal{X}_{02}\mathcal{X}_{01}\mathcal{X}_{01}\mathcal{X}_{01} + \frac{1}{\sqrt{3}} \mathcal{X}_{01}\mathcal{X}_{12}\mathcal{X}_{02}\mathcal{X}_{02} + \frac{1}{\sqrt{3}} \mathcal{X}_{12}\mathcal{X}_{02}\mathcal{X}_{12}\mathcal{X}_{12} \right) \right]
     \ \ \ ,
    \\
  \begin{gathered}\Qcircuit @R=0.2em @C=0.3em { & \multigate{3}{\controlsector{\mathbf{1}}{\mathbf{3}}{\mathbf{1}}{\mathbf{3}}} & \qw \\
     & \ghost{\controlsector{\mathbf{1}}{\mathbf{3}}{\mathbf{1}}{\mathbf{3}}} & \qw \\
     & \ghost{\controlsector{\mathbf{1}}{\mathbf{3}}{\mathbf{1}}{\mathbf{3}}} & \qw \\
     & \ghost{\controlsector{\mathbf{1}}{\mathbf{3}}{\mathbf{1}}{\mathbf{3}}} & \qw}\end{gathered} \quad &= \quad \exp\left[ -i \alpha \left( \frac{1}{3} \mathcal{X}_{01}\mathcal{X}_{02}\mathcal{X}_{01}\mathcal{X}_{01} + \frac{1}{\sqrt{3}} \mathcal{X}_{12}\mathcal{X}_{01}\mathcal{X}_{02}\mathcal{X}_{02} + \frac{1}{\sqrt{3}} \mathcal{X}_{02}\mathcal{X}_{12}\mathcal{X}_{12}\mathcal{X}_{12} \right) \right]
     \ \ \ ,
    \\
     \begin{gathered}\Qcircuit @R=0.2em @C=0.3em { & \multigate{3}{\controlsector{\mathbf{1}}{\mathbf{3}}{\mathbf{3}}{\mathbf{1}}} & \qw \\
     & \ghost{\controlsector{\mathbf{1}}{\mathbf{3}}{\mathbf{3}}{\mathbf{1}}} & \qw \\
     & \ghost{\controlsector{\mathbf{1}}{\mathbf{3}}{\mathbf{3}}{\mathbf{1}}} & \qw \\
     & \ghost{\controlsector{\mathbf{1}}{\mathbf{3}}{\mathbf{3}}{\mathbf{1}}} & \qw}\end{gathered}  \quad &= \quad \exp\left[-i \alpha \left( \frac{1}{3}
    \mathcal{X}_{01}\mathcal{X}_{01}\mathcal{X}_{01}\mathcal{X}_{01} -\frac{1}{\sqrt{3}} \mathcal{X}_{12}\mathcal{X}_{12}\mathcal{X}_{02}\mathcal{X}_{02} -\frac{1}{\sqrt{3}} \mathcal{X}_{02}\mathcal{X}_{02}\mathcal{X}_{12}\mathcal{X}_{12} \right) \right]
     \ \ \ ,
    \\
    \begin{gathered}\Qcircuit @R=0.2em @C=0.3em { & \multigate{3}{\controlsector{\mathbf{1}}{\mathbf{3}}{\overline{\mathbf{3}}}{\overline{\mathbf{3}}}} & \qw \\
     & \ghost{\controlsector{\mathbf{1}}{\mathbf{3}}{\mathbf{3}}{\mathbf{3}}} & \qw \\
     & \ghost{\controlsector{\mathbf{1}}{\mathbf{3}}{\mathbf{3}}{\mathbf{3}}} & \qw \\
     & \ghost{\controlsector{\mathbf{1}}{\mathbf{3}}{\mathbf{3}}{\mathbf{3}}} & \qw}\end{gathered}  \quad &= \quad \exp\left[-i \alpha \left(-\frac{1}{3}
    \mathcal{X}_{01}\mathcal{X}_{12}\mathcal{X}_{01}\mathcal{X}_{01} -\frac{1}{3} \mathcal{X}_{12}\mathcal{X}_{02}\mathcal{X}_{02}\mathcal{X}_{02} +\frac{1}{3} \mathcal{X}_{02}\mathcal{X}_{01}\mathcal{X}_{12}\mathcal{X}_{12} \right) \right]
    \label{eq:csecrot133bar3bar}
     \ \ \ ,
    \\
    \begin{gathered}\Qcircuit @R=0.2em @C=0.3em { & \multigate{3}{\controlsector{\mathbf{3}}{\mathbf{3}}{\mathbf{3}}{\mathbf{3}}} & \qw \\
     & \ghost{\controlsector{\mathbf{1}}{\mathbf{1}}{\mathbf{1}}{\mathbf{1}}} & \qw \\
     & \ghost{\controlsector{\mathbf{1}}{\mathbf{1}}{\mathbf{1}}{\mathbf{1}}} & \qw \\
     & \ghost{\controlsector{\mathbf{1}}{\mathbf{1}}{\mathbf{1}}{\mathbf{1}}} & \qw}\end{gathered}  \quad &= \quad \exp\left[-i \alpha \left(\frac{1}{3\sqrt{3}}
    \mathcal{X}_{01}\mathcal{X}_{02}\mathcal{X}_{12}\mathcal{X}_{01} +\frac{1}{3\sqrt{3}} \mathcal{X}_{12}\mathcal{X}_{01}\mathcal{X}_{01}\mathcal{X}_{02} +\frac{1}{3} \mathcal{X}_{02}\mathcal{X}_{12}\mathcal{X}_{02}\mathcal{X}_{12} \right) \right]
     \ \ \ ,
    \label{eq:csecrot3333}
    \\
    \begin{gathered}\Qcircuit @R=0.2em @C=0.3em { & \multigate{3}{\controlsector{\mathbf{3}}{\overline{\mathbf{3}}}{\mathbf{3}}{\overline{\mathbf{3}}}} & \qw \\
     & \ghost{\controlsector{\mathbf{1}}{\mathbf{1}}{\mathbf{1}}{\mathbf{1}}} & \qw \\
     & \ghost{\controlsector{\mathbf{1}}{\mathbf{1}}{\mathbf{1}}{\mathbf{1}}} & \qw \\
     & \ghost{\controlsector{\mathbf{1}}{\mathbf{1}}{\mathbf{1}}{\mathbf{1}}} & \qw}\end{gathered}  \quad &= \quad \exp\left[-i \alpha \left(\frac{1}{3}
    \mathcal{X}_{12}\mathcal{X}_{02}\mathcal{X}_{12}\mathcal{X}_{01} +\frac{1}{9} \mathcal{X}_{02}\mathcal{X}_{01}\mathcal{X}_{01}\mathcal{X}_{02} +\frac{1}{3} \mathcal{X}_{01}\mathcal{X}_{12}\mathcal{X}_{02}\mathcal{X}_{12} \right) \right]
     \ \ \ ,
    \\
    \begin{gathered}\Qcircuit @R=0.2em @C=0.3em { & \multigate{3}{\controlsector{\mathbf{3}}{\overline{\mathbf{3}}}{\overline{\mathbf{3}}}{\mathbf{3}}} & \qw \\
     & \ghost{\controlsector{\mathbf{1}}{\mathbf{1}}{\mathbf{1}}{\mathbf{1}}} & \qw \\
     & \ghost{\controlsector{\mathbf{1}}{\mathbf{1}}{\mathbf{1}}{\mathbf{1}}} & \qw \\
     & \ghost{\controlsector{\mathbf{1}}{\mathbf{1}}{\mathbf{1}}{\mathbf{1}}} & \qw}\end{gathered}  \quad &= \quad \exp\left[-i \alpha \left( \frac{1}{3}
    \mathcal{X}_{12}\mathcal{X}_{01}\mathcal{X}_{12}\mathcal{X}_{01} +\frac{1}{3\sqrt{3}} \mathcal{X}_{02}\mathcal{X}_{12}\mathcal{X}_{01}\mathcal{X}_{02} +\frac{1}{3\sqrt{3}} \mathcal{X}_{01}\mathcal{X}_{02}\mathcal{X}_{02}\mathcal{X}_{12} \right) \right]
         \ \ \ ,
    \label{eq:BoxBoxD}
\end{align}
where, when applied in the evolution of the Yang-Mills Hamiltonian, $\alpha$
will be determined by $g$ and $t$.
As enumerated in Table~\ref{tab:controlsectors133bar}, the 27 rotations present in the decomposition of Eq.~\eqref{eq:magneticCircuit} can be determined from these 8 through basic parity and conjugation transformations.
For example, the last rotation at the right of Eq.~\eqref{eq:magneticCircuit} can be defined from Eq.~\eqref{eq:csecrot133bar3bar} through a paired horizontal and vertical parity transformations.
Applying the link directionality structure of Fig.~\ref{fig:labels}, the horizontal parity transformation
affects the controls as $|\mathbf{C}_1\rangle |\mathbf{C}_2\rangle |\mathbf{C}_3\rangle |\mathbf{C}_4\rangle \rightarrow |\overline{\mathbf{C}}_3\rangle |\overline{\mathbf{C}}_4\rangle |\overline{\mathbf{C}}_1\rangle |\overline{\mathbf{C}}_2\rangle$ and affects the active space as $|\mathbf{R}_b\rangle |\mathbf{Q}_r\rangle |\mathbf{R}_t\rangle |\mathbf{Q}_\ell\rangle \rightarrow |\overline{\mathbf{R}}_b\rangle |\mathbf{Q}_\ell\rangle |\overline{\mathbf{R}}_t\rangle |\mathbf{Q}_r\rangle$.
Likewise, the vertical parity transformation affects the controls as $|\mathbf{C}_1\rangle |\mathbf{C}_2\rangle |\mathbf{C}_3\rangle |\mathbf{C}_4\rangle \rightarrow |\mathbf{C}_2\rangle |\mathbf{C}_1\rangle |\mathbf{C}_4\rangle |\mathbf{C}_3\rangle $ and affects the active space as $|\mathbf{R}_b\rangle |\mathbf{Q}_r\rangle |\mathbf{R}_t\rangle |\mathbf{Q}_\ell\rangle \rightarrow |\mathbf{R}_t\rangle |\overline{\mathbf{Q}}_r\rangle |\mathbf{R}_b\rangle |\overline{\mathbf{Q}}_\ell\rangle$.
For an example of the conjugation transformation, when implementing the third operator from the right-hand circuit of Eq.~\eqref{eq:magneticCircuit} in the $\overline{\mathbf{3}}^{\otimes 4}$ control sector, one employes the operator described in Eq.~\eqref{eq:csecrot3333} but with swapped mode indices $1\leftrightarrow 2$ performing the conjugation $\mathbf{3} \leftrightarrow \overline{\mathbf{3}}$ in both the control and active spaces.
Appendix~\ref{app:explicitlocalops133} provides an explicit enumeration of all 81 operators, transformed from those of Eqns.~\eqref{eq:boxboxD1}-\eqref{eq:BoxBoxD}, for the $\{\mathbf{1}, \mathbf{3}, \overline{\mathbf{3}}\}$ Trotterized magnetic time evolution.
Crucially, the construction of this local operator is applicable to lattices of any size when local irreps are truncated to single-index tensors.

The plaquette time evolution operator can be implemented through a collection of $e^{-i \alpha \mathcal{X}\mathcal{X} \mathcal{X} \mathcal{X}}$-type operators using the Givens rotations and a generalized CNOT operator controlling the application of a qutrit Pauli operator from Eq.~\eqref{eq:qutritXs} on the mode occupation of a second qutrit.
When expressing the two-qutrit Givens rotation between the same two modes in each qutrit, a second rotation can be applied to the spectator mode to remove inadvertent rotation,
\begin{equation}
  \begin{gathered}
    \Qcircuit @C = 0.2em @R = 0.6cm {
    &\multigate{1}{e^{-i \mathcal{X}_{jk} \otimes \mathcal{X}_{jk} \alpha}} & \qw \\
    &\ghost{e^{-i \mathcal{X}_{jk} \otimes \mathcal{X}_{jk} t}} & \qw
    }
  \end{gathered}
  =
  \begin{gathered}
    \Qcircuit @C = 0.2em @R = 0.6cm {
    &\multigate{1}{G_{jk}^{\mathcal{X}\mathcal{X}}(\alpha)} & \qw \\
    &\ghost{G_{jk}^{\mathcal{X}\mathcal{X}}(\alpha)} & \qw
    }
  \end{gathered}
  =
  \begin{gathered}
    \Qcircuit @C = 0.2em @R = 0.6cm {
    & \gate{X_{jk}} & \qw & \qcontrol{$\ell$} & \qw & \qcontrol{$\ell$} & \gate{X_{jk}} & \qw \\
    & \qcontrol{k} \qwx & \gate{G_{jk}^{\mathcal{X}} \left( \frac{\alpha}{2} \right)} & \gate{Y_{jk}} \qwx & \gate{G_{jk}^{\mathcal{X}}\left( \frac{\alpha}{2} \right)} & \gate{Y_{jk}} \qwx & \qcontrol{k} \qwx & \qw
    }
  \end{gathered}\ \ \ ,
  \label{eq:circuitTEVxx}
\end{equation}
where $\ell$ is the third mode, the complement of $j,k$.
In order to switch the relative modes acted upon in either qudit space, a pair of $X$ operators can be used to  change  basis,
\begin{equation}
  \begin{gathered}
    \Qcircuit @C = 0.2em @R = 0.3cm {
    &\multigate{1}{e^{-i \mathcal{X}_{jk} \otimes \mathcal{X}_{mn} \alpha}} & \qw \\
    &\ghost{e^{-i \mathcal{X}_{jk} \otimes \mathcal{X}_{mn} \alpha}} & \qw
    }
  \end{gathered}
  =
  \begin{gathered}
    \Qcircuit @C = 0.2em @R = 0.3cm {
    &\qw & \multigate{1}{G_{jk}^{\mathcal{X}\mathcal{X}}(\alpha)} & \qw & \qw \\
    &\gate{X_{s}} & \ghost{G_{jk}^{\mathcal{X}\mathcal{X}}(\alpha)} & \gate{X_s} & \qw
    }
  \end{gathered} \ \ \ ,
\end{equation}
where $X_s$ is the $X$ operator necessary to bring the modes of the second qutrit in line with those of the first $\{m,n\} = \{j,k\}$ for the duration of this operator's action.
This function is akin to the basis transformations commonly used in qubit implementations of multi-Pauli time evolutions.
Extending the tactics of Eq.~\eqref{eq:circuitTEVxx} to three qudits yields,
\begin{align}
\begin{gathered}
    \Qcircuit @C = 0.2em @R = 0.3cm {
    &\multigate{2}{G_{jk}^{\mathcal{X}\mathcal{X}\mathcal{X}}(\alpha)} & \qw \\
    &\ghost{G_{jk}^{\mathcal{X}\mathcal{X}\mathcal{X}}(\alpha)} & \qw \\
    &\ghost{G_{jk}^{\mathcal{X}\mathcal{X}\mathcal{X}}(\alpha)} & \qw
    }
  \end{gathered}
  &=
  \scalemath{0.9}{
\begin{gathered}
  \Qcircuit @R=0.2em @C=0.3em {
  & \gate{X_{jk}} & \qw & \qw & \qw & \qcontrol{$\ell$} & \qcontrol{$\ell$} & \qw & \qcontrol{$\ell$} & \qcontrol{$\ell$} & \qw & \qw & \gate{X_{jk}} & \qw \\
  & \qcontrol{k} \qwx & \gate{X_{jk}} & \qw & \qcontrol{$\ell$} & \qw \qwx & \qcontrol{$\ell$} \qwx & \qw & \qcontrol{$\ell$} \qwx & \qw \qwx & \qcontrol{$\ell$} & \gate{X_{jk}} & \qcontrol{k} \qwx & \qw \\
  & \qw & \qcontrol{k} \qwx & \gate{G_{jk}^{\mathcal{X}}\left(\frac{\alpha}{2} \right)} & \gate{Y_{jk}} \qwx & \gate{Y_{jk}} \qwx & \gate{Y_{jk}} \qwx & \gate{G_{jk}^{\mathcal{X}}\left( \frac{\alpha}{2} \right)} & \gate{Y_{jk}} \qwx & \gate{Y_{jk}} \qwx & \gate{Y_{jk}} \qwx & \qcontrol{k} \qwx & \qw & \qw
  }
\end{gathered}} \ \ \ , \nonumber \\
  &=
  \begin{gathered}
  \Qcircuit @R =0.2em @C=0.3em {
  & \gate{X_{jk}} & \qw & \qw & \qcontroldouble{$\ell$} & \qw & \qcontroldouble{$\ell$} & \qw & \gate{X_{jk}} & \qw \\
  & \qcontrol{k} \qwx & \gate{X_{jk}} & \qw & \qcontroldouble{$\ell$} \qwx & \qw & \qcontroldouble{$\ell$} \qwx & \gate{X_{jk}} & \qcontrol{k} \qwx & \qw \\
  & \qw & \qcontrol{k} \qwx & \gate{G_{jk}^{\mathcal{X}}\left(\frac{\alpha}{2} \right)} & \gate{Y_{jk}} \qwx & \gate{G_{jk}^{\mathcal{X}}\left(\frac{\alpha}{2} \right)} &
   \gate{Y_{jk}} \qwx & \qcontrol{k} \qwx & \qw & \qw
  }
  \end{gathered} \ \ \ ,
\end{align}
where the double-circled controls $\Qcircuit @R=0.2em @C=0.3em { & \qcontroldouble{$\ell$} & \qw }$ represent the inclusive-or for multicontrols, applying the target operation if the state $\ell$ is populated in any of the controlled subspaces.
For the $G^{\mathcal{X}\mathcal{X}\mathcal{X}\mathcal{X}}$ operator, 7 $\ell$-controlled operators will be used on either side to construct the inclusive-or-controlled
Pauli for removal of the rotation when the third state is populated in any of the first three qutrits
\begin{equation}
\begin{gathered}
    \Qcircuit @C = 0.2em @R = 0.3cm {
    &\multigate{3}{G_{jk}^{\mathcal{X}\mathcal{X}\mathcal{X}\mathcal{X}}(\alpha)} & \qw \\
    &\ghost{G_{jk}^{\mathcal{X}\mathcal{X}\mathcal{X}\mathcal{X}}(\alpha)} & \qw \\
    &\ghost{G_{jk}^{\mathcal{X}\mathcal{X}\mathcal{X}\mathcal{X}}(\alpha)} & \qw \\
    &\ghost{G_{jk}^{\mathcal{X}\mathcal{X}\mathcal{X}\mathcal{X}}(\alpha)} & \qw
    }
  \end{gathered}
  =
  \begin{gathered}
  \Qcircuit @R =0.2em @C=0.3em {
  & \gate{X_{jk}} & \qw & \qw & \qw & \qcontroldouble{$\ell$} & \qw & \qcontroldouble{$\ell$} & \qw & \qw & \gate{X_{jk}} & \qw \\
  & \qcontrol{k} \qwx & \gate{X_{jk}} & \qw & \qw & \qcontroldouble{$\ell$} \qwx & \qw & \qcontroldouble{$\ell$} \qwx& \qw & \gate{X_{jk}} & \qcontrol{k} \qwx & \qw \\
  & \qw & \qcontrol{k} \qwx & \gate{X_{jk}} & \qw & \qcontroldouble{$\ell$} \qwx & \qw & \qcontroldouble{$\ell$} \qwx & \gate{X_{jk}} & \qcontrol{k} \qwx & \qw & \qw \\
  & \qw & \qw & \qcontrol{k} \qwx & \gate{G_{jk}^{\mathcal{X}}\left(\frac{\alpha}{2} \right)} & \gate{Y_{jk}} \qwx & \gate{G_{jk}^{\mathcal{X}}\left(\frac{\alpha}{2} \right)} &
   \gate{Y_{jk}} \qwx & \qcontrol{k} \qwx & \qw & \qw & \qw
  }
  \end{gathered} \ \ \ ,
  \label{eq:circGXXXX}
\end{equation}
where one functional realization of this multi-controlled inclusive-or operation is
\begin{equation}
\begin{gathered}
  \Qcircuit @R=0.2em @C=0.3em {
  & \qcontroldouble{$\ell$}  & \qw \\
  & \qcontroldouble{$\ell$} \qwx  & \qw \\
  & \qcontroldouble{$\ell$} \qwx  & \qw \\
  & \gate{Y_{jk}} \qwx & \qw
  }
\end{gathered}
=
\begin{gathered}
  \Qcircuit @R=0.2em @C=0.3em {
  & \qw & \qw & \qcontrol{$\ell$} & \qw & \qcontrol{$\ell$} & \qcontrol{$\ell$} & \qcontrol{$\ell$} & \qw \\
  & \qw & \qcontrol{$\ell$} & \qw \qwx & \qcontrol{$\ell$} & \qw \qwx & \qcontrol{$\ell$}  \qwx & \qcontrol{$\ell$} \qwx & \qw \\
  & \qcontrol{$\ell$} & \qw \qwx & \qw \qwx & \qcontrol{$\ell$} \qwx & \qcontrol{$\ell$} \qwx & \qw \qwx & \qcontrol{$\ell$} \qwx & \qw \\
  & \gate{Y_{jk}} \qwx & \gate{Y_{jk}} \qwx & \gate{Y_{jk}} \qwx & \gate{Y_{jk}} \qwx & \gate{Y_{jk}} \qwx & \gate{Y_{jk}} \qwx & \gate{Y_{jk}} \qwx & \qw
  }
\end{gathered} \ \ \ .
\end{equation}
While this particular formulation comprised of two single-qutrit rotation operators is functionally clear and seems advantageous when considering $T$-costs for a potentially fault-tolerant implementation,
it is entirely expected that hardware-specific variations will be made to this circuit decomposition in the course of practical implementation.
Specifically, it is expected that different quantum architectures may offer unique techniques for implementing the isolated two-mode rotation when the third state is not populated in any of the four qutrits, as represents the core of this circuit decomposition.
Furthermore, a qubit embedding with the common Gray-code implementation of two-level unitaries discussed in Subsection~\ref{subsec:oneplaqscaling} may be advantageous for particular architectures.

In the above, a single qudit is used to capture the local Hilbert space of each link, demanding the two-dimensional hexagonal connectivity of Fig.~\ref{fig:connectivitydiagram} between qudit modes.
Section~\ref{subsec:embeddings} discusses the alternate opportunity to introduce two qudits per link representing the $(p,q)$ registers of tensor indices, with the advantage that the necessary connectivity of modes in each qudit is reduced to nearest-neighbor linear, similar in form to that of a link in SU(2).
The quantum circuits necessary for implementing the plaquette operator in this $(p,q)$ basis of the local Hilbert space can be straightforwardly generalized from that above.
In particular, the Hermitian combinations of Eq.~\eqref{eq:boxboxD1}-\eqref{eq:BoxBoxD} can be modified with the following substitutions
\begin{equation}
  \mathcal{X}_{01} \rightarrow \mathcal{X}_{01} \otimes \mathbb{I}
  \qquad
  \mathcal{X}_{12} \rightarrow \mathcal{X}_{01} \otimes \mathcal{X}_{01}
  \qquad
  \mathcal{X}_{02} \rightarrow \mathbb{I} \otimes \mathcal{X}_{01}
  \label{eq:substitutions2pq}
\end{equation}
where the $\{\mathbf{1}, \mathbf{3}, \overline{\mathbf{3}}\}$ basis within a single Hilbert space is traded for the pair of Hilbert spaces $|p\rangle \otimes |q\rangle$.
In words, the last substitution in Eq.~\eqref{eq:substitutions2pq} replaces a hermitian operator mixing qudit states 0,2 (irreps $\mathbf{1},\overline{\mathbf{3}}$) with an identity operator in the $p$-register and a mixing of the lowest two levels in the $q$-register.
The plaquette operator implemented in the $(p,q)$ basis requires only (correlated)
nearest-neighbor interactions within each of the two qudits on every link.
One can write a similar translation for the control operators
\begin{equation}
  \begin{gathered}\Qcircuit @R=0.2em @C=0.3em {
  & \qcontrol{$\mathbf{1}$} & \qw
  }\end{gathered} =
  \begin{gathered}\Qcircuit @R=0.2em @C=0.3em {
  & \qcontrol{0} & \qw
  }\end{gathered}
  \rightarrow
  \begin{gathered}\Qcircuit @R=0.2em @C=0.3em {
  & \qcontrol{0} & \qw \\
  & \qcontrol{0} \qwx & \qw
  }\end{gathered}
  \qquad
  \begin{gathered}\Qcircuit @R=0.2em @C=0.3em {
  & \qcontrol{$\mathbf{3}$} & \qw
  }\end{gathered} =
  \begin{gathered}\Qcircuit @R=0.2em @C=0.3em {
  & \qcontrol{1} & \qw
  }\end{gathered}
  \rightarrow
  \begin{gathered}\Qcircuit @R=0.2em @C=0.3em {
  & \qcontrol{1} & \qw \\
  & \qcontrol{0} \qwx & \qw
  }\end{gathered}
  \qquad
  \begin{gathered}\Qcircuit @R=0.2em @C=0.3em {
  & \qcontrol{$\overline{\mathbf{3}}$} & \qw
  }\end{gathered} =
  \begin{gathered}\Qcircuit @R=0.2em @C=0.3em {
  & \qcontrol{2} & \qw
  }\end{gathered}
  \rightarrow
  \begin{gathered}\Qcircuit @R=0.2em @C=0.3em {
  & \qcontrol{0} & \qw \\
  & \qcontrol{1} \qwx & \qw
  }\end{gathered}
\end{equation}
where the two qudit lines at the right represent the $|p\rangle$ and $|q\rangle$ registers from top to bottom.
While the basis with a single qudit per link demonstrated homogeneity in the type of operators, $G_{jk}^{\mathcal{X}\mathcal{X}\mathcal{X}\mathcal{X}}(t)$,  necessary for a Trotterized implementation of the Hermitian combination of the plaquette operator and its conjugate, the $(p,q)$ basis becomes a mix of operators $G_{jk}^{\mathcal{X}^{\otimes n}}(t)$ for $4 \leq n \leq 8$.
These larger operators can be expressed as generalizations of Eq.~\eqref{eq:circGXXXX}.  Explicit lists characterizing the set of operators for Trotterized plaquette implementation in both of these local bases are provided in Appendix~\ref{app:explicitlocalops133}.

\subsection{Comparison Between Local and Global Bases}
\label{subsec:localCWglobal}
\noindent
Subsection~\ref{subsec:133} presented
results  for the time evolution of the two-plaquette system computed using wavefunctions defined in
a global basis truncated in color space by
a local link basis of $\{ {\bf 1}, {\bf 3} , \overline{\bf 3}  \}$.
Above, technology was developed to address this and other systems using local controlled-plaquette operators.
It is valuable to assure correspondence between results of the same quantities with these two quite different approaches.

\begin{figure}[!ht]
  \begin{overpic}[width = 0.5\textwidth,percent]{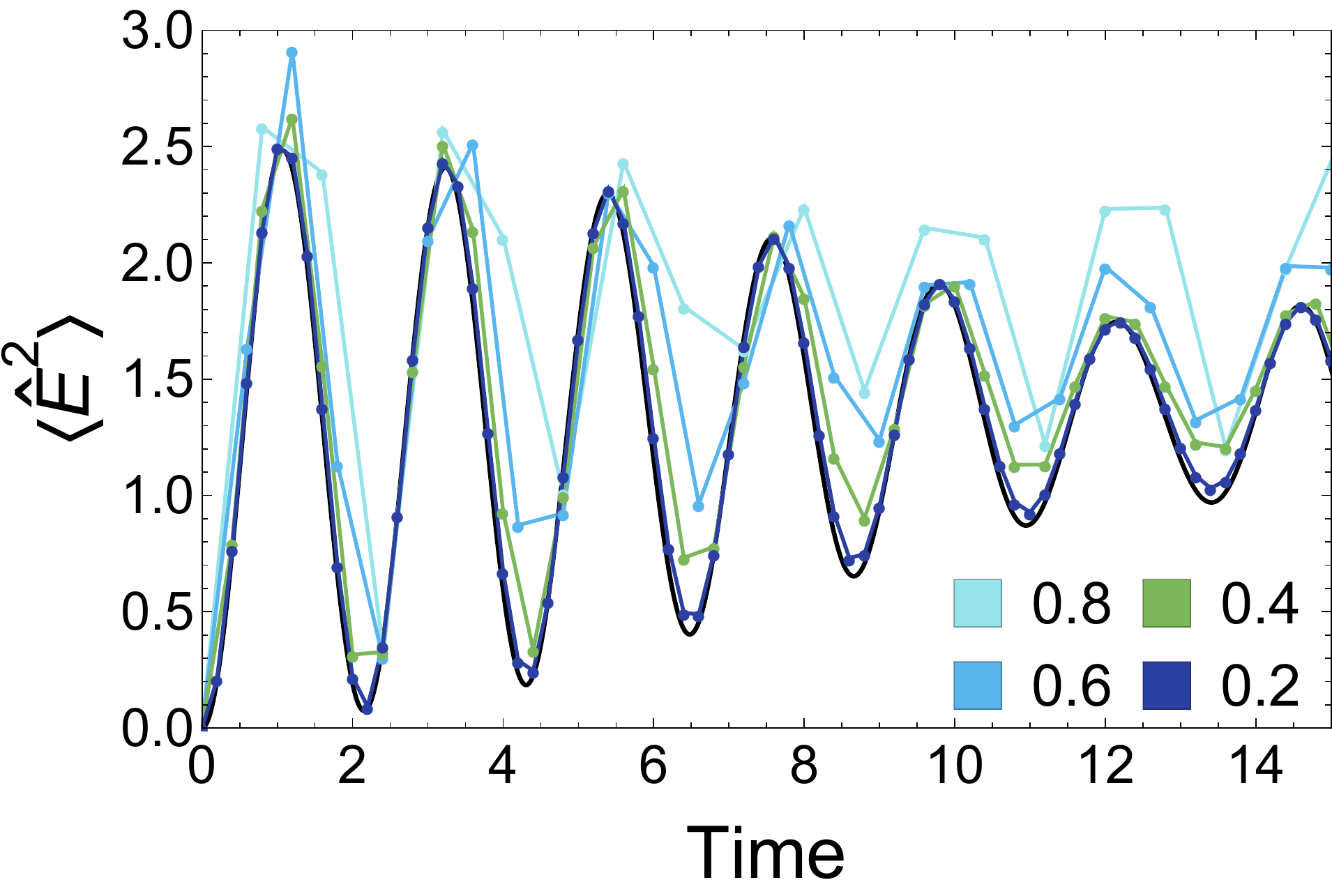}
    \put(92,58){\includegraphics[width=0.06\textwidth]{iconC1_bf.png}}
  \end{overpic}
  \caption{
Time evolution of the Casimirs $\sum\limits_a | {\bf E}^a |^2$ in the
two-plaquette system (with PBCs) with each link truncated to $\{ {\bf 1}, {\bf 3} , \overline{\bf 3} \}$
for $g=1$, initialized in the trivial vacuum.
The black curve is calculated in the global $++$ basis (the same curve as shown in the right panel of Fig.~\ref{fig:133energies} and Fig.~\ref{fig:1338truncgl}), while the points correspond to
Trotter evolution of the contributing controlled-plaquette operators in the local basis.
}
\label{fig:133trotevol}
\end{figure}
Figure~\ref{fig:133trotevol} shows the Trotterized time evolution of $\sum\limits_a | {\bf E}^a |^2$ in the chromo-electric field starting from the
trivial vacuum as a function of time with a local truncation of $\{ {\bf 1}, {\bf 3} , \overline{\bf 3}  \}$
for each link.
The black curve corresponds to those in the right panel of Fig.~\ref{fig:133energies} and Fig.~\ref{fig:1338truncgl}, obtained through the time evolution of the two-plaquette system using the four global basis states in
Eqs.~(\ref{eq:133basispp}) and (\ref{eq:1338basisppp}).
As the quantum circuits discussed above reproduce identically the unitary evolution of each component contribution to the Hamiltonian, the same curve is recovered from the local basis through Trotterized evolution leveraging the controlled-plaquette operator.
For example, the last operator in Eq.~(\ref{eq:BoxBoxD}) is implemented, schematically, as the following set of three controlled Givens rotations,
\begin{multline}
\exp\left[- i  \alpha
    \ \begin{pmatrix}
    \mathbf{3} & & \mathbf{3} \\
    &(\hat{\Box} + \hat{\Box}^\dagger)& \\
    \overline{\mathbf{3}} & & \overline{\mathbf{3}}
    \end{pmatrix}
 \right]
 \ \rightarrow \\
 \left(\exp \left[-i \frac{\alpha}{3}
    \mathcal{X}_{12}\mathcal{X}_{02}\mathcal{X}_{12}\mathcal{X}_{01} \right] \exp \left[-i\frac{\alpha}{9} \mathcal{X}_{02}\mathcal{X}_{01}\mathcal{X}_{01}\mathcal{X}_{02} \right] \exp \left[-i\frac{\alpha}{3} \mathcal{X}_{01}\mathcal{X}_{12}\mathcal{X}_{02}\mathcal{X}_{12}\right]\right)
    \otimes
    \Lambda_{1}\otimes \Lambda_{2} \otimes \Lambda_{1} \otimes \Lambda_{2}  \ \ \ ,
\end{multline}
and similar decompositions apply to the other contributing controlled operators.
As each unitary operator is associated with a physical transition of a plaquette, with coefficients determined from matrix elements between gauge-invariant states, the Trotterization preserves gauge invariance.
As usual, however, the lack of commutativity introduces discrepancies in the evolution, analogous to the higher dimension operators in the Symanzik action describing finite lattice spacing artifacts in lattice QCD calculations.

As anticipated, Fig.~\ref{fig:133trotevol} shows that the results of the Trotterized evolution in the local basis converges to a well-defined function as the Trotter step size is reduced toward zero that coincides with the result from evolution using the global basis with the same imposed color truncation.
This result, and others, provides a partial validation of the controlled-plaquette local basis construction for describing QCD dynamics, and thus a framework for a scalable implementation on quantum architectures.

\section{Technical Aspects for Simulating at Scale}
\noindent
The results of the previous sections inspire a discussion of technical hurdles that
can be anticipated on the path to simulations of SU(3) lattice gauge theory
at scale using local multiplet bases.
We focus on issues related to scalability in volume, circuit depth, and gauge field truncation, to potential hardware implementations, and to extensions to three dimensional spatial lattices.


\subsection{Scalability of Local Basis}
\label{subsec:scaling}
\noindent
This scalability discussion begins first with the total number of qubits required to express the gauge field; a diagrammatic notation parallel to that developed for large-$N_c$ scaling~\cite{tHooft:1973alw,Witten:1979kh} is then presented to study the number of gauge invariant vertices as well as physical states and matrix elements comprising the plaquette operator.
As na\"ively presented above, the latter quantity can be made to directly correspond to the number of Givens rotations and thus the quantum circuit depth for the implementation of each local plaquette operator.
In addition to their scaling, numerical values of these vertex and plaquette operator properties are provided to inform future practical implementations.

Qubit requirements are sensitive to the basis used to digitize and express the gauge field.
The most efficient use of a hardware Hilbert space is achieved through a global basis,
as discussed in Section~\ref{sec:globalbasis2p},
where the gauge variant space is removed through classical pre-processing and only the gauge invariant space is mapped onto quantum degrees of freedom.
As demonstrated above, neither the classical pre-processing nor the subsequent compilation for quantum implementation are expected to be scalable in global bases.
For this reason, local bases have been presented in Section~\ref{sec:localbasis},
trading an expanded Hilbert space for local operators that may be optimized and implemented equivalently throughout the lattice.
An initial mapping of Yang-Mills to local quantum registers, as first presented by Byrnes and Yamamoto~\cite{Byrnes:2005qx},
requires a large number of quantum registers to define the gauge field and include the flavor,
color and Dirac degrees of freedom of the quarks.
Each link would be described by
$|p, q, T_L, T_L^z, Y_L, T_R, T_R^z, Y_R \rangle$ with a quantum register of qubits (or qudits) associated with
each quantum number.
The number of qubits for each quantum number can be determined by the cutoff, $\Lambda_p$, defining a finite $(p,q)$ space.
The compression accomplished in this work, using the methods introduced in
Refs.~\cite{Banuls:2017ena,Klco:2019evd}, significantly reduces the required qubit requirements for a given lattice.
By integrating over the local color spaces, the number of registers per link is reduced from 8 to 2.
For a lattice of $L$ sites in each of $D$ spatial directions, the number of qubits required for SU(3) Yang-Mills is estimated to be
$\# \ qubits\sim 2 L^D \log_2(\Lambda_{p}+1)$.
For an $L=10$ lattice in $D = 3$ dimensions truncated at $\Lambda_p = \Lambda_q =  1$ and thus restricted to color irreps $\{ {\bf 1}, {\bf 3}, \overline{\bf 3}, {\bf 8} \}$, an estimate of
$\sim 2000$ logical qubits are required.

In the formulation of this paper, Gauss's law is implemented explicitly by the neighboring controls associated with the action of the plaquette operator
between link configurations and, in particular, the dependence on these controls of the non-vanishing matrix elements that transition between plaquette configurations.
Time evolution of a Yang-Mills wavefunction defined in a local basis can be accomplished by repeated applications
of these controlled-plaquette operators, parallelizable at separations of two plaquettes, and thus enjoys the volume independence of the circuit depth characteristic of locally-interacting theories~\cite{Lloyd1073}.

To estimate the scaling of quantum resources required for the implementation of each controlled-plaquette operator as the irrep truncation is raised,  it
is found to be convenient to work with the $(p,q)$ coupled register mapping.
With a truncation defined by the maximum number of upper and lower indices describing the highest dimension tensor within the active color space of each link, $\Lambda_{p,q}$,
it is essential to estimate the scaling as a function of increasing $\Lambda_{p,q}$.
A truncation that respects color parity, $\Lambda_{p}=\Lambda_{q}$ is employed.
The quasi-locality of the interactions ensures that further scaling with regard to this cutoff involves only (trivial)
factors of the spacetime volume.
The above analyses of the one- and two-plaquette systems indicate that, for the low-energy and low-energy-density sectors of calculations, contributions from color irreps high in the spectrum become exponentially suppressed beyond a coupling-dependent value.
If this suppression is maintained for extended lattices upon raising the tensor index truncation, $\Lambda_p$, will dominate over
the power-law scaling in the number of non-vanishing matrix elements defining the controlled-plaquette operators.
It is anticipated that analogous arguments will be established for localized high-energy density configurations that evolve forward in time through fragmentation and hadronization (processes that are important for nuclear and high-energy physics)
to configurations of low-lying final-state hadrons.

\begin{figure}[!ht]
    \centering
    \includegraphics[width = 0.6\textwidth]{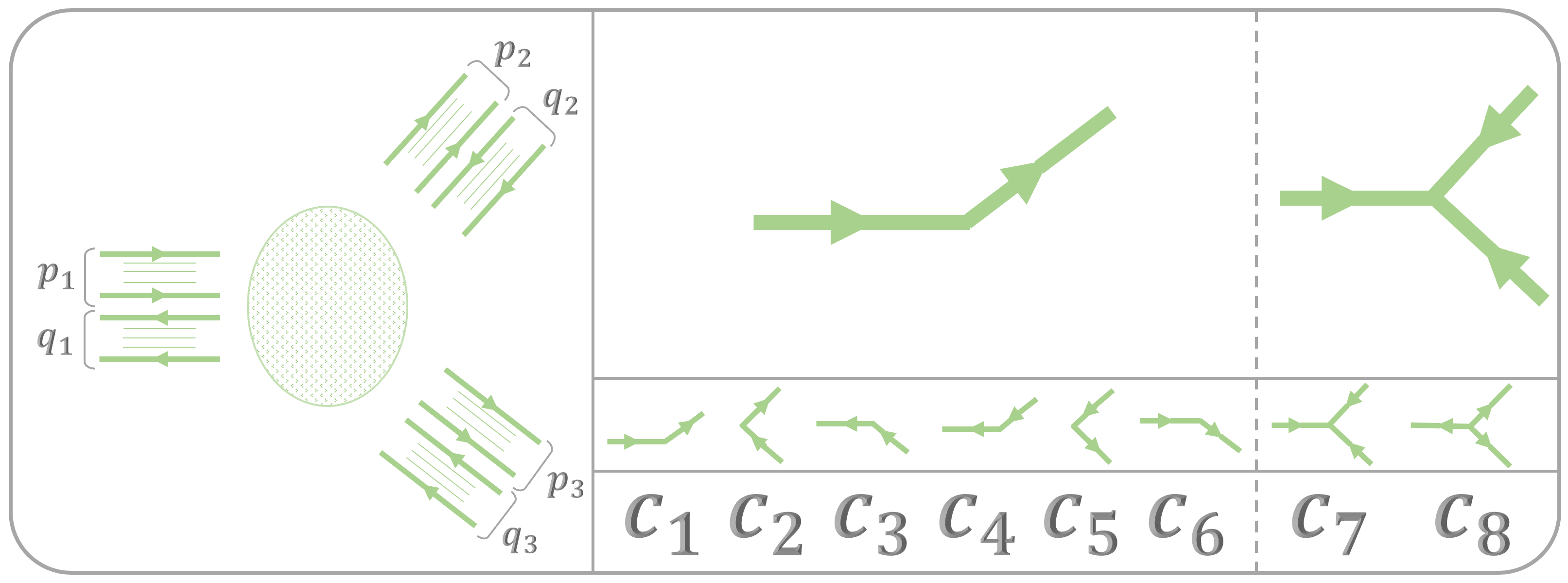}
  \caption{Diagrammatic representation of gauge invariant vertex contractions for the 3-pt vertex.  Arrows indicate the flow of tensor indices in the fundamental and anti-fundamental and external lines are contracted to produce irreducible representations in the local (p,q) link basis.}
  \label{fig:giVertices}
\end{figure}
To begin to establish the scaling of the number of controlled operations that will be required for time evolution using controlled-plaquette operators acting on the local basis, it is helpful to explore and quantify the
scaling of the 3-point vertex (or $2\rightarrow 1$ fusion).
This corresponds to multiplicities in the control of a single plaquette vertex.
The maximum possible growth of the Hilbert space for this three-point vertex is $\Lambda^6$, allowing a $(\Lambda +1)$-dimensional configuration space for each of the $(p,q)$ registers detailing the irrep tensor structure on each link.
However, not all vertex configurations satisfy Gauss's law, leading to a reduction to the number of physical vertices.
Inspired by the diagrammatic tactics of large-$N_c$ scaling calculations~\cite{tHooft:1973alw,Witten:1979kh}, consider directional lines expressing propagations of fundamental or anti-fundamental indices.
The left of Fig~\ref{fig:giVertices} shows the 3-point vertex characterized by $p$ fundamental indices and $q$ anti-fundamental indices as propagating left-right and right-left, respectively.
The external lines (conventionally absent in the large-$N_c$ analysis of color-singlet objects), can be mapped directly to irreducible representations defining the local link basis.
Whether or not the chosen external lines can be connected through a center  region constructed by gauge invariant index contractions determines whether the vertex is physical, or contains a singlet.
All  gauge invariant contractions relevant to the 3-point vertex are shown at the right of Fig.~\ref{fig:giVertices}.
The SU(3) structure provides two invariant tensors: the $\delta$-function, leading to contractions labeled $c_{1-6}$, and the Levi-Civita, leading to contractions $c_{7-8}$.
When implementing a local truncation in the $(p,q)$ basis, this appears as correlated constraints between the $c_i$'s e.g., $p_1 = c_1 + c_6 + c_7 \leq \Lambda_p$.
Every integer vector $\vec{c}$ refers to a physical 3-point vertex and a unique contraction pattern.
However, every unique contraction pattern does not provide a unique vertex in the local $(p,q)$ basis, as discussed in Section~\ref{sec:globalbasis2p}.
For example, the $\mathbf{8} \otimes \mathbf{8} \otimes \mathbf{8}$ vertex can be expressed by $\vec{c} = \{1,0,1,0,1,0,0,0\}, \{0,1,0,1,0,1,0,0\}, $ or $\{0,0,0,0,0,0,1,1\}$ through a set of three $\delta$-contractions or a pair of $\epsilon$-contractions.

Using this diagrammatic approach, the algebraic tensorial decomposition methods of Coleman~\cite{doi:10.1063/1.1704245} indicating that the number of three-point vertex completions for given $(p,q)_{2,3}$ is
\begin{equation}
 \dim(p_1,q_1)_{physical} = \sum_{i = 0}^{\min(p_2, q_3)} \sum_{j = 0}^{\min(p_3,q_2)} 1 + \min(p_2 -i, p_3-j) + \min(q_2 - j, q_3-i) \ \ \ ,
\end{equation}
or evaluations through computational packages e.g., {\tt SU-3-CG-Code} Mathematica code~\cite{Rowe_1999,Martins:2019sfj},
explicit calculations can be made of the unique physical vertices in the $(p,q)$ basis.
\begin{table}[!ht]
  \begin{tabular}{c|c||c|c}
  \hline
  \hline
    $p,q \le \Lambda_{p}$  & $\#$ of singlets & $p,q \le \Lambda_{p}$  & $\#$ of singlets\\
    \hline
0 & 1 & 9 & 182,803 \\
1 & 19 & 10 &  322,621\\
2 & 165 & 11 & 542,196\\
3 & 838 & 12 & 874,483\\
4 & 3,049 & 13 & 1,361,683\\
5 & 8,865 & 14 & 2,056,971\\
6 & 22,003 & 15 & 3,026,098\\
7 & 48,514 & 16 & 4,349,413\\
8 & 97,653 &  &\\
   \hline
   \hline
  \end{tabular}\ \
  \begin{tabular}{c|c||c|c}
  \hline
  \hline
    $p,q \le \Lambda_{p}$  & $\#$ of singlets & $p,q \le \Lambda_{p}$  & $\#$ of singlets\\
    \hline
0 & 1  &  6 & 1,739,833\\
1 & 82  &   7 & 5,080,226\\
2 &  1967 &  8 & 13,071,135\\
3 & 19,550  &  9  & 30,436,170\\
4 & 116,929  &  10 & 65,372,321 \\
5 & 504,932  &   11   &  131,352,884 \\
   \hline
   \hline
  \end{tabular}
  \caption{
  The number of singlets formed from three (left, $2\rightarrow 1$ fusion process) and four (right, $3\rightarrow 1$ fusion process) color irreps
  up to a cutoff in the number of upper and lower indices of $\Lambda_{p}$.
  }
  \label{tab:singletvertexscaling}
\end{table}
\begin{figure}[!ht]
\begin{overpic}[width = 0.47\textwidth,percent]{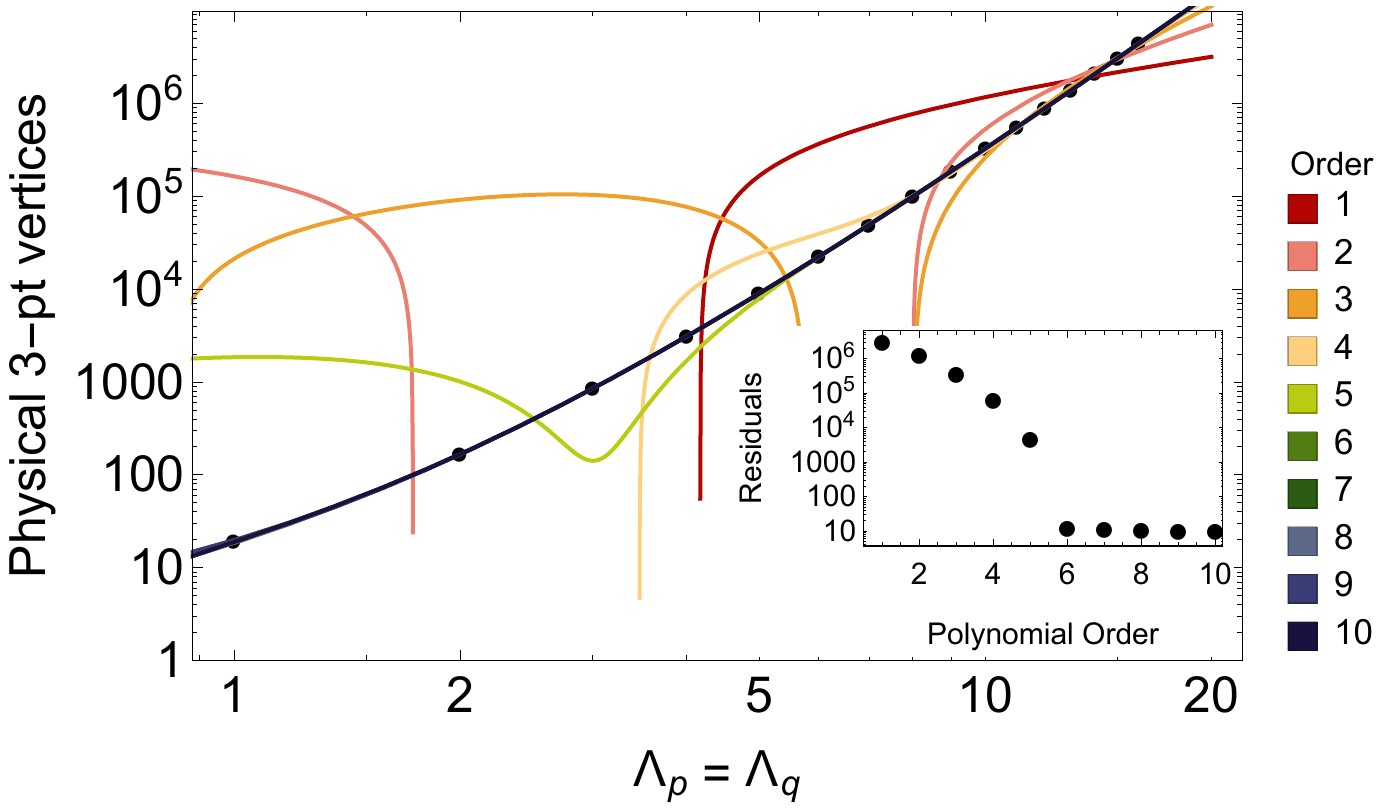}
    \put(82,50){\includegraphics[width=0.06\textwidth]{iconC1_bf.png}}
  \end{overpic}
\begin{overpic}[width = 0.47\textwidth,percent]{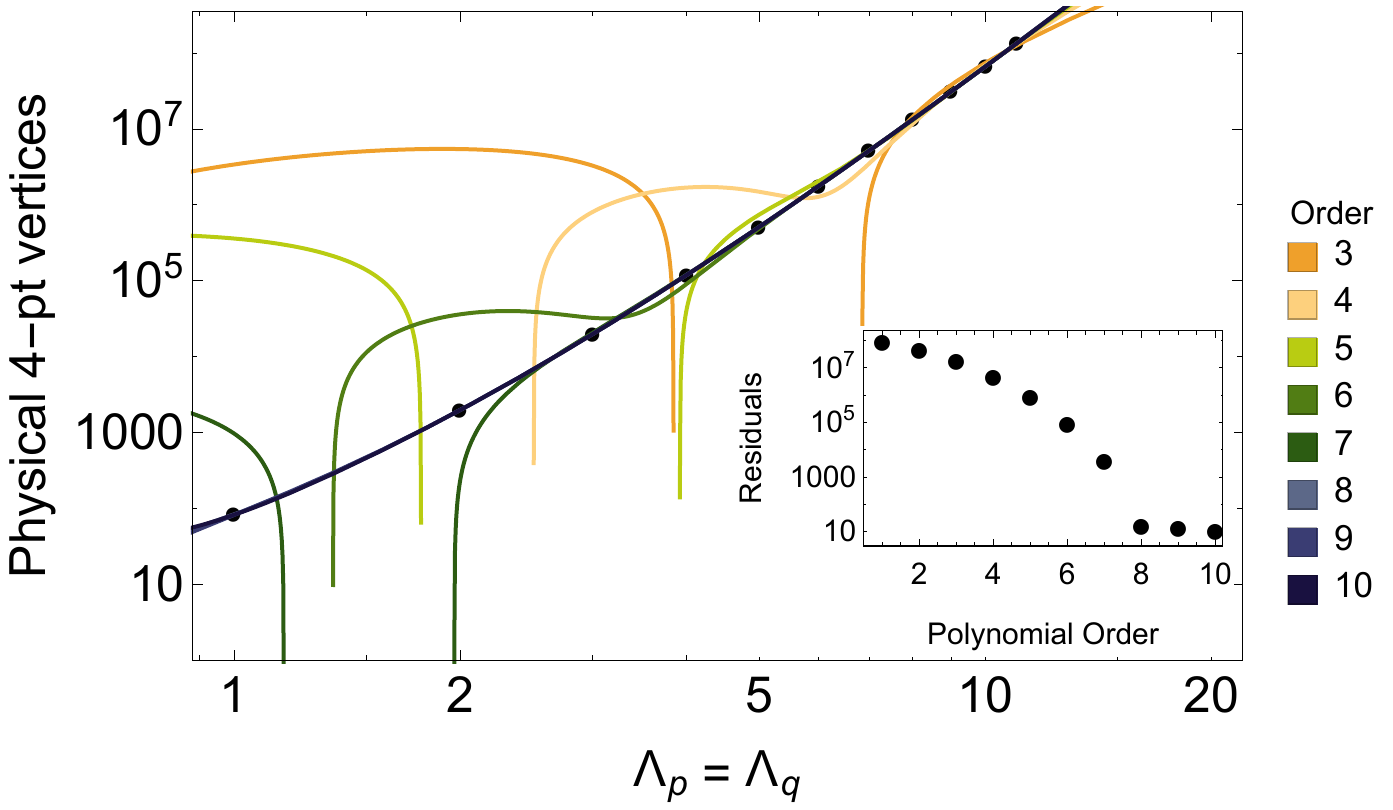}
    \put(82,50){\includegraphics[width=0.06\textwidth]{iconC1_bf.png}}
  \end{overpic}
  \caption{
The number of singlets in the product of three (left panel) and four (right panel)
color irreps as a function of the index cutoff   $\Lambda_{p}$.
The main panels show the number of singlets and polynomial fits for a range of orders,
$f(x) = \sum\limits_{n=0}^{n_{\rm max}} c_n x^n$
from $n_{\rm max}=1$ to $10$.
The inset panels shows the L2 norm of residuals in the fits.
}
\label{fig:3ptfits}
\end{figure}
Exact values computed for the fusion of modest-sized irreps are given at the left of Table~\ref{tab:singletvertexscaling}.
The left panel of Fig.~\ref{fig:3ptfits} shows the results of Table~\ref{tab:singletvertexscaling} along with polynomial fits and the associated residuals.
Significant reductions in fit residuals are found for  polynomials up to order $n=6$,
beyond which the inclusion of terms of higher degree do not improve the quality of the (single precision) fit to the exact results.
The results in Table~\ref{tab:singletvertexscaling} indicate $\Lambda_{p}^6$ scaling
 for the number of physical 3-point vertices below a local link truncation of $\Lambda_p$.
This asymptotic scaling is equivalent to that of the unconstrained Hilbert space of the locally-defined vertex, indicating that satisfying Gauss's law does not modify the asymptotic polynomial scaling of the physically relevant vertices.

Another quantity whose scaling is important is the number of plaquette control sectors.
This is lower bounded by the number of different products of four irreps that contains at least one singlet.
The unconstrained asymptotic scaling of the associated Hilbert space is $\Lambda_{p}^{8}$.
Using a similar array of techniques as those for the 3-point vertices above, explicit calculations furnish the results shown at the right of Table~\ref{tab:singletvertexscaling}.
As shown in the associated right panel of Fig.~\ref{fig:3ptfits}, the modest calculated values indicate a necessary and sufficient scaling exponent of $\Lambda_{p}^8$.
Once again, the Gauss's law constraint does not reduce the asymptotic polynomial scaling of the number of physical vertices.
The remaining stagnant residuals in both the cases of the 3-point and 4-point vertices shown in the insets of Fig.~\ref{fig:3ptfits} indicate that the Gauss's law constraint does, however, add an additional non-polynomial structure to the scaling, acting to reduce the physical dimensionality from the simple $\Lambda^{6,8}$ unconstrained values.
This strong scaling provides a potentially daunting backdrop to implementation on present and future quantum devices.
At low $\Lambda_p$ truncations, increasing the number of indices for color irreps by one in the dynamical local link basis can produce factors of $\sim10$ in the number of control sectors that are required to be implemented at each Trotter step of the plaquette time evolution.

To fully understand the circuit complexity of the local controlled plaquette operator, as discussed in Section~\ref{sec:localbasis}, with increasing gauge field truncation,
the total number of non-vanishing matrix elements within the physical subspace needs to be considered.
Explicit calculations of these properties are presented in Table~\ref{tab:plaquetteMEs}.
\begin{table}
  \begin{tabular}{cc||cc|c}
  \hline
  \hline
  $\Lambda_p = \Lambda_q$ & dimensions & physical states & matrix elements & elements/states \\
  \hline
    1 & $(\mathbf{1}, \mathbf{3})$ & 81 & 81 & 1 \\
    1 & $(\mathbf{1}, \mathbf{3},\mathbf{8})$ & 529 & 1,018 & 1.92 \\
    2 & $(\mathbf{1}, \mathbf{3},\mathbf{8}, \mathbf{6})$ & 5,937 & 19,594 & 3.30 \\
    2 & $(\mathbf{1}, \mathbf{3}, \mathbf{8},\mathbf{6}, \mathbf{15} )$ &  59,737 & 419,316 & 7.02 \\
    2 & $(\mathbf{1}, \mathbf{3},\mathbf{8}, \mathbf{6}, \mathbf{15}, \mathbf{27})$ & 139,317 & 1,049,931 & 7.54 \\
    3 & $(\mathbf{1}, \mathbf{3},\mathbf{8},\mathbf{6}, \mathbf{15}, \mathbf{27}, \mathbf{10})$ & 509,271 & 4,001,111 & 7.86 \\
    3 & $(\mathbf{1}, \mathbf{3},\mathbf{8},\mathbf{6}, \mathbf{15}, \mathbf{27}, \mathbf{10}, \mathbf{24})$ & 2,008,297 & 24,648,819 & 12.27  \\
    \hline
    \hline
  \end{tabular}
  \caption{
  Properties of the plaquette operator truncated in the local index $(p,q)$ basis and at intermediate truncations organized by dimension.  The number of physical states constituting the gauge-invariant basis of the plaquette operator,
  as well as the number of non-zero matrix elements within the physical subspace are presented.
  The ratio of these two quantities is shown in the right column. }
  \label{tab:plaquetteMEs}
\end{table}
While classically capturing dimensionalities for sufficiently high truncations to numerically constrain the scaling of the physical plaquette states is nontrivial, experience with the 3- and 4-point vertices suggests that an asymptotic polynomial scaling consistent with that of the unconstrained Hilbert space is likely.
This would result in a scaling of $\Lambda_p^{16}$.
Additionally, a clear lower bound of $\Lambda_p^{12}$ can be rationalized by the freedom of the diagonal 3-point vertices before determining physically viable values for the two remaining control links.
Crucially, the final step of determining the number of non-vanishing matrix elements within the physical space (connected to the time evolution circuit depth as discussed in Section~\ref{sec:localbasis}) does not contribute additional factors of $\Lambda_p$ to the asymptotic scaling of the number of physical plaquette states.
The irrep-locality of the plaquette operator produces a surface- rather than volume-type contribution to the dimension of physical states.
To see this, consider the active space of the plaquette operator contracting a $\mathbf{3}$ or $\overline{\mathbf{3}}$ with the irrep at each of the four active links.
As demonstrated in the connectivity diagram of Fig.~\ref{fig:connectivitydiagram}, an application of the fundamental or anti-fundamental is capable of producing only local nearest-neighbor transitions in the structure of a two-dimensional hexagonal lattice.
As such, the plaquette operator generates population in each of three new irreps for each link and thus a potential transition supported to $3^4 = 81$ different final states.
Many of the possible states generated will not satisfy Gauss's law at the four vertices, as enforced through the plaquette operator controls.
Thus, the number of non-zero matrix elements of the controlled plaquette operator is maximally a constant factor of 81 times larger than the total number of physical states.
The right column of Table~\ref{tab:plaquetteMEs} indicates that in practice, for low $\Lambda_p$-truncations, this number is significantly smaller than 81, its unconstrained upper bound.

While the scaling of the number of control structures and matrix elements in the controlled-plaquette operator
with cutoff in irrep space is a relatively high-order polynomial,
the operator is nearly local in space, extending over just a few links.
This remains the case, but involving more links, with the plethora of Hamiltonian improvements that could be implemented, for example, Refs.~\cite{Symanzik:1983dc,Luscher:1984xn,Lepage:1992xa,Alford:1995hw}.
Consequently, we expect that these operators can be determined using classical computing, and subsequently
applied repeatedly throughout the lattice volume.
With the anticipated color irrep localization of low-lying field configurations,
we do not anticipate that classical computing resources will impose a  limitation on defining the
Trotterized time evolution operator for the relevant range of lattice spacing (a range that remains to be quantified).
When increasing the gauge field truncation in this local multiplet basis,
high-order polynomial quantum resources are traded for improvement in the physical convergence.
The delocalization resulting from  reducing the lattice spacing will require controlled extrapolations
as the lattice scales are systematically removed.
To understand these features more clearly, and to be able to better estimate resource requirements,
even for modest-sized lattices, further calculations and simulations are required.
For example, the use of binary encodings, as has been used for the single plaquette in Subsection~\ref{subsec:oneplaqscaling}, may lead to slower polynomial growth, a subject of future investigations.

The  convergence in color space for low-lying states suggests that a low-dimension-color-irrep
EFT may exist.  We conjecture that plaquettes containing \enquote{high-energy links}, defined by their Casimir,
can be \enquote{integrated out} of the low-dimension-color-irrep space, with their effects reproduced by higher-dimension
gauge-invariant  operators in the Hamiltonian with coefficients determined by matching observables.
This possibility remains to be explored, and will be the subject of future work.
In order to perform precision calculations at scale, developing EFT techniques, such as this and those used to
make predictions from lattice QCD calculations, appear to be essential.

At this point it is worth commenting on instanton configurations that mediate transitions between distinct topological sectors in Yang-Mills theory
(for a review, see Ref.~\cite{Schafer:1996wv}).
Far from the center of an instanton, the field strength scales as $1/r^4$
from color fields that scale as $1/r^2$, and such configurations are expected to lead to
modifications to the na\"ive quantum resource requirements.
From a scaling perspective, these configurations are anticipated to introduce power-law structure in color space.
The resulting convergence in the presence of an instanton in 3-dimensional calculations is expected to be simply exponential in the number of qubits,
rather than a mix of single- and double-exponential convergence for the latticization and digitization, respectively, as was found in scalar field theory~\cite{Somma:2016:QSO:3179430.3179434,Jordan:2011ne,Jordan:2011ci,Macridin:2018gdw,Klco:2018zqz}.
Furthermore, motivated by the topological freezing effects experienced when updating gauge field configurations in Euclidean lattice QCD calculations~\cite{DelDebbio:2002xa,Brower:2003yx,DelDebbio:2004xh,Luscher:2010we,Schaefer:2010hu,Hasenbusch:2017fsd}, it will be important to understand the ability of quantum simulation time evolution and state preparation techniques to efficiently capture topological charge sectors.
Observed to be influential in this aspect for classical calculations, this feature further inspires the importance of thorough explorations of boundary conditions in simulation efficiency.
Reliable estimates of the impact of these configurations will only become possible
when 3-dim simulations become practical.
However, analysis of lattice QCD gauge-field configurations,
e.g., Ref.~\cite{Hackett:2018cel}, in particular in regions of topological charge density
may provide helpful information.

In this work, we have focused on controlled-plaquette operators for time evolution,
and have not presented an explicit formulation of a state preparation.
By confinement, the connected correlation functions of the vacuum are exponentially localized with a length scale
set by the mass gap.
As such, the techniques associated with exponentially convergent systematically-localizable operators and fixed-point
quantum circuits~\cite{Klco:2019yrb,Klco:2020aud} can, in principle, be implemented.
Classical computations of a lattice system containing at least a correlation length can be used to tune the parameters of a
link-based initialization quantum circuit, which can then be used throughout the larger volume of the quantum simulation.
This is, of course, limited by the lattice spacing, which if small enough would exceed classical computing resources.


\subsection{Hardware Implementation Exploratory Discussion}
\noindent
Focusing on the local controlled plaquette implementation due to its advantageous scaling,
two mappings of the color space into the Hilbert space(s) associated with each link have been considered.
The first, with a single quantum register or qudit per link, requires high connectivity among sets of 8 link registers and two-dimensional hexagonal connectivity within each.
The second, with a pair of quantum registers or qudits per link, simplifies the connectivity internal to the link space from 2D-hexagonal to a correlated set of one-dimensional hierarchies requiring only (correlated) nearest-neighbor raising and lowering operators within each qudit.
This further organization of two registers per link, one each for the upper and lower indices of the tensor describing the color irreps of the link, technically requires additional 16-register communication to implement the 8-register correlated ladder operators controlled by 8 neighboring registers.

\begin{figure}[!ht]
\begin{minipage}{0.5\textwidth}
  \includegraphics[width = 0.9\textwidth, valign = c]{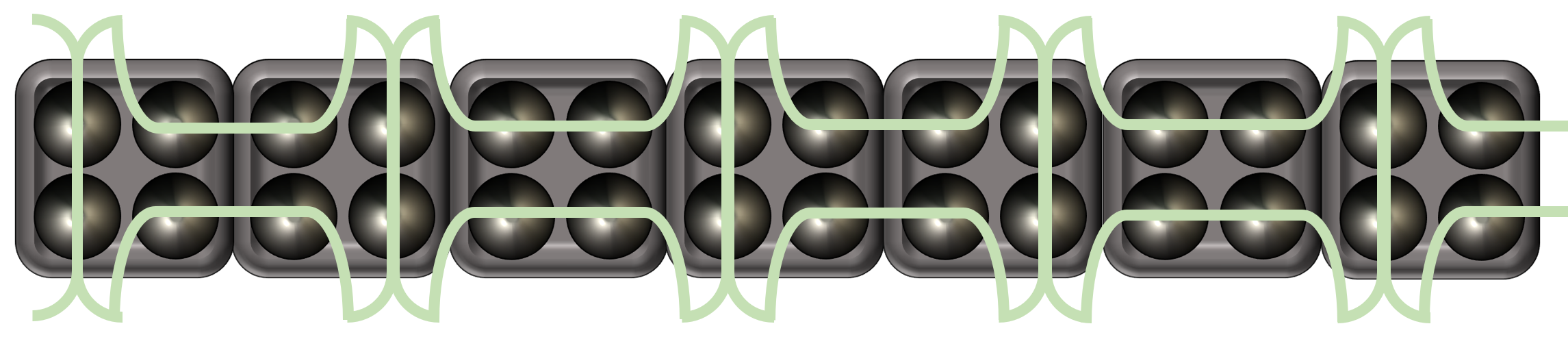}  \\
  \vspace{0.5cm}
  \includegraphics[width = 0.9\textwidth, valign = c]{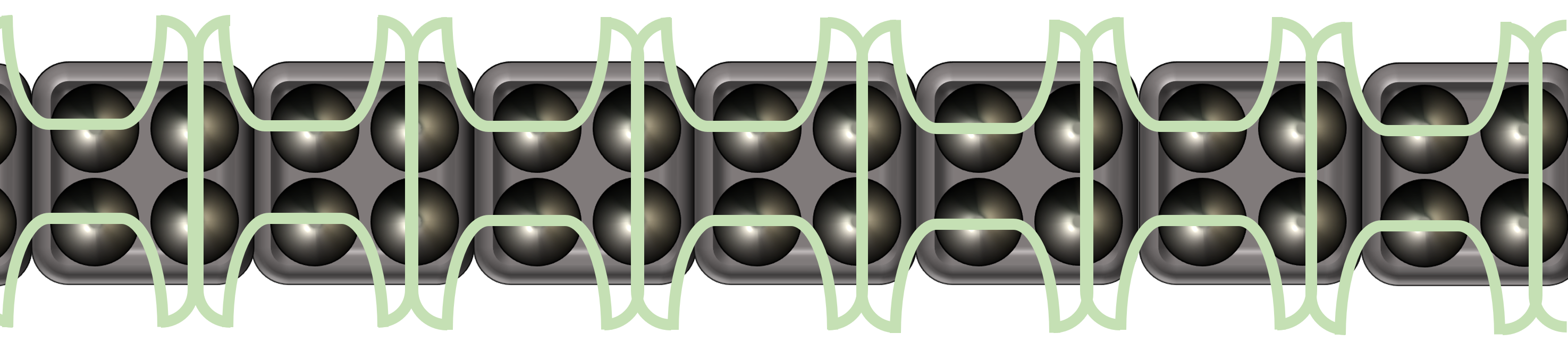}
\end{minipage}
  \includegraphics[width=0.3\textwidth, valign = c]{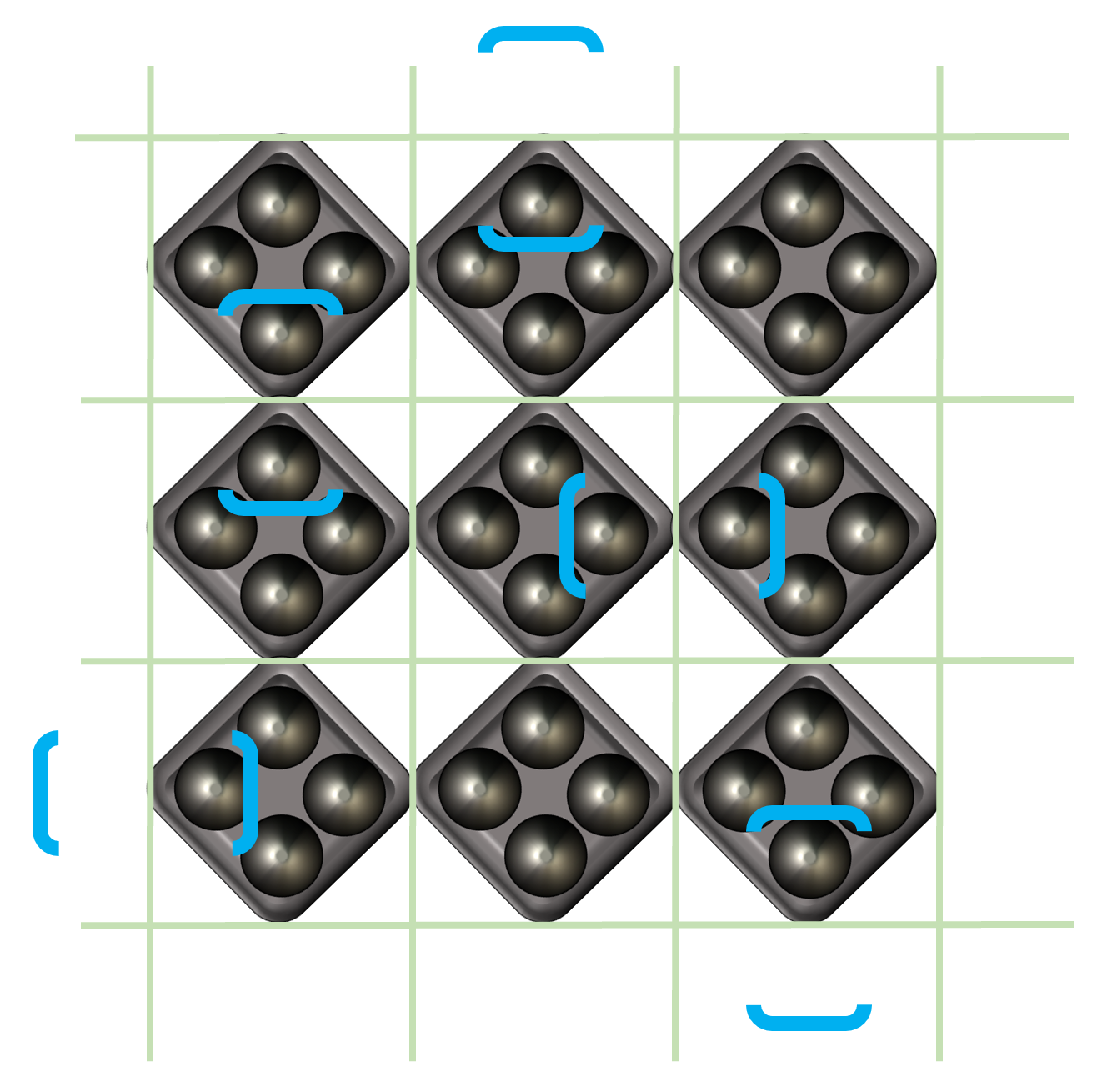}
\caption{
Plausible mappings of one (left panel)- and two (right panel)-dimensional SU(3) lattice gauge theory
onto quad-core SRF cavities utilizing the $(p,q)$ local basis.
The light green lines indicate the lattice structure.
At the left top, a one-dimensional plaquette string is illustrated in the $(p,q)$ basis with two qudits per link.
At the left bottom, the $(p,q)$ local basis is used only for the vertical links to homogenize the quad-core operations.
At the right, it is shown how an array of quad-core SRF cavities can be used to represent a
two-dimensional lattice of SU(3) gauge theory in the local $(p,q)$ basis with
blue brackets indicating the cavities used to represent the $(p,q)$ pair of qudits at local links.
}
\label{fig:quadcoredesigns}
\end{figure}

While the tactics presented here can be readily implemented on qubits, qutrits, and generally qudits,
they appear to be also suitable for the Superconducting Radio Frequency (SRF) cavity devices being advanced by Lawrence Livermore National Laboratory (LLNL)
and Fermi National Accelerator Laboratory  (FNAL)~\cite{Holland:2019zju,Romanenko:2018nut}.
Relatively large  cutoffs in color irrep space could be implemented for each link, even with today's cavities.
It would appear that a quantum communication fabric that connects eight nearest neighbor cavities
is sufficient to simulate a chain of plaquettes using controlled-plaquette operators, as detailed in this work.
The SRF cavities associated with each link and those of the four external control links would be engaged coherently.
This is somewhat more complicated than the standard hypercubic communication fabric used, for example, for lattice QCD calculations.

It may be profitable to develop SRF cavity systems with sufficient cavity interconnect and/or optimized control
to be able to implement
simulations of one- and two-plaquette SU(3) systems using the local basis (requiring 4 and 6 cavities, respectively, or twice these numbers if trading qudits for simplified intra-qudit connectivity as in the $(p,q)$ option of the local basis),
an extended one-dimensional chain of plaquettes, and a three-dimensional cube of plaquettes (requiring 12 cavities),
which would provide foundational steps toward simulations of QCD.
Diagrams are provided in Fig.~\ref{fig:quadcoredesigns} to demonstrate possible lattice connectivity through the use of quad-core SRF cavity architectures for one (left panels)- and two (right panel)-dimensional SU(3) lattices.
Two different embeddings for the one-dimensional plaquette string are imagined: the first with every link represented in the $(p,q)$ local basis with two cavities per link and the second with a mixture of the $(p,q)$ local basis on vertical links and the single-qudit local basis on horizontal links.
This mixture of local bases allows unitary operators for the magnetic time evolution to be translationally invariant among the quad-core cavities, and further emphasizes the flexibility in lattice structure amenable to hardware codesign.

Crucially, by working in localized bases, the plaquette operators designed and implemented in the process of the demonstrations suggested above remain relevant even as lattices are scaled to the infinite volume limit.
Looking further into the future, one could imagine $\sim 10^3$ SRF cavities with a
localized communication fabric and optimized controls
being used to simulate three-dimensional Yang-Mills theory on a $10\times 10\times 10$ spatial lattice
using or adapting
techniques demonstrated in this work, and their extension to higher dimensions.

Complementarily, recent advances in forming high-fidelity qudits within the hyperfine structure of trapped-ion systems,
for example Ref.~\cite{PhysRevResearch.2.033128},
indicate that exploratory calculations using the qutrit encodings presented above for the $\{\mathbf{1}, \mathbf{3}, \overline{\mathbf{3}}\}$ truncation may be implementable
on such systems in the near future.
In addition to their appealing connectivity, these systems of trapped ions are intriguingly capable of naturally generating interaction Hamiltonians that swap populations between qudit levels with a coupling constant determined by level-dependent CG sums.
While the local calculation of CG coefficients is not currently foreseen to be a limiting factor,
the presence of these natural interactions suggest a potential path forward for a hybrid digital-analog approach to constructing plaquette operators.

Looking forward toward future production-scale simulations, experience from classical lattice QCD calculations indicates that tuning QCD simulations, the lattice spacing and quark masses, will be required for each
set of calculations.
One could imagine that initial simple tunings in early productions
may involve calculations of the dynamics of one- and two-plaquette systems (and higher)
as benchmarks of device performance, both in execution time and fidelity.


\subsection{Higher Spatial Dimensions}
\noindent
Up to this point, strings of plaquettes residing in one spatial dimension have been considered.
In two spatial dimensions, four-point vertices are  required, though they can be
point-split into an expanded lattice of again only three-point vertices.
For systems in three dimensions, vertices involve the fusion of 6 links, two in each spatial dimension.
Integration over the gauge space at each vertex can be performed,
and links denoted by the irrep dimensionality, as in lower dimensions.
The controlled-plaquette operators will act in each of the spatial planes, controlled by
the total \enquote{external} color at each vertex, as dictated by Gauss's law.
Therefore, the plaquette control structure that has been developed in this work will be applicable to three dimensions, and no further structures are required.
However, at each vertex, the coherent sum over all irreps that can be formed from the external four links defines the control sectors of the operator.
This is reminiscent of the function provided by point-splitting, discussed extensively in the literature for such simulations, e.g., Refs.~\cite{PhysRevD.97.074511,Raychowdhury:2018tfj,Raychowdhury:2019iki},
here applied to the controlled plaquette operator.

This is an extra layer of complexity that has to be incorporated into the quantum circuits required to analyze the system.
Most of the extra layer is anticipated to be accomplished using classical computation, as it requires determining matrix elements of local objects, and does not increase in complexity with the system volume.
At the level of the quantum simulation, the increase of operator structure, corresponding to
an increase in the  number of link projectors (and the number of orientations), the number of quantum operations scales trivially with volume.

\section{Conclusions}
\label{sec:conc}
\noindent
The quantum simulation of lattice field theories offers a path toward computing dynamical,
non-equilibrium processes of importance for basic science and for advancing quantum technologies
that are inaccessible to classical simulation.
While classical simulations of field theories are sophisticated,
with ongoing improvements to
algorithms, workflows, hardware, infrastructure and community organization,
quantum simulations are at their very earliest stages.
Simulations of spin systems naturally map to quantum devices with quantum registers of qubits, and early
real-time calculations of elastic and in-elastic processes in low-dimensions
are being performed with present-day
quantum devices and quantum simulators.
Further,  powerful formal techniques are employed, such as tensor methods,
that are of benefit to both classical and quantum simulations.
As three of the four fundamental forces of nature are accurately described by quantum gauge field theories---describing the interactions and dynamics of quarks, gluons, electrons, electroweak bosons, and so forth---their simulation requires the inclusion of gauge fields.
Building upon a large body of work related to Hamiltonian formulations of non-Abelian quantum field theories,
and somewhat recent investigations related to their implementation on quantum devices,
we have investigated early steps along one of the possible paths forward (a \enquote{Trailhead}) for
quantum simulations of lattice SU(3) Yang-Mills gauge theory, with an eye toward QCD.
We anticipate that the results and insights gained in this work may be of benefit to the simulation of
other lattice field theories where local symmetries play a  central role and which are implemented, in part,
through the action of plaquette operators.
This includes algorithms that are relevant for quantum error correction for quantum hardware and computation.

In this work, the path toward implementation of QCD on quantum devices through the Kogut-Susskind Hamiltonian presented by
Byrnes and Yamamoto~\cite{Byrnes:2005qx} has been
adapted to reduce the qubit (qudit) requirements through integration over the local gauge space at each lattice site.
The continuous gauge field is digitized in the field conjugate representation by truncating the magnitude of the color-electric field that can be supported by any given link.
These local discrete Hilbert spaces can be captured through a basis of quantum registers distributed locally across the lattice in two main ways:
one using a mapping of irreps onto a qudit and one using two registers for the $(p,q)$ values defining tensor indices of SU(3) irreps for each link.
Applications of the plaquette operator then update each link in the plaquette, with amplitudes that depend upon irreps of the nearest neighbor links.
In contrast, a global basis can be defined by mapping symmetry-projected link configurations of the entire lattice onto the states of a quantum device, and the action of the Hamiltonian determined.
The resources required to implement local bases scale with the volume of the lattice, while those required for global
bases scale super-polynomially.
We examined, using classical simulation and IBM's superconducting quantum hardware,
the dynamics and mappings of a single SU(3) plaquette onto a quantum device, and are encouraged by
the exponential convergence of low-lying states and low-energy dynamics with increasing color-irrep cutoff.
Two-plaquette systems were  studied in detail using  global and local bases and are the simplest systems
that receive contributions from controlled-plaquette operators, the construction and implementation of which we have detailed.
While the low-lying states and time evolution of the single-plaquette systems can be efficiently
accommodated in a Hilbert space defined by qubits,
the structure of the link color-irrep Hilbert space and the action of the controlled-plaquette operator
lend themselves for embedding into qudit systems, such as qutrits or SRF cavity based systems.

The plaquette operator plays a central role in the simulation of lattice gauge theories, and the fidelity with which it
can be implemented across the Hilbert space of a given quantum device is a key measure of the ultimate quantum simulation fidelity.
Benchmarks for the performance of devices using both global bases and local bases may provide complementary information, the latter isolating the physical Hilbert space.
In addition to the one-plaquette benchmarks above, analogous benchmarks for the two-, three-, four-plaquette and higher-dimensional systems will also be valuable,
with the two-plaquette system providing a measure of the fidelity of the controlled-plaquette operator, the four-plaquette
system sensitive to the point-split four-vertex interaction, and the three-dimensional systems sensitive to 6-link vertices.
In near-term simulations, it is likely that a series of such benchmarks, starting from the single plaquette in the global basis, will be performed to identify preferred mappings onto device architectures and provide calibrations for simulations of larger systems.
Furthermore, such calibrations are expected to begin the process of tuning simulation parameters, such as the cutoff in color irrep space, balancing theory approximations with device performance for the array of lattices and couplings that will eventually be necessary to extract continuum quantities.

At these early stages of development, there is value in considering, quantifying, and comparing all potential implementations
of Yang-Mills gauge theories on quantum devices.
Beyond the time evolution operator, the finite mass gap in Yang-Mills theories suggests that systematically localizable quantum circuits may be realizable
to prepare the ground state and localized scattering states through the use of small-volume classical simulations.
Unfortunately, the asymptotic polynomial scaling of the required matrix elements, while sub-exponential, is sufficiently significant to expect that large-scale classical resources will be required for both state preparation and design of the controlled-plaquette operators.

Building upon previous frameworks,
the multiplet basis has here been further developed through local gauge-space integrations and circuit-level decomposition of local time evolution operators, showing promise for designing scalable quantum simulations of SU(3) Yang-Mills lattice gauge field theory.
Early examples of the proposed strategies have been concretely demonstrated, through implementation on a superconducting quantum architecture and through explicit enumeration of relevant operators, for one- and two-plaquette systems.
These simple systems are expected to guide near-term quantum simulations, inform future codesign, and provide calibration quantities for simulations at scale.

\begin{acknowledgments}
We would like to thank Silas Beane, Doug Beck, David Kaplan, Aidan Murran, and Alessandro Roggero for valuable discussions.
We acknowledge the use of IBM Quantum services for this work. The views expressed are those of the authors, and do not reflect the official policy or position of IBM or the IBM Quantum team.
Data presented throughout this manuscript is available upon email request.
AC was supported in part by Fermi National Accelerator Laboratory PO No. 652197.
NK is supported in part by the Walter Burke Institute for Theoretical Physics, and by the U.S. Department of Energy Office of Science, Office of Advanced Scientific Computing Research, (DE-SC0020290), and Office of High Energy Physics DE-ACO2-07CH11359.
MJS was supported in part by the U.S. Department of Energy, Office of Science, Office of Nuclear Physics, Inqubator for Quantum Simulation (IQuS) under Award Number DOE (NP) Award DE-SC0020970.
\end{acknowledgments}

\appendix

\section{Plaquette Matrix Elements}
\label{app:plaquetteMEs}

A derivation of plaquette operator matrix elements is presented after classical incorporation of the local vector indices.
Due to the single-link delocalization of the plaquette operator in a generically structured Hilbert space,
as discussed in Section~\ref{subsec:plaquetteop}, calculating matrix elements on a three plaquette lattice will be sufficient for evaluations of plaquette strings of arbitrary length.

Employing the labels of Fig.~\ref{fig:labels}, the wavefunction for the three-plaquette system with open boundary conditions is
\begin{multline}
    \Bigg|
    \begin{pmatrix}
     \mathbf{C}_1, \mathbf{R}_t, \mathbf{C}_3 \\
     \mathbf{Q}_\ell, \mathbf{Q}_r \\
     \mathbf{C}_2, \mathbf{R}_b, \mathbf{C}_4
   \end{pmatrix}
   \Bigg\rangle
   =
    \frac{1}{\dim(\mathbf{Q}_\ell) \dim (\mathbf{Q}_r)} \sum_{all} |\mathbf{C}_1, a, b\rangle |\mathbf{Q}_\ell, c, d\rangle | \mathbf{C}_2, e, f\rangle |\mathbf{R}_t, g, h\rangle |\mathbf{Q}_r, i, j\rangle  |\mathbf{R}_b, k, \ell\rangle |\mathbf{C}_3, m, n\rangle |\mathbf{C}_4, p, q \rangle \\
  \CG{\mathbf{C}_1}{b}{\overline{\mathbf{R}}_t}{g}{\overline{\mathbf{Q}}_\ell}{d}_{\Gamma_1}
  \CG{\mathbf{R}_t}{h}{\overline{\mathbf{C}}_3}{m}{\overline{\mathbf{Q}}_r}{j}_{\Gamma_2}
  \CG{\mathbf{C}_2}{f}{\overline{\mathbf{R}}_b}{k}{\mathbf{Q}_\ell}{c}_{\Gamma_3}
  \CG{\mathbf{R}_b}{\ell}{\overline{\mathbf{C}}_4}{p}{\mathbf{Q}_r}{i}_{\Gamma_4} \ \ \ ,
\end{multline}
where the sum is over all component indices,  characterized by color isospin and color hypercharge quantum numbers.
The normalization of this state has been determined through the application of the CG sum relation,
\begin{equation}
  \sum_{a, \alpha, a'} | \CG{\mathbf{R}_1}{a}{\mathbf{R}_2}{\alpha}{\mathbf{R}'}{a'}_{\Gamma}|^2 = \dim(\mathbf{R}') \qquad
  \ \ ,\ \
  \qquad  \mathbf{R}' \in \mathbf{R}_1 \otimes \mathbf{R}_2
  \ \ \ .
  \label{eq:CGsum}
\end{equation}
Application of the plaquette operator can be expressed in terms of four local link operators,
\begin{multline}
  \hat U_{\alpha, \beta}^{(\mathbf{3})} \hat U_{\beta, \gamma}^{(\mathbf{3})}
  \left( \hat U_{\gamma, \delta}^{(\mathbf{3})} \right)^\dagger
  \left( \hat U_{\delta, \alpha}^{(\mathbf{3})}\right)^\dagger
    \Bigg|
    \begin{pmatrix}
     \mathbf{C}_1, \mathbf{R}_t, \mathbf{C}_3 \\
     \mathbf{Q}_\ell, \mathbf{Q}_r \\
     \mathbf{C}_2, \mathbf{R}_b, \mathbf{C}_4
   \end{pmatrix}
   \Bigg\rangle
  \ = \
  \frac{1}{\dim(\mathbf{Q}_\ell) \dim (\mathbf{Q}_r)} \sum_{all} \\
  \CG{\mathbf{C}_1}{b}{\overline{\mathbf{R}}_t}{g}{\overline{\mathbf{Q}}_\ell}{d}_{\Gamma_1}
  \CG{\mathbf{R}_t}{h}{\overline{\mathbf{C}}_3}{m}{\overline{\mathbf{Q}}_r}{j}_{\Gamma_2}
  \CG{\mathbf{C}_2}{f}{\overline{\mathbf{R}}_b}{k}{\mathbf{Q}_\ell}{c}_{\Gamma_3}
  \CG{\mathbf{R}_b}{\ell}{\overline{\mathbf{C}}_4}{p}{\mathbf{Q}_r}{i}_{\Gamma_4} \\
  |\mathbf{C}_1, a, b\rangle
    \hat U_{\alpha, \delta}^{(\bar{\mathbf{3}})}|\mathbf{Q}_\ell, c, d\rangle
  |\mathbf{C}_2, e, f\rangle
    \hat U_{\delta, \gamma}^{(\bar{\mathbf{3}})}|\mathbf{R}_t, g, h\rangle
    \hat U_{\beta, \gamma}^{(\mathbf{3})}|\mathbf{Q}_r, i, j\rangle
    \hat U_{\alpha, \beta}^{(\mathbf{3})}|\mathbf{R}_b, k, \ell\rangle
  |\mathbf{C}_{3}, m, n\rangle
  |\mathbf{C}_{4}, p, q\rangle \ \ \ .
\end{multline}
Explicitly acting in the left and right spaces, the link operator functions as,
\begin{align}
    \hat U_{\ell r}^{(\mathbf{3})} |\mathbf{R}, a, b\rangle &=
      \hat U_{\ell r}^{(\mathbf{3})} | \overline{\mathbf{R}}, a\rangle \otimes |\mathbf{R}, b\rangle \ \ \ , \nonumber \\
  &= \sum_{\oplus \mathbf{R}', \vec{\Gamma}} \sqrt{\frac{\dim \mathbf{R}}{\dim \mathbf{R}'}} \  |\bar{\mathbf{R}}', a'\rangle
  \CGstar{\overline{\mathbf{R}}}{a}{\overline{\mathbf{3}}}{\ell}{\overline{\mathbf{R}}'}{a'}_{\Gamma_1}
  \otimes |\mathbf{R}, b\rangle
  \CGstar{\mathbf{R}}{b}{\mathbf{3}}{r}{\mathbf{R}'}{b'}_{\Gamma_2} \ \ \ , \nonumber
 \\
 &= \sum_{\oplus \mathbf{R}', \vec{\Gamma}} \sqrt{\frac{\dim \mathbf{R}}{\dim \mathbf{R}'}} \ |\mathbf{R}', a',b'\rangle \ \CG{\mathbf{R}}{a}{\mathbf{3}}{\ell}{\mathbf{R}'}{a'}_{\Gamma_1}
 \CGstar{\mathbf{R}}{b}{\mathbf{3}}{r}{\mathbf{R}'}{b'}_{\Gamma_2} \ \ \ ,
\end{align}
leading to a final state of,
\begin{multline}
  \hat{\Box}
      \Bigg|
    \begin{pmatrix}
     \mathbf{C}_1, \mathbf{R}_t, \mathbf{C}_3 \\
     \mathbf{Q}_\ell, \mathbf{Q}_r \\
     \mathbf{C}_2, \mathbf{R}_b, \mathbf{C}_4
   \end{pmatrix}
   \Bigg\rangle
  = \frac{1}{\dim(\mathbf{Q}_\ell) \dim (\mathbf{Q}_r)} \sum_{all}\sum_{\oplus \vec{\mathbf{R}}', \vec{\mathbf{Q}}', \vec{\Gamma}} \sqrt{\frac{\dim(\mathbf{R}_t) \dim(\mathbf{R}_b) \dim(\mathbf{Q}_\ell) \dim(\mathbf{Q}_r)}{\dim(\mathbf{R}'_t) \dim(\mathbf{R}'_b) \dim(\mathbf{Q}'_\ell) \dim(\mathbf{Q}'_r)}} \\
  \CG{\mathbf{C}_1}{b}{\overline{\mathbf{R}}_t}{g}{\overline{\mathbf{Q}}_\ell}{d}_{\Gamma_1}
  \CG{\mathbf{R}_t}{h}{\overline{\mathbf{C}}_3}{m}{\overline{\mathbf{Q}}_r}{j}_{\Gamma_2}
  \CG{\mathbf{C}_2}{f}{\overline{\mathbf{R}}_b}{k}{\mathbf{Q}_\ell}{c}_{\Gamma_3}
  \CG{\mathbf{R}_b}{\ell}{\overline{\mathbf{C}}_4}{p}{\mathbf{Q}_r}{i}_{\Gamma_4}
  \\
  \CG{\mathbf{R}_b}{k}{\mathbf{3}}{\alpha}{\mathbf{R}_b'}{k'}_{\Gamma_5}
  \CGstar{\mathbf{R}_b}{\ell}{\mathbf{3}}{\beta}{\mathbf{R}_b'}{\ell'}_{\Gamma_6}
  \\
  \CG{\mathbf{Q}_r}{i}{\mathbf{3}}{\beta}{\mathbf{Q}_r'}{i'}_{\Gamma_7}
  \CGstar{\mathbf{Q}_r}{j}{\mathbf{3}}{\gamma}{\mathbf{Q}_r'}{j'}_{\Gamma_8} \\
  \CG{\mathbf{R}_t}{g}{\overline{\mathbf{3}}}{\delta}{\mathbf{R}_t'}{g'}_{\Gamma_9}
  \CGstar{\mathbf{R}_t}{h}{\overline{\mathbf{3}}}{\gamma}{\mathbf{R}_t'}{h'}_{\Gamma_{10}}
  \\
  \CG{\mathbf{Q}_\ell}{c}{\overline{\mathbf{3}}}{\alpha}{\mathbf{Q}_\ell'}{c'}_{\Gamma_{11}}
  \CGstar{\mathbf{Q}_\ell}{d}{\overline{\mathbf{3}}}{\delta}{\mathbf{Q}_\ell'}{d'}_{\Gamma_{12}}
  \\
   |\mathbf{C}_1, a, b\rangle |\mathbf{Q}_\ell', c', d'\rangle | \mathbf{C}_2, e, f\rangle |\mathbf{R}_t', g', h'\rangle |\mathbf{Q}_r', i', j'\rangle  |\mathbf{R}_b', k', \ell'\rangle |\mathbf{C}_3, m, n\rangle |\mathbf{C}_4, p, q \rangle \ \ \ .
\end{multline}
Contraction with a final state produces the following  matrix elements,
\begin{multline}
  \Bigg\langle \begin{pmatrix}
     \mathbf{C}_1, \mathbf{R}_t', \mathbf{C}_3 \\
     \mathbf{Q}_\ell', \mathbf{Q}_r' \\
     \mathbf{C}_2, \mathbf{R}_b', \mathbf{C}_4
   \end{pmatrix} \Bigg|
   \hat{\Box}
   \Bigg| \begin{pmatrix}
     \mathbf{C}_1, \mathbf{R}_t, \mathbf{C}_3 \\
     \mathbf{Q}_\ell, \mathbf{Q}_r \\
     \mathbf{C}_2, \mathbf{R}_b, \mathbf{C}_4
   \end{pmatrix}
   \Bigg\rangle =
   \\ \frac{1}{\dim (\mathbf{Q}_\ell) \dim(\mathbf{Q}_r) \dim (\mathbf{Q}_\ell') \dim(\mathbf{Q}_r') } \sum_{all, \vec{\Gamma}} \sqrt{\frac{\dim(\mathbf{R}_t) \dim(\mathbf{R}_b) \dim(\mathbf{Q}_\ell) \dim(\mathbf{Q}_r)}{\dim(\mathbf{R}'_t) \dim(\mathbf{R}'_b) \dim(\mathbf{Q}'_\ell) \dim(\mathbf{Q}'_r)}}
   \\
  \CG{\mathbf{C}_1}{b}{\overline{\mathbf{R}}_t}{g}{\overline{\mathbf{Q}}_\ell}{d}_{\Gamma_1}
  \CG{\mathbf{R}_t}{h}{\overline{\mathbf{C}}_3}{m}{\overline{\mathbf{Q}}_r}{j}_{\Gamma_2}
  \CG{\mathbf{C}_2}{f}{\overline{\mathbf{R}}_b}{k}{\mathbf{Q}_\ell}{c}_{\Gamma_3}
  \CG{\mathbf{R}_b}{\ell}{\overline{\mathbf{C}}_4}{p}{\mathbf{Q}_r}{i}_{\Gamma_4}
  \\
  \CG{\mathbf{R}_b}{k}{\mathbf{3}}{\alpha}{\mathbf{R}_b'}{k'}_{\Gamma_5}
  \CGstar{\mathbf{R}_b}{\ell}{\mathbf{3}}{\beta}{\mathbf{R}_b'}{\ell'}_{\Gamma_6}
  \\
  \CG{\mathbf{Q}_r}{i}{\mathbf{3}}{\beta}{\mathbf{Q}_r'}{i'}_{\Gamma_7}
  \CGstar{\mathbf{Q}_r}{j}{\mathbf{3}}{\gamma}{\mathbf{Q}_r'}{j'}_{\Gamma_8} \\
  \CG{\mathbf{R}_t}{g}{\overline{\mathbf{3}}}{\delta}{\mathbf{R}_t'}{g'}_{\Gamma_9}
  \CGstar{\mathbf{R}_t}{h}{\overline{\mathbf{3}}}{\gamma}{\mathbf{R}_t'}{h'}_{\Gamma_{10}}
  \\
  \CG{\mathbf{Q}_\ell}{c}{\overline{\mathbf{3}}}{\alpha}{\mathbf{Q}_\ell'}{c'}_{\Gamma_{11}}
  \CGstar{\mathbf{Q}_\ell}{d}{\overline{\mathbf{3}}}{\delta}{\mathbf{Q}_\ell'}{d'}_{\Gamma_{12}}
  \\
  \CGstar{\mathbf{C}_1}{b}{\overline{\mathbf{R}}_t'}{g'}{\overline{\mathbf{Q}}_\ell'}{d'}_{\Gamma_{13}}
  \CGstar{\mathbf{R}_t'}{h'}{\overline{\mathbf{C}}_3}{m}{\overline{\mathbf{Q}}_r'}{j'}_{\Gamma_{14}}
  \CGstar{\mathbf{C}_2}{f}{\overline{\mathbf{R}}_b'}{k'}{\mathbf{Q}_\ell'}{c'}_{\Gamma_{15}}
  \CGstar{\mathbf{R}_b'}{\ell'}{\overline{\mathbf{C}}_4}{p}{\mathbf{Q}_r'}{i'}_{\Gamma_{16}} \ \ \ .
\end{multline}
This expression can be collected into four vertex factors
\begin{multline}
  \Bigg\langle \begin{pmatrix}
     \mathbf{C}_1, \mathbf{R}_t', \mathbf{C}_3 \\
     \mathbf{Q}_\ell', \mathbf{Q}_r' \\
     \mathbf{C}_2, \mathbf{R}_b', \mathbf{C}_4
   \end{pmatrix} \Bigg| \hat{\Box}
   \Bigg|
   \begin{pmatrix}
     \mathbf{C}_1, \mathbf{R}_t, \mathbf{C}_3 \\
     \mathbf{Q}_\ell, \mathbf{Q}_r \\
     \mathbf{C}_2, \mathbf{R}_b, \mathbf{C}_4
   \end{pmatrix} \Bigg\rangle = \\ \sqrt{\frac{\dim(\mathbf{R}_t) \dim(\mathbf{R}_b) }{\dim(\mathbf{R}'_t) \dim(\mathbf{R}'_b) \dim(\mathbf{Q}_\ell) \dim(\mathbf{Q}_r) \dim(\mathbf{Q}'_\ell)^3 \dim(\mathbf{Q}'_r)^3}}\\
   \sum  \CG{\mathbf{C}_1}{b}{\overline{\mathbf{R}}_t}{g}{\overline{\mathbf{Q}}_\ell}{d}_{\Gamma_1}
   \CG{\mathbf{R}_t}{g}{\overline{\mathbf{3}}}{\delta}{\mathbf{R}_t'}{g'}_{\Gamma_2}
   \CGstar{\mathbf{Q}_\ell}{d}{\overline{\mathbf{3}}}{\delta}{\mathbf{Q}_\ell'}{d'}_{\Gamma_3}
   \CGstar{\mathbf{C}_1}{b}{\overline{\mathbf{R}}_t'}{g'}{\overline{\mathbf{Q}}_\ell'}{d'}_{\Gamma_4}
   \\
   \sum
   \CG{\mathbf{R}_t}{h}{\overline{\mathbf{C}}_3}{m}{\overline{\mathbf{Q}}_r}{j}_{\Gamma_5}
   \CGstar{\mathbf{Q}_r}{j}{\mathbf{3}}{\gamma}{\mathbf{Q}_r'}{j'}_{\Gamma_6}
   \CGstar{\mathbf{R}_t}{h}{\overline{\mathbf{3}}}{\gamma}{\mathbf{R}_t'}{h'}_{\Gamma_7}
   \CGstar{\mathbf{R}_t'}{h'}{\overline{\mathbf{C}}_3}{m}{\overline{\mathbf{Q}}_r'}{j'}_{\Gamma_8}
   \\
   \sum
   \CG{\mathbf{C}_2}{f}{\overline{\mathbf{R}}_b}{k}{\mathbf{Q}_\ell}{c}_{\Gamma_9}
   \CG{\mathbf{R}_b}{k}{\mathbf{3}}{\alpha}{\mathbf{R}_b'}{k'}_{\Gamma_{10}}
   \CG{\mathbf{Q}_\ell}{c}{\overline{\mathbf{3}}}{\alpha}{\mathbf{Q}_\ell'}{c'}_{\Gamma_{11}}
   \CGstar{\mathbf{C}_2}{f}{\overline{\mathbf{R}}_b'}{k'}{\mathbf{Q}_\ell'}{c'}_{\Gamma_{12}}
   \\
   \sum
   \CG{\mathbf{R}_b}{\ell}{\overline{\mathbf{C}}_4}{p}{\mathbf{Q}_r}{i}_{\Gamma_{13}}
   \CGstar{\mathbf{R}_b}{\ell}{\mathbf{3}}{\beta}{\mathbf{R}_b'}{\ell'}_{\Gamma_{14}}
   \CG{\mathbf{Q}_r}{i}{\mathbf{3}}{\beta}{\mathbf{Q}_r'}{i'}_{\Gamma_{15}}
   \CGstar{\mathbf{R}_b'}{\ell'}{\overline{\mathbf{C}}_4}{p}{\mathbf{Q}_r'}{i'}_{\Gamma_{16}} \ \ \ .
\end{multline}
Using a phase convention of real CGs, thus equivalent in the left and right spaces, and a full-conjugation of the irreps in
4 of the 16 above CGs to standardize indices, yields the identification of plaquette matrix elements as written in Eq.~\eqref{eq:plaquetteArbMEs}.

Below is presented the collection of physical plaquette matrix elements of Eq.~\eqref{eq:plaquetteArbMEs} for the 8 unique control sectors of the local plaquette operator in the truncated qutrit space $\{\mathbf{1}, \mathbf{3}, \overline{\mathbf{3}}\}$.
\begin{table}[!ht]
\begin{tabular}{c|ccc}
\hline
\hline
$\controlsectorc{\mathbf{1}}{\mathbf{1}}{\mathbf{1}}{\mathbf{1}}$ & $\plaquettestateket{\mathbf{1}}{\mathbf{1}}{\mathbf{1}}{\mathbf{1}}$ & $\plaquettestateket{\overline{\mathbf{3}}}{\overline{\mathbf{3}}}{\mathbf{3}}{\mathbf{3}}$ & $\plaquettestateket{\mathbf{3}}{\mathbf{3}}{\overline{\mathbf{3}}}{\overline{\mathbf{3}}}$ \\
\hline
$\plaquettestatebra{\mathbf{1}}{\mathbf{1}}{\mathbf{1}}{\mathbf{1}}$ & $(0,0)$ & $(1,0)$ & $(0,1)$ \\
$\plaquettestatebra{\overline{\mathbf{3}}}{\overline{\mathbf{3}}}{\mathbf{3}}{\mathbf{3}}$ & $(0,1)$ & $(0,0)$ & $(1,0)$ \\
$\plaquettestatebra{\mathbf{3}}{\mathbf{3}}{\overline{\mathbf{3}}}{\overline{\mathbf{3}}}$ & $(1,0)$ & $(0,1)$ & $(0,0)$ \\
\hline
\hline
\end{tabular}
\begin{tabular}{c|ccc}
\hline
\hline
$\controlsectorc{\mathbf{1}}{\mathbf{1}}{\mathbf{3}}{\overline{\mathbf{3}}}$ & $\plaquettestateket{\mathbf{1}}{\mathbf{3}}{\mathbf{1}}{\mathbf{1}}$ & $\plaquettestateket{\overline{\mathbf{3}}}{\mathbf{1}}{\mathbf{3}}{\mathbf{3}}$ & $\plaquettestateket{\mathbf{3}}{\overline{\mathbf{3}}}{\overline{\mathbf{3}}}{\overline{\mathbf{3}}}$ \\
\hline
$\plaquettestatebra{\mathbf{1}}{\mathbf{3}}{\mathbf{1}}{\mathbf{1}}$ & $(0,0)$ & $(\frac{1}{3},0)$ & $(0,\frac{1}{\sqrt{3}})$ \\
$\plaquettestatebra{\overline{\mathbf{3}}}{\mathbf{1}}{\mathbf{3}}{\mathbf{3}}$ & $(0,\frac{1}{3})$ & $(0,0)$ & $(\frac{1}{\sqrt{3}},0)$ \\
$\plaquettestatebra{\mathbf{3}}{\overline{\mathbf{3}}}{\overline{\mathbf{3}}}{\overline{\mathbf{3}}}$ & $(\frac{1}{\sqrt{3}},0)$ & $(0,\frac{1}{\sqrt{3}})$ & $(0,0)$ \\
\hline
\hline
\end{tabular}
\end{table}
\begin{table}[ht]
\begin{tabular}{c|ccc}
\hline
\hline
$\controlsectorc{\mathbf{1}}{\mathbf{3}}{\mathbf{1}}{\mathbf{3}}$ & $\plaquettestateket{\mathbf{1}}{\overline{\mathbf{3}}}{\mathbf{3}}{\mathbf{3}}$ & $\plaquettestateket{\overline{\mathbf{3}}}{\mathbf{3}}{\overline{\mathbf{3}}}{\overline{\mathbf{3}}}$ & $\plaquettestateket{\mathbf{3}}{\mathbf{1}}{\mathbf{1}}{\mathbf{1}}$ \\
\hline
$\plaquettestatebra{\mathbf{1}}{\overline{\mathbf{3}}}{\mathbf{3}}{\mathbf{3}}$ & $(0,0)$ & $(\frac{1}{\sqrt{3}},0)$ & $(0,\frac{1}{3})$ \\
$\plaquettestatebra{\overline{\mathbf{3}}}{\mathbf{3}}{\overline{\mathbf{3}}}{\overline{\mathbf{3}}}$ & $(0,\frac{1}{\sqrt{3}})$ & $(0,0)$ & $(\frac{1}{\sqrt{3}},0)$ \\
$\plaquettestatebra{\mathbf{3}}{\mathbf{1}}{\mathbf{1}}{\mathbf{1}}$ & $(\frac{1}{3},0)$ & $(0,\frac{1}{\sqrt{3}})$ & $(0,0)$ \\
\hline
\hline
\end{tabular}
\begin{tabular}{c|ccc}
\hline
\hline
$\controlsectorc{\mathbf{1}}{\mathbf{3}}{\mathbf{3}}{\mathbf{1}}$ & $\plaquettestateket{\mathbf{1}}{\mathbf{1}}{\mathbf{3}}{\mathbf{3}}$ & $\plaquettestateket{\overline{\mathbf{3}}}{\overline{\mathbf{3}}}{\overline{\mathbf{3}}}{\overline{\mathbf{3}}}$ & $\plaquettestateket{\mathbf{3}}{\mathbf{3}}{\mathbf{1}}{\mathbf{1}}$ \\
\hline
$\plaquettestatebra{\mathbf{1}}{\mathbf{1}}{\mathbf{3}}{\mathbf{3}}$ & $(0,0)$ & $(-\frac{1}{\sqrt{3}},0)$ & $(0,\frac{1}{3})$ \\
$\plaquettestatebra{\overline{\mathbf{3}}}{\overline{\mathbf{3}}}{\overline{\mathbf{3}}}{\overline{\mathbf{3}}}$ & $(0,-\frac{1}{\sqrt{3}})$ & $(0,0)$ & $(-\frac{1}{\sqrt{3}},0)$ \\
$\plaquettestatebra{\mathbf{3}}{\mathbf{3}}{\mathbf{1}}{\mathbf{1}}$ & $(\frac{1}{3},0)$ & $(0,-\frac{1}{\sqrt{3}})$ & $(0,0)$ \\
\hline
\hline
\end{tabular}
\end{table}
\begin{table}[ht]
\begin{tabular}{c|ccc}
\hline
\hline
$\controlsectorc{\mathbf{1}}{\mathbf{3}}{\overline{\mathbf{3}}}{\overline{\mathbf{3}}}$ & $\plaquettestateket{\mathbf{1}}{\mathbf{3}}{\mathbf{3}}{\mathbf{3}}$ & $\plaquettestateket{\overline{\mathbf{3}}}{\mathbf{1}}{\overline{\mathbf{3}}}{\overline{\mathbf{3}}}$ & $\plaquettestateket{\mathbf{3}}{\overline{\mathbf{3}}}{\mathbf{1}}{\mathbf{1}}$ \\
\hline
$\plaquettestatebra{\mathbf{1}}{\mathbf{3}}{\mathbf{3}}{\mathbf{3}}$ & $(0,0)$ & $(\frac{1}{3},0)$ & $(0,-\frac{1}{3})$ \\
$\plaquettestatebra{\overline{\mathbf{3}}}{\mathbf{1}}{\overline{\mathbf{3}}}{\overline{\mathbf{3}}}$ & $(0,\frac{1}{3})$ & $(0,0)$ & $(-\frac{1}{3},0)$ \\
$\plaquettestatebra{\mathbf{3}}{\overline{\mathbf{3}}}{\mathbf{1}}{\mathbf{1}}$ & $(-\frac{1}{3},0)$ & $(0,-\frac{1}{3})$ & $(0,0)$ \\
\hline
\hline
\end{tabular}
\begin{tabular}{c|ccc}
\hline
\hline
$\controlsectorc{\mathbf{3}}{\mathbf{3}}{\mathbf{3}}{\mathbf{3}}$ & $\plaquettestateket{\mathbf{1}}{\overline{\mathbf{3}}}{\overline{\mathbf{3}}}{\mathbf{3}}$ & $\plaquettestateket{\overline{\mathbf{3}}}{\mathbf{3}}{\mathbf{1}}{\overline{\mathbf{3}}}$ & $\plaquettestateket{\mathbf{3}}{\mathbf{1}}{\mathbf{3}}{\mathbf{1}}$ \\
\hline
$\plaquettestatebra{\mathbf{1}}{\overline{\mathbf{3}}}{\overline{\mathbf{3}}}{\mathbf{3}}$ & $(0,0)$ & $(\frac{1}{3},0)$ & $(0,\frac{1}{3 \sqrt{3}})$ \\
$\plaquettestatebra{\overline{\mathbf{3}}}{\mathbf{3}}{\mathbf{1}}{\overline{\mathbf{3}}}$ & $(0,\frac{1}{3})$ & $(0,0)$ & $(\frac{1}{3 \sqrt{3}},0)$ \\
$\plaquettestatebra{\mathbf{3}}{\mathbf{1}}{\mathbf{3}}{\mathbf{1}}$ & $(\frac{1}{3 \sqrt{3}},0)$ & $(0,\frac{1}{3 \sqrt{3}})$ & $(0,0)$ \\
\hline
\hline
\end{tabular}
\end{table}
\begin{table}[ht]
\begin{tabular}{c|ccc}
\hline
\hline
$\controlsectorc{\mathbf{3}}{\overline{\mathbf{3}}}{\mathbf{3}}{\overline{\mathbf{3}}}$ & $\plaquettestateket{\mathbf{1}}{\mathbf{3}}{\mathbf{1}}{\overline{\mathbf{3}}}$ & $\plaquettestateket{\overline{\mathbf{3}}}{\mathbf{1}}{\mathbf{3}}{\mathbf{1}}$ & $\plaquettestateket{\mathbf{3}}{\overline{\mathbf{3}}}{\overline{\mathbf{3}}}{\mathbf{3}}$ \\
\hline
$\plaquettestatebra{\mathbf{1}}{\mathbf{3}}{\mathbf{1}}{\overline{\mathbf{3}}}$ & $(0,0)$ & $(\frac{1}{9},0)$ & $(0,\frac{1}{3})$ \\
$\plaquettestatebra{\overline{\mathbf{3}}}{\mathbf{1}}{\mathbf{3}}{\mathbf{1}}$ & $(0,\frac{1}{9})$ & $(0,0)$ & $(\frac{1}{3},0)$ \\
$\plaquettestatebra{\mathbf{3}}{\overline{\mathbf{3}}}{\overline{\mathbf{3}}}{\mathbf{3}}$ & $(\frac{1}{3},0)$ & $(0,\frac{1}{3})$ & $(0,0)$ \\
\hline
\hline
\end{tabular}
\begin{tabular}{c|ccc}
\hline
\hline
$\controlsectorc{\mathbf{3}}{\overline{\mathbf{3}}}{\overline{\mathbf{3}}}{\mathbf{3}}$ & $\plaquettestateket{\mathbf{1}}{\overline{\mathbf{3}}}{\mathbf{1}}{\overline{\mathbf{3}}}$ & $\plaquettestateket{\overline{\mathbf{3}}}{\mathbf{3}}{\mathbf{3}}{\mathbf{1}}$ & $\plaquettestateket{\mathbf{3}}{\mathbf{1}}{\overline{\mathbf{3}}}{\mathbf{3}}$ \\
\hline
$\plaquettestatebra{\mathbf{1}}{\overline{\mathbf{3}}}{\mathbf{1}}{\overline{\mathbf{3}}}$ & $(0,0)$ & $(\frac{1}{3 \sqrt{3}},0)$ & $(0,\frac{1}{3 \sqrt{3}})$ \\
$\plaquettestatebra{\overline{\mathbf{3}}}{\mathbf{3}}{\mathbf{3}}{\mathbf{1}}$ & $(0,\frac{1}{3 \sqrt{3}})$ & $(0,0)$ & $(\frac{1}{3},0)$ \\
$\plaquettestatebra{\mathbf{3}}{\mathbf{1}}{\overline{\mathbf{3}}}{\mathbf{3}}$ & $(\frac{1}{3 \sqrt{3}},0)$ & $(0,\frac{1}{3})$ & $(0,0)$ \\
\hline
\hline
\end{tabular}
\end{table} 
\section{One SU(2) Plaquette}
\label{app:onesu2plaquette}
\noindent
The SU(2) gauge theory of a single plaquette is interesting to study
in the context of SU(3)
as it allows a first glimpse of what may be expected for the behavior of
color fields associated with low-energy states in  larger systems.
Of course, SU(2) Hamiltonian gauge theory has been extensively studied in the
past, for example, Refs.~\cite{Chin:1985ua,Piekarewicz:1988mx}.
The Hamiltonian for SU(2) Yang-Mills theory for one plaquette
is similar to that for SU(3) in form,
\begin{eqnarray}
\hat H & = &
\frac{g^2}{2} \sum_{\rm a, links} |{\bf E}^a|^2
\ +\
\frac{1}{2 g^2}\ \left( 4 \ -\ \hat\Box - \hat\Box^\dagger\ \right)
\ \ \ \ .
\label{eq:su21plaq}
\end{eqnarray}
Working with normalized states that satisfy Gauss's law at each of the four vertices, and in the basis eigenstates of the Casimir operator, $|\chi_j\rangle$,
the Hamiltonian matrix in this basis is,
\begin{eqnarray}
H_{j,j^\prime}& = &
\frac{1}{2}g^2  j(j+1) \delta_{j,j^\prime}
\ +\
\frac{1}{g^2}
\left( 2 \delta_{j,j^\prime} - \delta_{j+1,j^\prime}- \delta_{j-1,j^\prime} \right)
\ \ \ ,
\label{eq:ham1j}
\end{eqnarray}
where $j, j^\prime \ge 0$.
Recognizing the form of the magnetic operator as the single-separation finite-difference
approximation to $\nabla^2$, and neglecting boundary and positivity issues, and taking the continuum limit in j-space,  the Hamiltonian can then be written as,
\begin{eqnarray}
\hat H & \rightarrow &
\frac{1}{2}g^2 \hat J^2 - \frac{1}{g^2} \nabla_j^2
\ \ \ ,
\label{eq:hamcont1}
\end{eqnarray}
acting in $j$-space, where $\hat J^2$ is the SU(2) Casimir operator.
The eigenstates of the low-lying wavefunctions  satisfy,
\begin{eqnarray}
\nabla_j^2\psi(j) \ +\ \left(g^2 E - \frac{1}{2} g^4 j(j+1) \right)\psi(j) = 0
\ \ \ ,
\label{eq:wave1}
\end{eqnarray}
a Weber type-A differential equation, with Parabolic Cylinder functions as solutions.
Asymptotically for large j, these  functions scale as
\begin{eqnarray}
\psi & \rightarrow &
e^{-\frac{g^2}{2\sqrt{2}} (j+\frac{1}{2})^2}
\ \ \ ,
\label{eq:psiasym}
\end{eqnarray}
demonstrating Gaussian convergence in field space.
Considerations have to be given to the discrete nature of the field, to the boundary conditions at $j=0$, and so forth, but for wavefunctions with support over many j-values,
the approximate dependence on large $j$ is expected to be given by Eq.~(\ref{eq:psiasym}).
This result is  encouraging as it suggests a rapid convergence in j-space for $j\gsim 4/g$,
which provides a  guide for the extent of field values in a finite dimensional Hilbert space defining the plaquette.  To obtain comparable fidelity for somewhat higher lying states, only modest increases in maximum field values are expected to be required.
This distribution resembles that of the low-lying field configurations in scalar field theory.

Extending this discussion to multi-plaquette systems, it is anticipated that the support of the field on any given link in the low-lying states will have a localized distribution that is similar to Eq.~(\ref{eq:psiasym}) up to polynomial corrections.  This remains to be verified by direct simulation. Further, the comparison between SU(2) and SU(3) is complicated by the non-linear behavior in color space of SU(3).  However, the expectation is that SU(3) color fields are also localized in irrep space, in a way that resembles the behavior found in SU(2).

\section{Benchmarks for Single Plaquette Time Evolution}
\label{app:benchmarks}
\noindent
It is important to establish robust benchmarks to guide future quantum simulations of gauge field theories.
Section~\ref{sec:singleplaquette} of the main text explores  bases within the multiplet digitization of
an SU(3) plaquette and their consequences for digital quantum simulation.
To complement,  subsection~\ref{subsec:benchmarks}
focuses on the effectiveness of simulations in capturing local maxima and minima in time evolution.
Tables~\ref{table:ElectricMax} and~\ref{table:ElectricMin} provided in this appendix show the numerical values in
Fig.~\ref{fig:minmaxbenchmark} of the main text.
\begin{table}[ht!]
	\centering
	\begin{tabular}{|c|c|c|c|c|c|c|}
		\hline
		\multicolumn{7}{|c|}{First Maximum of $\langle H_E \rangle $ for  $g=1$} \\
		\hline
		Basis Truncation & Exact &\  Trotter Steps\  &\  Order\  & \ CNOTs\  & \ Trotterized \ &\  {\tt Athens}\  \\
		\hline
		Global Basis $p,q\rightarrow \infty$ & 0.9389 & - & - & - & - & - \\
        Color Parity truncated at $\mathbf{3}$ & 0.7967 & - & - & 0 & - & \ $0.829\left({\tiny\begin{matrix}+0.039\\-0.045\end{matrix}}\right)$\  \\
		Global Basis $p,q \leq 1$ &\ \  0.8699 \ \ & 1 & 2 & 3 & 1.1602 & $1.06 \pm 0.08$ \\
		Global Basis $p,q \leq 1$ & 0.8699 & 2 & 2 & 6 & 0.9019 & $0.91 \pm 0.18$ \\
		Global Basis $p,q \leq 1$ & 0.8699 & 3 & 2 & 9 & 0.8837 & $0.9 \pm 0.4$ \\
		Global Basis $p,q \leq 1$ & 0.8699 & 4 & 2 & 12 & 0.8776 &  $1.0 \pm 0.5$ \\
		Color Parity truncated at $\mathbf{6}$ & 0.9296 & 1 & 1 & 5 & 4.2582 & $4.3 \pm 0.4 $ \\
		Color Parity truncated at $\mathbf{6}$ & 0.9296 & 2 & 1 & 10 & 1.8280 & $1.9 \pm 0.5$ \\
		\ Color Parity truncated at $\mathbf{6}$\  & 0.9296 & 1 & 2 & 7 & 0.8820 & $0.94 \pm 0.24$ \\
		\hline
	\end{tabular}
	\caption{
	The first maximum of $\langle H_E \rangle $ for  $g=1$
	in the time evolution of the trivial vacuum for a single plaquette.
	The columns of entries correspond to:
	(first) the basis truncation,
	(second) results of exact time evolution performed on a classical computer,
	(third) the number of Trotter steps,
	(fourth) the order of the Trotterization,
	(fifth) the number of CNOT's used on the {\tt Athens} quantum processor,
	(sixth) the results of Trotterized time evolution using a classical computer,
	(seventh) the results of Trotterized time evolution obtained from IBM's {\tt Athens} quantum processor.
		}
	\label{table:ElectricMax}
\end{table}
\begin{table}[ht!]
	\centering
	\begin{tabular}{ |c|c|c|c|c|c|c| }
		\hline
		\multicolumn{7}{|c|}{First Minimum of $\langle H_E \rangle $ for  $g=1$} \\
		\hline
		Basis Truncation & Exact &  \ Trotter Steps \ &\  Order\  & \ CNOTs\  & \ Trotterized\  & {\tt Athens} \\
		\hline
		Global Basis $p,q\rightarrow \infty$ & \ \ 0.0234 \ \ & - & - & - & - & - \\
        Color Parity truncated at $\mathbf{3}$ & 0.0000 & - & - & 0 & - & \ $0.0037 \left({\tiny\begin{matrix}+0.012\\-0.0033\end{matrix}}\right)$\  \\
		Global Basis $p,q \leq 1$ & 0.0096 & 1 & 2 & 3 & 0.0000 & $0.04 \pm 0.08$ \\
		Global Basis $p,q \leq 1$ & 0.0096 & 2 & 2 & 6 & 0.0803 & $0.11 \pm 0.1$ \\
		Global Basis $p,q \leq 1$ & 0.0096 & 3 & 2 & 9 & 0.0452 & $0.18 \pm 0.29$ \\
		Global Basis $p,q \leq 1$ & 0.0096 & 4 & 2 & 12 & 0.0140 & $0.09 \pm 0.24$  \\
		Color Parity truncated at $\mathbf{6}$ & 0.0206 & 1 & 1 & 5 & 2.782 & - \\
		Color Parity truncated at $\mathbf{6}$ & 0.0206 & 2 & 1 & 10 & 1.1840 & $1.1 \pm 0.7$ \\
		\ Color Parity truncated at $\mathbf{6}$\  & 0.0206 & 1 & 2 & 7 & 0.1555 & $0.25 \pm 0.2$ \\
		\hline
	\end{tabular}
	\caption{
		The first minimum of $\langle H_E \rangle $ for  $g=1$
	in the time evolution of the trivial vacuum for a singe plaquette.
	The columns of entries correspond to:
	(first) the basis truncation,
	(second) results of exact time evolution performed on a classical computer,
	(third) the number of Trotter steps,
	(fourth) the order of the Trotterization,
	(fifth) the number of CNOT's used on the {\tt Athens} quantum processor,
	(sixth) the results of Trotterized time evolution using a classical computer,
	(seventh) the results of Trotterized time evolution obtained from IBM's {\tt Athens} quantum processor.
In the simulations, if a minimum was not obtained in the time evolution, the seventh column contains a dash.
}
	\label{table:ElectricMin}
\end{table} 

\section{SU(3) Clebsch Gordan Coefficients}
\label{app:su3CG}

The methods proposed in the main text for the implementation of real time dynamics of SU(3) lattice gauge theory through the use of quantum devices currently require the classical calculation of CG tensorial projections in the contraction of SU(3) irreducible representations.
While the calculation of generic CG coefficients has been classified
as \#P-complete and thus a daunting dependency, the restriction to fixed rank (2 for SU(3)) has been identified as a sufficient condition for the existence of a polynomial time algorithm~\cite{Narayanan2006,de2006computation}.
As shown in Eq.~\eqref{eq:plaquetteArbMEs}, matrix elements of the local plaquette operator can be formulated through 9-R composite CG vertex factors.
These vertex factors and plaquette matrix elements are independent of the basis chosen to enumerate the states within each irrep.
Due to discrepancies we have
observed in the CG coefficients calculated
with publicly accessible codes  when multiplicities occur in tensor decomposition
(e.g., Ref~\cite{Alex:2010wi}),  an explicit set of 9-R symbols for irreps appearing at index truncation
$\Lambda_p = \Lambda_q = 1$
is provided in Eq.~\eqref{eq:explicit9Rs}.
\begin{align}
  \nineR{1}{3}{3}{\overline{3}}{1}{\overline{3}}{\overline{3}}{3}{1} &= 1 \qquad & \nineR{1}{3}{3}{\overline{3}}{1}{\overline{3}}{\overline{3}}{3}{8} &= 8 \qquad &
  \nineR{3}{\overline{3}}{1}{\overline{3}}{1}{\overline{3}}{8}{\overline{3}}{\overline{3}} &= 2\sqrt{2} \nonumber \\
  \nineR{3}{\overline{3}}{1}{3}{1}{3}{\overline{3}}{\overline{3}}{3}
  &= \sqrt{3} \qquad &
  \nineR{3}{3}{\overline{3}}{3}{1}{3}{\overline{3}}{3}{1}
  &= -1 \qquad &
  \nineR{3}{3}{\overline{3}}{3}{1}{3}{\overline{3}}{3}{8} &= 4 \label{eq:explicit9Rs}\\
  \nineR{3}{3}{\overline{3}}{\overline{3}}{1}{\overline{3}}{8}{3}{3}
  &= - \sqrt{6} \qquad &
  \nineR{8}{8}{8}{3}{1}{3}{3}{8}{3} &= \frac{3}{4}\left( \sqrt{5}+3\right)
  \qquad & \nineR{8}{8}{8}{\overline{3}}{1}{\overline{3}}{\overline{3}}{8}{\overline{3}}
  &= \frac{3}{4} \left( \sqrt{5} - 3\right) \nonumber
\end{align}
Importantly, note that the antisymmetric contribution within the $\mathbf{8}\otimes \mathbf{8} \rightarrow \mathbf{8}$  CG produces a factor of (-1) upon conjugation of the last 9-R symbol in this set.  

\section{\texorpdfstring{Local Operators for $\{\mathbf{1},\mathbf{3}, \overline{\mathbf{3}}\}$ Magnetic Time Evolution}{Explicit Local Operators}}
\label{app:explicitlocalops133}

In this appendix, we provide an explicit enumeration of the 81 operators in the $\{\mathbf{1}, \mathbf{3}, \overline{\mathbf{3}}\}$
local basis contributing to Trotterized magnetic time evolution,
see Eqns.~\eqref{eq:boxboxD1}-\eqref{eq:BoxBoxD}.
\begingroup
\setlength{\extrarowheight}{.5em}
  \begin{longtable}{@{}cc|c|cc@{}}
    \hline
    \hline
    \multicolumn{2}{c}{Control Sector} & & \multicolumn{2}{c}{Givens Rotation} \\
    $\{\mathbf{C}_1, \mathbf{C}_2, \mathbf{C}_3, \mathbf{C}_4\}$ & $\{\mathbf{C}_1, \mathbf{C}_2, \mathbf{C}_3, \mathbf{C}_4\}_{(p,q)}$ &  coefficient & $\{\mathbf{R}_b, \mathbf{Q}_r, \mathbf{R}_t, \mathbf{Q}_\ell\}$ & $\{\mathbf{R}_b, \mathbf{Q}_r, \mathbf{R}_t, \mathbf{Q}_\ell\}_{(p,q)}$ \\
    \hline
    \hline
    \multirow{3}{*}{$\{\mathbf{1}, \mathbf{1}, \mathbf{1}, \mathbf{1}\}$} & \multirow{3}{*}{$\{00000000 \} $}  &  $ 1 $ & $ \mathcal{X}_{01} \mathcal{X}_{01} \mathcal{X}_{02} \mathcal{X}_{02}  $ & $ \mathcal{X}_{01} \mathbb{I} \mathcal{X}_{01} \mathbb{I} \mathbb{I} \mathcal{X}_{01} \mathbb{I} \mathcal{X}_{01}  $ \\ & &  $ 1 $ & $ \mathcal{X}_{02} \mathcal{X}_{02} \mathcal{X}_{01} \mathcal{X}_{01}  $ & $ \mathbb{I} \mathcal{X}_{01} \mathbb{I} \mathcal{X}_{01} \mathcal{X}_{01} \mathbb{I} \mathcal{X}_{01} \mathbb{I}  $ \\ & &  $ 1 $ & $ \mathcal{X}_{12} \mathcal{X}_{12} \mathcal{X}_{12} \mathcal{X}_{12}  $ & $ \mathcal{X}_{01} \mathcal{X}_{01} \mathcal{X}_{01} \mathcal{X}_{01} \mathcal{X}_{01} \mathcal{X}_{01} \mathcal{X}_{01} \mathcal{X}_{01}  $ \\
\hline
\multirow{3}{*}{$\{\mathbf{1}, \mathbf{1}, \mathbf{3}, \overline{\mathbf{3}}\}$} & \multirow{3}{*}{$\{00001001 \} $}  &  $ \frac{1}{\sqrt{3}} $ & $ \mathcal{X}_{01} \mathcal{X}_{12} \mathcal{X}_{02} \mathcal{X}_{02}  $ & $ \mathcal{X}_{01} \mathbb{I} \mathcal{X}_{01} \mathcal{X}_{01} \mathbb{I} \mathcal{X}_{01} \mathbb{I} \mathcal{X}_{01}  $ \\& &  $ \frac{1}{3} $ & $ \mathcal{X}_{02} \mathcal{X}_{01} \mathcal{X}_{01} \mathcal{X}_{01}  $ & $ \mathbb{I} \mathcal{X}_{01} \mathcal{X}_{01} \mathbb{I} \mathcal{X}_{01} \mathbb{I} \mathcal{X}_{01} \mathbb{I}  $ \\& &  $ \frac{1}{\sqrt{3}} $ & $ \mathcal{X}_{12} \mathcal{X}_{02} \mathcal{X}_{12} \mathcal{X}_{12}  $ & $ \mathcal{X}_{01} \mathcal{X}_{01} \mathbb{I} \mathcal{X}_{01} \mathcal{X}_{01} \mathcal{X}_{01} \mathcal{X}_{01} \mathcal{X}_{01}  $ \\
\hline
\multirow{3}{*}{$\{\mathbf{1}, \mathbf{3}, \mathbf{1}, \mathbf{3}\}$} & \multirow{3}{*}{$\{00100010 \} $}  &  $ \frac{1}{3} $ & $ \mathcal{X}_{01} \mathcal{X}_{02} \mathcal{X}_{01} \mathcal{X}_{01}  $ & $ \mathcal{X}_{01} \mathbb{I} \mathbb{I} \mathcal{X}_{01} \mathcal{X}_{01} \mathbb{I} \mathcal{X}_{01} \mathbb{I}  $ \\& &  $ \frac{1}{\sqrt{3}} $ & $ \mathcal{X}_{02} \mathcal{X}_{12} \mathcal{X}_{12} \mathcal{X}_{12}  $ & $ \mathbb{I} \mathcal{X}_{01} \mathcal{X}_{01} \mathcal{X}_{01} \mathcal{X}_{01} \mathcal{X}_{01} \mathcal{X}_{01} \mathcal{X}_{01}  $ \\& &  $ \frac{1}{\sqrt{3}} $ & $ \mathcal{X}_{12} \mathcal{X}_{01} \mathcal{X}_{02} \mathcal{X}_{02}  $ & $ \mathcal{X}_{01} \mathcal{X}_{01} \mathcal{X}_{01} \mathbb{I} \mathbb{I} \mathcal{X}_{01} \mathbb{I} \mathcal{X}_{01}  $ \\
\hline
\multirow{3}{*}{$\{\mathbf{1}, \mathbf{3}, \mathbf{3}, \mathbf{1}\}$} & \multirow{3}{*}{$\{00101000 \} $}  &  $ \frac{1}{3} $ & $ \mathcal{X}_{01} \mathcal{X}_{01} \mathcal{X}_{01} \mathcal{X}_{01}  $ & $ \mathcal{X}_{01} \mathbb{I} \mathcal{X}_{01} \mathbb{I} \mathcal{X}_{01} \mathbb{I} \mathcal{X}_{01} \mathbb{I}  $ \\& &  $ -\frac{1}{\sqrt{3}} $ & $ \mathcal{X}_{02} \mathcal{X}_{02} \mathcal{X}_{12} \mathcal{X}_{12}  $ & $ \mathbb{I} \mathcal{X}_{01} \mathbb{I} \mathcal{X}_{01} \mathcal{X}_{01} \mathcal{X}_{01} \mathcal{X}_{01} \mathcal{X}_{01}  $ \\& &  $ -\frac{1}{\sqrt{3}} $ & $ \mathcal{X}_{12} \mathcal{X}_{12} \mathcal{X}_{02} \mathcal{X}_{02}  $ & $ \mathcal{X}_{01} \mathcal{X}_{01} \mathcal{X}_{01} \mathcal{X}_{01} \mathbb{I} \mathcal{X}_{01} \mathbb{I} \mathcal{X}_{01}  $ \\
\hline
\multirow{3}{*}{$\{\mathbf{1}, \mathbf{3}, \overline{\mathbf{3}}, \overline{\mathbf{3}}\}$} & \multirow{3}{*}{$\{00100101 \} $}  &  $ -\frac{1}{3} $ & $ \mathcal{X}_{01} \mathcal{X}_{12} \mathcal{X}_{01} \mathcal{X}_{01}  $ & $ \mathcal{X}_{01} \mathbb{I} \mathcal{X}_{01} \mathcal{X}_{01} \mathcal{X}_{01} \mathbb{I} \mathcal{X}_{01} \mathbb{I}  $ \\& &  $ \frac{1}{3} $ & $ \mathcal{X}_{02} \mathcal{X}_{01} \mathcal{X}_{12} \mathcal{X}_{12}  $ & $ \mathbb{I} \mathcal{X}_{01} \mathcal{X}_{01} \mathbb{I} \mathcal{X}_{01} \mathcal{X}_{01} \mathcal{X}_{01} \mathcal{X}_{01}  $ \\& &  $ -\frac{1}{3} $ & $ \mathcal{X}_{12} \mathcal{X}_{02} \mathcal{X}_{02} \mathcal{X}_{02}  $ & $ \mathcal{X}_{01} \mathcal{X}_{01} \mathbb{I} \mathcal{X}_{01} \mathbb{I} \mathcal{X}_{01} \mathbb{I} \mathcal{X}_{01}  $ \\
\hline
\multirow{3}{*}{$\{\mathbf{3}, \mathbf{3}, \mathbf{3}, \mathbf{3}\}$} & \multirow{3}{*}{$\{10101010 \} $}  &  $ \frac{1}{3 \sqrt{3}} $ & $ \mathcal{X}_{01} \mathcal{X}_{02} \mathcal{X}_{12} \mathcal{X}_{01}  $ & $ \mathcal{X}_{01} \mathbb{I} \mathbb{I} \mathcal{X}_{01} \mathcal{X}_{01} \mathcal{X}_{01} \mathcal{X}_{01} \mathbb{I}  $ \\& &  $ \frac{1}{3} $ & $ \mathcal{X}_{02} \mathcal{X}_{12} \mathcal{X}_{02} \mathcal{X}_{12}  $ & $ \mathbb{I} \mathcal{X}_{01} \mathcal{X}_{01} \mathcal{X}_{01} \mathbb{I} \mathcal{X}_{01} \mathcal{X}_{01} \mathcal{X}_{01}  $ \\& &  $ \frac{1}{3 \sqrt{3}} $ & $ \mathcal{X}_{12} \mathcal{X}_{01} \mathcal{X}_{01} \mathcal{X}_{02}  $ & $ \mathcal{X}_{01} \mathcal{X}_{01} \mathcal{X}_{01} \mathbb{I} \mathcal{X}_{01} \mathbb{I} \mathbb{I} \mathcal{X}_{01}  $ \\
\hline
\multirow{3}{*}{$\{\mathbf{3}, \overline{\mathbf{3}}, \mathbf{3}, \overline{\mathbf{3}}\}$} & \multirow{3}{*}{$\{10011001 \} $}  &  $ \frac{1}{3} $ & $ \mathcal{X}_{01} \mathcal{X}_{12} \mathcal{X}_{02} \mathcal{X}_{12}  $ & $ \mathcal{X}_{01} \mathbb{I} \mathcal{X}_{01} \mathcal{X}_{01} \mathbb{I} \mathcal{X}_{01} \mathcal{X}_{01} \mathcal{X}_{01}  $ \\& &  $ \frac{1}{9} $ & $ \mathcal{X}_{02} \mathcal{X}_{01} \mathcal{X}_{01} \mathcal{X}_{02}  $ & $ \mathbb{I} \mathcal{X}_{01} \mathcal{X}_{01} \mathbb{I} \mathcal{X}_{01} \mathbb{I} \mathbb{I} \mathcal{X}_{01}  $ \\& &  $ \frac{1}{3} $ & $ \mathcal{X}_{12} \mathcal{X}_{02} \mathcal{X}_{12} \mathcal{X}_{01}  $ & $ \mathcal{X}_{01} \mathcal{X}_{01} \mathbb{I} \mathcal{X}_{01} \mathcal{X}_{01} \mathcal{X}_{01} \mathcal{X}_{01} \mathbb{I}  $ \\
\hline
\multirow{3}{*}{$\{\mathbf{3}, \overline{\mathbf{3}}, \overline{\mathbf{3}}, \mathbf{3}\}$} & \multirow{3}{*}{$\{10010110 \} $}  &  $ \frac{1}{3 \sqrt{3}} $ & $ \mathcal{X}_{01} \mathcal{X}_{02} \mathcal{X}_{02} \mathcal{X}_{12}  $ & $ \mathcal{X}_{01} \mathbb{I} \mathbb{I} \mathcal{X}_{01} \mathbb{I} \mathcal{X}_{01} \mathcal{X}_{01} \mathcal{X}_{01}  $ \\& &  $ \frac{1}{3 \sqrt{3}} $ & $ \mathcal{X}_{02} \mathcal{X}_{12} \mathcal{X}_{01} \mathcal{X}_{02}  $ & $ \mathbb{I} \mathcal{X}_{01} \mathcal{X}_{01} \mathcal{X}_{01} \mathcal{X}_{01} \mathbb{I} \mathbb{I} \mathcal{X}_{01}  $ \\& &  $ \frac{1}{3} $ & $ \mathcal{X}_{12} \mathcal{X}_{01} \mathcal{X}_{12} \mathcal{X}_{01}  $ & $ \mathcal{X}_{01} \mathcal{X}_{01} \mathcal{X}_{01} \mathbb{I} \mathcal{X}_{01} \mathcal{X}_{01} \mathcal{X}_{01} \mathbb{I}  $ \\
\hline
\multirow{3}{*}{$\{\mathbf{1}, \mathbf{1}, \overline{\mathbf{3}}, \mathbf{3}\}$} & \multirow{3}{*}{$\{00000110 \} $}  &  $ \frac{1}{3} $ & $ \mathcal{X}_{01} \mathcal{X}_{02} \mathcal{X}_{02} \mathcal{X}_{02}  $ & $ \mathcal{X}_{01} \mathbb{I} \mathbb{I} \mathcal{X}_{01} \mathbb{I} \mathcal{X}_{01} \mathbb{I} \mathcal{X}_{01}  $ \\& &  $ \frac{1}{\sqrt{3}} $ & $ \mathcal{X}_{02} \mathcal{X}_{12} \mathcal{X}_{01} \mathcal{X}_{01}  $ & $ \mathbb{I} \mathcal{X}_{01} \mathcal{X}_{01} \mathcal{X}_{01} \mathcal{X}_{01} \mathbb{I} \mathcal{X}_{01} \mathbb{I}  $ \\& &  $ \frac{1}{\sqrt{3}} $ & $ \mathcal{X}_{12} \mathcal{X}_{01} \mathcal{X}_{12} \mathcal{X}_{12}  $ & $ \mathcal{X}_{01} \mathcal{X}_{01} \mathcal{X}_{01} \mathbb{I} \mathcal{X}_{01} \mathcal{X}_{01} \mathcal{X}_{01} \mathcal{X}_{01}  $ \\
\hline
\multirow{3}{*}{$\{\mathbf{3}, \mathbf{1}, \mathbf{3}, \mathbf{1}\}$} & \multirow{3}{*}{$\{10001000 \} $}  &  $ \frac{1}{3} $ & $ \mathcal{X}_{01} \mathcal{X}_{01} \mathcal{X}_{01} \mathcal{X}_{02}  $ & $ \mathcal{X}_{01} \mathbb{I} \mathcal{X}_{01} \mathbb{I} \mathcal{X}_{01} \mathbb{I} \mathbb{I} \mathcal{X}_{01}  $ \\& &  $ \frac{1}{\sqrt{3}} $ & $ \mathcal{X}_{02} \mathcal{X}_{02} \mathcal{X}_{12} \mathcal{X}_{01}  $ & $ \mathbb{I} \mathcal{X}_{01} \mathbb{I} \mathcal{X}_{01} \mathcal{X}_{01} \mathcal{X}_{01} \mathcal{X}_{01} \mathbb{I}  $ \\& &  $ \frac{1}{\sqrt{3}} $ & $ \mathcal{X}_{12} \mathcal{X}_{12} \mathcal{X}_{02} \mathcal{X}_{12}  $ & $ \mathcal{X}_{01} \mathcal{X}_{01} \mathcal{X}_{01} \mathcal{X}_{01} \mathbb{I} \mathcal{X}_{01} \mathcal{X}_{01} \mathcal{X}_{01}  $ \\
\hline
\multirow{3}{*}{$\{\mathbf{3}, \mathbf{1}, \mathbf{1}, \mathbf{3}\}$} & \multirow{3}{*}{$\{10000010 \} $}  &  $ \frac{1}{3} $ & $ \mathcal{X}_{01} \mathcal{X}_{02} \mathcal{X}_{01} \mathcal{X}_{02}  $ & $ \mathcal{X}_{01} \mathbb{I} \mathbb{I} \mathcal{X}_{01} \mathcal{X}_{01} \mathbb{I} \mathbb{I} \mathcal{X}_{01}  $ \\& &  $ -\frac{1}{\sqrt{3}} $ & $ \mathcal{X}_{02} \mathcal{X}_{12} \mathcal{X}_{12} \mathcal{X}_{01}  $ & $ \mathbb{I} \mathcal{X}_{01} \mathcal{X}_{01} \mathcal{X}_{01} \mathcal{X}_{01} \mathcal{X}_{01} \mathcal{X}_{01} \mathbb{I}  $ \\& &  $ -\frac{1}{\sqrt{3}} $ & $ \mathcal{X}_{12} \mathcal{X}_{01} \mathcal{X}_{02} \mathcal{X}_{12}  $ & $ \mathcal{X}_{01} \mathcal{X}_{01} \mathcal{X}_{01} \mathbb{I} \mathbb{I} \mathcal{X}_{01} \mathcal{X}_{01} \mathcal{X}_{01}  $ \\
\hline
\multirow{3}{*}{$\{\mathbf{3}, \mathbf{1}, \overline{\mathbf{3}}, \overline{\mathbf{3}}\}$} & \multirow{3}{*}{$\{10000101 \} $}  &  $ -\frac{1}{3} $ & $ \mathcal{X}_{01} \mathcal{X}_{12} \mathcal{X}_{01} \mathcal{X}_{02}  $ & $ \mathcal{X}_{01} \mathbb{I} \mathcal{X}_{01} \mathcal{X}_{01} \mathcal{X}_{01} \mathbb{I} \mathbb{I} \mathcal{X}_{01}  $ \\& &  $ -\frac{1}{3} $ & $ \mathcal{X}_{02} \mathcal{X}_{01} \mathcal{X}_{12} \mathcal{X}_{01}  $ & $ \mathbb{I} \mathcal{X}_{01} \mathcal{X}_{01} \mathbb{I} \mathcal{X}_{01} \mathcal{X}_{01} \mathcal{X}_{01} \mathbb{I}  $ \\& &  $ \frac{1}{3} $ & $ \mathcal{X}_{12} \mathcal{X}_{02} \mathcal{X}_{02} \mathcal{X}_{12}  $ & $ \mathcal{X}_{01} \mathcal{X}_{01} \mathbb{I} \mathcal{X}_{01} \mathbb{I} \mathcal{X}_{01} \mathcal{X}_{01} \mathcal{X}_{01}  $ \\
\hline
\multirow{3}{*}{$\{\overline{\mathbf{3}}, \mathbf{3}, \overline{\mathbf{3}}, \mathbf{3}\}$} & \multirow{3}{*}{$\{01100110 \} $}  &  $ \frac{1}{9} $ & $ \mathcal{X}_{01} \mathcal{X}_{02} \mathcal{X}_{02} \mathcal{X}_{01}  $ & $ \mathcal{X}_{01} \mathbb{I} \mathbb{I} \mathcal{X}_{01} \mathbb{I} \mathcal{X}_{01} \mathcal{X}_{01} \mathbb{I}  $ \\& &  $ \frac{1}{3} $ & $ \mathcal{X}_{02} \mathcal{X}_{12} \mathcal{X}_{01} \mathcal{X}_{12}  $ & $ \mathbb{I} \mathcal{X}_{01} \mathcal{X}_{01} \mathcal{X}_{01} \mathcal{X}_{01} \mathbb{I} \mathcal{X}_{01} \mathcal{X}_{01}  $ \\& &  $ \frac{1}{3} $ & $ \mathcal{X}_{12} \mathcal{X}_{01} \mathcal{X}_{12} \mathcal{X}_{02}  $ & $ \mathcal{X}_{01} \mathcal{X}_{01} \mathcal{X}_{01} \mathbb{I} \mathcal{X}_{01} \mathcal{X}_{01} \mathbb{I} \mathcal{X}_{01}  $ \\
\hline
\multirow{3}{*}{$\{\overline{\mathbf{3}}, \mathbf{3}, \mathbf{3}, \overline{\mathbf{3}}\}$} & \multirow{3}{*}{$\{01101001 \} $}  &  $ \frac{1}{3 \sqrt{3}} $ & $ \mathcal{X}_{01} \mathcal{X}_{12} \mathcal{X}_{02} \mathcal{X}_{01}  $ & $ \mathcal{X}_{01} \mathbb{I} \mathcal{X}_{01} \mathcal{X}_{01} \mathbb{I} \mathcal{X}_{01} \mathcal{X}_{01} \mathbb{I}  $ \\& &  $ \frac{1}{3 \sqrt{3}} $ & $ \mathcal{X}_{02} \mathcal{X}_{01} \mathcal{X}_{01} \mathcal{X}_{12}  $ & $ \mathbb{I} \mathcal{X}_{01} \mathcal{X}_{01} \mathbb{I} \mathcal{X}_{01} \mathbb{I} \mathcal{X}_{01} \mathcal{X}_{01}  $ \\& &  $ \frac{1}{3} $ & $ \mathcal{X}_{12} \mathcal{X}_{02} \mathcal{X}_{12} \mathcal{X}_{02}  $ & $ \mathcal{X}_{01} \mathcal{X}_{01} \mathbb{I} \mathcal{X}_{01} \mathcal{X}_{01} \mathcal{X}_{01} \mathbb{I} \mathcal{X}_{01}  $ \\
\hline
\multirow{3}{*}{$\{\mathbf{1}, \overline{\mathbf{3}}, \mathbf{1}, \overline{\mathbf{3}}\}$} & \multirow{3}{*}{$\{00010001 \} $}  &  $ \frac{1}{\sqrt{3}} $ & $ \mathcal{X}_{01} \mathcal{X}_{12} \mathcal{X}_{12} \mathcal{X}_{12}  $ & $ \mathcal{X}_{01} \mathbb{I} \mathcal{X}_{01} \mathcal{X}_{01} \mathcal{X}_{01} \mathcal{X}_{01} \mathcal{X}_{01} \mathcal{X}_{01}  $ \\& &  $ \frac{1}{3} $ & $ \mathcal{X}_{02} \mathcal{X}_{01} \mathcal{X}_{02} \mathcal{X}_{02}  $ & $ \mathbb{I} \mathcal{X}_{01} \mathcal{X}_{01} \mathbb{I} \mathbb{I} \mathcal{X}_{01} \mathbb{I} \mathcal{X}_{01}  $ \\& &  $ \frac{1}{\sqrt{3}} $ & $ \mathcal{X}_{12} \mathcal{X}_{02} \mathcal{X}_{01} \mathcal{X}_{01}  $ & $ \mathcal{X}_{01} \mathcal{X}_{01} \mathbb{I} \mathcal{X}_{01} \mathcal{X}_{01} \mathbb{I} \mathcal{X}_{01} \mathbb{I}  $ \\
\hline
\multirow{3}{*}{$\{\mathbf{1}, \overline{\mathbf{3}}, \overline{\mathbf{3}}, \mathbf{1}\}$} & \multirow{3}{*}{$\{00010100 \} $}  &  $ -\frac{1}{\sqrt{3}} $ & $ \mathcal{X}_{01} \mathcal{X}_{01} \mathcal{X}_{12} \mathcal{X}_{12}  $ & $ \mathcal{X}_{01} \mathbb{I} \mathcal{X}_{01} \mathbb{I} \mathcal{X}_{01} \mathcal{X}_{01} \mathcal{X}_{01} \mathcal{X}_{01}  $ \\& &  $ \frac{1}{3} $ & $ \mathcal{X}_{02} \mathcal{X}_{02} \mathcal{X}_{02} \mathcal{X}_{02}  $ & $ \mathbb{I} \mathcal{X}_{01} \mathbb{I} \mathcal{X}_{01} \mathbb{I} \mathcal{X}_{01} \mathbb{I} \mathcal{X}_{01}  $ \\& &  $ -\frac{1}{\sqrt{3}} $ & $ \mathcal{X}_{12} \mathcal{X}_{12} \mathcal{X}_{01} \mathcal{X}_{01}  $ & $ \mathcal{X}_{01} \mathcal{X}_{01} \mathcal{X}_{01} \mathcal{X}_{01} \mathcal{X}_{01} \mathbb{I} \mathcal{X}_{01} \mathbb{I}  $ \\
\hline
\multirow{3}{*}{$\{\mathbf{1}, \overline{\mathbf{3}}, \mathbf{3}, \mathbf{3}\}$} & \multirow{3}{*}{$\{00011010 \} $}  &  $ \frac{1}{3} $ & $ \mathcal{X}_{01} \mathcal{X}_{02} \mathcal{X}_{12} \mathcal{X}_{12}  $ & $ \mathcal{X}_{01} \mathbb{I} \mathbb{I} \mathcal{X}_{01} \mathcal{X}_{01} \mathcal{X}_{01} \mathcal{X}_{01} \mathcal{X}_{01}  $ \\& &  $ -\frac{1}{3} $ & $ \mathcal{X}_{02} \mathcal{X}_{12} \mathcal{X}_{02} \mathcal{X}_{02}  $ & $ \mathbb{I} \mathcal{X}_{01} \mathcal{X}_{01} \mathcal{X}_{01} \mathbb{I} \mathcal{X}_{01} \mathbb{I} \mathcal{X}_{01}  $ \\& &  $ -\frac{1}{3} $ & $ \mathcal{X}_{12} \mathcal{X}_{01} \mathcal{X}_{01} \mathcal{X}_{01}  $ & $ \mathcal{X}_{01} \mathcal{X}_{01} \mathcal{X}_{01} \mathbb{I} \mathcal{X}_{01} \mathbb{I} \mathcal{X}_{01} \mathbb{I}  $ \\
\hline
\multirow{3}{*}{$\{\overline{\mathbf{3}}, \overline{\mathbf{3}}, \overline{\mathbf{3}}, \overline{\mathbf{3}}\}$} & \multirow{3}{*}{$\{01010101 \} $}  &  $ \frac{1}{3} $ & $ \mathcal{X}_{01} \mathcal{X}_{12} \mathcal{X}_{01} \mathcal{X}_{12}  $ & $ \mathcal{X}_{01} \mathbb{I} \mathcal{X}_{01} \mathcal{X}_{01} \mathcal{X}_{01} \mathbb{I} \mathcal{X}_{01} \mathcal{X}_{01}  $ \\& &  $ \frac{1}{3 \sqrt{3}} $ & $ \mathcal{X}_{02} \mathcal{X}_{01} \mathcal{X}_{12} \mathcal{X}_{02}  $ & $ \mathbb{I} \mathcal{X}_{01} \mathcal{X}_{01} \mathbb{I} \mathcal{X}_{01} \mathcal{X}_{01} \mathbb{I} \mathcal{X}_{01}  $ \\& &  $ \frac{1}{3 \sqrt{3}} $ & $ \mathcal{X}_{12} \mathcal{X}_{02} \mathcal{X}_{02} \mathcal{X}_{01}  $ & $ \mathcal{X}_{01} \mathcal{X}_{01} \mathbb{I} \mathcal{X}_{01} \mathbb{I} \mathcal{X}_{01} \mathcal{X}_{01} \mathbb{I}  $ \\
\hline
\multirow{3}{*}{$\{\overline{\mathbf{3}}, \mathbf{1}, \overline{\mathbf{3}}, \mathbf{1}\}$} & \multirow{3}{*}{$\{01000100 \} $}  &  $ \frac{1}{\sqrt{3}} $ & $ \mathcal{X}_{01} \mathcal{X}_{01} \mathcal{X}_{12} \mathcal{X}_{02}  $ & $ \mathcal{X}_{01} \mathbb{I} \mathcal{X}_{01} \mathbb{I} \mathcal{X}_{01} \mathcal{X}_{01} \mathbb{I} \mathcal{X}_{01}  $ \\& &  $ \frac{1}{3} $ & $ \mathcal{X}_{02} \mathcal{X}_{02} \mathcal{X}_{02} \mathcal{X}_{01}  $ & $ \mathbb{I} \mathcal{X}_{01} \mathbb{I} \mathcal{X}_{01} \mathbb{I} \mathcal{X}_{01} \mathcal{X}_{01} \mathbb{I}  $ \\& &  $ \frac{1}{\sqrt{3}} $ & $ \mathcal{X}_{12} \mathcal{X}_{12} \mathcal{X}_{01} \mathcal{X}_{12}  $ & $ \mathcal{X}_{01} \mathcal{X}_{01} \mathcal{X}_{01} \mathcal{X}_{01} \mathcal{X}_{01} \mathbb{I} \mathcal{X}_{01} \mathcal{X}_{01}  $ \\
\hline
\multirow{3}{*}{$\{\overline{\mathbf{3}}, \mathbf{1}, \mathbf{1}, \overline{\mathbf{3}}\}$} & \multirow{3}{*}{$\{01000001 \} $}  &  $ -\frac{1}{\sqrt{3}} $ & $ \mathcal{X}_{01} \mathcal{X}_{12} \mathcal{X}_{12} \mathcal{X}_{02}  $ & $ \mathcal{X}_{01} \mathbb{I} \mathcal{X}_{01} \mathcal{X}_{01} \mathcal{X}_{01} \mathcal{X}_{01} \mathbb{I} \mathcal{X}_{01}  $ \\& &  $ \frac{1}{3} $ & $ \mathcal{X}_{02} \mathcal{X}_{01} \mathcal{X}_{02} \mathcal{X}_{01}  $ & $ \mathbb{I} \mathcal{X}_{01} \mathcal{X}_{01} \mathbb{I} \mathbb{I} \mathcal{X}_{01} \mathcal{X}_{01} \mathbb{I}  $ \\& &  $ -\frac{1}{\sqrt{3}} $ & $ \mathcal{X}_{12} \mathcal{X}_{02} \mathcal{X}_{01} \mathcal{X}_{12}  $ & $ \mathcal{X}_{01} \mathcal{X}_{01} \mathbb{I} \mathcal{X}_{01} \mathcal{X}_{01} \mathbb{I} \mathcal{X}_{01} \mathcal{X}_{01}  $ \\
\hline
\multirow{3}{*}{$\{\overline{\mathbf{3}}, \mathbf{1}, \mathbf{3}, \mathbf{3}\}$} & \multirow{3}{*}{$\{01001010 \} $}  &  $ -\frac{1}{3} $ & $ \mathcal{X}_{01} \mathcal{X}_{02} \mathcal{X}_{12} \mathcal{X}_{02}  $ & $ \mathcal{X}_{01} \mathbb{I} \mathbb{I} \mathcal{X}_{01} \mathcal{X}_{01} \mathcal{X}_{01} \mathbb{I} \mathcal{X}_{01}  $ \\& &  $ -\frac{1}{3} $ & $ \mathcal{X}_{02} \mathcal{X}_{12} \mathcal{X}_{02} \mathcal{X}_{01}  $ & $ \mathbb{I} \mathcal{X}_{01} \mathcal{X}_{01} \mathcal{X}_{01} \mathbb{I} \mathcal{X}_{01} \mathcal{X}_{01} \mathbb{I}  $ \\& &  $ \frac{1}{3} $ & $ \mathcal{X}_{12} \mathcal{X}_{01} \mathcal{X}_{01} \mathcal{X}_{12}  $ & $ \mathcal{X}_{01} \mathcal{X}_{01} \mathcal{X}_{01} \mathbb{I} \mathcal{X}_{01} \mathbb{I} \mathcal{X}_{01} \mathcal{X}_{01}  $ \\
\hline
\multirow{3}{*}{$\{\overline{\mathbf{3}}, \mathbf{3}, \mathbf{1}, \mathbf{1}\}$} & \multirow{3}{*}{$\{01100000 \} $}  &  $ \frac{1}{3} $ & $ \mathcal{X}_{01} \mathcal{X}_{01} \mathcal{X}_{02} \mathcal{X}_{01}  $ & $ \mathcal{X}_{01} \mathbb{I} \mathcal{X}_{01} \mathbb{I} \mathbb{I} \mathcal{X}_{01} \mathcal{X}_{01} \mathbb{I}  $ \\& &  $ \frac{1}{\sqrt{3}} $ & $ \mathcal{X}_{02} \mathcal{X}_{02} \mathcal{X}_{01} \mathcal{X}_{12}  $ & $ \mathbb{I} \mathcal{X}_{01} \mathbb{I} \mathcal{X}_{01} \mathcal{X}_{01} \mathbb{I} \mathcal{X}_{01} \mathcal{X}_{01}  $ \\& &  $ \frac{1}{\sqrt{3}} $ & $ \mathcal{X}_{12} \mathcal{X}_{12} \mathcal{X}_{12} \mathcal{X}_{02}  $ & $ \mathcal{X}_{01} \mathcal{X}_{01} \mathcal{X}_{01} \mathcal{X}_{01} \mathcal{X}_{01} \mathcal{X}_{01} \mathbb{I} \mathcal{X}_{01}  $ \\
\hline
\multirow{3}{*}{$\{\mathbf{3}, \mathbf{3}, \mathbf{1}, \overline{\mathbf{3}}\}$} & \multirow{3}{*}{$\{10100001 \} $}  &  $ \frac{1}{3} $ & $ \mathcal{X}_{01} \mathcal{X}_{12} \mathcal{X}_{12} \mathcal{X}_{01}  $ & $ \mathcal{X}_{01} \mathbb{I} \mathcal{X}_{01} \mathcal{X}_{01} \mathcal{X}_{01} \mathcal{X}_{01} \mathcal{X}_{01} \mathbb{I}  $ \\& &  $ -\frac{1}{3} $ & $ \mathcal{X}_{02} \mathcal{X}_{01} \mathcal{X}_{02} \mathcal{X}_{12}  $ & $ \mathbb{I} \mathcal{X}_{01} \mathcal{X}_{01} \mathbb{I} \mathbb{I} \mathcal{X}_{01} \mathcal{X}_{01} \mathcal{X}_{01}  $ \\& &  $ -\frac{1}{3} $ & $ \mathcal{X}_{12} \mathcal{X}_{02} \mathcal{X}_{01} \mathcal{X}_{02}  $ & $ \mathcal{X}_{01} \mathcal{X}_{01} \mathbb{I} \mathcal{X}_{01} \mathcal{X}_{01} \mathbb{I} \mathbb{I} \mathcal{X}_{01}  $ \\
\hline
\multirow{3}{*}{$\{\mathbf{3}, \overline{\mathbf{3}}, \mathbf{1}, \mathbf{1}\}$} & \multirow{3}{*}{$\{10010000 \} $}  &  $ \frac{1}{\sqrt{3}} $ & $ \mathcal{X}_{01} \mathcal{X}_{01} \mathcal{X}_{02} \mathcal{X}_{12}  $ & $ \mathcal{X}_{01} \mathbb{I} \mathcal{X}_{01} \mathbb{I} \mathbb{I} \mathcal{X}_{01} \mathcal{X}_{01} \mathcal{X}_{01}  $ \\& &  $ \frac{1}{3} $ & $ \mathcal{X}_{02} \mathcal{X}_{02} \mathcal{X}_{01} \mathcal{X}_{02}  $ & $ \mathbb{I} \mathcal{X}_{01} \mathbb{I} \mathcal{X}_{01} \mathcal{X}_{01} \mathbb{I} \mathbb{I} \mathcal{X}_{01}  $ \\& &  $ \frac{1}{\sqrt{3}} $ & $ \mathcal{X}_{12} \mathcal{X}_{12} \mathcal{X}_{12} \mathcal{X}_{01}  $ & $ \mathcal{X}_{01} \mathcal{X}_{01} \mathcal{X}_{01} \mathcal{X}_{01} \mathcal{X}_{01} \mathcal{X}_{01} \mathcal{X}_{01} \mathbb{I}  $ \\
\hline
\multirow{3}{*}{$\{\mathbf{3}, \mathbf{3}, \overline{\mathbf{3}}, \mathbf{1}\}$} & \multirow{3}{*}{$\{10100100 \} $}  &  $ -\frac{1}{3} $ & $ \mathcal{X}_{01} \mathcal{X}_{01} \mathcal{X}_{12} \mathcal{X}_{01}  $ & $ \mathcal{X}_{01} \mathbb{I} \mathcal{X}_{01} \mathbb{I} \mathcal{X}_{01} \mathcal{X}_{01} \mathcal{X}_{01} \mathbb{I}  $ \\& &  $ -\frac{1}{3} $ & $ \mathcal{X}_{02} \mathcal{X}_{02} \mathcal{X}_{02} \mathcal{X}_{12}  $ & $ \mathbb{I} \mathcal{X}_{01} \mathbb{I} \mathcal{X}_{01} \mathbb{I} \mathcal{X}_{01} \mathcal{X}_{01} \mathcal{X}_{01}  $ \\& &  $ \frac{1}{3} $ & $ \mathcal{X}_{12} \mathcal{X}_{12} \mathcal{X}_{01} \mathcal{X}_{02}  $ & $ \mathcal{X}_{01} \mathcal{X}_{01} \mathcal{X}_{01} \mathcal{X}_{01} \mathcal{X}_{01} \mathbb{I} \mathbb{I} \mathcal{X}_{01}  $ \\
\hline
\multirow{3}{*}{$\{\overline{\mathbf{3}}, \overline{\mathbf{3}}, \mathbf{1}, \mathbf{3}\}$} & \multirow{3}{*}{$\{01010010 \} $}  &  $ -\frac{1}{3} $ & $ \mathcal{X}_{01} \mathcal{X}_{02} \mathcal{X}_{01} \mathcal{X}_{12}  $ & $ \mathcal{X}_{01} \mathbb{I} \mathbb{I} \mathcal{X}_{01} \mathcal{X}_{01} \mathbb{I} \mathcal{X}_{01} \mathcal{X}_{01}  $ \\& &  $ \frac{1}{3} $ & $ \mathcal{X}_{02} \mathcal{X}_{12} \mathcal{X}_{12} \mathcal{X}_{02}  $ & $ \mathbb{I} \mathcal{X}_{01} \mathcal{X}_{01} \mathcal{X}_{01} \mathcal{X}_{01} \mathcal{X}_{01} \mathbb{I} \mathcal{X}_{01}  $ \\& &  $ -\frac{1}{3} $ & $ \mathcal{X}_{12} \mathcal{X}_{01} \mathcal{X}_{02} \mathcal{X}_{01}  $ & $ \mathcal{X}_{01} \mathcal{X}_{01} \mathcal{X}_{01} \mathbb{I} \mathbb{I} \mathcal{X}_{01} \mathcal{X}_{01} \mathbb{I}  $ \\
\hline
\multirow{3}{*}{$\{\overline{\mathbf{3}}, \overline{\mathbf{3}}, \mathbf{3}, \mathbf{1}\}$} & \multirow{3}{*}{$\{01011000 \} $}  &  $ -\frac{1}{3} $ & $ \mathcal{X}_{01} \mathcal{X}_{01} \mathcal{X}_{01} \mathcal{X}_{12}  $ & $ \mathcal{X}_{01} \mathbb{I} \mathcal{X}_{01} \mathbb{I} \mathcal{X}_{01} \mathbb{I} \mathcal{X}_{01} \mathcal{X}_{01}  $ \\& &  $ -\frac{1}{3} $ & $ \mathcal{X}_{02} \mathcal{X}_{02} \mathcal{X}_{12} \mathcal{X}_{02}  $ & $ \mathbb{I} \mathcal{X}_{01} \mathbb{I} \mathcal{X}_{01} \mathcal{X}_{01} \mathcal{X}_{01} \mathbb{I} \mathcal{X}_{01}  $ \\& &  $ \frac{1}{3} $ & $ \mathcal{X}_{12} \mathcal{X}_{12} \mathcal{X}_{02} \mathcal{X}_{01}  $ & $ \mathcal{X}_{01} \mathcal{X}_{01} \mathcal{X}_{01} \mathcal{X}_{01} \mathbb{I} \mathcal{X}_{01} \mathcal{X}_{01} \mathbb{I}  $ \\
\hline
\hline \\
\multicolumn{5}{l}{} \\[-1cm]
\caption{Local operators in the 27 control sectors of the generic plaquette operator with irrep truncation on each link of $\{\mathbf{1}, \mathbf{3}, \overline{\mathbf{3}}\}$ mapped to qutrit levels $\{0, 1, 2\}$ or $(p,q)$-qutrit pair levels of $\{(0,0), (1,0), (0,1)\}$.}
\label{tab:plaquetteops133bar}
\end{longtable}
\endgroup

\begin{table}[ht]
\setlength{\extrarowheight}{.5em}
  \begin{tabular}{c|cc}
    \hline
    \hline
    $\{\mathbf{C}_1, \mathbf{C}_2\}$ &  coefficient & Givens Operator \\
    \hline
    \multirow{3}{*}{$\{\mathbf{1}, \mathbf{1}\} $} & $ 1 $ & $ \mathcal{X}_{01} \mathcal{X}_{01} \mathcal{X}_{02} \mathcal{X}_{02}  $ \\  & $ 1 $ & $ \mathcal{X}_{02} \mathcal{X}_{02} \mathcal{X}_{01} \mathcal{X}_{01}  $ \\  & $ 1 $ & $ \mathcal{X}_{12} \mathcal{X}_{12} \mathcal{X}_{12} \mathcal{X}_{12}  $ \\
\hline
\multirow{3}{*}{$\{\mathbf{1}, \mathbf{3}\} $} & $ \frac{1}{3} $ & $ \mathcal{X}_{01} \mathcal{X}_{02} \mathcal{X}_{01} \mathcal{X}_{01}  $ \\  & $ \frac{1}{\sqrt{3}} $ & $ \mathcal{X}_{02} \mathcal{X}_{12} \mathcal{X}_{12} \mathcal{X}_{12}  $ \\  & $ \frac{1}{\sqrt{3}} $ & $ \mathcal{X}_{12} \mathcal{X}_{01} \mathcal{X}_{02} \mathcal{X}_{02}  $ \\
\hline
\multirow{3}{*}{$\{\mathbf{3}, \mathbf{3}\} $} & $ \frac{1}{3 \sqrt{3}} $ & $ \mathcal{X}_{01} \mathcal{X}_{02} \mathcal{X}_{12} \mathcal{X}_{01}  $ \\  & $ \frac{1}{3} $ & $ \mathcal{X}_{02} \mathcal{X}_{12} \mathcal{X}_{02} \mathcal{X}_{12}  $ \\  & $ \frac{1}{3 \sqrt{3}} $ & $ \mathcal{X}_{12} \mathcal{X}_{01} \mathcal{X}_{01} \mathcal{X}_{02}  $ \\
\hline
\multirow{3}{*}{$\{\mathbf{3}, \overline{\mathbf{3}}\} $} & $ \frac{1}{3} $ & $ \mathcal{X}_{01} \mathcal{X}_{12} \mathcal{X}_{02} \mathcal{X}_{12}  $ \\  & $ \frac{1}{9} $ & $ \mathcal{X}_{02} \mathcal{X}_{01} \mathcal{X}_{01} \mathcal{X}_{02}  $ \\  & $ \frac{1}{3} $ & $ \mathcal{X}_{12} \mathcal{X}_{02} \mathcal{X}_{12} \mathcal{X}_{01}  $ \\
\hline
\multirow{3}{*}{$\{\mathbf{3}, \mathbf{1}\} $} & $ \frac{1}{3} $ & $ \mathcal{X}_{01} \mathcal{X}_{01} \mathcal{X}_{01} \mathcal{X}_{02}  $ \\  & $ \frac{1}{\sqrt{3}} $ & $ \mathcal{X}_{02} \mathcal{X}_{02} \mathcal{X}_{12} \mathcal{X}_{01}  $ \\  & $ \frac{1}{\sqrt{3}} $ & $ \mathcal{X}_{12} \mathcal{X}_{12} \mathcal{X}_{02} \mathcal{X}_{12}  $ \\
\hline
\end{tabular}
\quad
\begin{tabular}{c|cc}
\multirow{3}{*}{$\{\overline{\mathbf{3}}, \mathbf{3}\} $} & $ \frac{1}{9} $ & $ \mathcal{X}_{01} \mathcal{X}_{02} \mathcal{X}_{02} \mathcal{X}_{01}  $ \\  & $ \frac{1}{3} $ & $ \mathcal{X}_{02} \mathcal{X}_{12} \mathcal{X}_{01} \mathcal{X}_{12}  $ \\  & $ \frac{1}{3} $ & $ \mathcal{X}_{12} \mathcal{X}_{01} \mathcal{X}_{12} \mathcal{X}_{02}  $ \\
\hline
\multirow{3}{*}{$\{\mathbf{1}, \overline{\mathbf{3}}\} $} & $ \frac{1}{\sqrt{3}} $ & $ \mathcal{X}_{01} \mathcal{X}_{12} \mathcal{X}_{12} \mathcal{X}_{12}  $ \\  & $ \frac{1}{3} $ & $ \mathcal{X}_{02} \mathcal{X}_{01} \mathcal{X}_{02} \mathcal{X}_{02}  $ \\  & $ \frac{1}{\sqrt{3}} $ & $ \mathcal{X}_{12} \mathcal{X}_{02} \mathcal{X}_{01} \mathcal{X}_{01}  $ \\
\hline
\multirow{3}{*}{$\{\overline{\mathbf{3}}, \overline{\mathbf{3}}\} $} & $ \frac{1}{3} $ & $ \mathcal{X}_{01} \mathcal{X}_{12} \mathcal{X}_{01} \mathcal{X}_{12}  $ \\  & $ \frac{1}{3 \sqrt{3}} $ & $ \mathcal{X}_{02} \mathcal{X}_{01} \mathcal{X}_{12} \mathcal{X}_{02}  $ \\  & $ \frac{1}{3 \sqrt{3}} $ & $ \mathcal{X}_{12} \mathcal{X}_{02} \mathcal{X}_{02} \mathcal{X}_{01}  $ \\
\hline
\multirow{3}{*}{$\{\overline{\mathbf{3}}, \mathbf{1}\} $} & $ \frac{1}{\sqrt{3}} $ & $ \mathcal{X}_{01} \mathcal{X}_{01} \mathcal{X}_{12} \mathcal{X}_{02}  $ \\  & $ \frac{1}{3} $ & $ \mathcal{X}_{02} \mathcal{X}_{02} \mathcal{X}_{02} \mathcal{X}_{01}  $ \\  & $ \frac{1}{\sqrt{3}} $ & $ \mathcal{X}_{12} \mathcal{X}_{12} \mathcal{X}_{01} \mathcal{X}_{12}  $ \\
\hline
    \hline
  \end{tabular}
  \caption{Local operators (isolated from Table~\ref{tab:plaquetteops133bar} for convenience) in the nine control sectors of the plaquette operator on the two-plaquette lattice with PBCs and irrep truncation on each link of $\{\mathbf{1}, \mathbf{3}, \overline{\mathbf{3}}\}$ mapped to qutrit levels $\{0, 1, 2\}$.}
\end{table}

\FloatBarrier

\bibliography{bibsyrae}

\end{document}